\documentclass[prb,aps,preprint]{revtex4}
\usepackage{amsmath}
\usepackage{graphicx}
\usepackage{lmodern}
\usepackage{color}
\usepackage{amssymb}
\newcommand{\beq} {\begin{equation}}
\newcommand{\eeq} {\end{equation}}
\newcommand{\bea} {\begin{eqnarray}}
\newcommand{\eea} {\end{eqnarray}}
\newcommand{\be} {\begin{equation}}
\newcommand{\ee} {\end{equation}}
\renewcommand{\(}{\left(}
\renewcommand{\)}{\right)}
\renewcommand{\[}{\left[}
\renewcommand{\]}{\right]}

\renewcommand{\phi}{\varphi}

\begin{document}
\title {Charge-density-wave order with momentum $(2Q,0)$ and $(0,2Q)$ within the spin-fermion model: continuous and discrete symmetry breaking,
preemptive composite order, and relation to pseudogap in hole-doped cuprates}
\author{Yuxuan Wang and Andrey Chubukov}
\affiliation{~Department of Physics, University of Wisconsin, Madison, WI 53706, USA}
\date{\today}
\begin{abstract}
 We analyze charge order in hole-doped cuprates within the the spin-fermion model. We show that magnetically-mediated interaction, which is
  known to
 give rise to $d$-wave superconductivity and charge order with momentum along zone diagonal,
  also gives rise to charge order with momenta
  $Q_x =(2Q,0)$ and $Q_y =(0,2Q)$ consistent with the experiments.
  We show that  an instability towards $\Delta_k^Q = \langle c^\dagger_{{\bf k}+{\bf Q}} c_{{\bf k}-{\bf Q}}\rangle$ with ${\bf Q} = Q_x$ or $Q_y$
  is a threshold phenomenon, but the dimensionless spin-fermion coupling is above the threshold, if the magnetic correlation length $\xi$ exceeds a
   certain critical value. At a critical $\xi$, the onset temperature for the charge order terminates at a quantum-critical point distant from the
   magnetic
   one.  We argue that the charge order with $Q_x$ or $Q_y$
   changes sign under ${\bf k} \to {\bf k} + (\pi,\pi)$, but $|\Delta_k^Q| \neq |\Delta_{k+(\pi,\pi)}^Q|$. In real space,
   such an order has both bond and site components, the bond one is larger.
   We further argue that $\Delta_k^Q$ and $\Delta_{-k}^Q$ are not equivalent, and their symmetric and antisymmetric combinations describe, in real space,
    incommensurate density modulations and incommensurate bond current, respectively.  We derive Ginzburg-Landau functional for four-component $U(1)$
    order parameters $\Delta^Q_{\pm k}$ with ${\bf Q} = Q_x$ or $Q_y$ and analyze it first in mean-field theory and then beyond mean-field.  Within
    mean-field we  find two types of CDW states, I and II, depending on system parameters.
       In state  I, density and current modulations emerge with the same ${\bf Q} = Q_x$ or $Q_y$, breaking $Z_2$
    lattice rotational symmetry, and differ in phase
       by $\pm \pi/2$. The selection of $\pi/2$ or $-\pi/2$ additionally breaks $Z_2$ time-reversal symmetry, such that
        the total order parameter manifold is $U(1) \times Z_2 \times Z_2$. In state II
        density and current modulations emerge with  different ${\bf Q}$ and the order
        parameter manifold is $U(1) \times U(1) \times Z_2$, where in the two realizations of state II $Z_2$ corresponds to either lattice rotational or
         time-reversal symmetry breaking.  We extend the analysis beyond mean-field and argue that discrete symmetries get broken before long-range charge order sets in. For state I,
which, we argue, is related to hole-doped cuprates, we show that, upon lowering the temperature, the system first breaks $Z_2$ lattice
rotational symmetry ($C_4 \to C_2$) at $T= T_n$ and develops a nematic order, then breaks $Z_2$ time-reversal symmetry at $T_t < T_n$ and locks the
relative phase between  density and current fluctuations, and finally breaks $U(1)$ symmetry of a common phase of even and odd components of
$\Delta^Q_{k}$
at $T= T_{\rm cdw} < T_t < T_n$ and develops a true charge order. We argue that at a mean-field $T_{\rm cdw}$ is smaller than superconducting $T_{\rm sc}$, but
  preemptive composite order lifts $T_{\rm cdw}$ and reduces $T_{\rm sc}$ such that at large $\xi$ charge order develops prior to superconductivity.
 We obtain the full phase diagram  and present quantitative comparison of our results with ARPES data for hole-doped cuprates.
\end{abstract}
\maketitle
\tableofcontents

\newpage
\section{Introduction}
  Intensive experimental studies of hole-doped cuprates over the last few years have provided strong indications that the
  pseudogap region
  is a state (or even a set of states) with broken symmetry.
  First,
   X-ray and  neutron scattering data on La$_{1.875}$Ba$_{0.125}$CuO$_4$ strongly indicate~\cite{labacuo,stripes} that
  lattice  rotational
   symmetry is broken from $C_4$ down to $C_2$ below a certain temperature $T^* (x)$. Evidence for rotational symmetry breaking has been also found in neutron scattering data on YBCO (Ref.
   [\onlinecite{hinkov}]) and in  STM data on Bi$_2$Sr$_2$CaCu$_2$O$_{8+\delta}$, at energies comparable to $T^*$ (Ref.\ [\onlinecite{davis,davis_1}]).
   Second, measurements of the Kerr angle
 at optical frequencies detected a polar Kerr effect~\cite{kerr}, and polarized elastic neutron scattering measurements
 detected
 an intra-unit cell magnetic order~\cite{bourges, greven}.
  The onset temperatures for the Kerr effect and for intra-cell magnetic order are not equal,
   but roughly follow the same doping dependence as  $T^{*} (x)$.
     The most natural
 interpretation of these two measurements would be that time-reversal symmetry is broken, although the
  absence of a sign change of a Kerr signal under the change of the direction of the applied magnetic field
   raises a possibility that the Kerr effect may be a non-reciprocal phenomenon, in which case it should be
   associated with the breaking of mirror symmetries.  Recent optical experiments in the terahertz regime has found~\cite{armitage} a non-zero
 linear birefringence, which was also interpreted as the result of the breaking of mirror symmetries {\it and } of $C_4$ lattice rotational symmetry.
 The temperature dependence of the onset of a linear birefringence in YBCO closely follows the one
  for Kerr signal.

   Third,
  X-ray measurements on YBCO (Refs.\ [\onlinecite{ybco,ybco_1}]), Bi$_2$Sr$_{2-x}$La$_x$CuO$_{6+\delta}$ (Ref.\ [\onlinecite{X-ray}]), and
    Bi$_2$Sr$_2$CaCu$_2$O$_{8+\delta}$ (Ref.\ [\onlinecite{X-ray_1}])  detected a static incommensurate charge density-wave (CDW) order with momenta ${\bf Q}_x =(2Q,0)$ and/or ${\bf Q}_y = (0,2Q)$,  and $2Q$ was determined to be equal
     to the distance between neighboring hot spots --points where the Fermi surface (FS) intersects with the magnetic Brillouin zone
     boundary~\cite{X-ray,X-ray_1}. The observed order is not long-ranged, but this well may be due to pinning by impurities~\cite{mark_last,steve_last}.
   Earlier NMR measurements~\cite{wu,mark} and more recent sound velocity measurements~\cite{ultra}
 in a magnetic field $H$  found a true CDW order  at $H \geq 20 T$.
  Quantum oscillation measurements~\cite{suchitra} and measurements of Hall and Seebeck coefficients~\cite{taillefer} were interpreted
    as feedbacks effect from the CDW order on fermions.
  The onset temperature $T_{\rm cdw}(x)$ of the CDW  order  was found to be smaller than
      $T^*(x)$ but follow a similar doping dependence.
      Fourth,
      ARPES
      measurements  deep under the superconducting dome have found~\cite{zxshen,zxshen_0,Kaminski} a change of system behavior at a certain doping, and were interpreted
       as evidence for the existence of a quantum-critical point (QCP) at $x = x_{\rm cr}$, at which a new order emerges.
       It is tempting to associate this emerging order with CDW.

   These and other experimental data~\cite{basov,alloul} pose a  challenge to the theory.  System behavior in the metallic region
    outside the pseudogap
    can be reasonably well described within a theoretical framework that fermions interact by exchanging quanta of  collective excitations.
    One proposal along these lines~\cite{grilli}, is that these excitations are charge fluctuations enhanced by phonons
    (a similar set of ideas has been recently displayed for Fe-pnictides~\cite{kontani}).
    An incommensurate CDW order with ${\bf Q}$ along $x$ or $y$ directions in the momentum space is a natural
     part of this scenario, and studies of a true and fluctuating CDW order within  a microscopic Hubbard-Holstein model and using a more general reasoning of
      frustrated phase separation mechanism did indeed find~\cite{castellani} a CDW QCP at around optimal doping, identified the pseudogap temperature with the onset of CDW order~\cite{extra_gr} and obtained a number of
      features in Raman scattering~\cite{gr_raman}, STM~\cite{gr_stm}, and ARPES~\cite{gr_arpes}, consistent with the experimental data in hole-doped cuprates~\cite{review_features}. Furthermore, the  residual momentum-dependent repulsive interaction mediated by charge critical fluctuations was argued
 to give rise to $d$-wave superconducting instability, although additional interaction component, for fermions in antinodal regions, had
  to be included to match the experimental angular variation of the $d$-wave gap~\cite{varlamov}.
    An alternative proposal  is that relevant collective excitations  are
     spin fluctuations, peaked at or near antiferromagnetic momenta $(\pi,\pi)$.
   The corresponding spin-fluctuation approach~\cite{scalapino,pines,acf,acs} naturally explains $d$-wave symmetry of the superconducting state
   and yields a non-Fermi liquid
   behavior of fermionic self-energy and optical conductivity~\cite{hartnol,maslov} in a rather wide frequency ranges, even when magnetic correlation length is only a few lattice spacings. This approach does describe precursors to magnetism~\cite{tremblay,joerg,sedrakyan} and accounts reasonably well for the
    phase diagram of electron-doped cuprates~\cite{wang_el}, where pseudogap behavior is very likely a crossover behavior due to magnetic precursors~\cite{el_prec}.
    At the same time, until recently, spin-fluctuation approach was believed to be incapable to describe charge order and symmetry breaking in the pseudogap phase of hole-doped cuprates. Other explanations of charge order/symmetry breaking have been proposed, including loop-current order~\cite{varma} or $d$-density-wave (current) order~\cite{kotliar,sudip}.
    Other widely discussed scenarios of the pseudogap associate pseudogap behavior with  precursors  to either Mott physics~\cite{lee,millis,tremblay_1,ph_ph,rice}, or  superconductivity~\cite{emery,randeria,mike_last}.

   The spin-fluctuation scenario was revitalized by Metlitski and Sachdev~\cite{ms}  who found that the spin-mediated interaction is attractive not only
   in
   the $d$-wave superconducting
    channel but also in the $d$-wave charge channel, at momenta ${\bf Q}_d=2{\bf k}_{\rm hs}$, where ${\bf k}_{\rm hs}$ is the momentum of one of hot spot on
    a FS
       and $2{\bf k}_{\rm hs}=(\pm 2Q,\pm 2Q)$ are directed along one of Brillouin zone diagonals.
    In real space, an instability in a $d$-wave charge channel implies a charge bond order, for which
    $\langle c^\dagger({\bf r}+{\bf a}) c({\bf r})\rangle$ acquires an ${\bf r}$-dependent component of different sign for ${\bf a}$ along $x$ and $y$ directions,
    while  $\langle c^\dagger({\bf r}) c({\bf r})\rangle$ remains unperturbed.
     The analysis of
    CDW instability
    with ${\bf Q}_d=2{\bf k}_{\rm hs}$ within spin-fluctuation approach was extended by Efetov, Meier, and P\'epin~\cite{efetov}, who argued that the
    pseudogap behavior may be the consequence of the competition between
     bond order and superconductivity (in their scenario, the modulus of the combined SC/CDW ``super-vector" order parameter emerges at $T^*$ but its direction gets fixed along the SC ``axis" only at a  smaller $T_{\rm sc}$).
      The ``super-vector" scenario is appealing from theory perspective and allows one to explain some experimental
     data~\cite{subir_3,efetov_2}. However, it has three
     discrepancies with the experiments. First, the momenta $2{\bf k}_{\rm hs}$ are directed  along
     one of the two Brillouin
     zone diagonals, while CDW momentum detected by resonant X-ray scattering~\cite{X-ray,X-ray_1} and in STM~\cite{davis,davis_1}
      is along horizontal or vertical axis in momentum space (${\bf Q} = {\bf Q}_x = (2Q,0)$ or ${\bf Q} = {\bf Q}_y =
     (0,2Q)$.
     Second, bond-order instability is close to superconducting $T_{\rm sc}$, but is below $T_{\rm sc}$ (Refs.\ [\onlinecite{ms,efetov}]), while
     experiments see the development of charge order
   above superconducting $T_{\rm sc}$. Third,  bond order with momentum ${\bf Q}_d=2{\bf k}_{\rm hs}$ does not break time-reversal
   or mirror symmetries and therefore
     does not explain Kerr, neutron
   scattering, and magneto-electric birefringence experimennts~\cite{kerr,bourges,greven,armitage}.

   In this paper we present a different  scenario for the pseudogap due to spin-fluctuation exchange.
    We  argue that magnetically-mediated interaction yields an attraction in
    the  CDW channel for incoming  momenta ${\bf Q}_x$ and ${\bf Q}_y$, and, when magnetic correlation length is large enough,
     gives rise to a CDW instability
    at
    a nonzero temperature
 $T_{\rm cdw}$. That such critical temperature exists is not guaranteed a'priori, despite that, as we show below, there are logarithms in the perturbation theory. The reason is that magnetically-mediated interaction is dynamical, and the gap equation
    is an integral equation in frequency. For the latter, the summation of the leading logarithms does not necessarily give rise an
    instability~\cite{finn,wang}, and one
      one has to go beyond the leading logarithmic approximation to verify whether or not the interaction exceeds a certain finite  threshold.
       We show that for CDW with ${\bf Q} = Q_x$ or $Q_y$, the interaction is above the threshold, and the
       linearized gap equation, or, more accurately, the set of coupled
      equations for
      $\Delta_{k}^{Q} = \langle c^\dagger_{\bf k+Q} c_{\bf k- Q}\rangle$ and $\Delta^{Q}_{{k}+(\pi,\pi)}$,
     does have a solution at a finite $T = T_{\rm cdw}$.
We compute $T_{\rm cdw}$ first in Eliashberg-type calculations and then by treating the effects of thermal bosonic fluctuations beyond Eliashberg theory, and compare $T_{\rm cdw}$ with $T_{\rm sc}$ obtained using the same procedures. In Eliashberg calculation, we show that $T_{\rm cdw}$ and $T_{\rm sc}$ are finite at $\xi = \infty$ at $T_{\rm sc} > T_{\rm cdw}$. With more accurate treatment of thermal fluctuation (equivalently, the contribution from zero bosonic Matsubara frequency)
 we find that the ratio  $T_{\rm sc}/T_{\rm cdw}$ approaches one at infinite $\xi$, i.e.,
  $T_{\rm sc}$ and $T_{\rm cdw}$ must be quite close at large $\xi$. We also analyze non-ladder diagrams and show that they are small numerically.

     The CDW order parameter $\Delta_{k}^{Q}$ changes sign under momentum shift by $(\pi,\pi)$, as the bond order does, but it also has a
     non-zero on-site (a true CDW) component $\langle c^\dagger (r) c(r)\rangle = f(r)$ because $|\Delta_{k}^{Q}| \neq |\Delta^{\bf Q}_{\bf k + (\pi,\pi)}|$.
     This agrees with the structure of the charge order extracted from  STM and X-ray data~\cite{davis_1,X-ray_last}. Because on-site component of
     $\Delta^{\bf Q}_{\bf k}$
     is non-zero (albeit small) we will be calling this order a CDW, primarily to distinguish it from a true bond order with diagonal ${\bf Q}_d =
     (2Q,\pm 2Q)$, for which, by symmetry,
     $\Delta^{{\bf Q}_d}_{\bf k}  = -\Delta^{{\bf Q}_d}_{\bf k + (\pi,\pi)}$.

     We analyze the structure CDW order in detail, first in mean-field approximation and then by going beyond mean-field.
      Within mean-field,  we first assume that $\Delta^{Q_x}_k$ and $\Delta^{Q_y}_k$ are even functions in $k$
       and
       discuss the interplay between CDW orders with $Q_x$ and $Q_y$.
     The linearized equations for both CDW orders have solution at the same $T=T_{\rm cdw}$. What happens at a smaller $T$ depends on how the two orders
      $\Delta^{Q_x}$ and $\Delta^{Q_y}$
     interact with each other. We show that the interaction is repulsive, i.e., the two orders tend to repel each other. If  the repulsion is weak,
      the two orders appear simultaneously and with the same amplitude, and the system develops a checkerboard
     order. If the repulsion is strong enough, it becomes energetically advantageous for a system to spontaneously break lattice rotational symmetry from $C_4$
     down to $C_2$ and  develop
     CDW with only  $Q_x$ or $Q_y$. In the real space, such an order has the  form of stripes, e.g., $\langle c^\dagger (r) c(r)\rangle \propto \cos(2 Q r_y)$ with ${\bf Q} = Q_y = (0,2Q)$.
      To understand which type of CDW order develops, we derive the Ginzburg-Landau action to order $(\Delta^Q_k)^4$ and analyze its form. We find that the repulsion is strong enough such that the system
      prefers to
      break $C_4$ symmetry down to $C_2$ and  develop a stripe order. This is consistent with STM data~\cite{davis_1}.
      A different scenario for CDW order with ${\bf Q} = (2Q,0)$ and $(0,2Q)$ has been proposed recently~\cite{efetov_3}, in which CDW  is induced by superconducting fluctuations. In that scenario, CDW emerges as a checkerboard order

\begin{figure}
\includegraphics [width=0.5\columnwidth]{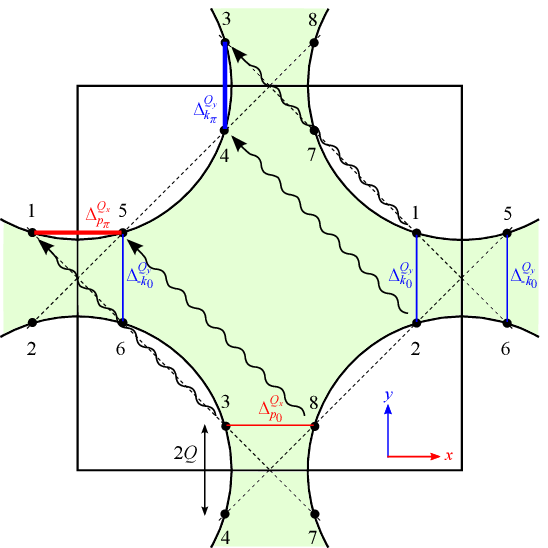}
\caption{The Fermi surface, Brillouin zone and magnetic Brillouin zone (dashed line). Hot spots are defined as intersections of the FS with magnetic
Brillouin zone. The hot spot pairs 1-2 and 3-4 denotes the CDW pairing we consider. They are coupled through the antiferromagnetic exchange interaction
peaked at momentum $(\pi, \pi)$, as shown by the dashed arrows.}
\label{fig3}
\end{figure}

      We next take a more careful look at the dependence of $\Delta_{k}^{Q}$
      on the center of mass momentum $k$. CDW order with, say,  ${\bf Q} = Q_y$ can be constructed out of hot fermions with
      ${\bf k} \approx {\bf k}_0 = (\pi - Q,0)$ and ${- \bf k}_0 = (-\pi +Q,0)$
      (pairs 1-2 and 5-6 in Fig.\ \ref{fig3}). The CDW order parameters $\Delta_{k_0}^Q$ and $\Delta_{-k_0}^Q$ are not identical because
       $2{\bf k}_0$ is not a reciprocal lattice vector. As a result,
      $\Delta_{k}^Q$ with ${\bf k} \approx {\bf k}_0$ generally  has two components -- one is  even in $k$ and
      the other is odd (e.g., $\Delta_{k}^{Q_y} \propto \cos k_x$ and $\Delta_{k}^Q \propto \sin k_x$, respectively).
        In contrast, for charge order with diagonal ${\bf Q}_d= 2{\bf k}_{\rm hs}$, only the even in $k$ solution
       is possible because  the center of mass momentum is at ${\bf k}_0=(\pi,0$) or $(0,\pi)$ and ${\bf k}_0$ and $-{\bf k}_0$ are equivalent points.

  We show that the even component $\Delta_{1,k}^{Q_y} = (\Delta_{k}^{Q_y} + \Delta_{-k}^{Q_y})/2 $ represents a variation of site and  bond charge densities (a  variation of the cite density is
  $\delta \rho (r) \propto \cos 2Q r_y$),  while the  odd
  component $\Delta_{2, k}^{Q_y} = (\Delta_{k}^{Q_y} - \Delta_{- k}^{Q_y})/2$  represents a fermionic current $j_x (r) \propto \sin 2Q r_y$   This current gives rise to a non-zero
  orbital magnetic field $H_z \propto \cos 2Q r_y$
  and, by definition, breaks time-reversal symmetry (TRS).
     This, however, does not lead to orbital ferromagnetism as $\int H_z dV$ vanishes.

                 We compute $T_{\rm cdw}$  for even and odd components and show that
                $T_{\rm cdw}$
   for the even component is larger, in agreement with Ref.\ [\onlinecite{subir_2,subir_4}], but
   the one for the odd component is a close second. We  derive the GL model
    for four $U(1)$ CDW fields, $\Delta_{1,k}^{Q_x}$, $\Delta_{2,k}^{Q_x}$, $\Delta_{1,k}^{Q_y}$, and $\Delta_{2,k}^{Q_y}$.
    We argue that at low $T$ both density and charge components are generally non-zero, and the
      system develops a CDW order of one of two types,
     depending on the interplay between system parameters. We label the corresponding
      ordered states as states I and II,
       In the state I, density and current modulations emerge with the same ${\bf Q}$ (either $Q_x$ or $Q_y$) via a continuous second-order transition.
        Such an order spontaneously breaks $C_4$ lattice rotational symmetry down to $C_2$, like in the case when only $\Delta_{1,k}^{Q}$ was set to be non-zero. The density and the current component with a given $Q$ are both non-zero at low enough $T$ and the
         phase difference between them  is locked at  $\pm \pi/2$.
        The order parameter in this state breaks $Z_2$ lattice rotational symmetry and $U(1)$ symmetry of the common phase of the two order parameters, and
         breaks an additional
        $Z_2$  symmetry by selecting the relative phase to be either $\pi/2$ or $-\pi/2$.  It is natural to associate this additional $Z_2$ symmetry with time reversal (TR), which is then explicitly broken in the state I.

        In the state II incommensurate density and current modulations emerge with different ${\bf Q}$ via first-order transition.
       There are two realizations of state II: in the first all four CDW components are non-zero and have equal magnitudes, while relative phases between
         $\Delta_{1,k}^{Q_x}$  and $\Delta_{2,k}^{Q_x}$ and between $\Delta_{1,k}^{Q_y}$  and $\Delta_{2,k}^{Q_y}$ are, simultaneously, either $\pi/2$ or $-\pi/2$. This is a $C_4$-symmetric checkerboard state with order parameter manifold $U(1) \times U(1) \times Z_2$, where $Z_2$ is associated with TR.
         In the second realization, only one density and one current components are non-zero, e.g., $\Delta_{1,k}^{Q_x}$  and $\Delta_{2,k}^{Q_y}$. Such an order breaks
          $C_4$ lattice symmetry down to $C_2$, but does not additionally break TR symmetry because the phases of  $\Delta_{1,k}^{Q_x}$  and $\Delta_{2,k}^{Q_y}$
           are uncorrelated. The order parameter is again $U(1) \times U(1) \times Z_2$, with $Z_2$ now associated with lattice rotational symmetry.

         We extend the analysis of the Ginzburg-Landau (GL) action beyond mean-field by applying Hubbard-Stratonovich transformation to collective
   variables and analyzing the resulting action within saddle-point approximation, in close similarity to the analysis of the nematic order in Fe-pnictides~\cite{rafael}. We specifically focus on the state I, which in mean-field emerges via a continuous transition.
    We show that discrete symmetries get broken before long-range charge order sets in. We show that, upon lowering the temperature, the system first breaks $Z_2$ lattice
rotational symmetry ($C_4 \to C_2$) at $T= T_{n}$ and develops a nematic order. Below $T_{n}$,  $\langle|\Delta_{i,\bf k}^{Q_y}|^2\rangle$ becomes non-equal to $\langle|\Delta_{i,\bf k}^{Q_x}|^2\rangle$ ($i =1,2$), while
    $\langle \Delta_{i,\bf k}^{Q_{x,y}}\rangle =0$, i.e., density and current modulations do not develop,
       $\langle\delta \rho(r)\rangle = \langle j_x (r)\rangle=0$.   Such a nematic order has been discussed in series of recent publications on the cuprates~\cite{lederer} and Fe-pnictides~\cite{rafael}.  Then, at a smaller $T_t  \leq T_{n}$, another composite order parameter $\Upsilon \propto
       \langle\Delta_1^{Q_{y}} (\Delta_2^{Q_{y}})^*\rangle \rangle$ becomes non-zero (for the order with ${\bf Q} = Q_y$), while still $\langle \Delta_{i,\bf k}^{Q_{x,y}}\rangle =0$.
      Under time reversal, $\Upsilon$ transforms into $-\Upsilon$, hence this composite order breaks TRS.
         This order can be understood as the locking of a relative phase $\psi$ of $\Delta_{1,\bf k}^{Q_{y}}$ and $\Delta_{2,\bf k}^{Q_{y}}$ at $\psi = \pi/2$ or $\psi = - \pi/2$ without the locking of the common phase of $\Delta_{1,\bf k}^{Q_{y}}$ and $\Delta_{2,\bf k}^{Q_{y}}$. The emergence of a preemptive composite order which breaks time-reversal symmetry has been verified in Ref.\ [\onlinecite{tsvelik}] using a different computational technique.
      Finally, below $T_{\rm cdw} < T_t$ the system breaks $U(1)$ symmetry of the common phase  and the system develops a true CDW order (a quasi-long-range order in 2D).
      Within our theory, we identify the temperatures $T_n$ and $T_t$ as the experimental pseudogap temperature $T^*$.

 The existence of the preemptive order is the crucial element in our scenario.  Without it,  CDW instability would be subleading to $d$-wave superconductivity
 and to bond order with diagonal ${\bf Q}_d = (2Q, \pm 2Q)$ as in mean-field approximation both have larger onset temperatures than $T_{\rm cdw}$.
 However, superconducting order parameter and  order parameter for bond charge order  have only one, even in $k$, component, and for these two there is no preemptive
   instability which would break time-reversal symmetry. Moreover, neither superconductivity nor bond order break $C_4$ symmetry. For bond order
      this is the consequence of the fact that  bond orders with $(2Q,2Q)$ and $(2Q,-2Q)$
      only weakly  interact with each other  because in a fourth-order square diagram for the interaction term some fermions are necessary far away from the FS.
       As a result, the two orders appear simultaneously and form a  checkerboard-type structure.
       If the system parameters  are such that $T_{n}$ gets larger than the onset temperature for superconductivity/bond-order,
        the first instability upon  lowering of $T$ is into a state with a composite CDW order with $Q_x (Q_y)$.
          Once composite order forms, it reconstructs fermionic excitations and tends reduce the onset temperatures for superconductivity/bond-order because
   composite charge order and superconductivity/bond-order compete for the FS. At the same time, a composite CDW order
 increases the susceptibility for the primary CDW fields and hence increases $T_{\rm cdw}$,  much like a spin-nematic order in Fe-pnictides increases the
 Neel temperature of SDW order~\cite{rafael}.  An increase  of $T_{\rm cdw}$ compared to the onset of superconductivity/bond-order becomes even stronger once we
 include into consideration 2D fluctuation effects because composite order only breaks discrete Ising symmetry, while near-degenerate $d$-wave
 superconductivity and bond order form weakly anisotropic $O(4)$ model, in which $T_{\rm sc}$ is strongly
 reduced by fluctuations from $O(4)$  manifold.

 The two transitions at $T_{n}$ and $T_{\rm cdw}$ have been also found in the scenario~\cite{efetov_3} that CDW order is due to strong superconducting fluctuations,
  but in that case CDW order has only an even in $k$ component and there is no intermediate $T$ range where $C_4$ symmetry and/or
   TRS are broken.

 We next consider doping evolution of $T_{\rm cdw}$ and the interplay between charge order and superconductivity at various dopings.
 We argue that $T_{\rm cdw}$ decreases when magnetic correlation length $\xi$ decreases
        and vanishes at some finite $\xi$, setting up a charge QCP at some distance away from the magnetic instability (see Fig.\ \ref{phases}(a,b)).
 A similar doping dependence holds  for the onset temperature for bond order with diagonal ${\bf Q}$, as we also demonstrate.
 The ideas about a non-magnetic QCP at around optical doping have been presented in earlier publications~\cite{ccm,varma,varma_2}, in our theory we found such QCP in microscopic calculations.
 The onset temperatures of nematic and TRS-breaking composite orders follow the same doping dependence as $T_{\rm cdw}$. Within saddle-point Hubbard-Stratonovich theory, $T_{n}$ and $T_t$ merge with $T_{\rm cdw}$ at some small $T$ below which the system undergoes a single first-order CDW transition~\cite{Fernandes_13}. Whether this holds beyond
    saddle-point approximation remains to be seen, but in any case near the critical $\xi$, $T_{\rm sc}$ is higher than both $T_{\rm cdw}$ and $T_{n}$,
  and at larger dopings (smaller $\xi$) only
  superconducting order develops.  The precise location of the CDW QCP will likely by affected by superconductivity, as it generally happens when one order
  develops under the umbrella of another~\cite{umbrella,vvc,fs}.

    We assume that charge QCP exists and combine the doping dependencies of
         $T_{\rm sc}$, $T_{\rm cdw}$, $T_{n}$, and $T_t$   into the full phase diagram, which we show in Fig.\ \ref{phases}(c).  We conjecture
          that the reduction of
         $T_{\rm sc}$ in the
underdoped regime is primarily the result of a direct competition between superconductivity and composite CDW order, while
a reduction due to fluctuations between superconductivity and bond order~\cite{efetov} plays a secondary role.
 We emphasize that in our model
  superconductivity and CDW order are produced by the same underlying spin-fluctuation exchange interaction, and in this respect they are,
  in the terminology of Refs.\ [\onlinecite{fra_kiv,steve_k}],
   intertwined
  rather than competing orders.
  The situation is again similar to the one for underdoped pnictides where superconductivity and SDW orders are also intertwine orders as they
   originate from the same 4-fermion pair-hopping interaction~\cite{vvc,fs}.

   We compare our theoretical phase diagram with the one for hole-doped cuprates and present quantitative comparison of our theory with ARPES data,
   including Fermi arcs in the normal state~\cite{mike_arc} and the doping evolution of the spectral function at low $T$, when the systems moves from a pure
   superconducting  state into a state where superconductivity and charge order co-exist.  We argue that the agreement with the data is quite good, but to
   describe the evolution of the ARPES dispersion along the cuts closer to zone diagonals one needs to go beyond what we did so far and solve for the CDW
   order parameter $\Delta^Q_k$ for $k$ rather far away from the mid-point between hot spots.

The structure of the paper is the following. In the next section we consider the model. In Sec.\ \ref{sec:2_1} we
 analyze the onset of CDW order with momentum $Q_x = (2Q,0)$ and $Q_y = (0,2Q)$ at near-infinite magnetic correlation length, when spin-fluctuation mediated interaction in the strongest and fermionic self-energy is large and cannot be neglected.  We
present our solution of the ladder set of
equations for the CDW order parameter first to logarithmical accuracy and then beyond the logarithmical approximation.
We show that the CDW problem belongs to a class of threshold problems, however the value of the coupling in our case is above the threshold.
 We compute $T_{\rm cdw}$ first in Eliashberg-type calculations and then by treating the effects of thermal bosonic fluctuations beyond Eliashberg theory, and compare $T_{\rm cdw}$ with $T_{\rm sc}$ obtained using the same procedures.  We also analyze non-ladder diagrams and present non-linear equation for CDW order parameter.
    In Sec.\ \ref{sec:new} we expand near the ladder solution, show that the solution corresponds to the minimum
of the effective action within the CDW subset, and discuss the interplay between CDW and superconducting and bond-order instabilities.
In Sec.\ \ref{sec:3} we discuss the structure of the CDW solution within mean-field approximation.
 We first approximate CDW order parameters $\Delta_{k}^{Q} = \langle c^\dagger_{{\bf k}+{\bf Q}} c_{{\bf k}-{\bf Q}}\rangle$ by
 $\Delta^{Q_x}$ and  $\Delta^{Q_y}$ in hot regions
  and analyze the interplay between CDW orders with $Q_x$ and $Q_y$.
   We show that CDW order breaks lattice rotational $C_4$ symmetry down to $C_2$ and develops in the
form of stripes.  We then show that CDW order $\Delta_{k}^{Q}$ actually
 has two components, one is even under ${\bf k} \to - {\bf k}$ and the other is odd
($\Delta_{1,k}^Q$ and $\Delta_{2,k}^Q$, respectively). Both are $U(1)$ fields, and the odd component changes sign under time-reversal. We derive GL
functional for four coupled CDW order parameters
 $\Delta_{1,k}^{Q_x}$, $\Delta_{2,k}^{Q_x}$, $\Delta_{1,k}^{Q_y}$, and $\Delta_{2,k}^{Q_y}$ and
 argue that either state I or state II is realized at low $T$, depending on the interplay between the two input parameters.
  We show that in the state I, CDW order still breaks $C_4$ lattice symmetry down to $C_2$, and, in addition, the phases of
    $\Delta_{1,k}^{Q_y} = |\Delta_1|e^{i\phi_1}$ and $\Delta_{2,k}^{Q_y} = |\Delta_2|e^{i\phi_2}$ differ by $\phi_1-\phi_2 = \pm \pi/2$. The selection $\pi/2$ or $-\pi/2$  breaks TRS.
 In Sec.\ \ref{sec:4} we analyze GL action for the state I beyond mean-field, by introducing collective variables (bi-products of $\Delta_{1,2}$)
  and search for non-zero expectation values of these variables within saddle-point approximation.
 We argue that CDW order
 develops in three stages, via two intermediate phases,  one with pure nematic order and another with additional breaking of TRS.
   In Sec.\ \ref{sec:5} we consider the interplay between composite CDW orders, a true CDW order,  and superconductivity, and
  obtain
  the phase diagram of hole-doped cuprates as a function of hole doping whose increase we identify with the decrease of a magnetic correlation length $\xi$.
  Here we show that $T_{\rm cdw}$, $T_{n}$, and $T_t$ decrease with decreasing $\xi$ and vanish at (the same) finite $\xi$ setting up a CDW quantum-critical point at some distance from a quantum-critical point associated with the onset of a magnetic order.
  In Sec.\ \ref{sec:6} we compare
   our results with the ARPES data both above and below $T_{\rm sc}$.  We present our conclusions in Sec.\ \ref{sec:7}.
    The discussion on several technical issues is moved into Appendices.  For completeness, in  Appendix~\ref{app:b} we
     also discuss
   the doping dependence of the onset temperature for bond order and the corresponding phase diagram.

In our consideration we approximate the electronic structure and collective spin excitations as two-dimensional, i.e., neglect fermionic and bosonic dispersions along $k_z$ direction.  We believe that the essential physics is captured within 2D treatment, although a coherent interlayer tunneling maybe important for the stabilization of the stripe phase~\cite{raghu_a}. We also assume that near CDW instability the system remains a metal, albeit with strong incoherence caused by quantum criticality.
A development of stripe CDW order from a quantum antiferromagnet in the strong coupling regime has been recently considered in Ref.\ [\onlinecite{arun_new}].

\section{The model}
\label{sec:2}

 We use the same spin-fermion model as in earlier  studies of magnetically-mediated
 $d$-wave superconductivity~\cite{scalapino,acf} and
 non-Fermi liquid physics  outside of pseudogap region\cite{acs,ms}.
 The model describes  low-energy fermions with the FS shown in Fig.\ \ref{fig3} and with 4-fermion interaction  mediated by soft spin collective excitations  peaked at or near $(\pi,\pi)$.  We focus on hot regions on the FS, for which shifting ${\bf k}_F$ by ${\bf k}_F + (\pi,\pi)$ keeps a
 fermion near the FS, and
  expand fermionic dispersion near a hot spot as $\epsilon_k = v_{F,{\bf k}} (k_\perp + \kappa k^2_\parallel/k_F)$, where $v_{F,\bf k}$ is a Fermi velocity at
  a hot spot, $k_\parallel$ is a deviation from a hot spot along the FS, and dimensionless $\kappa$
   specifies the curvature of the FS.  The Fermi velocities at hot spots ${\bf k}_1$, ${\bf k}_2$ and ${\bf k}_3 = {\bf k}_1 + (\pi,\pi)$, ${\bf k}_4 =
   {\bf k}_2 + (\pi,\pi)$ in Fig.\ \ref{fig3} are ${\bf v}_{F,{\bf k}_1} = (v_x, v_y)$ ${\bf v}_{F,{\bf k}_2} = (v_x, -v_y)$, ${\bf v}_{F,{\bf k}_3} =
   (-v_y, -v_x)$, and ${\bf v}_{F,{\bf k}_4} = (-v_y, v_x)$.
   The amplitude of the Fermi velocity $v_{F,{\bf k}} = \sqrt{v^2_x + v^2_y}$ and the value of $\kappa$ are the same at all hot spots.
   The effective 4-fermion interaction is mediated by soft spin excitations is
   \beq
    {H}_{int} = {\bar g} \chi(q) c^{\dagger}_{k+q,\alpha} {\bf \sigma}_{\alpha\beta} c_{k,\beta}
   c^{\dagger}_{p-q,\gamma} {\bf \sigma}_{\gamma\delta} c_{p,\delta},
  \label{as_1}
   \eeq
  where $k = ({\bf k}, \omega_m)$, $q = ({\bf q, \Omega_m})$, $\omega_m (\Omega_m)$ are fermionic (bosonic) Matsubara frequencies, and
  \beq
  \chi(q) = \chi ({\bf q}, \Omega_m) = \frac{1}{{\bf q}^2 + \xi^{-2} + \gamma |\Omega_m|},
  \label{as_1_1}
  \eeq
   where the last term in the denominator
    is the Landau damping with the coefficient $\gamma = 4 {\bar g}/(\pi |v^2_y-v^2_x|)$ (Ref.\ [\onlinecite{acs}]).
     The Landau damping  contains the same ${\bar g}$ as in (\ref{as_1})  because Landau damping of collective spin excitations
      also originates from the spin-fermion interaction.

 Following earlier works~\cite{acs,ms,efetov}, we assume that the coupling ${\bar g}$ is small compared to the Fermi energy $E_F =v_F k_F/2$ and focus on
 instabilities which occur at energies well below $E_F$ and at
 $\xi^{-1} \geq 0$, before the system becomes magnetically ordered.
 One such instability is towards a $d$-wave superconductivity~\cite{scalapino,pines,acf,wang}. It involves fermions from
 hot and lukewarm regions on the FS
 (with the self-energy $\Sigma(k_\parallel, \omega_m) \propto \sqrt{\omega_m}$ and  $\Sigma(k_\parallel, \omega_m) \propto
   \omega_m/|k_\parallel|$, respectively), and, taken alone (i.e., without competition with CDW order)  occurs at $T_{\rm sc} = T_{\rm sc} (\xi)$, which
   is
   non-zero at all $\xi$
   and interpolates between $T_{\rm sc} (\xi) \approx 0.04 {\bar g}$ at large $\xi$, with weak dependence on $v_x/v_y$, (Refs.\ [\onlinecite{wang,ar_mike}]),
  and BCS-like result $T_{\rm sc} (\xi) \sim ({\bar g}/\lambda^2) e^{-1/\lambda}$,  at smaller $\xi \ll E_F/{\bar g}$, when
  dimensionless coupling $\lambda = 3 {\bar g}/(4\pi v_F \xi^{-1})$ is small (see panel b in Fig.\ \ref{phases}).  Another instability, considered in
  [\onlinecite{ms,efetov}], is towards a $d$-wave charge bond order with diagonal momentum $2 k_{hs} = (2Q , \pm 2Q)$, where $k_{hs} = (Q, \pi \pm Q)$.
  This instability develops at $T_{\rm bo} (\xi)$, which is smaller than $T_{\rm sc} (\xi)$ at any non-zero $\kappa$,  although rather close to it at $\xi \to \infty$ (Refs.\
  [\onlinecite{ms,efetov}]). We analyze the form of $T_{\rm bo} (\xi)$ in (see Appendix~\ref{app:b}) where we show that $T_{\rm bo} (\xi)$  vanishes at a certain $\xi$, when $\lambda \log {[E_F/({\bar g} \kappa)]} = O(1)$.

  Our goal is to analyze another CDW channel, the one with momentum $Q_x = (2Q,0)$ or $Q_y =(0,2Q)$. This pairing involves fermions from hot/lukewarm regions
   1-2, 3-4, etc. in Fig.\ \ref{fig3}.
    The analysis of a potential CDW instability involving these fermions is a bit tricky,
    because fermions in  the two regions connected by  $(\pi,\pi)$ (e.g., regions 1-2 and 3-4 in Fig.\ \ref{fig3}) are different in
    the sense
    that parallel (antiparallel) velocities are $v_x$ ($v_y$) in the first set and $-v_y$ ($-v_x$) in the second.
      Accordingly, the CDW order parameter $\Delta_{k}^Q = \langle c^\dagger_{k+Q,\alpha} c_{k-Q,\alpha}\rangle$ does not
       obey
       a particular symmetry relation under $k \to {k} + (\pi,\pi)$,  and one has to solve the $2\times 2$  coupled set of equations
       for $\Delta_{k}^Q$ and $\Delta_{k+(\pi,\pi)}^Q$.

 In the next two sections we consider CDW instability  with momentum ${\bf Q}$ along either $x$ or $y$ axis near the onset of SDW order, when  the magnetic
 correlation length $\xi$ is near-infinite.  We consider what happens at smaller $\xi$ later in Sec.\ \ref{sec:5}.
We perform our analysis in two stages. In the next section (Sec.\ III) we solve the set of linearized gap equations for $\Delta_{k}^Q$ and $\Delta_{k+\pi}^Q$ within
the ladder approximation
 and show that this set has  non-trivial solution at a non-zero $T_{\rm cdw}$.   In Sec.\ \ref{sec:new}
 we
 re-derive the same set by
   performing Hubbard-Stratonovich transformation from original fermions to bosonic CDW variables and show that diagrammatic ladder approximation is
   equivalent to saddle point approximation in Hubbard-Stratonovich approach. We expand around Hubbard-Stratonovich solution within the CDW subset and
   show
   that saddle-point solution is the minimum of the effective action, i.e., fluctuations around saddle point solution do not diverge.
   We then discuss the interplay between our CDW  and superconductivity/bond-order.  We argue that our CDW order and the other orders
    can be treated within a generic Ginzburg-Landau functional. Taken alone, each order is stable and is not destroyed by fluctuations. This internal
    stability implies that the system develops the order which sets up at the highest $T$.
    At a mean-field level, $T_{\rm sc} \geq T_{\rm bo} > T_{\rm cdw}$, hence superconductivity develops first. However, we show in Sec.\ \ref{sec:4} that
    composite CDW order develops at $T_{n} > T_{\rm cdw}$ and this temperature well may exceed $T_{\rm sc}$. Once this happens, composite charge order provides negative feedback on superconducting/bond order, reducing the corresponding mean-field onset
    temperatures, and gives positive feedback on $T_{\rm cdw}$ which in some parameter range becomes larger than $T_{\rm sc}$.

\section{The onset of CDW order with momentum $(2Q,0)$ and $(0,2Q)$ at $\xi = \infty$}
\label{sec:2_1}

 Borrowing notations from superconductivity, we will be calling the  equations for $\Delta_{{k}}^Q$ and $\Delta_{{k} + {(\pi, \pi)}}^Q$ as ``gap"
 equations.
 We will first solve for the onset of CDW instability in the ladder approximation without discussing its validity and later show that non-ladder
 contributions to the gap equation are small numerically.
 We start with hot regions 1-2 and 3-4 in Fig.\ \ref{fig3}.
These two pairs forms a closed set for the CDW gap equations, and pairing in other hot regions should simply follow due to symmetry.

\begin{figure}
\includegraphics[width=0.5\columnwidth]{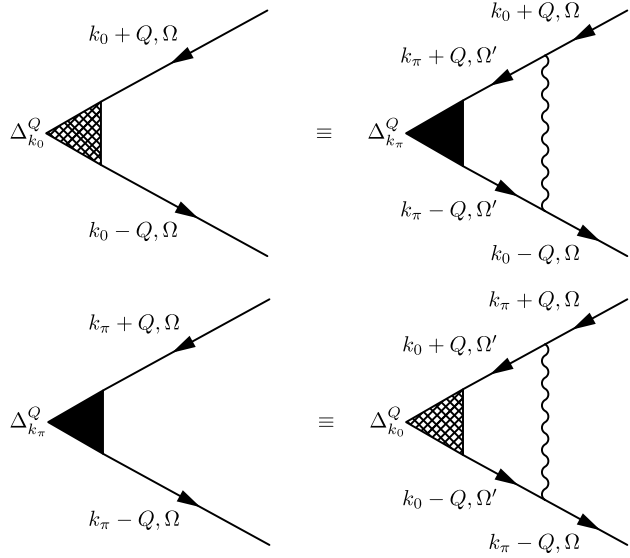}
\caption{The set of linearized equations for CDW  vertices constructed out of fermions near hot spots. The momenta $k_0$ and $k_\pi = k_0 + (\pi,\pi)$ are
in between the two neighboring hot spots either along $x$ or along $y$  direction, chosen such that $k_0 \pm Q$ and $k_\pi \pm Q$ are right at the hot
spots. The solid lines are full fermionic propagators, the wavy lines represent spin-mediated interaction peaked at $(\pi,\pi)$.}
\label{fig5}
\end{figure}

   The gap equations in the ladder approximation are shown in  Fig.\ \ref{fig5}.
 The  CDW order parameters   $\Delta_{{k}}^Q$ and $\Delta_{{k} + \pi}^Q \equiv \Delta_{{k} + {(\pi, \pi)}}^Q$
  are expressed via each other, and one needs to solve a  set of two coupled gap
 equations to find an instability. Each equation is, in general, an integral equation in both frequency and momentum. Besides, fermionic propagators in
 the
 r.h.s.\ of the gap equations contain the self-energy $\Sigma(k,\Omega_m)$ which is large and depends on frequency $\Omega_m$ and on the
 deviation of an internal fermion from a corresponding hot spot in the direction along the FS
  (Refs.\ [\onlinecite{acs,ms}]).
  We present the calculation of $T_{\rm cdw}$  in which  we  keep only frequency dependence in the fermionic self-energy and in the gap functions, i.e., approximate
   $\Sigma(k,\Omega_m) \approx\Sigma(k_{\rm hs},\Omega_m)\equiv\Sigma_{\Omega_m}$, $\Delta_k^{Q}(\Omega_m)\approx\Delta_{k_0}^{Q}(\Omega_m)$ when $k$ is near
   $k_0$, and
   $\Delta_k^{Q}(\Omega_m)\approx\Delta_{k_{\pi}}^{Q}(\Omega_m)$ when $k$ is near $k_{\pi}$.  Such an approximation has been verified~[\onlinecite{wang}]
   to be a valid one the calculation of a superconducting $T_{\rm sc}$ and we expect it to be valid also for a CDW instability. The full gap equations,
    with momentum-dependent CDW order parameters and momentum-dependent self-energy  are presented in Appendix \ref{app:a}.

  In analytical form the set of the two linearized integral equations in frequency for $\Delta_{k_0}^{Q}(\Omega_{m})$ and $\Delta_{k_\pi}^{Q}(\Omega_{m})$ is
\begin{align}
\Delta_{k_0}^{Q}(\Omega_{m})&=\frac{3\bar g T_{\rm cdw}}{4\pi^2}\sum_{m'}\int\frac{dx~dy}{[i{\tilde \Sigma} ({\Omega_{m'}})-v_x y+v_y x][i{\tilde
\Sigma} ({\Omega_{m'}})+v_x y+v_y x]}\frac{\Delta_{k_\pi}^{Q}(\Omega_{m'})}{x^2+y^2+\gamma|\Omega_{m}-\Omega_{m'}|}\label{Eq1a} \\
\Delta_{k_\pi}^{Q}(\Omega_{m})&=\frac{3\bar g T_{\rm cdw}}{4\pi^2} \sum_{m'}\int\frac{dx~dy}{[i{\tilde \Sigma} ({\Omega_{m'}})
-v_x x+v_y y][i{\tilde \Sigma} ({\Omega_{m'}})-v_x x-v_y y]}\frac{\Delta_{k_0}^{Q}(\Omega_{m'})}{x^2+y^2+\gamma|\Omega_{m}-\Omega_{m'}|}
\label{Eq1b}
\end{align}
where
$\gamma=4\bar g/(\pi (v_y^2-v_x^2))$, $x$ and $y$ are momentum deviations from the corresponding hot spots, ${\tilde \Sigma} (\Omega_m) = \Omega_m +
\Sigma (\Omega_m)$,
and the fermionic self-energy $\Sigma (\Omega_m)$  is the solution of the self-consistent equation
\beq
\Sigma (\Omega_m) = \frac{3\bar g T}{4\pi} \sum_{\Omega_{m'}} \int dy  \frac{{\rm sgn}( \Omega_{m'})}{\sqrt{y^2+
\gamma |\Omega_{m}-\Omega_{m'}|}}\frac{1}{\sqrt{y^2+\gamma |\Omega_{m}-\Omega_{m'}|}+|{\tilde \Sigma} (\Omega_m')|/v_F} \label{frac_0}.
\eeq
  The fermionic Green's function $G(k, \Omega_m)$
  is
  related to
  ${\tilde \Sigma} (\Omega_m)$ as
  \beq
  G^{-1} (k, \Omega_m) = i {\tilde \Sigma} (\Omega_m) - \epsilon_k
  \label{tuac_3}
  \eeq
  where $\epsilon_k$ is fermionic dispersion which in Eqs.\ (\ref{Eq1a}) and (\ref{Eq1b}) we expanded around hot spots.
 For hot fermions in regions 1 and 2 in Fig.\ \ref{fig3}, $k_0$ is near $(0,\pi)$, in which case $v_x < v_y$.

At $T=0$, the summation over frequency in (\ref{frac_0}) can be replaced by the integration. The equation for the fermionic self-energy becomes
 \beq
\Sigma (\Omega_m) =  \frac{3\bar g }{8\pi^2}\int d \Omega_{m'} \int dy  \frac{{\rm sgn}( \Omega_{m'})}{\sqrt{y^2+
\gamma |\Omega_{m}-\Omega_{m'}|}}\frac{1}{\sqrt{y^2+\gamma |\Omega_{m}-\Omega_{m'}|}+|{\tilde \Sigma} (\Omega_m')|/v_F} \label{frac_01}
\eeq
One can easily make sure that $\Sigma (\Omega_m) \propto {\rm sgn}(\Omega_{m'})\sqrt{|\Omega_{m'}|}$ at small enough frequencies.
Earlier calculations~\cite{acs,ms} of $\Sigma (\Omega_m)$ neglected self-energy in the r.h.s.\ of (\ref{frac_01}). In this approximation,
 the equation for $\Sigma(\Omega_m)$
is no longer
of
self-consistent form, and integration over $y$ and over $\Omega_{m'}$ yields
\beq
\Sigma (\Omega_m) =  {\rm sgn}(\Omega_{m'})\sqrt{\omega_0|\Omega_{m'}|}
\label{frac_02}
\eeq
where $\omega_0 = 9 {{\bar g}}/(16 \pi ) \times[(v_y^2-v_x^2)/v^2_F]$.   By order of magnitude, $\omega_0$ coincides with the spin-fermion coupling
constant ${\bar g}$. For consistency with previous works, below we will use $\omega_0$ as the overall scale for the self-energy.

When the self-energy is kept in the r.h.s of (\ref{frac_01}) the self-energy at frequencies $|\Omega_m| < \omega_0$ retains the same functional form as in
(\ref{frac_02}), but with the extra prefactor
\beq
\Sigma (\Omega_m) =  A~{\rm sgn}(\Omega_{m'})\sqrt{\omega_0|\Omega_{m'}|}
\label{frac_022}
\eeq
where, to a good numerical accuracy, $A = 2/3$.

To get an insight where the instability comes from, consider momentarily  the case $v_x =0$, $v_y = v_F$. Then Fermi velocities at the two hot spots near
$(\pi,0)$ (points 1 and 2 in Fig.\ \ref{fig3}) are antiparallel
 to each other, while the ones at the two hot spots near $(0,\pi)$ (points 3 and 4 in Fig.\ \ref{fig3}) are parallel.
  In this limit, Eqs.\ (\ref{Eq1a}) and (\ref{Eq1b}) reduce to
 \begin{align}
\Delta_{k_0}^{Q}(\Omega_{m})&=\frac{3\bar g T_{\rm cdw}}{4\pi^2}\sum_{m'}\int\frac{dx~dy}{\left[i{\tilde \Sigma} ({\Omega_{m'}})+v_F
x\right]^2}\frac{\Delta_{k_\pi}^{Q}(\Omega_{m'})}{x^2+y^2+\gamma|\Omega_{m}-\Omega_{m'}|}\label{Eq1aa} \\
\Delta_{k_\pi}^{Q}(\Omega_{m})&=- \frac{3\bar g T_{\rm cdw}}{4\pi^2} \sum_{m'}\int\frac{dx~dy}{\left[{\tilde \Sigma}^2 ({\Omega_{m'}})
+(v_F y)^2\right]}
\frac{\Delta_{k_0}^{Q}(\Omega_{m'})}{x^2+y^2+\gamma|\Omega_{m}-\Omega_{m'}|},
\label{Eq1bb}
\end{align}

In earlier large $N$ calculations of $T_{\rm sc}$ (Refs.\ [\onlinecite{acf,acs,ms}]), the dependence of the bosonic propagator on the momentum transverse to the FS
 (i.e., the $x$ dependence in the last term in Eq.\ \ref{Eq1aa}) and the $y$ dependence in Eq.\ (\ref{Eq1bb})
 was neglected. If we used the same approximation here, we would obtain no CDW instability because
 the integral over $x$ in Eq.\ (\ref{Eq1aa}) would vanish.
However, in our case the bosonic propagator does depend on $x$ and its poles are in both upper and lower half-planes of complex x.  As a consequence, the
momentum integration over $x$ in
 the r.h.s.\ of (\ref{Eq1aa}) gives a non-zero result even if we assume that Fermi velocities at points 3 and 4 in Fig.\ \ref{fig3} are parallel.
 Furthermore, because at $\Omega_m < \omega_0$,
 $|{\tilde \Sigma} ({\Omega_m})|/v_F \approx \Sigma (\Omega_m) \sim ({\bar g} |\Omega_m|/v^2_F)^{1/2} \sim (\gamma |\Omega_m|)^{1/2}$, the poles in the
 bosonic propagator are located at $x$ comparable to that of the double pole.  As a result, the result of the integration over $x$ is comparable to what
 one would get from integrating over $x$ in the two fermionic Green's functions, if the poles there were in different half-planes of $x$.  In other words,
 the fact that the velocities at the hot spots at points 3 and 4 in Fig.\ \ref{fig3} are parallel  does not make the r.h.s.\ of Eq.\ (\ref{Eq1aa})
 parametrically smaller compared to the case in Eq.\ (\ref{Eq1bb}), where the velocities at the two hot spots are anti-parallel and the momentum integral
 over the product of the two Green's function already gives a non-zero result. In Eq.\ (\ref{Eq1bb}), the contributions from  the poles in the Green's
 function and in the bosonic propagator are of the same order and just add up in the overall prefactor.

We now return back to Eqs.\ (\ref{Eq1a}) and (\ref{Eq1b}) for $\Delta_{k_0}^Q (\Omega_m)$ and $\Delta_{k_\pi}^Q (\Omega_m)$. We first integrate over $x$ in Eq.\ (\ref{Eq1a}) and obtain
\begin{align}
\Delta_{k_0}^{Q}(\Omega_{m})=-\frac{3\bar g
T_{\rm cdw}}{8\pi}\sum_{m'}\int_0^{\infty}\frac{dy}{\sqrt{y^2+\gamma|\Omega_{m}-\Omega_{m'}|}}\frac{\Delta_{k_\pi}^Q(\Omega_{m'})}{v_x^2y^2+
\left(v_y\sqrt{y^2+\gamma|\Omega_{m}-\Omega_{m'}|}+{\tilde \Sigma}({\Omega_{m'}})\right)^2}.
\label{ch_1}
\end{align}
Introducing then
$z=y/\sqrt{\gamma|\Omega_{m'}|}$ and $\phi=\arctan(v_x/v_y)$ and using zero-temperature form of $\Sigma (\Omega_m) \approx (2/3) {\rm
sgn}(\Omega_{m'})\sqrt{\omega_0|\Omega_{m'}|}$, we re-write (\ref{ch_1}) as
\begin{align}
\Delta_{k_0}^{Q}(\Omega_{m})=&-\frac{3\cos2\phi}{8} T_{\rm cdw} \sum_{m'}\frac{\Delta_{k_\pi}^{Q}(\Omega_{m'})}{|\Omega_{m'}|}\int_{0}^{\infty}\frac{dz}{\sqrt{z^2+|1-\Omega_{m}/\Omega_{m'}|}}\nonumber\\
&\times\frac{1}{z^2\sin^2\phi+\left[\sqrt{z^2+|1-\Omega_{m}/\Omega_{m'}|}\cos\phi+(1/4)\cos2\phi(1+\sqrt{{\Omega_{m'}}/{\omega_0}})\right]^2}.
\label{k_0}
\end{align}
Integrating over $x$ in Eq.\ (\ref{Eq1b}) and re-writing the result in the same variables $z$ and $\phi$ we obtain
\begin{align}
\Delta_{k_\pi}^{Q}(\Omega_{m})=&-\frac{3\cos2\phi}{8}T_{\rm cdw}\sum_{m'}\frac{\Delta_{k_0}^{Q}(\Omega_{m'})}{|\Omega_{m'}|}\int_{0}^{\infty}\frac{dz}{\sqrt{z^2+|1-\Omega_{m}/\Omega_{m'}|}}\nonumber\\
&\times\frac{1}{z^2\cos^2\phi+\left[\sqrt{z^2+|1-\Omega_{m}/\Omega_{m'}|}\sin\phi+(1/4)\cos2\phi(1+\sqrt{{\Omega_{m'}}/{\omega_0}})\right]^2}.
\label{kpi}
\end{align}
The value of $\phi$ depends on the geometry of the FS. When Fermi velocities at hot spots  1 and 2 in Fig.\ \ref{fig3}  are nearly antiparallel, and the
ones
 at hot spots 3 and 4 are nearly parallel, we have $\phi \approx 0$. When Fermi velocities at hot spots 1 and 2 are nearly perpendicular to each other, we
 have
  $\phi \approx \pi/4$. For the FS as in hole-doped cuprates, $\phi$ is non-zero, but small numerically.

The negative signs in the right hand sides of (\ref{k_0}) and (\ref{kpi}) imply that the solution is only possible when $\Delta_{k_0}^{Q}(\Omega_{m})$ and
$\Delta_{k_\pi}^{Q}(\Omega_{m})$ have opposite signs, i.e., CDW order parameter $\Delta_{k}^{Q}(\Omega_{m})$ changes sign under ${\bf k} \to {\bf k} +
(\pi,\pi)$.
 This does not imply, however, that the order has only a bond component $\langle c^\dagger ({\bf r+a}) c({\bf r})\rangle$ (Ref.~[\onlinecite{ms}]).
In our case,  $\Delta_{k_0}^{Q}(\Omega_{m})$ and $\Delta_{k_\pi}^{Q}(\Omega_{m})$ differ in magnitude, and
both on-site and bond components are present.
For the on-site charge density we have
\begin{align}
\langle c^\dagger ({\bf r}) c({\bf r})\rangle=\sum_{\bf k,Q} \Delta_{k}^{Q} e^{i\bf Qr} = \sum_{Q} \left(\Delta^Q_{k_0} + \Delta^Q_{k_\pi}\right) e^{i\bf Qr}
\end{align}

\subsection{Gap equations to logarithmic accuracy}

\subsubsection{Neglecting frequency dependencies of $\Delta_{k_0}^{\bf Q} (\Omega_m)$ and $\Delta_{k_\pi}^{\bf Q} (\Omega_m)$}

 As a first pass on Eqs.\ (\ref{k_0}) and (\ref{kpi}) we approximate gap functions as frequency-independent constants
 $\Delta_{k_0}^{\bf Q}$ and $\Delta_{k_\pi}^{\bf Q}$,
  set the lower limit of
 integration
 over internal fermionic frequency to $\pi T_{\rm cdw}$, and neglect the dependence on external frequency in (\ref{k_0}) and (\ref{kpi}).
   Evaluating the integrals with two fermionic and one bosonic propagators, we find that they are
 logarithmically singular, no matter
  what $\phi$ is. To logarithmical accuracy, we obtain
\begin{align}
\Delta_{k_0}^Q=&-S_1(\phi)\log\frac{\omega_0}{T_{\rm cdw}}\Delta_{k_\pi}^{Q},\nonumber\\
\Delta_{k_\pi}^Q=&-S_2(\phi)\log\frac{\omega_0}{T_{\rm cdw}}\Delta_{k_0}^Q.
\label{k0pilog}
\end{align}
where
\begin{align}
S_1(\phi)=&\frac{3\cos2\phi}{8\pi}\int_{0}^{\infty}\frac{dz}{\sqrt{z^2+1}}\frac{1}{z^2\sin^2\phi+\left(\sqrt{z^2+1}\cos\phi+(1/4)\cos2\phi\right)^2}\nonumber\\
S_2(\phi)=&\frac{3\cos2\phi}{8\pi}\int_{0}^{\infty}\frac{dz}{\sqrt{z^2+1}}\frac{1}{z^2\cos^2\phi+\left(\sqrt{z^2+1}\sin\phi+(1/4)\cos2\phi\right)^2}.
\end{align}
 We emphasize that these two functions remain finite even if we set $\phi = 0$ (i.e., set the velocities at hot spots 3 and 4 in  Fig.\ \ref{fig3} to be
  parallel to each other). Note also that $S_1$ and $S_2$  depend on the ratio of $v_x/v_y$ but not on $\bar g$, which cancels out between the overall factor in the spin-fermion
 interaction and in the Landau damping.  This cancellation is typical for an instability mediated by a massless Landau-overdamped collective
 mode~\cite{ar_mike}.
 \begin{figure}
\includegraphics[width=0.5\columnwidth]{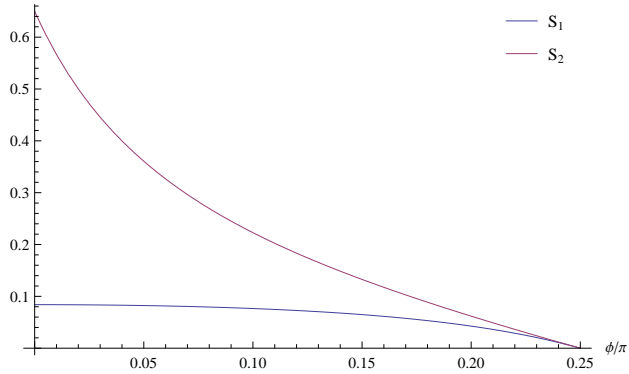}
\caption{The integrals $S_1$ and $S_2$ as functions of the angle $\phi$. Both integrals vanish at $\phi = \pi/4$ because at this $\phi$ the Landau damping
coefficient diverges.}
\label{Fig5}
\end{figure}

Evaluating $S_1$ and $S_2$ numerically, we found (see Fig.\ \ref{Fig5}) that $S_2 > S_1>0$ as long as $\phi<\pi/4$.
In the limit when $\phi =0$ (i.e., when the velocities at hot spots 1 and 2 are antiparallel, and the ones at hot spots 3 and 4 are
parallel),
we have $S_1=0.084$
 and $S_2=0.650$
 ($\sqrt{S_1 S_2} = 0.234$).  At non-zero $\phi$, the values
   of $S_1$ and $S_2$ are closer to each other.  That $S_2 >S_1$ implies that
 the CDW order parameter  in the region with nearly antiparallel Fermi velocities (region 1-2 in our case) is smaller than in the  region with nearly
 parallel velocities (region 3-4 in our case).  Solving the set
 (\ref{k0pilog})  we  immediately obtain that
  the linearized gap equation has a non-zero solution at
  \beq
  1 = S_1 S_2 \left(\log \frac{\omega_0}{T_{\rm cdw} }\right)^2
  \label{tuac_5}
  \eeq
 i.e., at  $T_{\rm cdw}  \sim \omega_0 e^{-1/\sqrt{S_1 S_2}}$. This
 $T_{\rm cdw}$  is of the same order of magnitude as superconducting $T_{\rm sc}$, which at the onset of SDW order is also of order $\omega_0$ (Refs. \
 [\onlinecite{acf,wang}]).
 Right at $T= T_{\rm cdw}$ we have from (\ref{k0pilog})
 \beq
 \Delta^Q_{k_0} = - \Delta^Q_{k_\pi} \sqrt{\frac{S_2}{S_1}}.
\label{tuac_7}
 \eeq

\subsubsection{Keeping frequency dependencies of  $\Delta_{k_0}^{Q} (\Omega_m)$ and $\Delta_{k_\pi}^{Q} (\Omega_m)$}

Eq.\ (\ref{tuac_5}) has been obtained within the approximations that (i) CDW order parameters $\Delta_{k_0}$ and $\Delta_{k_\pi}$ do not depend on
frequency, and
 (ii) one can neglect the dependence on external $\Omega$ in the r.h.s.\ of Eqs.\ (\ref{k_0}) and (\ref{kpi}).
   For a more accurate treatment, we need to keep the frequency dependence in $\Delta_{k_0}^Q (\Omega_m)$ and in $\Delta_{k_\pi}^Q (\Omega_m)$.

 Plugging Eq.\ (\ref {kpi}) into Eq.\ (\ref {k_0}) we get rid of $\Delta_{k_\pi}^Q (\Omega_m)$ and obtain integral equation for $\Delta_{k_0}^{Q}(\Omega)$ in
 the form,
\begin{align}
\Delta_{k_0}^Q(\Omega_m)= T\sum_{m'} I(\Omega_m,\Omega_{m'}) \Delta_{k_0}^Q(\Omega_{m'}),
\label{gapper}
\end{align}
where the kernel $I(\Omega_m,\Omega_{m'})$ depends on both external and internal frequency.

We first analyze the pairing susceptibility. For this we add the source term $\Delta_0$ to the right hand side of Eq.\ (\ref{gapper}) and
 search for the divergence of
$\Delta_{k_0}^Q/\Delta_0$ at $T = T_{\rm cdw}$.
At first order of iterations we replace
$\Delta_{k_0}^Q(\Omega_{m'})$ by $\Delta_0$ in the integral part and obtain
$\Delta_{k_0}^Q(\Omega_m=\pi T)=\Delta_0(1+\frac{1}{2}S_1S_2\log^2\frac{\omega_0}{T})$.
This is the same result as we had before, except for the additional  $1/2$ factor which is due to the requirement that, when we compute $\sum_{m'}
I(\Omega_m,\Omega_{m'})$,  the internal frequency must be much larger than the external one in order to obtain $\log^2$ correction.
If subsequent iterations would transform $1 + 0.5 S_1S_2\log^2\frac{\omega_0}{T})$ into $1/(1 -0.5 S_1S_2\log^2\frac{\omega_0}{T})$, as it happens in BCS theory, we would obtain $T_{\rm cdw}$ similar to Eq.\ (\ref{tuac_5}). However,  we found that in our case
 the series of $\log^2$ terms actually sum up into a power-law form
   $\Delta_{k_0}^{Q}=\Delta_0\cosh\left(\sqrt{S_1S_2}\log\frac{\omega_0}{T}\right)\sim\Delta_0\left(\frac{\omega_0}{T}\right)^{\sqrt{S_1S_2}}$. The implication is that, within the logarithmic approximation,  the ratio
    $\Delta_{k_0}^{Q}/\Delta_0$ does not diverge at any finite $T$, i.e., Eq.\ (\ref{tuac_5}) is an artifact of neglecting frequency dependence of $\Delta^Q$.
    A similar situation holds in the superconducting channel. There, previous works have found \cite{acf,wang,moon} that the
    instability
     does develop, but to detect it one has to go beyond logarithmic accuracy, solve the full integral equation in frequency and search whether or not
     there is an instability at $T >0$. This is what we do next.

\subsection{Beyond logarithmic approximation}

\subsubsection{Pairing susceptibility at $T=0$}

     The first step in the analysis is to consider $T=0$, when the summation over Matsubara frequencies can be replaced by integration.
      At $T=0$, the lower cut-off of the logarithm is set by frequency rather than by $T$, hence within the logarithmic approximation
       \beq
       \Delta_{k_0}^{Q} (\Omega_m), \Delta_{k_\pi}^Q (\Omega_m) \propto \Delta_0\left(\frac{\omega_0}{|\Omega_m|}\right)^{\sqrt{S_1S_2}}.
       \label{asp_1}
       \eeq
      We now verify whether
     the actual  pairing susceptibility at $T=0$ behaves similarly to (\ref{asp_1})  or changes sign, at least in some frequency ranges. The latter would indicate
      that that the normal state is unstable, hence $T_{\rm cdw}$ is non-zero.

       A similar strategy has been applied to superconducting problem at the onset of SDW order, when magnetic $\xi = \infty$~\cite{acf,wang}. We briefly
  discuss how it worked there and then apply it to our case.
    Like in our case,  the summation of the leading logarithms for the superconducting problem  does not lead to the instability and instead yields for
    the
    superconducting order parameter $\Delta_{\rm sc} (\Omega_m) \propto \Delta_0 \left(\frac{\omega_0}{|\Omega_m|}\right)^{\alpha_0}$, where $\Delta_0$ is
    again a source term and $\alpha_0$ is some positive number of order one~\cite{comm_A}.
       The solution of the full integral equation for $\Delta_{\rm sc} (\Omega_m)$ at $T=0$ also yields  a power-law form $\Delta_{\rm sc} (\Omega_m)
       \propto \Delta_0 \left(\frac{\omega_0}{|\Omega_m|}\right)^{\alpha}$ at $\Omega_m < \omega_0$, like in (\ref{asp_1}), however $\alpha$
         turns out to be a complex number. In this situation, there are two solutions, one with $\alpha$, another with $\alpha^*$.
      The linear combination of these two solutions
      yields oscillating
      \beq
      \Delta_{k_0}^Q \propto \Delta_0 \cos[{\rm Im}(\alpha)\log\omega+\theta]/|\Omega_m|^{{\rm Re} \alpha}
      \label{asp_2}
      \eeq
       with a
       ``free" phase variable $\theta$.  The presence of $\theta$ plays the crucial role when the analysis is extended to finite $T$ and the source term $\Delta_0$ is set to
       zero.
       The power-law behavior at a finite $T$ exists in the frequency range between
       $\omega_0$ and $T$ and has to satisfy  boundary conditions at the two edges. This requires two adjustable parameters. The temperature is one of
       them
       and the phase $\theta$ is the other one. Solving for the two boundary conditions requires care, but the result is that, very likely, they can be
        satisfied at a non-zero $T$, i.e., at this $T$ the linearized gap equation has a solution.  Whether this is the actual $T_{\rm sc}$ is a more subtle issue as there may exist some other
           solution of the linearized gap equation, with different
          behavior at small frequencies. In any case, however, the fact that Eq.\ (\ref{asp_2}) is the solution of the linearized gap equation
           at a finite $T$ implies that $T_{\rm sc}$ must be finite.
          From this perspective,  the fact that the exponent $\alpha$ is complex is a sufficient condition for the existence of the pairing
          instability at a finite $T_{\rm sc}$ in the
         quantum-critical regime.

 We follow the same strategy for the CDW case.
  We keep the frequency dependencies of $\Delta_{k_0}^{Q}$ and $\Delta_{k_\pi}^{Q}$  in Eqs.\ (\ref{k_0}) and (\ref{kpi}), solve these two equations as
  integral equations in frequency, search for a power-law solution and analyze whether or not the exponent is complex. To shorten the presentation, we only
  consider the limiting case $\phi=0$, and replace the soft upper cutoff at $\omega_0$ with a hard one. With this simplification, we obtain, replacing the sums by integrals
\begin{align}
\Delta_{k_0}^{Q}(\Omega_{m})=&\frac{-3}{16 \pi}\int _{-\omega_0}^{\omega_0} \frac{d
\Omega_{m'}~\Delta_{k_\pi}^{Q}(\Omega_{m'})}{|\Omega_{m'}|}\int_{0}^{\infty}\frac{dz}{\sqrt{z^2+
|1-\Omega_{m'}/\Omega_{m}|}\left(\sqrt{z^2+|1-\Omega_{m'}/\Omega_{m}|}+1/4\right)^2}\label{Eq2a}\\
\Delta_{k_\pi}^{Q}(\Omega_{m})=&\frac{-3}{16\pi}\int_{-\omega_0}^{\omega_0}\frac{d
\Omega_{m'}~\Delta_{k_0}^{Q}(\Omega_{m'})}{|\Omega_{m'}|}\int_{0}^{\infty}\frac{dz}{\sqrt{z^2+|1
-\Omega_{m'}/\Omega_{m}|}}\frac{1}{z^2+1/16},\label{Eq2b}
\end{align}
We search for the solution in the form
 $\Delta_{k_0}^Q,~\Delta_{k_\pi}^Q\sim|\Omega_{m}|^{-\alpha}$. One can easily verify that convergence of integrals requires $0< {\rm Re}\ \alpha<0.5$.
 Substituting this trial solution into (\ref{Eq2a}) and (\ref{Eq2b}) we find after simple algebra a self-consistency condition
 \begin{align}
g_{\rm eff}  F(\alpha)G(\alpha) =1,
\label{cond}
\end{align}
 where
 $g_{\rm eff}  = 9/64$ is the universal dimensionless coupling constant for our quantum-critical problem and
\begin{align}
F(\alpha)=&\frac{1}{2\pi}\int_{-\infty}^{\infty}\frac{d\omega}{|\omega|^{1+\alpha}}\int_{0}^{\infty}\frac{dz}{\sqrt{z^2+|1-1/\omega|}\left(\sqrt{z^2+|1-1/\omega|}+1/4\right)^2}\nonumber\\
G(\alpha)=&\frac{1}{2\pi}\int_{-\infty}^{\infty}\frac{d\omega}{|\omega|^{1+\alpha}}\int_{0}^{\infty}\frac{dz}{\sqrt{z^2+|1-1/\omega|}}\frac{1}{z^2+1/16},
\end{align}

We solved Eq.\ (\ref{cond}) for $\alpha$ and found that the solution is a complex number: $\alpha = 0.288 \pm 0.185 i$.
The corresponding eigenfunction has the form
 $\Delta_{k_0}^Q,~\Delta_{k_\pi}^Q\sim \cos[0.185 \log{|\Omega_{m}|} + \theta]/|\Omega_m|^{-0.288}$, where $\theta$ is a
 free
  phase factor. Like we said, this is the sufficient condition for a CDW instability at a non-zero $T_{\rm cdw}$.

\subsubsection{The computation of $T_{\rm cdw}$}

Because the effective coupling $g_{\rm eff}$ is of order one, the only relevant energy scale in the quantum-critical regime is $\omega_0$, hence $T_{\rm cdw}$ must be
 of order $\omega_0$.  From this perspective, the estimate of $T_{\rm cdw}$ in Eq.\ (\ref{tuac_5}) is actually not far off as it also gives $T_{\rm cdw}$ of order
$\omega_0$.  To get the right ratio of  $T_{\rm cdw}/\omega_0$, one need to solve the set of the two linearized gap equations numerically.

There is one caveat, however, associated with the special role of zero bosonic Matsubara frequency term in the gap equation. Indeed, the frequency sum in
each of
Eqs.\ (\ref{Eq2a}) and (\ref{Eq2b}) contains the term with $\Omega_{m'} = \Omega_m$. For this particular Matsubara frequency the integral over $z$ diverges
logarithmically, as $\log \xi$, and, if there was no counter-term, $T_{\rm cdw}$ would vanish at $\xi = \infty$.

This issue is known in the superconducting problem~\cite{acf,ar_mike,finn,msv}. The term with zero Matsubara frequency represents scattering with zero
frequency transfer and a finite momentum transfer and from this perspective acts like an ``impurity". The logarithmical divergence of the integral over $dz$ in
(\ref{Eq2a}) and (\ref{Eq2b})  implies that ``impurity"  has an infinite strength at $\xi = \infty$.  Still, for an $s$-wave
superconductor, the contribution to $T_{\rm sc}$ from impurities must
vanish  by Anderson theorem.  To see this vanishing in our formalism, one needs to do calculations more accurately than we did so far and re-evaluate
fermionic self-energy  at a finite $T$, as it also contains a $\log \xi$ contribution coming from zero bosonic Matsubara frequency.
 For an $s$-wave superconductor, the two contributions cancel each other.  For other cases, the situation is
 less obvious.  For $p$-wave pairing, the divergent terms coming from zero bosonic Matsubara
frequency do not cancel out within the Eliashberg approximation and this eventually gives rise
to first-order superconducting transition~\cite{finn}.
 For $d$-wave pairing,  earlier calculations within spin-fermion model used the Eliashberg approximation, in which
  the momentum integration is factorized -- the one
transverse
to the FS is performed over the two fermionic propagators, while the one along the FS is performed over the bosonic propagator.
  Within this approximation,
 the contributions from zero Matsubara frequency to the pairing vertex and to the
fermionic self-energy either
completely cancel out in the gap equation~\cite{acf,msv}, when fermionic self-energy is approximated by its value at a hot spot, or the divergent terms cancel out and the remaining non-divergent terms give rise to a modest
reduction of $T_{\rm sc}$, when the momentum dependence of the
fermionic self-energy on ${\bf k}$ along the FS is kept (Ref.\ [\onlinecite{ar_mike}]). On the other hand, in our solution of the CDW problem, it was crucial to go beyond Eliashberg approximation and include the contributions  from the poles in the bosonic
propagator  in the integration  along and transverse to the FS. (We recall that, for interactions mediated by collective modes of fermions, there is no
small parameter to justify Eliashberg approximation, except for special cases near three dimensions~\cite{senthil,max_very_last}).
To see whether or not the cancellation of the divergent contributions from zero Matsubara frequency holds in our case we have
to keep the summation over Matsubara frequencies not only in the equations for $\Delta_{k}^Q$ but also in the self-energy.
  For definiteness, we consider the case $\phi =0$, when Fermi
 velocities
 at one pair of hot spots are anti-parallel to each other and at the other are parallel. Like before, we neglect momentum dependencies of
 $\Delta^Q_{k_0}$,
 $\Delta^Q_{k_\pi}$ and the fermionic self-energy, i.e., approximate these quantities by their values at hot spots.  The inclusion of the $T$ dependence of the self-energy modifies Eqs.
  (\ref{Eq2a}) and
 (\ref{Eq2b}) to
  \begin{align}
\Delta_{k_0}^{Q}(\Omega^*_{m})=-\frac{3 T^*_{\rm cdw}}{8}\sum_{|\Omega^*_{m'}|<1} \Delta_{k_\pi}^{Q}(\Omega^*_{m'})\int_{0}^{\infty}&\frac{dy^*}{\sqrt{(y^*)^2+
|\Omega^*_{m}-\Omega^*_{m'}|}}\nonumber\\
&\times\frac{1}{\left(\sqrt{(y^*)^2+|\Omega^*_{m}-\Omega^*_{m'}|}+\frac{3}{8}|{\tilde \Sigma}^* (\Omega^*_{m'}|\right)^2},\label{Eq2aa}\\
\Delta_{k_\pi}^{Q}(\Omega^*_{m})=-\frac{3T^*_{\rm cdw}}{8}\sum_{|\Omega^8_{m'}|<1}
\Delta_{k_0}^{Q}(\Omega^*_{m'})\int_{0}^{\infty}&\frac{dy^*}{\sqrt{(y^*)^2+|\Omega^*_m -\Omega^*_{m'}|}}\frac{1}{(y^*)^2+\frac{9}{64}|{\tilde \Sigma}^*
(\Omega_{m'})|^2},\label{Eq2bb}
\end{align}
where ${\tilde \Sigma}^* (\Omega^*_m) = \Omega^*_m + \Sigma^* (\Omega^*_m)$, and
\begin{align}
 \Sigma^* (\Omega^*_m) = T^*_{\rm cdw} \sum_{|\Omega^*_{m'}|<\omega_0} \int_{0}^{\infty} dy^*  \frac{{~{\rm sgn}} (\Omega^*_{m'})}{\sqrt{(y^*)^2+
|\Omega_{m}-\Omega_{m'}|}}\frac{1}{\sqrt{(y^*)^2+|\Omega_{m}-\Omega_{m'}|}+\frac{3}{8}|{\tilde \Sigma}^* (\Omega_{m'})|} \label{Eq2cc}
\end{align}
In (\ref{Eq2aa})-(\ref{Eq2cc}) we used rescaled variables $\Omega^*_m = \Omega_m/\omega_0$, $\Sigma^* = \Sigma/\omega_0$, $T_{\rm cdw}^* = T_{\rm cdw}/\omega_0$ and $y^* = y/(\gamma \omega_0)^{1/2} = 2\pi v_F
 y/(3{\bar g})$.
 Let's first analyze the equation for the self-energy $\Sigma^*$. We recall that at $T=0$ we have $\Sigma^* (\Omega^*) = \Omega^* + (2/3)\sqrt{|\Omega^*|} {~{\rm sgn}}(\Omega^*)$.  At a finite $T$,
  the leading contribution to the sum in the r.h.s of (\ref{Eq2cc}) is $\log \xi$
    from the term with $\Omega^*_{m'} = \Omega^*_m$.
    Restricting with only this term and neglecting bare $\Omega^*$ (i.e., neglecting
 the difference between ${\tilde \Sigma}^*$ and $\Sigma^*$), we obtain from (\ref{Eq2cc})
  \beq
  \Sigma^* (\Omega^*_m) \approx \frac{8T^*_{\rm cdw} L {~{\rm sgn}} (\Omega^*_{m})}{3|\Sigma^* (\Omega^*_m)|}
  \label{frac_1}
  \eeq
  where $L = \log {\xi}$.  Solving (\ref{frac_1}) we obtain
  \beq
  \Sigma^* (\Omega^*_m) \approx \left(\frac{8 T^*_{\rm cdw} L}{3}\right)^{1/2} {{\rm sgn}} (\Omega^*_{m}) + ...
  \label{frac_2}
  \eeq
 where dots stand for terms of order one.  This formula is valid when $T^*_{\rm cdw} L$  is a large number.

 Substituting this $\Sigma^*(\Omega_m^*)$ into the first two equations and assuming that relevant $\Omega^*_m - \Omega^*_{m'}$ and  $y$  are of order one
 (we later verify this),
  we pull out $\Sigma^* (\Omega^*_m)$ and after integration over $y$  obtain
  \begin{align}
\Delta_{k_0}^{Q}(\Omega^*_{m}) \approx&-\frac{8 T^*_{\rm cdw}}{3}\sum_{|\Omega^*_{m'}|<1} \frac{\Delta_{k_\pi}^{Q}(\Omega^*_{m'})}{(\Sigma^*)^2}
\log{\frac{(\Sigma^*)^2}{|\Omega^*_{m}-\Omega^*_{m'}|}} + \ldots\nonumber\\
\Delta_{k_\pi}^{Q}(\Omega^*_{m}) \approx&-\frac{8T^*_{\rm cdw}}{3}\sum_{|\Omega^*_{m'}|<1} \frac{\Delta_{k_0}^{Q}(\Omega^*_{m'})}{(\Sigma^*)^2}
\log{\frac{(\Sigma^*)^2}{|\Omega^*_{m}-\Omega^*_{m'}|}} + \ldots,
  \label{frac_3}
\end{align}
 where $(\Sigma^*)^2 =(\Sigma^* (\Omega^*_m))^2 = 8 T^*_{\rm cdw} L/3$.
 We see that, as long as we neglect non-logarithmic terms,  $\Delta_{k_0}^{Q} = - \Delta_{k_\pi}^{Q}$. As a result, in this approximation CDW with
  $Q_x/Q_y$ has pure $d$-wave
  form-factor and in real space represents a bond order, just like CDW with diagonal $Q$.  A cite  component (a true CDW) appears
 once we include corrections to (\ref{frac_3}) (labeled as dots in (\ref{frac_3}), and is small in $1/L$.  Such a  structure of the charge
  order parameter is  consistent with Refs. [\onlinecite{subir_2,subir_4,bill}] and with the form of $\Delta_k^Q$ extracted from
 recent measurements~\cite{davis_1,X-ray_last}.

 The reason why $\Delta_{k_0}^{Q}$ and $\Delta_{k_\pi}^{Q}$ become almost equal by magnitude in spite of the difference in the arrangements of Fermi
 velocities in regions 1-2 and 3-4 in Fig.\ \ref{fig3} is that at large $T^*L$, the $\tilde\Sigma$ term in the fermionic propagator $G^{-1} (k, \omega) = i
 \tilde\Sigma (\Omega) - {\bf v}\cdot {\bf k}$ becomes  larger than the ${\bf v}\cdot {\bf k}$ term.
   Then the difference between nesting and
 anti-nesting becomes almost irrelevant as each fermionic  propagator can be
  approximated  by $G^{-1} (k, \omega) = i \tilde\Sigma (\Omega)$.

  The r.h.s.\ of each of the two Eqs.\ (\ref{frac_3}) contains $1/L$ coming from $(\Sigma^*)^2$ in the denominator and the logarithmical term in the
  numerator due to
  the presence of zero bosonic Matsubara frequency term in the summation over $\Omega^*_{m'}$. The logarithmical terms in the numerator and denominator
  exactly
   cancel each other, i.e., at this level there is no information what $T_{\rm cdw}$ is. To obtain $T_{\rm cdw}$ one has to keep terms with $\Omega_{m}
   \neq \Omega_{m'}$ in the formulas for
    $\Delta_{k_0}^{Q}(\Omega^*_{m}),~\Delta_{k_\pi}^{Q}(\Omega^*_{m})$ and
    for $\Sigma(\Omega^*_m)$.
       We follow the strategy used in Eliashberg-type
     theories and introduce ${\bar \Delta}_{k_0}^{Q}(\Omega^*_{m}) =  \Omega_m^*\Delta_{k_0}^{Q}(\Omega^*_{m})/\tilde\Sigma^* (\Omega_m^*)$ and ${\bar
     \Delta}_{k_\pi}^{Q}(\Omega^*_{m}) =  \Omega_m^*\Delta_{k_\pi}^{Q} (\Omega_m^*)/\tilde\Sigma^* (\Omega_m^*)$. Substituting this into any of the two equations
     in (\ref{frac_3}) and using  $\Delta_{k_0}^{Q} = - \Delta_{k_\pi}^{Q}$ we obtain [${\bar \Delta} (\Omega_m^*) \equiv {\bar
     \Delta}_{k_0}^{Q}(\Omega^*_{m})$]
     \beq
     {\bar \Delta}(\Omega^*_{m}) = \lambda T^*_{\rm cdw} \sum_{|\Omega^*_{m'}|<1} \log{\frac{(\Sigma^*)^2}{|\Omega^*_{m}-\Omega^*_{m'}|}} \left(\frac{{\bar
     \Delta}(\Omega^*_{m'})}{|\Omega^*_{m'}|} - \frac{{\bar \Delta}(\Omega^*_{m})}{\Omega^*_{m}} \frac{\Omega^*_{m'}}{|\Omega^*_{m'}|}\right)
\label{frac_4}
\eeq
where we defined $\lambda = 8/(3 |\Sigma^*|) = \sqrt{8/(3T^*_{\rm cdw} L)}$.
The second term in the last bracket in the r.h.s.\ of this equation (the one with ${\bar \Delta}(\Omega^*_{m})$) comes from the self-energy, once we
 express $\Delta_{k_0}^{Q}$ and $\Delta_{k_\pi}^{Q}$ in the l.h.s.\ of  the two equations in (\ref{frac_3}) via ${\bar \Delta}$.
 We see that term with zero Matsubara frequency vanishes, in agreement with what we found a few lines above (the term inside the parenthesis in the r.h.s.
 of (\ref{frac_4}) vanishes when $\Omega_{m'} = \Omega_m$), and the value of $T_{\rm cdw}$ is determined by
 the contributions from non-zero boisonic Matsubara frequencies.
  The vanishing  of $\Omega_{m'} = \Omega_m$ term is similar to what happens in an $s-$wave superconductor. However, contrary to an $s$-wave case, here
  the effective coupling $\lambda$ in (\ref{frac_4}) does depend on $L$.

  Eq.\ (\ref{frac_4}) has been studied in the context of color superconductivity~\cite{son} and of the pairing mediated by collective excitations in $D=3$
  (Refs.\ [\onlinecite{joerg,tsvelik,moon,senthil,max_very_last}]) and we just borrow the result:  at weak coupling
 (small $\lambda$) $T_{\rm cdw}$ is determined by
 \beq
 \log \frac{(\Sigma^*)^2}{T^*_{\rm cdw}} \sim \frac{1}{\lambda^{1/2}}
 \eeq
 Solving this equation, we obtain, in actual units
 \beq
 T_{\rm cdw} \sim \omega_0 \frac{(\log L)^4}{L}
 \label{frac_5}
 \eeq
 We see that thermal fluctuations reduce the CDW instability temperature by a factor $(\log L)^4/L$ compared to
 what we would obtain by using zero-temperature gap equation and just setting $T$ as the lower cutoff.

\subsection{The interplay between $T_{\rm cdw}$ and $T_{\rm sc}$}

There is another consequence of strong effect of thermal fluctuations -- the onset temperature for CDW order with  $Q_x/Q_y$ becomes
almost indistinguishable from the onset temperatures for superconductivity and for bond order with diagonal $Q$. Indeed, the equation for superconducting
$T_{\rm sc}$
 is the same as Eq.\ (\ref{Eq2bb}) if we replace $\Delta^Q_{k_0} \to \Delta_{\rm sc}$ and $\Delta^Q_{k_\pi} \to -\Delta_{\rm sc}$. To logarithmic accuracy,
 this leads to the same gap equation for ${\bar \Delta}_{\rm sc} (\Omega_m) = \Delta_{\rm sc}(\Omega^*_{m}) (\Omega_m)/\Sigma^* (\omega_m)$ as Eq.
 (\ref{frac_4}). This is an expected result as
  superconducting problem and CDW problem with $Q_x/Q_y$ differ in the interplay between the directions of Fermi velocities  in the two hot regions connected by $(\pi,\pi)$. For superconducting problem the velocities at hot spots at ${\bf k}$ and $-{\bf k}$ are strictly antiparallel, while for CDW
  they are almost antiparallel in one hot region and almost parallel in the other.  In a situation
  when the frequency dependent
  term in the fermionic propagator becomes parameterically larger than ${\bf v}\cdot {\bf k}$,  the difference in the directions of Fermi velocities
  becomes irrelevant and superconducting and CDW onset temperatures are both given by Eq.\ (\ref{frac_5}) to leading order in $L = \log \xi$, and differ
  only
  in the subleading terms, which are small in $1/L$, i.e.,
  \beq
T_{\rm sc} = T_{\rm cdw} \left(1 + f(1/L)\right),~~~f(0) =0.
\label{frac_7}
\eeq

In real quasi-3D systems, the logarithm remains finite even when magnetic $\xi = \infty$.  In this situation, both $T_{\rm cdw}$ and $T_{\rm sc}$ remain finite
 at the onset of CDW order (i.e., on a phase diagram they both cross $T_{\rm sdw}$ line at finite $T$) and both are of order $\omega_0$. Still,
  if $L$ is large enough,  $T_{\rm cdw}$ is close to $T{\rm sc}$, and the relative difference between the two temperatures is parametrically small in $1/L$.

\begin{figure}
 \includegraphics[width=0.5\columnwidth]{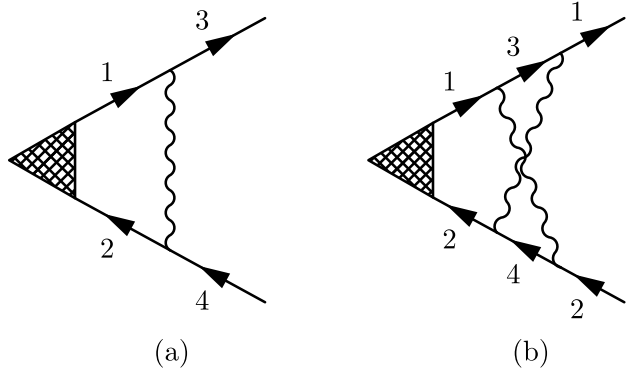}
\caption{The one-loop ladder  diagram and the two-loop non-ladder diagram for the renormalization of the CDW vertex. Both diagrams are logarithmical and, parameter-wise, of the same order, however
 the numerical prefactor for the two-loop diagram is much smaller. }
\label{fig10_a}
\end{figure}

\subsection{The role of non-ladder diagrams for $T_{\rm cdw}$}

So far, in our consideration of the onset temperature for CDW instability, we analyzed the set of gap equations within ladder approximation, i.e., used
the
same type of diagrams as in BSC/Eliashberg theory of superconductivity, only in particle-hole channel.  The ladder
approximation is justified either at weak coupling, when the kernel in the gap equation is logarithmically singular and ladder diagrams contain higher
powers of logarithms than non-ladder diagrams, or in the Eliashberg theory, when the coupling is not small but bosons, which mediate pairing, are slow modes
compared to electrons.  In this last case, non-ladder diagrams for the pairing vertex are small in Eliashberg parameter.  Neither of this approximations can be
justified in our case -- the effective coupling is of order one and Eliashberg parameter is also of order one because the interaction in the CDW channel
is mediated by collective modes of electrons, which have the same Fermi velocity as fermions themselves. Eliashberg parameter can be artificially made
small by extending theory to large $N$ (Refs.\ [\onlinecite{acs,ms,senthil}]), but we do not use this extension in our analysis.

In the absence of any small parameter, non-ladder diagrams for the CDW order parameter $\Delta_{k}^Q$ are of the same order as the ladder ones, and
one has to check whether they can significantly affect the result for $T_{\rm cdw}$.  We do a simple check to verify this. Namely, we return back to the
logarithmical approximation and check how much the prefactor of the logarithm obtained from one-loop ladder diagram changes when we include two-loop
non-ladder diagrams.
 Specifically, we compare the prefactors from the two diagrams shown in Fig.\ \ref{fig10_a}.
 The one-loop diagram has been analyzed above, see Eqs.\ (\ref{k0pilog}).  To logarithmical accuracy it yields
 \beq
 \sqrt{S_1 S_2}  \Delta^{Q}_{k_0} \log{\frac{\omega_0}{T}}
 \label{frac_8}
 \eeq
 where we used the fact that $\Delta^Q_{k_\pi} = - \Delta^Q_{k_0} \sqrt{S_2/S_1}$. For the case when Fermi velocities at hot spots 1 and 2 in
 Fig.\ \ref{fig3} are anti-parallel, $\sqrt{S_1 S_2} =0.234$.
 For the same geometry the two-loop diagram in Fig.\ \ref{fig10_a} yields, to logarithmic accuracy,
 \beq
 0.024 \Delta^{Q}_{k_0} \log{\frac{\omega_0}{T}}
 \label{frac_9}
 \eeq
We see that the prefactor in Eq.\ (\ref{frac_9}) is ten times smaller than in Eq.\ (\ref{frac_8}), i.e., at least at this level of consideration, the
two-loop non-ladder diagram in Fig.\ \ref{fig10_a} only contributes a small correction to the one-loop ladder diagram.  We take this result as an indication that the ladder
approximation
 for the CDW order, while not justified parametrically, is  reasonably well justified numerically.

\section{The stability of the CDW solution}
\label{sec:new}

 The existence of the solution of the gap equation for CDW order parameter by itself does not guarantee that there is an instability towards an CDW state.
 To prove that the
  system is truly unstable one has to verify that the solution with a non-zero CDW order parameter
  corresponds to a minimum of the Free energy, rather than a maximum.
 For a single order parameter $\Delta$, bilinear in fermionic operators a way to verify the stability is to use
   Hubbard-Stratonovich (HS) transformation, introduce $\Delta$ as a real HS field,  integrate out fermions, and expand the effective action
   ${S}_{\rm eff} (\Delta)$ in powers of $\Delta$. The expansion generally has the form
   \beq
   {S}_{\rm eff} (\Delta) = \alpha \Delta^2 + \beta \Delta^4.
\label{suac_1}
\eeq
 where $\alpha = a (T -T_0)$ and $\beta>0$ must be positive for a continuous transition.
  The saddle-point solution $\partial {S}_{\rm eff}/\partial \Delta$ yields a conventional result $\Delta^2 = -\alpha/(2\beta)$. Expanding around
  saddle point to quadratic order in fluctuations and evaluating fluctuating contribution to the Free energy one finds that fluctuations increase
  ${S}_{\rm eff}$ and the Gaussian integral over fluctuations of $\Delta$ nicely converges, i.e., the saddle point solution is a stable minimum.
 This simple reasoning, however, implies that $a>0$.  If $a$ was negative, saddle-point solutions with
  $\langle\Delta\rangle =0$ at $T > T_0$ and $\langle\Delta\rangle \neq 0$ at $T < T_0$ would
   correspond to a maximum rather than a minimum of the effective action. One can formally convert these states into minima, but for this one has to
   transform the integration contour over $\Delta$ from  real to imaginary axis.

 In our case  there  are two CDW orders, $\Delta_{k_0}^Q$ and $\Delta_{k_\pi}^Q$ hence
 one has to introduce two HS fields. We show below that the saddle point solution for $\Delta_{k_0}^Q -\Delta_{k_\pi}^Q$ is along real axis, while the
 saddle point solution for $\Delta_{k_0}^Q + \Delta_{k_\pi}^Q$ is along imaginary axis.  Given that solutions along real and imaginary axis have very
 different physical meaning in the case of a single field, one has to perform a more detailed analysis of fluctuations around the saddle-point solution to
 verify whether in our case
    a disordered state is stable at $T > T_{\rm cdw}$ and the ordered state is stable at $T < T_{\rm cdw}$.

    Another complication in our case is associated with the fact that there are several directions of fluctuations around the CDW solution. The system can
    fluctuate in  the CDW subset (i.e., within the plane set by  $\Delta_{k_0}^Q$ and $\Delta_{k_\pi}^Q$),  but it also can fluctuate
    in the directions of different orders, including superconductivity and bond order.  All these fluctuations must be included in the full analysis of
    stability of CDW order~\cite{efetov_private}.

    We perform the stability analysis in several stages.
    First, we analyze the solution of the set of non-linear ladder equations and show that the mean-field solution with a non-zero $\Delta^Q_k$ appears continuously below $T_{\rm cdw}$. Then we analyze fluctuations around the mean-field solution in the $\Delta_{k_0}^Q$ and
    $\Delta_{k_\pi}^Q$ plane. Finally, we discuss the interplay with bond order and superconductivity.

To simplify the analysis, below we neglect the complications associated with the frequency dependence of $\Delta^Q_k (\Omega)$ and with pair-breaking effects
 of thermal fluctuations, i.e., approximate $\Delta^Q_k (\Omega)$ by $\Delta^Q_k$ as set $T$ as the lower cutoff of $T=0$ formulas.  Within this approximation,
$T_{\rm cdw} = \omega_0 e^{-1/\sqrt{S_1 (0) S_2 (0)}}$.  The inclusion of  the frequency dependence of $\Delta^Q_k$ and thermal fluctuations will complicate the analysis
  but not change the conclusions.

\subsection{The non-linear gap equations at $T < T_{\rm cdw}$}

 The set of  non-linear gap equations for CDW order with $Q_x/Q_y$ has been obtained in Ref. ~[\onlinecite{private}] and we reproduced their formula.
   For completeness, we  briefly outline the details of our derivation. We again assume that Fermi velocities at hot spots 1 and 2 in Fig.\ \ref{fig3}) are anti-parallel, while the
 velocities of hot spots 3 and 4 are parallel. In regions 1 and 2 a non-zero $\Delta^Q_{k_0}$ acts in the same way as superconducting order
 parameter, i.e., $(v_F y)^2$ is replaced by $(v_F y)^2 + (\Delta^Q_{k_0})^2$.  In regions 3 and 4 a non-zero $\Delta^Q_{k_\pi}$ just shifts quasiparicle dispersions by $\pm \Delta^Q_{k_\pi}$, and the new fermionic operators which
 diagonalize the
  quadratic form, are $(c_3 +c_4)/\sqrt{2}$ and $(c_3-c_4)/\sqrt{2}$. In both regions we have normal Green's functions of original fermions $\langle T c_i
  c^\dagger_i\rangle$
  ($i =1,2,3,4$) and ``anomalous" Green's functions $\langle Tc_1 c^\dagger_2\rangle$ and $\langle Tc_3 c^\dagger_4\rangle$.
   Combining the contributions to the ladder renormalizations of
  $\Delta^Q_{k_0}$ and $\Delta^Q_{k_\pi}$ from diagrams with two normal and two anomalous Green's functions, we obtain, restoring momentarily the frequency
   dependence of $\Delta^Q_k$:
\begin{align}
\Delta_{k_0}^{Q}(\Omega_{m})=\frac{3\bar g T}{4\pi^2}&\sum_{m'}\int \frac{dx~dy}{x^2+y^2+\gamma|\Omega_{m}-\Omega_{m'}|}\ \nonumber \\
&\times\frac{\Delta_{k_\pi}^{Q}(\Omega_{m'})}{\left[i{\tilde \Sigma} ({\Omega_{m'}})+v_F x + \Delta_{k_\pi}^{Q}(\Omega_{m})\right] \left[i{\tilde \Sigma}
({\Omega_{m'}})+v_F x - \Delta_{k_\pi}^{Q}(\Omega_{m})\right]}  \label{Eq1ee} \\
\Delta_{k_\pi}^{Q}(\Omega_{m})=- \frac{3\bar gT}{4\pi^2}& \sum_{m'}\int
\frac{dx~dy}{x^2+y^2+\gamma|\Omega_{m}-\Omega_{m'}|}\frac{\Delta_{k_0}^{Q}(\Omega_{m'})}{{\tilde \Sigma}^2 ({\Omega_{m'}})
+(v_F y)^2 + [\Delta_{k_0}^Q(\Omega_{m'})]^2}.
\label{Eq1ff}
\end{align}
For definiteness, we set both $\Delta_{k_0}^{Q}(\Omega_{m})$ and $\Delta_{k_\pi}^{Q}(\Omega_{m})$ to be real.

At $T$  slightly below $T_{\rm cdw}$, one can  expand the r.h.s.\ of (\ref{Eq1ee}) and (\ref{Eq1ff})
  in powers  $\Delta^Q_{k_0}$ and $\Delta^Q_{k_\pi}$. Approximating now $\Delta^Q_k (\Omega)$ by frequency-independent values and restricting with the
logarithmic approximation, we obtain to order $\Delta^3$,
  \begin{align}
  \Delta_{k_0}^Q=&- S_1(0) \Delta_{k_\pi}^{Q} \left[\log\frac{\omega_0}{T} + \frac{\pi C_1}{4T}(\Delta_{k_\pi}^{Q})^2\right] ,\nonumber\\
\Delta_{k_\pi}^Q=&-S_2(0) \Delta_{k_0}^Q \left[\log\frac{\omega_0}{T} + \frac{\pi C_2}{4T}(\Delta_{k_0}^{Q})^2\right].
\label{k0pilog_1}
\end{align}
where $C_1 = 0.43$ and $C_2 = 9.03$.
 Eliminating $\Delta^Q_{k_\pi}$ from these equations we obtain
 \beq
  (\Delta_{k_0}^Q)^2 =-\frac{\alpha}{\beta},
  \label{suac_2}
  \eeq
  where
  \begin{align}
  \alpha =& 1 - S_1 (0) S_2 (0) \log^2{\omega_0/T} =a (T-T_{\rm cdw}); \nonumber \\
  a=& 2 \sqrt{S_1 (0) S_2 (0)}/T_{\rm cdw} >0,~~~ T_{\rm cdw} = \omega_0 e^{-1/\sqrt{S_1 (0) S_2 (0)}}
  \label{suac_3}
  \end{align}
  and
  \beq
  \beta = \[C_2+C_1\frac{S_2(0)}{S_1(0)}\]\frac{\pi S_1(0)S_2(0)}{4T_{\rm cdw}}\log\frac{\omega_0}{T_{\rm cdw}}.
 \label{suac_4}
  \eeq
 We see that $\beta >0$, hence the CDW transition is continuous.

\subsection{Fluctuations within the CDW subset}
\label{ivb}

  To analyze fluctuations within the CDW subset, we derive the effective action in terms of fields $\Delta_{k_0}^Q$ and $\Delta_{k_\pi}^Q$, show that the
  saddle-point solution is equivalent to the solution that we found by summing up ladder diagrams, and then expand $\Delta_{k_0}^Q$ and
  $\Delta_{k_\pi}^Q$ around the saddle-point solution and analyze the stability of the effective action.

  To make presentation easier to follow, we temporarily replace the actual momentum and frequency-dependent spin-mediated interaction
  $3 {\bar g} \chi (q, \Omega)$ by a constant ${\bar \chi}$. We restore the actual momentum and frequency dependence of the interaction in the final
  formulas for the effective action.

Consider for definiteness the ordering with ${\bf Q} = Q_y$, between regions 1-2 and 3-4. The 4-fermion interaction, which provides the glue for
CDW, is
   \beq
   H'= {\bar \chi} c^{\dagger}_{k_0-Q}c_{k_0+Q}c^{\dagger}_{k_\pi+Q}c_{k_\pi-Q}+{ h.c.}
   \label{tuac_1}
   \eeq
    We define   ${\rho}_{k_0}=c^{\dagger}_{k_0+Q}c_{k_0-Q}$, and ${\rho}_{k_\pi}=c^{\dagger}_{k_\pi+Q}c_{k_\pi-Q}$, and rewrite  4-fermion interaction
    as
\begin{align}
H'=\bar \chi \left(\bar{\rho}_{k_0}{\rho}_{k_\pi}+\bar{\rho}_{k_\pi}{\rho}_{k_0}\right) = \frac{\bar \chi}2
\left(\bar{\rho}_{k_0}+\bar{\rho}_{k_\pi}\right)\left({\rho}_{k_0}+{\rho}_{k_\pi}\right)-\frac{\bar
\chi}{2}\left(\bar{\rho}_{k_0}-\bar{\rho}_{k_\pi}\right)\left({\rho}_{k_0}-{\rho}_{k_\pi}\right).
\label{aa_1}
\end{align}
The Free energy $F = - T \log Z$ and the partition function is $Z = \prod_k \int d c^\dagger_k d c_k e^{-(H_0 + H')/T}$.

We use the HS identities~\cite{Fernandes_13,Fern_13_a},
\begin{align}
\exp{\(\frac{\bar\chi}{2}{\bar z_+} z_-\)} &= \int \frac{d {\Delta}_+ d {\bar\Delta}_{+}}{2\pi\bar\chi} \exp{\[-\frac{|{\Delta}_+|^2}{2\bar\chi} +
\frac{i}{2}\( {\Delta}_+ z_+ + {\bar\Delta}_{+} {\bar z_+}\)\]}
\nonumber\\
\exp{\(\frac{\bar\chi}{2}{\bar z_-} z_-\)} &=  \int \frac{d {\Delta}_- d {\bar\Delta}_{-}}{2\pi\bar\chi} \exp{\[-\frac{|{\Delta}_-|^2}{2\bar\chi}
+\frac{1}{2}\( {\Delta}_- z_- + {\bar\Delta}_{-} {\bar z_-}\)\]}
\label{ywfri1}
\end{align}
where ${\Delta}$'s are in general  complex fields, and apply these identities to  $z_+ =
{\rho}_{k_0} + {\rho}_{k_\pi}=c^{\dagger}_{k_0+Q}c_{k_0-Q} + c^{\dagger}_{k_\pi+Q}c_{k_\pi-Q}$
 and $z_- = {\rho}_{k_0} - {\rho}_{k_\pi}=c^{\dagger}_{k_0+Q}c_{k_0-Q} - c^{\dagger}_{k_\pi+Q}c_{k_\pi-Q}$ to
  decouple  bilinear terms in $\rho$ in (\ref{aa_1}).
  The partition function is now $Z = \prod_k \int d c^\dagger_k d c_k d {\Delta}_{-} d {\bar\Delta}_{-} d{\Delta}_+ d{\bar\Delta}_+  e^{-{
  S}(c^\dagger,c, {\Delta}_-,{\Delta}_+)}$ and the action
  is now quadratic in fermionic fields. Integration over fermionic variables can be carried out explicitly and we obtain
  \beq
  Z = \int d {\Delta}_{-} d {\bar\Delta}_{-} d{\Delta}_{+} d{\bar\Delta}_{+}  e^{-{S}_{\rm eff} ({\Delta}_{-},
  {\bar\Delta}_{-},{\Delta}_{+},{\bar\Delta}_{+})}.
  \eeq
    We analyze the action in the saddle-point approximation and consider fluctuations around
   saddle point solutions.  Because ${\Delta}_{+}$ and ${\Delta}_{-}$ couple linearly to $z_+$ and $z_-$, non-zero saddle-point solutions for
   ${\Delta}_{+}$ and/or ${\Delta}_{-}$ imply that
    the corresponding $z_+$ and $z_-$ are also non-zero: $z_+  = i
 {\Delta}_{+}
 /\bar\chi$ and $z_- =
  {\Delta}_{-}/\bar\chi$.
     In our notations then, $\Delta^Q_{k_0} = \langle \bar\chi c^{\dagger}_{k_0+Q}c_{k_0-Q}\rangle = ( i {\Delta}_{+}+{\Delta}_{-} )/2$ and
     $\Delta^Q_{k_\pi} = \langle c^{\dagger}_{k_\pi+Q}c_{k_\pi-Q}\rangle =
    (i {\Delta}_{+}-{\Delta}_{-} )/2$.
    (this does {\it not} mean that $\Delta_{k_0}^Q$ and $\Delta_{k_\pi}^Q$ are related by complex conjugation, since
    $\Delta_{+}$ and $\Delta_{-}$ are in general complex.)
    For our CDW solution
    $\Delta_{k_0}^Q\neq\pm\Delta_{k_\pi}^Q$, hence we expect non-zero saddle-point values of both ${\Delta}_{+}$ and ${\Delta}_{-}$.

 \subsubsection{Fluctuations at $T > T_{\rm cdw}$}

\begin{figure}
 \includegraphics[width=0.5\columnwidth]{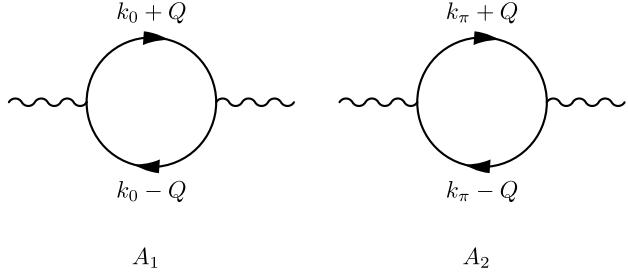}
\caption{The  diagrams for the coefficients $A_1$ and $A_2$  in the effective action, Eq.\ (\ref{ch_10}).}
\label{Fig11}
\end{figure}

     We first consider the situation at $T > T_{\rm cdw}$, when we expect that the minimum of the effective action corresponds to ${\Delta}_{-} =
     {\Delta}_{+} =0$.
   Integrating out  fermions and expanding the result to quadratic order in ${\Delta}_{+}$ and ${\Delta}_{-}$ we obtain the effective action in the form
 \begin{align}
{S}_{\rm eff}=  \frac{1}{2} \left[ A |{\Delta}_{+}|^2 - B |{\Delta}_{-}|^2 -i C ({\Delta}_{+}
{\bar\Delta}_{-}+{\bar\Delta}_{+}{\Delta}_{-})\right]
\label{y_3_2}
\end{align}
where
\begin{align}
& A =  \frac{1}{\bar \chi} + \frac{A_1 + A_2}{2},~~~B =   -\frac{1}{\bar \chi} + \frac{A_1 + A_2}{2},~~~C = \frac{A_1 - A_2}{2},
\end{align}
and
\begin{align}
A_1=&-\sum_{k,\omega}\frac{1}{G_{k_0+Q}^{-1}G_{k_0-Q}^{-1}} \nonumber \\
A_2=&-\sum_{k,\omega}\frac{1}{G_{k_\pi+Q}^{-1}G_{k_\pi-Q}^{-1}},
\label{ch_10}
\end{align}
where the Green's functions were introduced in (\ref{tuac_3}).
We show diagrammatic expressions for $A_1$ and $A_2$ in Fig.\ \ref{Fig11}. The overall negative signs in Eq.\ (\ref{ch_10}) are due to the presence of fermion
loops. Evaluating $A_1$ and $A_2$, we find that they are both positive.

The action in (\ref{y_3_2}) contains two terms with real prefactors of different sign ($A$ and $B$ terms) and one term with imaginary prefactor.
This apparently leads to some uncertainty as for a single field $\eta$ fluctuations described by ${S}_{\rm eff} = D |\eta|^2$ converge if $D >0$
and diverge if $D <0$.
In our case, the prefactor for  $|{\Delta}_{+}|^2$ term is positive and the one for $|{\Delta}_{-}|^2$ is negative, i.e, without the coupling provided by the $C$
term in  (\ref{y_3_2})
 fluctuations of ${\Delta}_{-}$ field would diverge. We will show, however that the $C$ term makes fluctuations of both  ${\Delta}_{+}$ and ${\Delta}_{-}$
 convergent at $T > T_{\rm cdw}$.

 To prove this, we first notice that the effective action in (\ref{y_3_2}) can be written in the form
\begin{align}
{S}_{\rm eff}=  \frac{1}{2} \left[ A \left({\bar\Delta}_{+} - i \frac{C}{A}
{\bar\Delta}_{-}\right)\left({\Delta}_{+}-i\frac{C}{A}{\Delta}_{-}\right)  + \left(\frac{C^2 -AB}{A}\right) |{\Delta}_{-}|^2 \right]
\label{y_4_1}
\end{align}
Expressing complex ${\Delta}_{+}$ and ${\Delta}_{-}$ as ${\Delta}_{+} = x+iy$, ${\Delta}_{-} = z+iw$, where $x,y,z,w$ are all real, we re-write
 \beq
 \left({\bar\Delta}_{+} - i \frac{C}{A} {\bar\Delta}_{-}\right)\left({\Delta}_{+}-i\frac{C}{A}{\Delta}_{-}\right)
 \eeq
  as $(x+ izC/A )^2 + (y+i wC/A )^2$. Introducing then  $x + izC/A = {\bar x}$ and $y + iwC/A = {\bar y}$ as new variables, we find that integration over
  ${\Delta}_{+}$ reduces to
  \beq
  \int d {\bar x}~d{\bar y}~e^{-(A/2) ({\bar x}^2 + {\bar y}^2)}
\eeq
This integral obviously converges.  Integrating then over $\bar x$ and $\bar y$ before integrating over $z$ and $w$ we obtain that fluctuating part of the
effective action
 reduces to
 \begin{align}
{S}_{\rm eff}=  \left(\frac{C^2 -AB}{2A}\right) |{\Delta}_{-}|^2
\label{y_4_1_1}
\end{align}
The prefactor $(C^2 - AB)/(2A)$ is
\beq
\frac{C^2 -AB}{2A} = \frac{1}{\bar \chi} \frac{1 - ({\bar \chi} A_1) (\bar \chi  A_2)}{2 + {\bar \chi} (A_1 + A_2)}
\label{tuac_4}
\eeq
The combinations
${\bar \chi} A_1$ and ${\bar \chi} A_2$ in the
 numerator  have the same forms as the kernels in the gap equations for $\Delta^Q_{k_0}$ and $\Delta^Q_{k_\pi}$. To see this we note that
${\bar \chi}A_{1,2}$ is the product of magnetically mediated interaction and two fermionic propagators with momentum difference $2Q$.
Restoring frequency and momentum dependence of ${\bar \chi} = 3 {\bar g} \chi (q, \Omega)$ and evaluating ${\bar \chi}A_{1,2}$ with logarithmic accuracy,
we
 obtain

 \begin{align}
 {\bar \chi} A_1 &= \frac{3\bar g }{8\pi^3}\int\frac{dx~dy~d \Omega_{m'}}{\left[i{\tilde \Sigma} ({\Omega_{m'}})-v_x y+v_y x\right]\left[i{\tilde \Sigma}
 ({\Omega_{m'}})+v_x y+v_y x\right]}\frac{
 1}{x^2+y^2+\gamma|\Omega_{m}-\Omega_{m'}|}\nonumber\\
  &= -S_1(\phi)\log\frac{\omega_0}{T}, \label{Eq1atu} \\
  {\bar \chi} A_2 &= \frac{3\bar g }{8\pi^3}\int\frac{dx~dy~d \Omega_{m'}}{\left[i{\tilde \Sigma} ({\Omega_{m'}})
-v_x x+v_y y\right]\left[i{\tilde \Sigma} ({\Omega_{m'}})-v_x x-v_y y\right]}\frac{
1}{x^2+y^2+\gamma|\Omega_{m}-\Omega_{m'}|},
\nonumber\\ &= -S_2(\phi)\log\frac{\omega_0}{T}. \label{Eq1btu}
\end{align}
Hence
\beq
\frac{C^2 -AB}{2A} \propto \left(1 - {\bar \chi}^2 A_1 A_2\right) = 1 - S_1 (\phi) S_2 (\phi) \left(\log \frac{\omega_0}{T}\right)^2
\label{tuac_6}
\eeq
Comparing this with Eq.\ (\ref{tuac_5}) we immediately find that $(C^2-AB)/(2A) = a (T - T_{\rm cdw})$, and the prefactor $a$ is positive.
This obviously implies that disordered state is stable at $T > T_{\rm cdw}$.

Another variable, whose  fluctuations are convergent, is ${\Delta}_{+}-i\frac{C}{A}{\Delta}_{-}$. The prefactor of the
corresponding term in the effective action equals $A/2$ and remains positive at $T_{\rm cdw}$. Hence the combination ${\Delta}_{+}-i\frac{C}{A}{\Delta}_{-}$,
does not acquire a non-zero value even when $C^2 -AB$ becomes negative and the field ${\Delta}_{-}$
 condenses.  Because
$\Delta^Q_{k_0}/\Delta^Q_{k_\pi} =
({\Delta}_{-} + i {\Delta}_{+})/({\Delta}_{-}-i{\Delta}_{+})$, the condition ${\Delta}_{+} -i\frac{C}{A}{\Delta}_{-} =0$ together with ${\bar \chi}^2 A_1
A_2
=1$ yields
\beq
\Delta^Q_{k_0} = - \Delta^Q_{k_\pi} \sqrt{\frac{S_2}{S_1}}.
\label{tuac_7_1}
\eeq
This is exactly the same as Eq.\ (\ref{tuac_7}), which we obtained by summing up ladder diagrams.

We now pause momentarily and summarize what we just did. We re-expressed the effective action (\ref{y_3_2}) as in (\ref{y_4_1}), shifted
 variables ${\rm Re} {\Delta}_{+}$  and ${\rm Im} {\Delta}_{+}$ into the complex plane by adding to them $i (C/A) {\rm Re}{\Delta}_{-}$ and $i(C/A) {\rm
 Im
 {\Delta}_{-}}$, respectively, and then integrated first over ${\rm Re} {\Delta}_{+}$  and ${\rm Im} {\Delta}_{+}$
 along the direction parallel to real axis, and then over ${\rm Re} {\Delta}_{-}$  and ${\rm Im} {\Delta}_{-}$.  We found that all Gaussian
 integrals are convergent at $T > T_{\rm cdw}$, i.e., the disordered state appears as stable at $T > T_{\rm cdw}$.

\begin{figure}
\includegraphics[width=0.5\columnwidth]{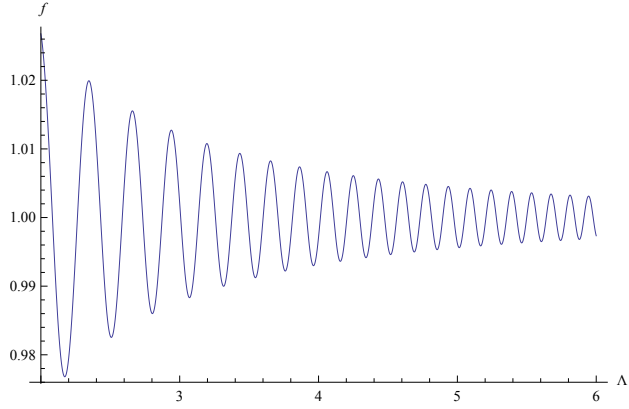}
\caption{The scaling function $f(\Lambda/A, C^2/AB, A/B)$ from Eq. (\ref{june27}) is plotted as a function of $\Lambda$ for fixed $A=B=1$, $C=2$.
The function has a $\Lambda-$dependent oscillating component, but clearly converges to $f=1$ at large $\Lambda$.}
\label{fig:june27}
\end{figure}

  We could, however, combine the three
 terms in (\ref{y_3_2}) differently,
  by keeping ${\Delta}_{+}$ as one variable and shifting ${\Delta}_{-}$ by a term proportional to ${\Delta}_{+}$. This way, we re-write (\ref{y_3_2}) as
  \begin{align}
{S}_{\rm eff}=  \frac{1}{2} \left[ -B \left({\bar\Delta}_{-} + i \frac{C}{B} {\bar\Delta}_{+}\right)\left({\Delta}_{-}
+i\frac{C}{B}{\Delta}_{+}\right)  - \left(\frac{C^2 -AB}{A}\right) |{\Delta}_{+}|^2 \right]
\label{y_4_1_a}
\end{align}
 Shifting now real and imaginary parts of ${\Delta}_{-}$ by $-i (C/B) {\rm Re}{\Delta}_{+}$ and $-i(C/B) {\rm Im {\Delta}_{+}}$, respectively, and
 integrating first over
 ${\rm Re} {\Delta}_{-}$  and ${\rm Im} {\Delta}_{-}$ parallel to real axis and then over ${\rm Re} {\Delta}_{+}$  and ${\rm Im} {\Delta}_{+}$ we obtain
 two divergent Gaussian integrals. Taken at a face value, this would imply that fluctuations of ${\Delta}_{-}$ and ${\Delta}_{+}$ diverge at $T > T_{\rm
 cdw}$, when $C^2 -AB >0$.  In reality, however, fluctuations do not diverge even if we integrate this way. To see this, one has to keep the limits of integration finite and
  set them to infinity only at the end of the calculation. One can then explicitly verify that
 in this computational scheme the  Gaussian
 integral
 $\int d {\Delta}_+ d {\Delta}_{-} e^{-{S}_{\rm eff}}$ with ${S}_{\rm eff}$ as in (\ref{y_4_1_a})
  yields the same result as we found before.   Specifically, the integral $W = \int d {\bar \Delta}_+ d {\bar \Delta}_- d ({\bar \Delta}_+)^* d ({\bar \Delta}_-)^*  exp{-S_{eff}}$,
  obtained by integrating over $d {\bar \Delta}_+ d ({\bar \Delta}_+)^*$ first, yields
  \beq
  W = \frac{4\pi^2}{C^2 -AB}
  \eeq
  If, instead, we integrate first over $d {\bar \Delta}_- d ({\bar \Delta}_-)^*$  and then over $d {\bar \Delta}_+ d ({\bar \Delta}_+)^*$, but each time will
   keep the limits of integration finite, from $-\Lambda$ to $\Lambda$, we obtain
   \beq
  W = \frac{4\pi^2}{C^2 -AB} f(\Lambda/A, C^2/AB, A/B)
  \label{june27}
  \eeq
  In Fig.\ \ref{fig:june27} we plot  $f$ as a function of $\Lambda$ for a fixed set of $A,B,C$. We see that $f$ has an oscillating component, but
   clearly tends to one when $\Lambda$ gets larger.
  The conclusion here is that, no matter in which the integration is done,
  the disordered state is stable at $T > T_{cdw}$ and becomes unstable at $T < T_{\rm cdw}$ when $C^2 - AB$ changes sign.

We discuss additional technical aspects  of the evaluation of the partition function for complex effective action in  Appendix \ref{app:c1}.

\subsubsection{Fluctuations at $T < T_{\rm cdw}$}
\label{sec:2aa}

\begin{figure}
\includegraphics[width=0.5\columnwidth]{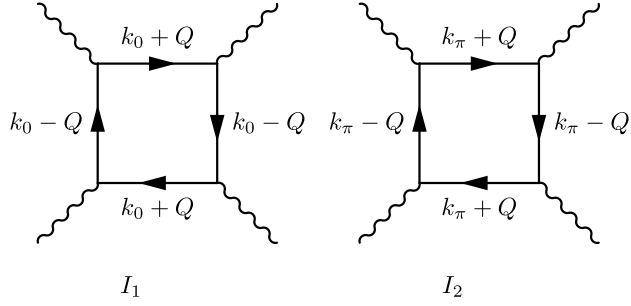}
\caption{The diagrammatic representation of the prefactors $I_1$ and $I_2$ in Eq.\ \ref{sat1}.}
\label{2}
\end{figure}

The HS analysis can be straightforwardly extended to $T < T_{\rm cdw}$, however to perform it we need to
expand the effective action ${S}_{\rm eff}[\Delta_-,\bar\Delta_-,\Delta_+,\bar\Delta_+]$
up to quartic terms. Applying HS transformation and expanding to fourth order in $\Delta$ we obtain
\begin{align}
{S}_{\rm eff}=  &\frac{1}{2} \left[ A |{\Delta}_{+}|^2 - B |{\Delta}_{-}|^2 -i C ({\Delta}_{+}
{\bar\Delta}_{-}+{\bar\Delta}_{+}{\Delta}_{-})\right]\nonumber\\
&-\frac{1}{16}I_1\[(\bar\Delta_-+i\bar\Delta_+)(\Delta_-+i\Delta_+)\]^2-\frac{1}{16}I_2\[(\bar\Delta_--i\bar\Delta_+)(\Delta_--i\Delta_+)\]^2,
\label{sat1}
\end{align}
The coefficients $I_1$ and $I_2$ are given by square diagrams made out fermions and are shown in Fig.\ \ref{2}. In analytical form,
\begin{align}
I_1=&-\frac{1}{2}\sum_{k,\omega}\frac{1}{G_{k_0+Q}^{-2}G_{k_0-Q}^{-2}} \nonumber \\
I_2=&-\frac{1}{2}\sum_{k,\omega}\frac{1}{G_{k_\pi+Q}^{-2}G_{k_\pi-Q}^{-2}}.
\label{sat2}
\end{align}
The integrals $I_1$ and $I_2$ are evaluated in Appendix \ref{app:extra}
using linear dispersion
around hot spots. In this approximation,
 $I_1$ is negative and finite and $I_2 =0$. Expanding further the dispersion relation, we find that $I_2$ is also non-zero, but is smaller than $I_1$. The discussion below does not rely on the precise numerical values of $I_1$ and $I_2$, and we keep both $I_1$ and $I_2$ as finite.

 Differentiating (\ref{sat1}) with respect to $\Delta_+$ we find from $\frac{\delta S}{\delta \bar\Delta_+} =0$ the relation
  $\Delta_+ = i {\tilde \lambda} \Delta_-$, with the prefactor renormalized from $\lambda = C/A$  by a non-zero $\Delta_-$:
\begin{align}
\tilde\lambda = \frac{C}{A}+ \frac{1}{4A }\[I_1(1-\lambda)^3-I_2(1+\lambda)^3\]|\Delta_-|^2
\label{sat4}
\end{align}
 Solving then the other saddle-point equation $\frac{\delta S}{\delta \bar\Delta_-} =0$
 and using  $\Delta_+ /(i \Delta_-) = {\tilde \lambda} = (\Delta^Q_{k_0} + \Delta^Q_{k_\pi})/(\Delta^Q_{k_\pi} - \Delta^Q_{k_0})$,
  we obtain
\begin{align}
&\frac{1}{\tilde\chi}\Delta_{k_0}^Q+A_1\Delta_{k_\pi}^Q+2I_1|\Delta_{k_\pi}^Q|^2\Delta_{k_\pi}^Q =0\nonumber\\
&\frac{1}{\tilde\chi}\Delta_{k_\pi}^Q+A_2\Delta_{k_0}^Q+2I_2|\Delta_{k_0}^Q|^2\Delta_{k_0}^Q =0.
\end{align}
Restoring the frequency and momentum dependence of $\tilde\chi$, like we did before, we immediately find that this set is analogous to  Eq.\ (\ref{k0pilog_1}),
 which we obtained by expanding in $\Delta^Q_k$ in the set ladder gap equations (it is important to keep $I_2$ for this comparison).
  The equivalence shows that the set of ladder gap equations is equivalent to
  saddle-point of the effective action.

We next replace $\Delta_+$ in Eq.\ (\ref{sat1})  by its saddle-point value and expand ${S}_{\rm eff}[\Delta_-]$ in powers of $\Delta_-$.
We obtain
\begin{align}
S_{\rm eff}[\Delta_-]=\(\frac{C^2-AB}{2A}\)|\Delta_-|^2+\beta'|\Delta_-|^4.
\label{sat5}
\end{align}
 where, we remind, $\frac{C^2-AB}{2A} \propto (T - T_{\rm cdw})$ and
 \beq
 \beta' = -\frac{1}{16}\[(1-\lambda)^4I_1+(1+\lambda)^4I_2\],
 \label{j_1}
 \eeq
 where, again, $\lambda=C/A$ and hence, $(1+\lambda)/(1-\lambda)=\sqrt{A_1/A_2}$ when $C^2-AB$ is close to zero.
 Because $I_1$ is negative, $I_2$ is much smaller than $|I_1|$, and $\lambda$ is also small,
 it follows from (\ref{j_1}) that $\beta'>0$, as expected.
We verified that, if we restore the frequency and momentum dependence of  $\tilde\chi$, the order parameters $\Delta^Q_{k_0}$ and $\Delta^Q_{k_0}$, obtained by minimizing (\ref{sat5}) with respect to $\Delta_-$ and using (\ref{sat4}) to obtain $\Delta_+$,
are equivalent to the solution of the non-linear ladder equations (\ref{suac_2}). Gaussian fluctuations around the HS solution can be obtained by usual means and indeed show that the saddle-point solution is a minimum with respect to variations of $\Delta_-$.

In Appendix~\ref{app:c2} we present an alternative derivation of Eq.\ (\ref{sat5}), using another HS formalism, in which the saddle points for $\Delta_+$ and $\Delta_-$, are both located along the real axis.

\subsection{The interplay between CDW order and superconductivity/diagonal bond order}

So far, we considered only fluctuations within the CDW subset.  The discussion in the preceding section shows that within this subset the CDW solution is
is a local minimum and fluctuations are convergent. The order parameters $\Delta^Q_{k_0}$ and $\Delta^Q_{k_\pi}$ are proportional to each other and the
effective action can be expressed in terms of one of them, which we label $\Delta_{\rm cdw}$:
\beq
{S}_{\rm eff} = \alpha_{\rm cdw} |\Delta_{\rm cdw}|^2 +
\beta_{\rm cdw} |\Delta_{\rm cdw}|^4 + ...
\eeq
 with $\alpha_{\rm cdw} = a (T - T_{\rm cdw})$ and $a >0$, $\beta_{\rm cdw} >0$.

There are indeed also fluctuations in the other directions, including the direction of $d$-wave superconductivity and diagonal bond order.
There fluctuations are longitudinal ones for CDW order and describe the change of ${S}_{\rm eff}$ when the magnitude of $\Delta_{\rm cdw}$
decreases and the
 magnitude of superconducting order of diagonal bond order increases.   To describe these fluctuations we extend the GL expansion of the
 effective action to include the competing channels.  To avoid complex formulas we only consider superconducting channel with order parameter $\Delta_{\rm
 sc}$.

 A straightforward analysis shows that the full effective action has the form
 \beq
{S}_{\rm eff} = \alpha_{\rm cdw} |\Delta_{\rm cdw}|^2 + {\beta_{\rm cdw}} |\Delta_{\rm cdw}|^4 + \alpha_{\rm sc} |\Delta_{\rm sc}|^2 +
{\beta_{\rm sc}} |\Delta_{\rm sc}|^4 + \beta_m  |\Delta_{\rm cdw}|^2  |\Delta_{\rm sc}|^2 + ...
\label{apac_26}
\eeq
where dots stand from higher-order terms, $\beta_i >0$, and $\alpha_{\rm sc} = a (T - T_{\rm sc})$.
The effective action of this form has been presented in Ref.\ [\onlinecite{deban}].
 The prefactor $a$ doesn't have to be the same as for
CDW order
 but can be adjusted to match that of $\alpha_{\rm cdw}$  by rescaling the magnitude of $\Delta_{\rm sc}$.

We know from the analysis in Sec.\ \ref{sec:2_1} that the instability temperature in the superconducting channel is close to $T_{\rm cdw}$, but still
larger than $T_{\rm cdw}$ (see Eq.\ (\ref{frac_7})).
 Analyzing the effective action (\ref{apac_26}) within mean-field theory, we find that immediately below $T_{\rm sc}$ only
 superconducting order emerges, while CDW order emerges at a lower $T = {\bar T}_{\rm cdw}$:
 \beq
 {\bar T}_{\rm cdw} = T_{\rm cdw} \frac{\beta_{\rm sc} - \beta_m \frac{T_{\rm sc}}{T_{\rm cdw}}}{\beta_{\rm sc} - \beta_m},
 \label{apac_27}
\eeq
provided that two conditions are met\cite{maxim,vvc,fs}
  \bea
&&\beta_{\rm sc} \beta_{\rm cdw} > \beta^2_m \nonumber\\
&& \beta_{\rm sc} > \beta_m \frac{T_{\rm sc}}{T_{\rm cdw}}
 \label{apac_28}
\eea
The first condition makes certain that the mixed state has lower energy than either of the two pure states and the second one guarantees
 that ${\bar T}_{\rm cdw}>0$.  If any these two conditions is not met, the system remains in a pure superconducting
 state down to $T=0$ and CDW order does not develop.

 We show below  that beyond mean-field the situation is more involved and the first instability upon lowering $T$ can actually happen within CDW
 subset,  before superconducting order or bond order with diagonal ${\bf Q}$ develop.  The reason is
  that the manifold for the
  CDW order parameter  includes one or two additional discrete $Z_2$ symmetries, depending on the actual structure of the CDW order.
    We demonstrate this in Sec.\ \ref{sec:3}. In Sec.\ \ref{sec:4} we show that composite charge orders, associated with there $Z_2$ symmetries, develop
   at temperatures larger than $T_{\rm cdw}$. Given that $T_{\rm cdw}$ is close to $T_{\rm sc}$, the onset temperature for composite charge order likely  exceeds $T_{\rm sc}$. Once a $Z_2$ composite orders sets in, it gives a negative feedback on superconductivity and reduces $T_{\rm sc}$, and, at the same time,
   increases the susceptibility for the primary CDW order and hence enhances $T_{\rm cdw}$. It the enhanced $T_{\rm cdw}$ becomes larger than the
    reduced  $T_{\rm sc}$, the same GL analysis as we just did below shows that $T_{\rm sc}$ is further reduced and whether it develops in co-existence of CDW at a lower $T$ depends on the same conditions as in (\ref{apac_28}).

\section{The structure of charge order: Mean-field analysis}
\label{sec:3}

In previous sections we considered CDW order with momentum either ${Q}_x=(2Q,0)$ or ${Q}_y=(0,2Q)$ and assumed that
 $\Delta_k^Q$ with a given $Q$ is a single $U(1)$ field, i.e., that $\Delta_{k_0}^Q = \Delta_{-k_0}^Q$.

In reality, the CDW order can emerge with either only $Q_x$ or $Q_y$, or with both momenta,  and also  $\Delta_{k_0}^Q$ and $\Delta_{-k_0}^Q$
  are in general not identical because ${\bf k}_0$ is not
   a high-symmetry point in the Brillouin zone. Indeed, by construction, the order parameter satisfies $(\Delta_k^{Q})^* = \Delta_k^{-Q}$. This condition implies that an incommensurate charge order parameter has an overall
 phase factor associated with the breaking of $U(1)$ symmetry, but does not specify how $\Delta_{k}^Q$  changes under $k \to -k$.
  For set 1-2 and 3-4 in Fig.\ \ref{fig3}, relevant $k$ are near $k_0= (\pi-Q,0)$ and $k_\pi = (-Q,\pi)$.
  For the set  5-6 and 7-8 in Fig.\ \ref{fig3}, relevant $k$ are near ${\bar k}_0 = -k_0 \equiv (\pi +Q,0)$ and ${\bar k}_\pi = - k_\pi$.
  As long as typical $|k-k_0|$ are smaller than $2Q$ (i.e., as long as $T_{\rm cdw}$ is smaller than, roughly, $E_F |Q/\pi|$),
  the two regions are weakly connected and at zero-order approximation can be considered independent on each other, in which case the gap
 equation does not distinguish between the solutions  for $\Delta_{k}^Q$, which are even and which are odd under $k \to -k$.
   One can easily check~\cite{subir_2} that under time reversal
  $\Delta_{k}^Q \to \Delta_{-k}^Q$, hence the odd in $k$ solution changes sign under time-reversal, and its emergence
   therefore implies that CDW order breaks time-reversal symmetry.
  We emphasize that the possibility to have two types of solutions is specific to CDW order with $Q_y$ ($Q_x$). For a charge order with a diagonal $Q$,
  the
  center of mass momentum is at $k=0$, and only an even in $k$ solution is possible.

 We label the even in $k$ solution as $\Delta^Q_1$ and the odd in $k$ solution as $\Delta^Q_2 \propto {\rm sgn} {(k)}$.
   We will show later in this Section that in real space $\Delta_1^Q$ describes an incommensurate  cite or bond charge density modulation,
       while $\Delta_2^Q$ describes an incommensurate  bond current.

 Combining two different ${\bf Q}$ with two components of $\Delta_k^Q$  for a given $Q$, we find   that
  the full order parameter for CDW order has two components: $\Delta_1^{Q_x}, \Delta_2^{Q_x}, \Delta_1^{Q_y}$, and $\Delta_2^{Q_y}$.
 In this Section, we obtain  the effective action for 4-component CDW order parameter and analyze it in the mean-field approximation. In Sec.\ \ref{sec:4}, we
  preform the analysis beyond mean-field and study preemptive composite orders.

 \subsection{Truncated effective action: stripe vs checkerboard order}
\label{sec:3a}

\begin{figure}
\includegraphics[width=0.5\columnwidth]{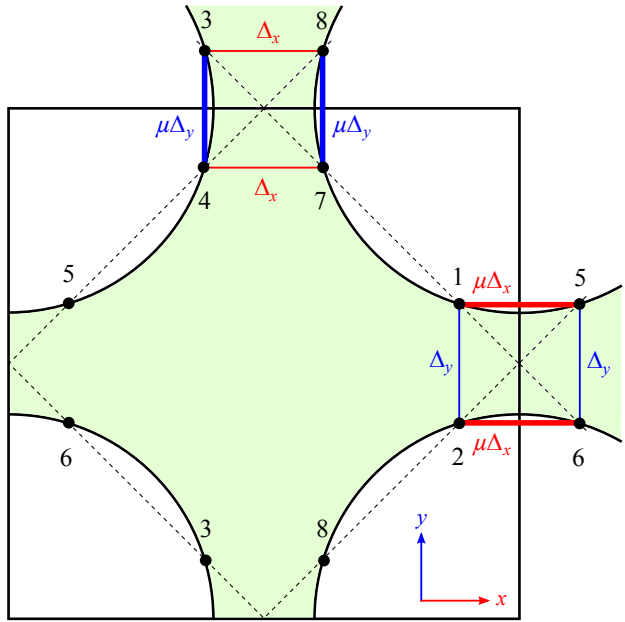}
\caption{The two order parameters responsible for stripe or checkerboard order.}
\label{Fig9}
\end{figure}

As a first pass on the structure of $\Delta^Q_k$, we assume that $\Delta^Q_k$ has only an even-in-$k$ component $\Delta_1^Q$
 (i.e., that $\Delta_{k_0}^Q=\Delta_{-k_0}^Q$) and analyze the GL model for two-component order parameter $\Delta_x = \Delta_{k_0}^{Q_x}$ and
  $\Delta_y = \Delta_{k_0}^{Q_y}$,
  subject to $\Delta^{Q_x}_{k_\pi} = \mu \Delta_x$ and $\Delta^{Q_x}_{k_\pi} = \mu \Delta_x$. Our goal here is
   address the issue whether CDW order develops simultaneously with both $Q_x$ and $Q_y$, in which case it preserves the underlying lattice $C_4$ symmetry and  gives rise to checkerboard charge order in the real space, or with either $Q_x$ or $Q_y$,
    in which case it spontaneously breaks $C_4$ symmetry down to $C_2$ and gives rise to stripe order
  The order with  $Q_x$ corresponds to CDW between fermions in regions 1-2 and 3-4 in Fig.\ \ref{fig3} and the order with $Q_y$
   corresponds to CDW between fermions in regions 1-5 and 3-8.
   We introduce $\Delta_x$ and $\Delta_y$ as two HS fields,
     integrate over fermions, and obtain the effective action
      ${S}_{\rm eff} (\Delta_x, \Delta_y)$.
       The prefactors for $|\Delta_y|^2$ and $|\Delta_x|^2$ are identical by symmetry, and the
 full action to order $\Delta^2$ is
 \begin{align}
{S}_{\rm eff}^{(2)}=\alpha \left(|\Delta_x|^2 + |\Delta_y|^2\right)
\end{align}

\begin{figure}
\includegraphics[width=0.4\columnwidth]{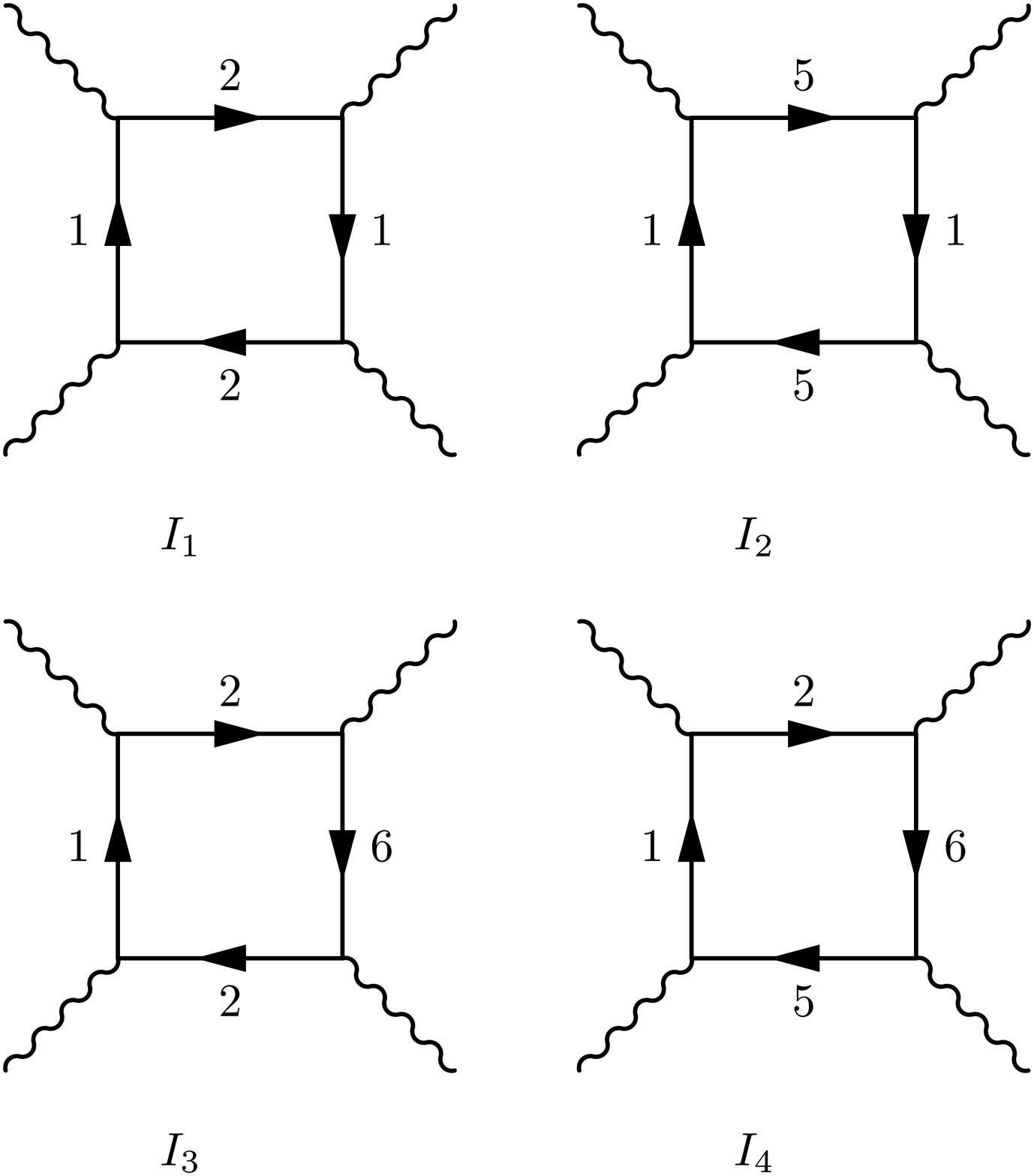}
\caption{The  diagrammatic representation of the quartic terms in the effective action.}
\label{Fig10}
\end{figure}

Extending the result to fourth order in $\Delta$, we obtain
      \begin{align}
      {S}_{\rm eff} (\Delta_x, \Delta_y) = \alpha \left(\Delta^2_x + \Delta^2_y\right) + \beta \left(\Delta^4_x + \Delta^4_y\right) + 2 \beta_m \Delta^2_x\Delta^2_y,
      \label{eq:2}
      \end{align}
  where, we remind, $\alpha = a(T-T_{\rm cdw})$, $a>0$.
 At a mean-field level, the  effective action (\ref{eq:2}) gives rise to a  checkerboard order when $\beta> \beta_m$, and to a stripe order when $\beta_m > \beta$.
 The coefficients $\beta$ and $\beta_m$ are expressed via the
  square diagrams with four fermionic propagators as
\begin{align}
\beta=&-2(I_1+\mu^4I_2)\nonumber\\
\beta_m=&-2\mu^2(2I_3+I_4),
\end{align}
where $\mu$ is the ratio $\Delta_{k_\pi}^{Q_y}/\Delta_{k_0}^{Q_y}$, which, we remind, is $-\sqrt{S_1/S_2}$ (see Eqs.\ (\ref{tuac_7}, \ref{tuac_7_1})).
 The terms  $I_i$ are the convolutions of four fermionic propagators
\begin{align}
I_1\equiv&-\frac{1}{2}\int G_1^2G_2^2\nonumber\\
I_2\equiv&-\frac{1}{2}\int G_1^2G_5^2\nonumber\\
I_3\equiv&-\int G_1G_5^2G_6\nonumber\\
I_4\equiv&-\int G_1G_2G_5G_6.
\label{Is}
\end{align}
We show $I_i$ graphically  in Fig.\ \ref{Fig10}, using the notations from Fig.\ \ref{Fig9}.
The overall minus sign in every line in (\ref{Is}) is due to the presence of a  fermionic loop.
 The abbreviations for the Green's function as $G_1\equiv G(\omega_{m},{\bf k_1}+(k_x,k_y))$, etc, and the integrals are performed over running frequency $\omega_{m}$ and  momenta $k_x$ and $k_y$.
  The integrals  $I_1$ and $I_2$ have been already introduced in Sec.\ \ref{sec:2}.

We evaluate $I_1$-$I_4$ in Appendix \ref{app:extra}, by expanding to linear order in the deviations from hot spots,
 and here quote the results. We obtain
\begin{align}
I_1&= - \frac{1}{16\pi^2v_x^2v_y}\frac{1}{\Lambda} \nonumber \\
I_2&=0 \nonumber \\
I_3&= - \frac{1}{16\pi^2 v^2_x v_y}\frac{1}{\Lambda} \log{\frac{\omega_0}{\Lambda}} \nonumber \\
I_4&= - \frac{1}{32 v_x v_y}\frac{1}{T}
\label{new_eq_1}
\end{align}
Using these results  we find that the prefactors $\beta$ and $\beta_m$ in Eq.\ (\ref{xy}) are given by
\begin{align}
\beta&=-2(I_1+\mu^4I_2)=\frac{1}{8\pi v_x^2 v_y}\frac{1}{\Lambda},\nonumber\\
\beta_m&=-2\mu^2(2I_3+I_4)
\approx \frac{\mu^2}{16v_xv_y}\frac{1}{T}.
\end{align}
 Because $\mu>1$, we have at low $T \ll \Lambda v_x$, $\beta_m \gg \beta$, i.e., the system chooses the stripe order in which only $\Delta_x$ or $\Delta_y$ emerges.
 Such an order  spontaneously breaks $C_4$ lattice rotational symmetry.

Phenomenological arguments for stripe charge order in hole-underdoped cuprates have been displayed earlier~\cite{steve_k}, and our microscopic analysis is consistent with earlier works.  We caution, however, that more accurate treatment is needed  when CDW order emerges either from a pre-existing superconducting state, or in an applied magnetic field.
Both a superconducting gap and a magnetic field cut the divergencies in $I_3$ and $I_4$ at low $T$, and it may happen that in this situation
 $\beta$ becomes larger than $\beta_m$, in which case the checkerboard CDW order develops.

\subsection{The full effective action}
\label{sec:3b}

We next
 analyze the effective action for the full 4-component CDW order parameter.  We split
  $\Delta_{k}^Q$ into even and odd parts as
\begin{align}
\Delta_{k}^{Q}=\Delta_{1,k}^{Q}+\Delta_{2,k}^{Q}~{\rm sgn}(k).
\label{ywfri3}
\end{align}
 and, to shorten notations, define $\Delta_1^{Q_x}$, $\Delta_2^{Q_x}$, $\Delta_{1}^{Q_y}$ and $\Delta_{2}^{Q_y}$
  as $\Delta_1^x$, $\Delta_2^x$, $\Delta_1^y$ and $\Delta_2^y$, respectively. The four order parameters transform differently under translation along $x$ and $y$
   directions in real space, lattice rotation by $\pi/2$, and inversion of time. We list the symmetry properties of the four $\Delta$'s in Table \ref{table:1}.

 \begin{table}
 \caption{The symmetry properties of the four order parameters $\Delta_1^x$, $\Delta_2^x$,  $\Delta_1^y$, and $\Delta_2^y$ under translation,
  $C_4$ lattice rotation, and time reversal.}
 \begin{ruledtabular}
 \begin{tabular}{ccccc}
 & $\Delta_1^x$ & $\Delta_2^x$ & $\Delta_1^y$ & $\Delta_2^y$ \\ \hline
 Translation along  $x$ & $\Delta_1^x e^{2i Q_x x}$ &  $\Delta_2^x e^{2i Q_x x}$ & $\Delta_1^y$ & $\Delta_2^y$ \\
  Translation along $y$ & $\Delta_1^x$ &  $\Delta_2^x$ & $\Delta_1^y e^{2i Q_y y}$ & $\Delta_2^ye^{2i Q_y y}$ \\
  $C_4$ lattice rotation & $\Delta_1^y$ & $\Delta_2^y$ & $\Delta_1^x$ & $\Delta_2^x$ \\
  Time reversal &$\Delta_1^x$ & -$\Delta_2^x$ & $\Delta_1^y$ & -$\Delta_2^y$
 \end{tabular}
 \end{ruledtabular}
 \label{table:1}
 \end{table}

We again use HS transformation from the action written in terms of fermionic operators to the action in terms of collective bosonic variables and obtain the
 prefactors for quadratic and quartic terms in $\Delta_i^j$ by integrating over the loops made out of two and four fermions, respectively.
The full analysis is somewhat involved and to give insights what CDW configurations may emerge we first
 approximate the CDW order parameters by their values at hot spots, which in technical terms implies that we approximate $c^\dagger c ~\Delta$ vertices in the square diagrams for the prefactors for $\Delta^4$ terms by their values at hot spots.
   Following the same steps as in the previous subsection, we
 obtain the effective action in the form
  \begin{align}
{S}_{\rm eff}=&\alpha(|\Delta_1^x|^2+|\Delta_1^y|^2+|\Delta_2^x|^2+|\Delta_2^y|^2)
+\beta\left\{|\Delta_1^x|^4+|\Delta_1^y|^4+|\Delta_2^x|^4+|\Delta_2^y|^4\right.\nonumber\\
&\left.+\[(\Delta_1^x)^*\Delta_2^x\]^2+\[(\Delta_2^x)^*\Delta_1^x\]^2+4|\Delta_1^x|^2|\Delta_2^x|^2+\[(\Delta_1^y)^*\Delta_2^y\]^2+\[(\Delta_2^y)^*\Delta_1^y\]^2+4|\Delta_1^y|^2|\Delta_2^y|^2\right\}\nonumber\\
&+2\bar\beta_m\left\{\[|\Delta_1^x|^2-|\Delta_2^x|^2\]\[|\Delta_1^y|^2-|\Delta_2^y|^2\]+\[\Delta_1^x(\Delta_2^x)^*-(\Delta_1^x)^*\Delta_2^x\]\[\Delta_1^y(\Delta_2^y)^*-(\Delta_1^y)^*\Delta_2^y\]\right\}\nonumber\\
&+2\tilde\beta_m(|\Delta_1^x|^2+|\Delta_2^x|^2)(|\Delta_1^y|^2+|\Delta_2^y|^2).
\label{ywth1}
\end{align}
where
 $\beta=-2(I_1+\mu^4I_2)$, $\bar\beta_m=-2\mu^2I_4$ and $\tilde\beta_m=-4\mu^2I_3$. For $\Delta_2^i =0$, Eq.\ (\ref{ywth1}) reduces to (\ref{eq:2}) with
 $\beta_m = \bar\beta_m + \tilde\beta_m$.
The expressions for $I_i$ are presented in (\ref{new_eq_1}). For these $I_i$,  all $\beta$'s are positive and ${\bar \beta}_m$ and ${\tilde \beta}_m$ well exceed $\beta$ because corresponding $I_i$ are larger and also because $\mu$ is larger than one.  The ratio of ${\bar\beta}_m/{\tilde \beta}_m$ does not depend on $\mu$ and is given by $I_4/I_3$. At low $T$ this ratio is large, but at $T= T_{\rm cdw}$ it is generally of order one. To account for all possible phases, we
 will treat  $\bar\beta_m$ and $\tilde\beta_m$ as the two parameters of comparable strength, but will keep $\tilde\beta_m,\bar\beta_m\gg\beta$.

We parametrize the four fields $\Delta_1^x$, $\Delta_2^x$, $\Delta_1^x$ and $\Delta_2^y$ as
 \begin{align}
 \Delta_1^x=|\Delta| \cos\theta \cos\phi_1 ~e^{i\psi_1}   &,~~~\Delta_2^x= |\Delta| \sin\theta \cos\phi_2 ~e^{i\psi_2}  ,\nonumber\\
 \Delta_1^y=|\Delta| \cos\theta \sin\phi_1 ~e^{i\bar\psi_1}  &,~~~\Delta_2^y= |\Delta| \sin\theta \sin\phi_2 ~e^{i\bar\psi_2},
 \label{ywmo2}
 \end{align}
 where all angles are taken between 0 and $\pi/2$.
 Plugging this into Eq.\ (\ref{ywth1}) and varying over $\psi$  we find that the action is minimized when
 \begin{align}
 \psi_1-\psi_2=\frac\pi2,&~~~\bar\psi_1-\bar\psi_2=\frac\pi2,\nonumber\\
 {\rm or}~~~\psi_1-\psi_2=-\frac\pi2,&~~~\bar\psi_1-\bar\psi_2=-\frac\pi2.
 \label{ywmo1}
 \end{align}
 This condition ``locks" the phase difference between $\Delta_1$'s and $\Delta_2$'s for CDW order parameters along the two directions of ${\bf Q}$
 to be simultaneously either $\pi/2$ or both $-\pi/2$. Plugging Eqs.\ (\ref{ywmo2}, \ref{ywmo1}) back into Eq.\ (\ref{ywth1}) we obtain
 \begin{align}
{S}_{\rm eff}=&\alpha|\Delta|^2+\beta|\Delta|^4\[(\cos^2\theta\cos^2\phi_1+\sin^2\theta\cos^2\phi_2)^2+(\cos^2\theta\sin^2\phi_1+\sin^2\theta\sin^2\phi_2)^2\]\nonumber\\
&+\frac{\bar\beta_m+\tilde\beta_m}2|\Delta|^4\(\cos^2\theta\sin2\phi_1-\sin^2\theta\sin2\phi_2\)^2-\frac{\bar\beta_m-\tilde\beta_m}2|\Delta|^4\sin^22\theta\sin^2(\phi_1+\phi_2)]\nonumber\\
\approx&~\alpha|\Delta|^2+\frac{\bar\beta_m+\tilde\beta_m}2|\Delta|^4\(\cos^2\theta\sin2\phi_1-\sin^2\theta\sin2\phi_2\)^2\nonumber\\
&-\frac{\bar\beta_m-\tilde\beta_m}2|\Delta|^4\sin^22\theta\sin^2(\phi_1+\phi_2)],
\label{ywmo3}
\end{align}
where in the last line we have used the approximation $\tilde\beta_m, \bar\beta_m\gg\beta$.

\begin{figure}
\includegraphics[width=0.5\columnwidth]{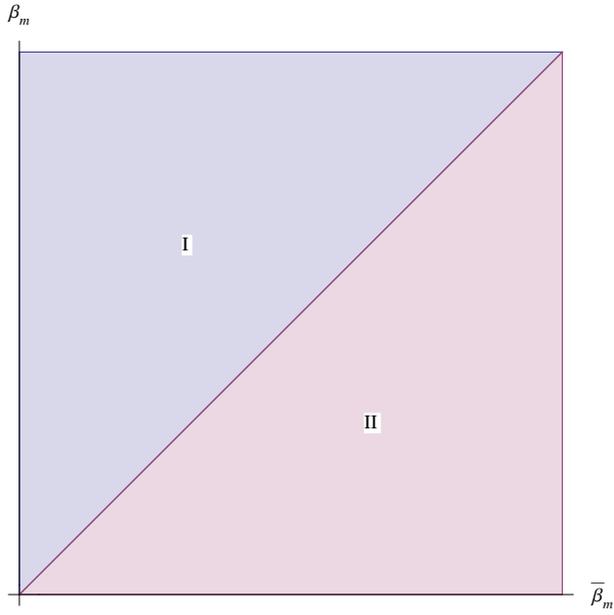}
\caption{States I and II in the parameter space of $\tilde\beta_m$ and $\bar\beta_m$.}
\label{ywtu3}
\end{figure}

The structure of the CDW order is now obtained by varying this action over $\phi_1$, $\phi_2$, and $\theta$. We found two types of states, I and II, depending on the interplay between $\tilde\beta_m$ and $\bar\beta_m$ (see \ref{ywtu3}).

 For $\tilde\beta_m>\bar\beta_m$,  we find that the minimum of ${S}_{\rm eff}$ corresponds to
$\phi_1=\phi_2=0,{\rm or}~\phi_1=\phi_2=\frac\pi2$, and arbitrary $\theta$.
The implication is that CDW order develops either with $Q_x$ or $Q_y$, i.e., is in the stripe form, like we found before.
 We see, however, that both $\Delta_1$ and $\Delta_2$ develop, in general, and the relative phase between the two is $\pm \pi/2$.
 We label this state as state I.
The relative magnitude of $\Delta_1$ and $\Delta_2$ is arbitrary at this level of consideration, but we show in the next subsection that it gets fixed when we include the $k$ dependence of $c^\dagger c \Delta_{k}^Q$ vertices.

For $\tilde\beta_m < \bar\beta_m$, the action (\ref{ywmo3}) is minimized when
$\theta=\pi/4$ and $\phi_1+\phi_2=\pi/2$. In terms of $\Delta$'s this implies $|\Delta^x_1| = |\Delta^y_2|$ and $|\Delta^y_1| = |\Delta^x_2|$.
 We label this state as state II.
The relative phases of $\Delta^x_1$ and $\Delta^x_2$ and of $\Delta^y_1$ and $\Delta^y_2$ are again fixed at either $\pi/2$ or $-\pi/2$ (with the same value for $x$ and $y$ components), but the relative phase of $\Delta^x_1$ and  $\Delta^y_1$ and the  relative magnitude  of $\Delta^x_1$ and $\Delta^x_2$ remain arbitrary at this level of consideration. We show in the next subsection that the relative magnitude gets fixed once we include the $k-$dependence of vertices, but the relative phase of $\Delta^x_1$ and  $\Delta^y_1$ still remains arbitrary.

  In the next two subsections we present a
more detailed study on states I and II.

\subsection{Properties of state I}
\label{sec:3c}

Suppose for definiteness that the ordering is along $Q_y$, i.e., $\Delta_1^y$ and $\Delta_2^y$ become non-zero below $T_{\rm cdw}$.
 The corresponding $S_{\rm eff}$ from (\ref{ywth1}) is
  \begin{align}
{S}_{\rm eff}=&\alpha(|\Delta_1^y|^2+|\Delta_2^y|^2)
+\beta\left(|\Delta_1^y|^4+|\Delta_2^y|^4
+\[(\Delta_1^y)^*\Delta_2^y\]^2+\[(\Delta_2^y)^*\Delta_1^y\]^2+4|\Delta_1^y|^2|\Delta_2^y|^2\right) \nonumber \\
=&\alpha(|\Delta_1^y|^2+|\Delta_2^y|^2) + \beta\(|\Delta_1^y|^2+|\Delta_2^y|^2\right)^2
+\beta\((\Delta_1^y)^*\Delta_2^y+\Delta_1^y(\Delta_2^y)^*\)^2
\label{ywth1_n}
\end{align}
As we already said, at this level,
while the phase difference of $\Delta_1^y$ and $\Delta_2^y$ is locked to be $\pm\pi/2$, the
relative magnitude of $\Delta_1^y$ and $\Delta^y_2$ can be arbitrary, only
$|\Delta_1^y|^2+|\Delta_2^y|^2$ is specified by minimizing Eq.\ \ref{ywth1_n}.
In fact, in this approximation one can easily make sure that the fields $\Delta^y_{k_0} = \Delta^y_1 + \Delta^y_2$ and
$\Delta^y_{-k_0} = \Delta^y_1 - \Delta^y_2$ decouple -- the first line in (\ref{ywth1_n}) exactly reduces to
\beq
{S}_{\rm eff}=\frac{1}{2} \left[\alpha(|\Delta^y_{k_0}|^2+|\Delta^y_{-k_0}|^2)
+\beta\left(|\Delta^y_{k_0}|^4+|\Delta^y_{-k_0}|^4\right)\right].
\label{ss_1}
\eeq
One implication of this equivalence is that in the hot spot approximation the fact that the phases of $\Delta^y_1$ and $\Delta^y_2$ are locked at $\pm \pi/2$
 does not have a physical consequence in the sense that the parameter manifold is $U(1) \times U(1) \times Z_2$, where
 the two $U(1)$'s are the two completely decoupled order parameters at $k$ and $-k$, and
  $Z_2$ symmetry is for the choice between
 $Q_x$ and $Q_y$, and there is no additional $Z_2$ component associated with the two choices for the phase locking.
 However, as we will see below, the decoupling between $\Delta^y_{k_0}$ and $\Delta^y_{-k_0}$ is  the artifact of
 the approximation of $\Delta_k^{Q_x}$ by its value at ${\bf k} = {\bf k}_0 = (\pi - Q, 0)$.
Once we go beyond this approximation, $\Delta^y_{k}$ and $\Delta^y_{-k}$ become coupled and $Z_2$ symmetry associated with the two choices of phase locking
becomes physically relevant part of the order parameter manifold.

To account for the effects due to momentum dependence of $\Delta^y_k$ we  adopt a simple ``toy model" approach and assume that odd and even components of $\Delta_k^{y}$ behave near $k_0$ as
\begin{align}
\Delta_{1,k}^{y} = \Delta^{y}_1 \frac{\cos k}{\cos k_0}, ~~ \Delta_{2,k}^{y}~{\rm sgn}(k) = \Delta^{y}_2 \frac{\sin k}{\sin k_0}.
\label{yw211}
\end{align}
 where $k$ is along $x$ direction.  The correspondent momentum dependence then appears in the vertices in 2-fermion and 4-fermion loop diagrams for $\alpha$ and $\beta$ terms.
  Re-evaluating the GL action with these vertices we obtain
 \begin{align}
{S}_{\rm eff} =
&\alpha_1 |\Delta_1^y|^2+ \alpha_2 |\Delta_2^y|^2+ \beta_1 |\Delta_1^y|^4 + \beta_2 |\Delta_2^y|^4 \nonumber\\
&+ 2 \beta_3 |\Delta_
1^y|^2|\Delta_2^y|^2 + \beta_3  \left((\Delta_1^y)^*\Delta_2^y+\Delta_1^y(\Delta_2^y)^*\right)^2,
\label{seff3}
\end{align}
where
\begin{align}
\alpha_{1,2}=\alpha-J_{\alpha_{1,2}},~~\beta_{1,2,3}=\beta+J_{\beta_{1,2,3}}.
\label{apac_01}
\end{align}
Here  $\alpha$ and $\beta$ are GL coefficients in the approximation $\Delta_{k}^{Q_y} = \Delta_{k_0}^{Q_y}$ and the corrections
 $J_{\alpha_{1,2}}$ and
$J_{\beta_{1,2,3}}$ are given by
\begin{align}
& J_{\alpha_1} = \int_{-\Lambda}^\Lambda dk \(\frac{\cos^2 k}{\cos^2 k_0}-1\), ~~J_{\alpha_2} = \int dk \(\frac{\sin^2 k}{\sin^2 k_0}-1\), \nonumber\\
& J_{\beta_1} = \int_{-\Lambda}^\Lambda dk \(\frac{\cos^4 k}{\cos^4 k_0}-1\),  ~~ J_{\beta_2} = \int dk \(\frac{\sin^4 k}{\sin^4 k_0}-1\)\nonumber\\
& J_{\beta_3} = \int_{-\Lambda}^\Lambda dk \(\frac{\sin^2 k \cos^2 k}{\sin^2 k_0 \cos^2 k_0}-1\).
\label{apac_1}
\end{align}
 where the integration  extends to a finite range $\Lambda$ around hot spots.
Expanding in (\ref{apac_1}) in $k-k_0$ we obtain
\begin{align}
& J_{\alpha_1} =  -2\epsilon \frac{\cos 2k_0}{\cos^2 k_0}, ~~J_{\alpha_2} =   2\epsilon \frac{\cos 2k_0}{\sin^2 k_0},
\label{apac_222}\\
& J_{\beta_1} =  2\epsilon \left(3 \cot^2 k_0 -1\right),~~ J_{\beta_2} = 2\epsilon \left(3 \tan^2 k_0 -1\right),\nonumber\\
 &J_{\beta_3} =   \epsilon \left(\cot^2 k_0 + \tan^2 k_0  -6\right)
\label{apac_2}
\end{align}
where $\epsilon = \int_{-\Lambda}^\Lambda (k-k_0)^2 dk >0$.
  In  Bi$_2$Sr$_{2-x}$La$_x$CuO$_{6+y}$, $\pi - k_0 \approx 0.255\pi$ (Ref.\ [\onlinecite{X-ray}]) hence
$|\tan k_0| \approx 1.03$ and $\cos 2 k_0 \approx -0.03$. From Eq.\ (\ref{apac_222}) we then find that $\alpha_2>\alpha_1$, i.e., the renormalized mean-field
CDW transition temperature for the even component, $T_{cdw,e} = T_e$ is larger than that for the odd component $T_{cdw,o} = T_o$. This agrees with  Refs.\ [\onlinecite{subir_2,subir_4}]. We note, however, that the two are still very close to the original $T_{\rm cdw}$ because $J_{\alpha_1}$ and $J_{\alpha_2}$ are very small numerically.
 A complimentary approach how to go beyond hot spot treatment is presented in Appendix \ref{app:d}.  It also leads to $\alpha_2 \geq \alpha_1$.
We also have
 \bea
  \beta_1 - \beta_3 &=& \epsilon \left(4 - (\tan k_0)^2 + 5 (\cot k_0)^2\right) \nonumber\\
  \beta_2 - \beta_3 &=& \epsilon \left(4 - (\cot k_0)^2 + 5 (\tan k_0)^2\right)
    \label{apac_7}
  \eea
For $|\tan k_0| \approx |\cot k_0| \approx 1$, $\beta_1 \approx \beta_2 > \beta_3$.

Analyzing the effective action (\ref{seff3}) in the mean-field approximation
we observe that a relative phase between $\Delta^y_1 = |\Delta^y_1|e^{i\psi_1}$ and $\Delta^y_2 = |\Delta^y_2|e^{i\psi_2}$ is locked at
  $\pm \pi/2$, like in the case of a constant $\Delta_k^x$. In other words,  if $\Delta^x_1$ is real, $\Delta^x_2$ should
  be imaginary.  From $\partial {S}_{\rm eff}/\partial \Delta^y_1 = \partial {S}_{\rm eff}/\partial
 \Delta^y_2 =0$
 we obtain
   \begin{align}
   &\Delta^y_1 \left(\alpha_1 + 2 \beta_1 |\Delta^y_1|^2 + 2 \beta_3 |\Delta^y_2|^2\right) =0 \nonumber \\
   & \Delta^y_2 \left(\alpha_2 + 2 \beta_2 |\Delta^y_2|^2 + 2 \beta_3 |\Delta^y_1|^2\right) =0
   \label{apac_3}
   \end{align}
   Assuming that both orders are non-zero, we obtain from (\ref{apac_3})
   \begin{align}
   |\Delta^y_1|^2 =& \frac{1}{2} \frac{\alpha_2 \beta_3 - \alpha_1 \beta_2}{\beta_1 \beta_2 - \beta^2_3} \nonumber \\
   |\Delta^y_2|^2 =& \frac{1}{2} \frac{\alpha_1 \beta_3 - \alpha_2 \beta_1}{\beta_1 \beta_2 - \beta^2_3}.
 \label{apac_4}
   \end{align}
 An elementary analysis shows that this solution is a minimum of the effective action when $\beta_1 \beta_2 > \beta^2_3$.
 In our case
 \beq
 \beta_1 \beta_2 - \beta^2_3 = \frac{16 \beta \epsilon}{(\sin 2k_0)^2} >0
  \label{apac_5}
  \eeq
  i.e., this condition is satisfied.
  The temperature at which $\Delta^x_2$ acquires a non-zero value is
  \beq
  T_{co} = T_o \frac{\beta_1 - \beta_3}{\beta_1 - \beta_3 \frac{T_o}{T_e}} \approx T_{o}
  \label{apac_6}
  \eeq
 Below this temperature both order parameters acquire non-zero values and the relative phase $\psi_1 - \psi_2$ is either $\pi/2$ or $-\pi/2$.
  The broken symmetry in the phase when both $\Delta^y_1$ and $\Delta^y_2$ are non-zero is $U(1) \times Z_2$, where continuous $U(1)$ corresponds to
   the common phase ${\bar \phi}_1 + {\bar \phi}_2$ of $\Delta^y_1$ and $\Delta^y_2$ and Ising $Z_2$ corresponds to the choice $\pi/2$ or $-\pi/2$ for the relative phase.

 What happens at lower $T$ depends on the sign of $\beta_2 - \beta_3 >0$, and the two orders
 co-exist down to $T=0$. Interestingly, when $\beta_2 < \beta_3$, there is another temperature
 \beq
{\bar T}_{co} = T_o \frac{\beta_3 - \beta_2}{\beta_3 - \beta_2 \frac{T_o}{T_e}} < T_o
  \label{apac_9}
  \eeq
at which  $\Delta^y_1$ disappears and at  smaller $T$ only $\Delta^y_2$ is non-zero.

It is also instructive to re-write the effective action (\ref{seff3}) in terms of the original CDW order parameters $\Delta^y_{k_0}$ and $\Delta^y_{-k_0}$
 at the hot spots. From Eqs.\ (\ref{ywfri3},\ref{yw211}) we have $\Delta_{1}^y=(\Delta_{k_0}^y+\Delta_{-k_0}^y)/2$ and $\Delta_{2}^y=(\Delta_{k_0}^y-\Delta_{-k_0}^y)/2$, which is the same relation as in the hot spot approximation. Plugging them into Eq.\ (\ref{seff3}), we obtain,
    \begin{align}
{S}_{\rm eff} =
&\left(\frac{\alpha_1 + \alpha_2}{4}\right) \left(|\Delta^y_{k_0}|^2+ |\Delta^y_{-k_0}|^2\right) +
\left(\frac{\alpha_1 - \alpha_2}{4}\right) \left(\Delta^y_{k_0} (\Delta^y_{-k_0})^* +  \Delta^y_{-k_0} (\Delta^y_{k_0})^*\right) \nonumber\\
&+ \left(\frac{3\beta_3}{8} + \frac{\beta_1 + \beta_2}{16} \right) \left(|\Delta^y_{k_0}|^2+ |\Delta^y_{-k_0}|^2\right)^2   \nonumber\\
&- \left(\frac{3\beta_3}{8} - \frac{\beta_1 + \beta_2}{16} \right) \left(\Delta^y_{k_0} (\Delta^y_{-k_0})^* +  \Delta^y_{-k_0} (\Delta^y_{k_0})^*\right)^2 \nonumber \\
 & + \frac{\beta_1 - \beta_2}{8} \left(|\Delta^y_{k_0}|^2+ |\Delta^y_{-k_0}|^2\right) \left(\Delta^y_{k_0} (\Delta^y_{-k_0})^* +  \Delta^y_{-k_0} (\Delta^y_{k_0})^*\right)\nonumber\\
&+\frac{\beta_3}{4} \left(\Delta^y_{k_0} (\Delta^y_{-k_0})^* -  \Delta^y_{-k_0} (\Delta^y_{k_0})^*\right)^2.
\label{seff3_a}
\end{align}

 For momentum-independent vertices, $\Delta^y_{k_0}$ and $\Delta^y_{-k_0}$ decouple in the effective action (\ref{ss_1}). However, we see that the fields $\Delta^y_k$ and $\Delta^y_{-k_0}$ now interact with each other. To make this more clearly visible, let's neglect small differences between
 $\alpha$ and $\alpha_{1,2}$ and between $\beta_1$ and $\beta_{2}$, but keep a larger difference between $\beta_1$ and $\beta_3$.  In this approximation,
 the effective action reduces to
   \begin{align}
{S}_{\rm eff} =
&\frac{\alpha}{2} \left(|\Delta^y_{k_0}|^2+ |\Delta^y_{-k_0}|^2\right) + \left(\frac{3\beta_3 + \beta_1}{8}\right) \left(|\Delta^y_{k_0}|^4+ |\Delta^y_{-k_0}|^4\right) \nonumber \\
& + \left(\frac{\beta_1 - \beta_3}{8}\right) \left(6|\Delta^y_{k_0}|^2  |\Delta^y_{-k_0}|^2 + \left(\Delta^y_{k_0} (\Delta^y_{-k_0})^* -  \Delta^y_{-k_0} (\Delta^y_{k_0})^*\right)^2\right)
\label{seff3_b}
\end{align}
This can be equivalently re-expressed as
   \begin{align}
{S}_{\rm eff} =
&\frac{\alpha}{2} \left(|\Delta^y_{k_0}|^2+ |\Delta^y_{-k_0}|^2\right) + \frac{\beta_1}{4} \left(|\Delta^y_{k_0}|^2+ |\Delta^y_{-k_0}|^2\right)^2 \nonumber \\
& + \left(\frac{3\beta_3 - \beta_1}{8}\right) \left(|\Delta^y_{k_0}|^2 - |\Delta^y_{-k_0}|^2\right)^2 -\left(\frac{\beta_1 - \beta_3}{8}\right)   \left(i\left(\Delta^y_{k_0} (\Delta^y_{-k_0})^* -  \Delta^y_{-k_0} (\Delta^y_{k_0})^*\right)\right)^2
\label{seff3_c}
\end{align}
 The advantage of this last expression is that it clearly shows that, for $\beta_1 > \beta_3$,  ${S}_{\rm eff}$ is reduced when $\Delta^y_{k_0}$ and $\Delta^y_{-k_0}$ appear together, and $\Delta^y_{k_0} = \Delta^y_1 \pm i |\Delta^y_2|$ and $\Delta^y_{-k_0} = \Delta^y_1 \mp i |\Delta^y_2|$  because then the last term in (\ref{seff3_c}) becomes $- ((\beta_1 - \beta_3)/2) |\Delta^y_1||\Delta^y_{2}|$.
 At the same time, the prefactor for the ``nematic" term $\left(|\Delta^y_{k_0}|^2 - |\Delta^y_{-k_0}|^2\right)^2$ is positive, which implies that $|\Delta^y_{k_0}|$ and
 $|\Delta^y_{-k_0}|$ must be equal.  This holds when $\Delta^y_1$ and $\Delta^y_2$ are orthogonal to each other.

\subsubsection{Physical properties of the coexistence state}
\label{5c1}

We now consider physical properties of the coexistence state, when both even and odd CDW order parameters are non-zero.
The generic condition $(\Delta_{k}^Q)^* = \Delta_{k}^{-Q}$ imposes the constraint that $\Delta^Q_1$ must be even in $Q$ and $\Delta^Q_2$ must be odd in
$Q$.  We then re-express the $\Delta_{k}^Q$  at hot spots 1-2 and 3-4 as
   (for ${\bf Q} = \pm Q_y$)
\begin{align}
\Delta_{k}^Q=\Delta_1\pm i \Delta_2 {~\rm sgn} (k) {~\rm sgn} (Q)
\label{apac:3}
\end{align}
where $\Delta_1$ and $\Delta_2$ are numbers.
 This $\Delta^Q_k$  breaks time reversal symmetry because under time reversal $\Delta_{k}^Q$ transforms into $(\Delta_{-k}^{-Q})^* = \Delta^Q_{-k}$.  The
  choice of relative sign in (\ref{apac:3}) specifies one of two non-equivalent solutions which transform into each other under time-inversion.
  On the other hand, the parity is not broken as under parity operation $\Delta^Q_k$ transforms into $\Delta^{-Q}_{-k} = \Delta^Q_k$.
   Note that the  order
  parameter (\ref{apac:3}) is similar, but not equivalent, to incommensurate complex $d$-density wave order proposed in~[\onlinecite{inc_1,inc_2}].

\begin{figure}
 \includegraphics[width=0.3\columnwidth]{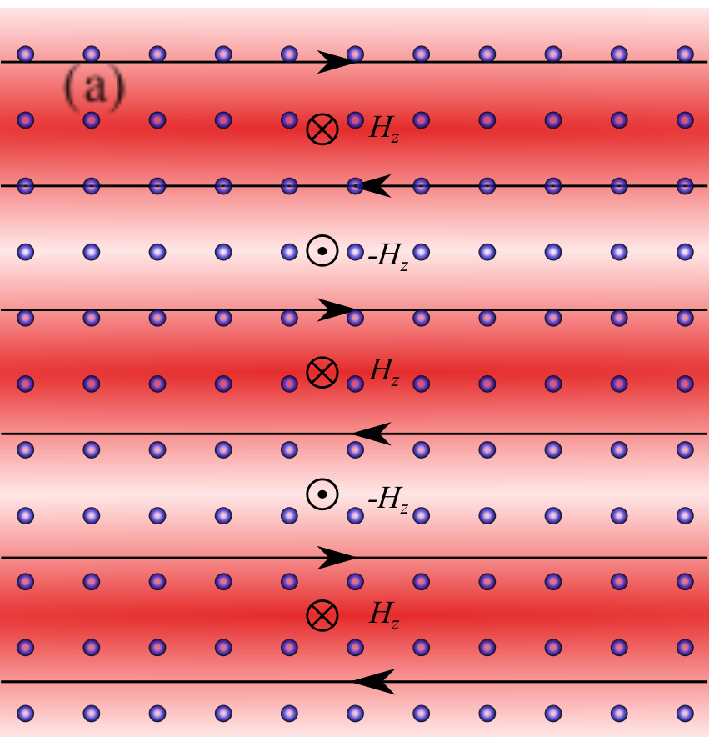}~~~
 \includegraphics[width=0.3\columnwidth]{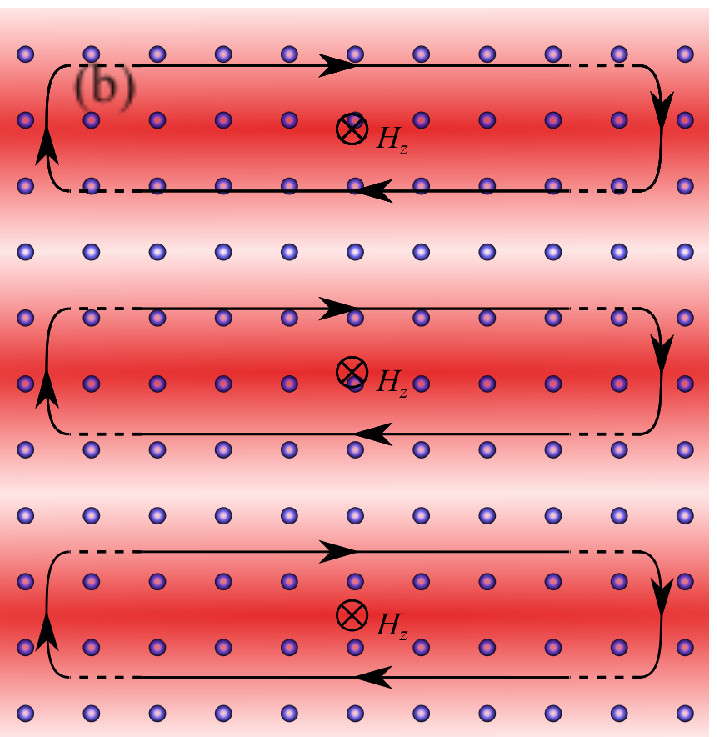}
 \caption{Panel (a): The structure of density and current modulations in the state I.
   The regions of higher and lower fermionic density are shown by darker and lighter color, respectively. The direction of the current is shown by arrows.
    The current vanishes when the density fluctuation has either the highest or the lowest value.  An oscillating current gives rise to an oscillating
    magnetic field, whose values are shown by dots and crosses. Panel (b): Current loops, formed by connecting oscillating currents in the bulk via the regions of
     higher local charge density on the surface.}
 \label{loopcurr_1}
 \end{figure}

Converting to real space, we find that the  term $\Delta_1$ corresponds to an incommensurate modulation of local charge and bond density in $y$
direction
\begin{align}
\delta\rho(r)=&{\rm Re}\langle c^\dagger(r)c(r)\rangle=\sum_{k}\langle c^\dagger(k+Q_y)c(k-Q_y)\rangle e^{i(k+Q_y)r}e^{-i(k-Q_y)r}+h.c.\propto\Delta_1\cos 2Q r_y,\nonumber\\
\delta\rho(r,a_x)=&{\rm Re}\left[\langle c^\dagger(r+a_x/2)c(r-a_x/2)\rangle\right] \propto \Delta_1\cos 2Q r_y \cos{k_0 a_x} \approx - \Delta_1\cos 2Q r_y, \nonumber \\
\delta\rho(r,a_y)=&{\rm Re}\left[\langle c^\dagger(r+a_y/2)c(r-a_y/2)\rangle\right] \propto \Delta_1\cos 2Q \left(r_y +a_y/2\right) \approx \Delta_1\cos 2Q r_y.
\end{align}
 The term $\Delta_2$ corresponds to an incommensurate  bond current, which flows along $x$ direction and has incommensurate modulation in $y$ direction
(see Fig.\ \ref{loopcurr_1} (a))
 \begin{align}
j_x(r)=&{\rm Re}\left[i\langle c^\dagger(r-a_x/2)c(r+a_x/2)\rangle\right]\nonumber\\
=&{\rm Re}\left[i\sum_{k}\langle c^\dagger(k+Q_y) c(k-Q_y)\rangle e^{i(k+Q_y)(r-a_x/2)}e^{-i(k-Q_y)(r+a_x/2)}+(Q_y\rightarrow-Q_y)\right]\nonumber\\
=&{\rm Re}\left[i\sum_k(\Delta_k^{Q} e^{2iQr_y}+\Delta_k^{-Q} e^{-2iQr_y})e^{-ik a_x}\right]\nonumber\\
=&2{\rm Re}\left[i\Delta_1\cos(2Qr_y)\sum_{k}e^{-ik a_x}-i|\Delta_2|\sin(2Qr_y)\sum_k{\rm sgn}(k)e^{-ik a_x}\right]\nonumber\\
\propto&|\Delta_2|\sin 2Q r_y\sin k_0 a_x~=|\Delta_2|\sin 2Q r_y\sin Q a_x.
\end{align}
 Note that the bond current modulation is in anti-phase with the density modulation.
  An incommensurate bond current in turn creates an incommensurate
  magnetic field  $H_z (r) \propto |\Delta_2| \cos {2Q r_y}$. This, however, does not lead to orbital ferromagnetism as the
   total magnetic field,
  integrated over the volume of the system, vanishes: $(1/V)\int H_z dV =0$.
To be more precise,  current lines have to close at
 the boundary of a sample, and it is natural to expect that they  close through the regions of excess charge, as shown in Fig \ref{loopcurr_1} (b).  This
  gives rise to
    a set of loop currents with circulation along the same direction, which do create a uniform magnetic field. However, a uniform field  scales as the area
    of the sample rather than its volume and vanishes in the thermodynamic limit. This  is very different from
     triangular loop currents proposed in Ref.\ [\onlinecite{varma}].

 \subsection{Properties of state  II}
\label{5d}
We recall that in hot spot approximation, the minimum of the effective action, Eq.\ (\ref{ywmo3}) for ${\bar \beta}_m > {\tilde \beta}_m$ (state II) is at
$\theta=\pi/4$ and $\phi_1+\phi_2=\pi/2$, which in terms of $\Delta$'s implies that $|\Delta^x_1| = |\Delta^y_2|$ and $|\Delta^y_1| = |\Delta^x_2|$.
The relative phases between $\Delta^x_1$ and $\Delta^x_2$ and between $\Delta^y_1$ and $\Delta^y_2$ are either both $\pi/2$ or $-\pi/2$.  Substituting these relations into (\ref{ywmo3}) and using $\Delta^y_1$ and $\Delta^y_2$ as two variables, we obtain the same effective action (\ref{ywth1_n}) as for the state I, namely
\beq
{S}_{\rm eff} = \alpha(|\Delta_1^y|^2+|\Delta_2^y|^2) + \beta(\left|\Delta_1^y|^2+|\Delta_2^y|^2\right)^2
 \eeq
Like for the state I, the relative magnitude between $\Delta^y_1$ and $\Delta^y_2$ is not fixed in the hot spot approximation, and to find the actual CDW ordering
 one has to include the $k-$dependence of $c^\dagger c ~\Delta^Q_k$  vertices. This gives rise to an effective action similar to Eq.\ (\ref{seff3}), however for the state II the full effective action is more involved as all four CDW components are different from zero.  We will not discuss the full form of the action because we believe that the state II is less relevant to the cuprates than the state I, and rather describe two potential realizations of the freezing of the relative magnitude of
 $|\Delta_1^y|$ and $|\Delta_2^y|$, i,e., the breaking of the freedom associated with the realization of $\phi_1 + \phi_2 = \pi/2$. One obvious choice is
$\phi_1 = \phi_2 = \pi/4$, another is $\phi_1 = \pi/2$, $\phi_2=0$ or $\phi_1 = 0$, $\phi_2=\pi/2$.

  \subsubsection{$\theta=\pi/4$, $\phi_1=\pi/4$, $\phi_2=\pi/4$}
\label{5d1}
In this case all four CDW components, $\Delta_1^x$, $\Delta_2^x$, $\Delta_1^y$, $\Delta_2^y$, develop with the same magnitude $\Delta$.
In real space this order corresponds to a checkerboard type incommensurate charge density modulation and incommensurate current in both $x$ and $y$ directions. We show this in Fig.\ \ref{yw30} (a). The order parameter manifold is $U(1) \times U(1) \times Z_2$, where two continuous $U(1)$ symmmetries are associated with the phases of $\Delta^x_1$ and $\Delta^y_1$, and the Ising $Z_2$ is associated with the relative phase between $\Delta_1$ and $\Delta_2$, which is $\pi/2$ or $-\pi/2$, simultaneously for $x$ and $y$ components. This CDW order preserves $C_4$ lattice rotational symmetry but breaks time-reversal symmetry.

Plugging $\theta=\pi/4$, $\phi_1=\pi/4$, $\phi_2=\pi/4$  into Eq.\ (\ref{ywmo3}) and again neglecting $\beta$ compared to ${\bar \beta}_m$ and
 ${\tilde\beta}_m$,   we find the effective action Eq.\ (\ref{ywth1}) in the form
\begin{align}
{S}_{\rm eff}=&\alpha\Delta^2-\frac{\bar\beta_m-\tilde\beta_m}{2}\Delta^4.
\end{align}
Because ${\bar \beta}_m - {\tilde\beta}_m >0$, the transition is first order. It occurs at some $T$ larger than mean-field $T_{\rm cdw}$ at which $\alpha$ changes sign.

\subsubsection{$\theta=\pi/4$, $\phi_1=\pi/2$, $\phi_2=0$}
\label{5d2}

\begin{figure}
 \includegraphics[width=0.3\columnwidth]{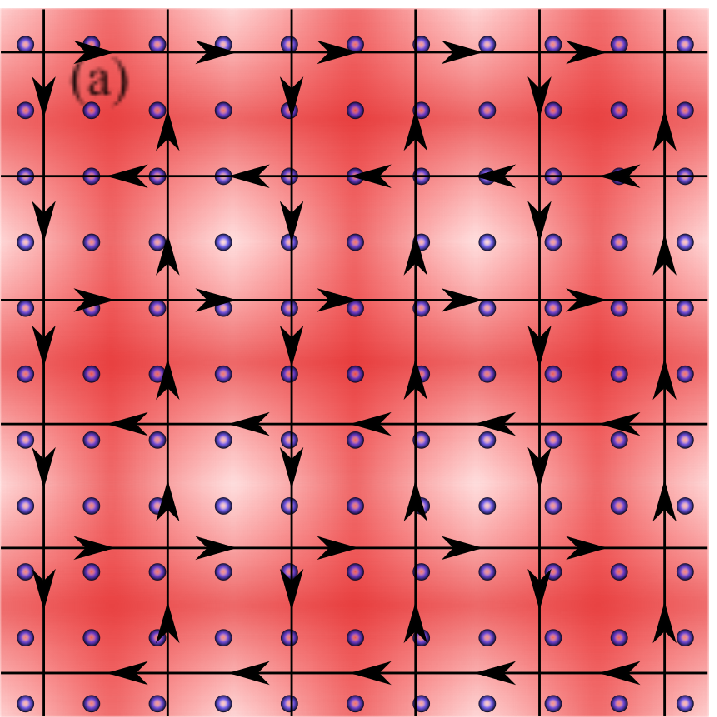}~~~
 \includegraphics[width=0.3\columnwidth]{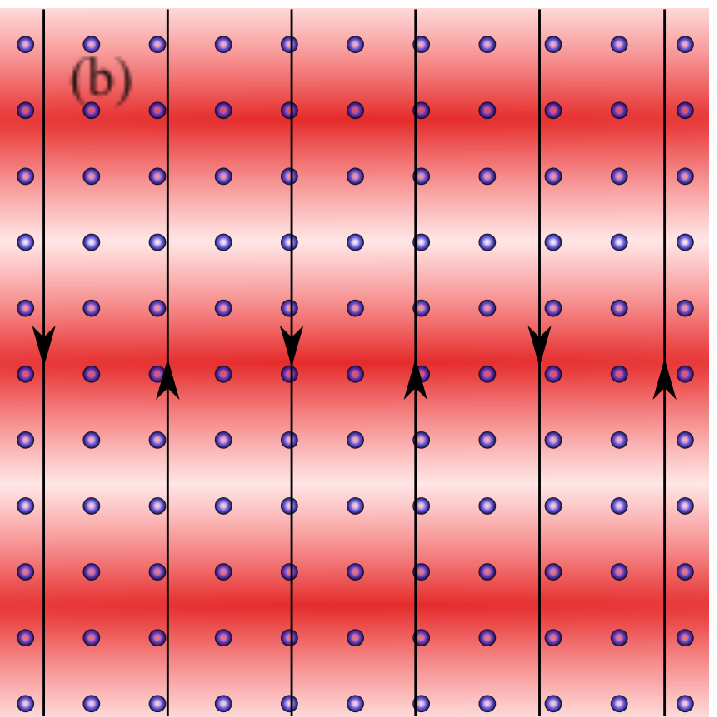}
 \caption{Two possible real space structures of charge order in the state II. Panel (a): a checkerboard charge density order together with oscillating currents along both $x$
 and $y$ directions. Panel (b): a stripe charge density order together with an oscillating current along orthogonal direction.}
 \label{yw30}
 \end{figure}

In this case, $\Delta^Q_1$ develops along $y$ direction and $\Delta^Q_2$ develops along $x$ direction, i.e., $|\Delta^y_1| = |\Delta^x_2| \neq 0$ and
$|\Delta^y_2| = |\Delta^x_1| = 0$.
 In real space this order corresponds to incommensurate charge density modulations in the $x$ direction and incommensurate current in $y$ direction.
 We show this in Fig.\ \ref{yw30} (b).
 Such an order breaks two $U(1)$ phase symmetries and breaks $C_4$ lattice symmetry down to $C_2$ such that the order parameter manifold is $U(1) \times U(1) \times Z_2$, where $Z_2$ corresponds to $C_4 \to C_2$. However the order parameter manifold
 does not has additional $Z_2$ component, which one would associate with time-reversal symmetry, because only $\Delta_1$ or $\Delta_2$ appear along a particular direction of ${\bf Q}$. Indeed, $\Delta_2$ changes sign under time reversal, but this change is absorbed into $U(1)$ phase symmetry.

The effective action for non-zero $\Delta^y_1$ and $\Delta^x_2$ is obtained from (\ref{ywth1}):
\begin{align}
{S}_{\rm eff}=&\alpha(|\Delta_1^y|^2+|\Delta_2^x|^2)
- 2({\bar \beta}_m - {\tilde\beta}_m) |\Delta_1^y|^2|\Delta_2^x|^2
\label{ywth1_2}
\end{align}
Because ${\bar \beta}_m - {\tilde\beta}_m >0$, the transition is again first order,
 into a state in which $\Delta^x_1$ and $\Delta^y_2$ have equal magnitudes.

 \section{Preemptive composite CDW order}
\label{sec:4}

We now go beyond mean-field theory and discuss potential preemptive transitions, when a discrete Ising symmetry gets broken at a higher $T$ than a continuous phase symmetry.  We focus on the state I because in this state transitions are second order and the analysis of preemptive instabilities can be carried out within the GL expansion. This state is also more favorable for applications to hole-doped cuprates as phase transitions there are likely continuous ones.
 We will briefly discuss a potential preemptive order in the state II at the end of this section. We remind that the order parameter manifold in the state I
  is $U(1) \times Z_2 \times Z_2$, where one $Z_2$ is associated with the breaking of $C_4$ lattice rotational symmetry down to $C_2$ and another $Z_2$ is associated with Ising symmetry breaking associated with the relative phase between even and odd in $k$ components of $\Delta^Q_k$ with a given $Q$. The lattice $Z_2$ symmetry
   is broken by strong interactions (${\tilde \beta}_m$ and ${\bar \beta}_m$ terms in the action), while $Z_2$ associated with the relative phase is broken by weaker interactions, of order $\beta \ll {\tilde \beta}_m, {\bar \beta}_m$.

 Below we discuss two preemptive composite orders associated with the breaking of each of the two discrete Ising $Z_2$ symmetries without breaking of the
 $U(1)$ phase symmetry.  We analyze composite orders within ``stage II" HS formalism, by which we mean that we introduce HS fields associated with $Z_2$ composite orders, apply HS transformation to effective action written in terms of $\Delta$ fields to decouple $\Delta^4$ terms, integrate over $\Delta$ fields and
  analyze the effective action for composite fields in the saddle point approximation. A similar procedure was applied in the study of a preemptive nematic order in Fe-pnictides~\cite{rafael} and of a preemptive translational symmetry breaking in doped graphene~\cite{Fern_13_a}.

The saddle-point approximation for the action in terms of composite order parameters is valid when fluctuations around saddle-point solution are weak.
  This is the case when the number of components of the primary field $\Delta$ field  is large. The HS approach assumes that the original model can be safely extended to large number of field components $M \gg1$ in the sense that the results obtained in the controlled analysis at large $M$ are at least qualitatively correct for the original model with $M \sim 1$. We will perform large $M$ calculation below and show that composite orders associated with the breaking of each
      of $Z_2$ symmetries in our case emerge at a higher $T$ than the one at which the primary field orders.
       A caveat here is that in 2D a primary field with $M \geq 3$ does not order down to $T=0$ (Ref.\ [\onlinecite{polyakov}]), hence a breaking of a $Z_2$ symmetry at any non-zero $T$ is a preemptive order. Whether the actual system shows the same behavior depends on the type of the problem.  For Fe-pnictides, the (magnetic) order parameter is a three- component unit vector, and it indeed does not order down to $T=0$ in 2D, like in large $M$ approximation. In quasi-2D systems, the primary field does order, but at a very low $T$, which for weak coupling along the third direction is certainly smaller than a finite critical $T$ at which $Z_2$ symmetry gets broken. In our case, however, the primary field is a two-component unit vector (a $U(1)$ field), and the temperature at which the primary field orders in a quasi-2D system is finite and tends to Berezinskii-Kosterlitz-Thouless temperature in the 2D limit.  Whether this temperature is still smaller than the one at which composite order develops is a'priori unclear and cannot be addressed within HS-based, large $M$  analysis. Fortunately,  the emergence of  preemptive composite orders has been verified within the approach specifically designed for a $U(1)$ primary field~\cite{tsvelik}.
        We use the result of~[\onlinecite{tsvelik}] as a verification that for the issue of a preemptive order a model with a two-component primary field is not qualitatively different from  models with larger number of field components and proceed with the HS-based analysis.

 \subsection{A nematic transition}

We first discuss whether the breaking of $C_4$ lattice rotational symmetry down to $C_2$ can occur before the continuous $U(1)$ phase symmetry gets broken.
For this discussion, the presence of the two components of $\Delta_k^Q$ with a given $Q$ does not play a role (the analysis of the truncated and full GL functional yield the same results with regard to $C_4$ breaking in the ordered state I). To  simplify presentation, we then analyze the truncated GL functional with only $\Delta_1$ component present.  Our analysis of a preemptive nematic order will closely follow that in Ref.\ [\onlinecite{rafael}], but we also discuss the stability of the nematic phase.

The effective action for coupled order parameters $\Delta_x = \Delta_1^{Q_x}$ and $\Delta_y = \Delta_1^{Q_y}$ is presented in  Eq.\ (\ref{eq:2}). Adding gradient terms and rescaling, we re-express (\ref{eq:2}) as
\begin{align}
{ S} (\Delta_x,\Delta_y) =&\alpha(|\Delta_x|^2+|\Delta_y|^2)+ |\partial_{\mu} \Delta_x|^2 + |\partial_{\mu} \Delta_y|^2 + \nonumber \\
& \frac{1}{2}\left(|\Delta_x|^2+|\Delta_y|^2\right)^2 -\frac{\beta^*}{2} \left(|\Delta_x|^2-|\Delta_y|^2\right)^2
\label{sef3}
\end{align}
where, in comparison with (\ref{eq:2}),  $\alpha^* = a^* (T - T_{\rm cdw})$ with $a^* = a/(\beta + \beta_m)$, and $\beta^* = (\beta_m - \beta)/(\beta_m + \beta)$.
 Because both $\beta_m$ and $\beta$ are positive and $\beta_m > \beta$, we have $0<\beta^* <1$.
 In principle, one should also include frequency dependence of the $\Delta$ fields  add the dynamical Landau damping $\gamma |\omega_m|$ term to
 $\alpha_q$, but to analyze the transition at a finite $T$ it is sufficient to consider only
 thermal fluctuations, i.e., the ones coming from $\omega_m =0$.

We extend each $\Delta$ field to $M\gg1$ components and  re-write ${ S} (\Delta_x,\Delta_y)$ as
 \begin{align}
{ S} (\Delta_x,\Delta_y) =&\sum_{i=1}^M \left(\alpha \left(|\Delta_{x,i}|^2+|\Delta_{y,i}|^2\right)+ |\partial_{\mu} \Delta_{x,i}|^2 +
|\partial_{\mu} \Delta_{y,i}|^2\right) + \nonumber \\
& \frac{1}{2M}\left(\sum_{i=1}^M\left(|\Delta_{x,i}|^2+|\Delta^2_{y,i}|^2\right)^2 \right)
-\frac{\beta^*}{2M}\left(\sum_{i=1}^M\left(|\Delta_{x,i}|^2-|\Delta_{y,i}^2|\right)\right)^2
\label{seff3m}
\end{align}
We  introduce two HS fields: ${\psi}$, conjugated to $i(|\Delta_{x,i}|^2+|\Delta_{y,i}|^2)$,
and $\upsilon$, conjugated to $|\Delta_{x,i}|^2-|\Delta_{y,i}|^2$,  as
\begin{align}
\exp\left(-\sum_{i=1}^M\left(|\Delta_{x,i}|^2+|\Delta_{y,i}|^2\right)^2/(2M)\right)&= \sqrt{\frac{M}{2\pi}}\int d{\psi}e^{\frac{-M
{\psi}^2}{2}}\exp\left[i{\psi}\left(\sum_{i=1}^M\left(|\Delta_{x,i}|^2+|\Delta^2_{y,i}|^2\right)\right)\right]\nonumber\\
\exp\left(\sum_{i=1}^M \left(|\Delta_{x,i}|^2-|\Delta_{y,i}|^2\right)^2/(2M)\right)&= \sqrt{\frac{M}{2\pi \beta^*}}\int d\upsilon e^{-\frac{M
v^2}{2\beta^*}}\exp\left[\upsilon \left(\sum_{i=1}^M\left(|\Delta_{x,i}|^2-|\Delta_{y,i}^2|\right)\right)\right]
\end{align}
Substituting these integrals into the the partition function $I = \int d \Delta_x d \Delta_y e^{-{ S} (\Delta_x, \Delta_y)}$ and integrating over
$\Delta_x$ and $\Delta_y$, we
 obtain  $I \propto \int d \psi d \upsilon e^{-M{S}_{\rm eff} (\psi, \upsilon)}$, where
\begin{align}
{S}_{\rm eff}[{\psi},\upsilon]= \frac{{\psi}^2}{2}+\frac{\upsilon^2}{2\beta^*}+ \int\frac{d^2q}{4\pi^2} \log\[(\alpha +
q^2-i{\psi})^2-\upsilon^2\].
\label{seff4}
\end{align}
The extremum of ${S}_{\rm eff}$ is obtained from $\partial {S}_{\rm eff}/\partial {\psi} =0$ and $\partial { S}_{\rm
eff}/\partial \upsilon =0$.
This gives two equations
\begin{align}
\frac{\partial{S}_{\rm eff}}{\partial{\psi}}=& \psi-2i \int\frac{d^2q}{4\pi^2}\frac{\alpha +q^2 -i {\psi}}{(\alpha + q^2 - i
{\psi})^2-\upsilon^2}=0\label{effphi}\\
\frac{\partial {S}_{\rm eff}}{\partial\upsilon}=&\frac{\upsilon}{\beta^*}-2\int\frac{d^2q}{4\pi^2}\frac{\upsilon}{(\alpha +q^2 -i
{\psi})^2-\upsilon^2}=0.\label{effups}
\end{align}
 The solution exists for an imaginary ${\psi} = i \psi_0$.

 We follow Ref.\ [\onlinecite{rafael}] and introduce  $r\equiv\alpha+\psi_0$ and $x\equiv q^2+r$. The primary fields get ordered when $r$ changes sign and
 becomes negative. This doesn't happen in 2D, as long as $T >0$. Replacing $\psi_0$ by $r -\alpha$, we obtain from Eq.\ (\ref{effups}):
\begin{align}
r=&\alpha+\frac{1}{2\pi}\int_{r}^{\Lambda}\frac{dx~x}{x^2-\upsilon^2}=\alpha+\frac{1}{2\pi}\log\frac{\Lambda}{\sqrt{r^2-\upsilon^2}}, \nonumber \\
\upsilon=& \upsilon \frac{\beta^*}{2\pi}\int_{r}^{\infty}\frac{dx}{x^2-\upsilon^2} = \upsilon\frac{\beta^*}{2\pi} \coth^{-1}\frac{r}{\upsilon}.
\label{n_1}
\end{align}

\subsubsection{The solution $\upsilon=0$}

The set of equations (\ref{n_1}) obviously allows a ``trivial" solution $\upsilon =0$.
We have then
\begin{align}
r = \alpha + \frac{1}{2\pi} \log{\frac{\Lambda}{r}}.
\label{n_2}
\end{align}
One can easily check the stability of this solution by verifying how the effective action changes when one moves along the
 trajectory which passes through a saddle point.  For $\upsilon$ this implies shifting from $\upsilon =0$ along the real axis, for $\psi$ this
 implies shifting along the {\it real} axis from ${\psi} = i \psi_0 = i(r-\alpha)$, where $r$  is the solution of (\ref{n_2}).
 Introducing $\psi = i \psi_0 + \delta {\psi}$ and $\upsilon \equiv \delta \upsilon$, substituting into the action, and expanding to
  second order in $\delta {\psi}$ and $\delta \upsilon$, we obtain
  \begin{align}
{S}_{\rm eff} ({\psi}, \upsilon) = {S}_{\rm eff}(i\psi_0,0) + \frac{\left(\delta {\psi}\right)^2}{2} \left(1 + \frac{1}{2\pi r}\right)
 + \frac{\left(\delta \upsilon\right)^2}{2\beta^*} \left(1 - \frac{\beta^*}{2\pi r}\right)
\label{n_3}
\end{align}
We see that the prefactor for $\left(\delta {\psi}\right)^2$ is definitely positive, i.e ${S}_{\rm eff}$ definitely increases along the
trajectory
 on Fig.\ 1.  The prefactor for  $\left(\delta \upsilon\right)^2$ term is positive as long as $r > \beta^*/2\pi$.
 Combining this with Eq.\ (\ref{n_2}), one finds that this holds when $\alpha > \alpha_{\rm cr}$, where
 \begin{align}
 \alpha_{\rm cr} = \frac{\beta^*}{2\pi} - \frac{1}{2\pi} \log{\frac{2\pi \Lambda}{\beta^*}}
 \label{n_4}
 \end{align}
 The condition $\alpha > \alpha_{\rm cr}$ implies that $T > T_{\rm cr}$, where $T_{\rm cr} = T_{\rm cdw} + \alpha_{\rm cr}/a$.

 \subsubsection{The solution with $\upsilon \neq 0$}

\begin{figure}
\includegraphics[width=0.5\columnwidth]{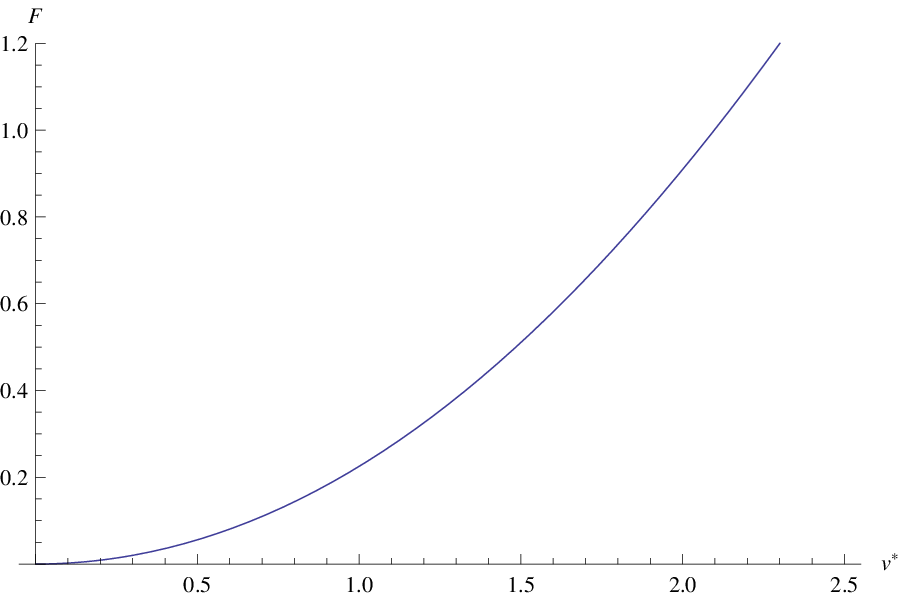}
\caption{ The plot of $F(\upsilon^*)$ from Eq.\ (\ref{n_6}) as a function of $\upsilon^*$ with $\beta^*$=0.3.}
\label{wed_4}
\end{figure}

 Solving the set of saddle-point equations for $\upsilon \neq 0$, we obtain
 \beq
 r = \frac{\beta^*}{2\pi}  \frac{\upsilon^*}{\tanh\upsilon^*}
 \label{n_5}
 \eeq
 where $\upsilon^*\equiv{\pi\upsilon}/{\beta^*}$. The equation on $\upsilon^*$ takes the form
 \begin{align}
F(\upsilon^*) &=\frac{2\pi}{\beta^*} \left(\alpha_{\rm cr} - \alpha\right) \nonumber \\
F(\upsilon^*) &=
1 - \frac{\upsilon^*}{\tanh\upsilon^*}+ \frac{1}{\beta^*} \log\frac{\sinh\upsilon^*}{\upsilon^*}
\label{n_6}
\end{align}
where, we remind, $\alpha = a (T - T_{\rm cdw})$
Expanding the l.h.s.\ of (\ref{n_6}) at small $\upsilon^*$, we obtain
\beq
\frac{(\upsilon^*)^2}{6} \left(\frac{1}{\beta^*} -2 \right) = \frac{2\pi}{\beta^*} \left(\alpha_{\rm cr} - \alpha\right) \propto \left(T_{\rm cr} - T \right)
\label{n_7}
\eeq
We see that, if $2 \beta^* <1$,  $\upsilon^*$ gradually increases when $T$ becomes smaller than $T_{\rm cr}$. To put it simply,
 the solution of the non-linear saddle-point
 equation shows that the order in $\upsilon$ emerges below $T_{\rm cr}$, as it is expected for a continuous, second-order transition.
We see the same behavior from Fig.\ \ref{wed_4} where we  plot
$F(\upsilon^*)$ from (\ref{n_6}) as a function of $\upsilon^*$.

 Let's now see what we get if we expand near the saddle point solution.  Introducing ${\psi} = i\psi_0 + \delta {\psi}$ and
 $\upsilon = (\beta^*/2\pi) \upsilon^* + \delta \upsilon$, where, for definiteness, $\upsilon^*$ is the positive solution of Eq.\ (\ref{n_6}), and
 expanding to second order in $\delta {\psi}$ and $\delta \upsilon$, we obtain after a straightforward algebra
  \begin{align}
  {S}_{\rm eff} ({\psi}, \upsilon) = {S}_{\rm eff}(i\psi_0, \beta^*\upsilon^*/\pi) + A
  \left(\delta {\psi}\right)^2 - B \left(\delta \upsilon\right)^2 - 2 i C \left(\delta {\psi}\right) \left(\delta \upsilon\right)
\label{n_8}
\end{align}
where
 \begin{eqnarray}
 &&A = \frac{1}{4 \beta^* \upsilon^*} \left(2\beta^* \upsilon^* + \sinh{2 \upsilon^*}\right) \nonumber \\
 &&B =  \frac{1}{4 \beta^* \upsilon^*} \left(\sinh{2 \upsilon^*} - 2 \upsilon^*\right) \nonumber \\
 &&C =   \frac{1}{2 \beta^* \upsilon^*} \sinh^2{\upsilon^*}
\label{n_9}
\end{eqnarray}
Obviously, $A, B$, and $C$ are positive for $\upsilon^* \neq 0$.

Eq.\ (\ref{n_8}) has the same form as Eq.\ (\ref{y_3_2}) in the main text and Eq.\ (\ref{a71}) in Appendix \ref{app:b}.
Like we did there, we re-express  $ {S}_{\rm eff} ({\psi}, \upsilon)$ in (\ref{n_8}) as
\begin{align}
 {S}_{\rm eff} ({\psi}, \upsilon) = {S}_{\rm eff}(i\psi_0, \beta^*\upsilon^*/\pi) + A
  \left(\left(\delta {\psi}\right) - i \frac{C}{A} \left(\delta \upsilon\right) \right)^2 + \frac{C^2-AB}{A} \left(\delta \upsilon\right)^2
\label{n_10}
\end{align}
The contour has to be chosen such that the variable $\left(\delta {\psi}\right) - i \frac{C}{A} \left(\delta \upsilon\right)$ is real, i.e we integrate
over $\delta \psi$ parallel to real axis.

As we already know, the condition that the saddle-point is the minimum of the action along the integration contour is $C^2-AB >0$.
Substituting the expressions from (\ref{n_9}), we find
\begin{align}
C^2 -AB = \frac{\sinh^2{\upsilon^*}}{4 \beta^* (\upsilon^*)^2} I (\upsilon^*)
\label{n_11}
\end{align}
where
\begin{align}
I (\upsilon^*) = \frac{1}{\beta^*} \left(\frac{\upsilon^*}{\tanh{\upsilon^*}} -1\right) - \left(\frac{\upsilon^*}{\tanh{\upsilon^*}} -
\left(\frac{\upsilon^*}{\sinh{\upsilon^*}}\right)^2 \right)
\label{n_12}
\end{align}
The condition that the saddle-point is the minimum of the action along the integration contour is then $I (\upsilon^*) >0$.
Expanding at small $\upsilon^*$ we obtain
\begin{align}
I (\upsilon^*) = \frac{(\upsilon^*)^2}{3} \left(\frac{1}{\beta^*} -2\right)
\label{n_14}
\end{align}
We see that $I (\upsilon^*) >0$ when $2 \beta^*<1$. This is the same condition as in Eq.\ (\ref{n_7}).  One can easily verify that when $0<2\beta^*<1$,
$I(\upsilon^*)$ is positive for all values of $\upsilon^*$.

\subsubsection{First-order transition at $1/2 < \beta^* < 1$}

\begin{figure}
\includegraphics[width=0.5\columnwidth]{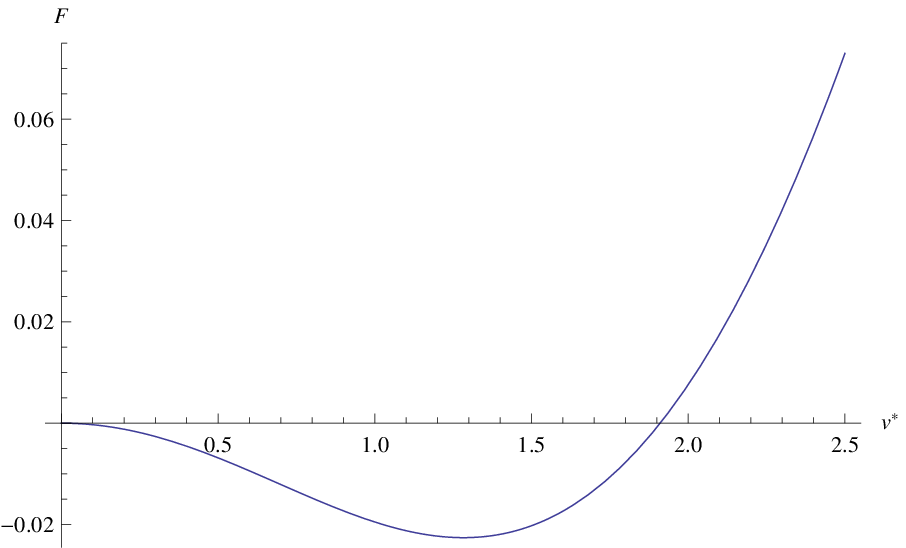}
\caption{ The plot of $F(\upsilon^*)$ from Eq.\ (\ref{n_6}) as a function of $\upsilon^*$ with $\beta^*$=0.55.}
\label{yw191}
\end{figure}

 For larger $\beta^*$, the prefactors in (\ref{n_14}) and (\ref{n_7}) are negative.  The analysis of the full saddle-point solution, Eq.\ (\ref{n_6})
  shows that, as $\alpha$ gets smaller, the saddle-point solution (i.e., the solution of (\ref{n_6}) first emerges at a finite $\upsilon^*_{\rm cr}$, i.e.,
  the
   transition is first order (see Fig.\ \ref{yw191}, in which we plot $F(\upsilon^*)$ from (\ref{n_6})  vs $\upsilon^*$ for $1/2 <\beta^* <1$.)

 As $\alpha$ gets smaller, two saddle-point solutions appear, one at $\upsilon^* > \upsilon^*_{\rm cr}$, another at $\upsilon^* < \upsilon^*_{\rm cr}$.
 By obvious reasons, the solution with $\upsilon^* > \upsilon^*_{\rm cr}$ is expected to be stable, while the one at $\upsilon^* < \upsilon^*_{\rm cr}$ is
 expected
  to be unstable. We see from Fig.\ \ref{yw191} that the solution at $\upsilon^* > \upsilon^*_{\rm cr}$ corresponds to $d F(\upsilon^*)/d \upsilon^* >0$, and the
  solution
    at $\upsilon^* <\upsilon^*_{\rm cr}$ corresponds to $d F(\upsilon^*)/d \upsilon^* <0$. Now, evaluate
    \begin{align}
    \frac{d F(\upsilon^*)}{d \upsilon^*} = \frac{\beta^*_1}{\beta^*} \left(\frac{\upsilon^*}{\tanh{\upsilon^*}} -1\right) -
    \left(\frac{\upsilon^*}{\tanh{\upsilon^*}} - \left(\frac{\upsilon^*}{\sinh{\upsilon^*}}\right)^2 \right)
    \label{n_15}
    \end{align}
    Comparing this with (\ref{n_12}), we see that
    \beq
     \frac{d F(\upsilon^*)}{d \upsilon^*} \equiv I(\upsilon^*)
     \label{n_16}
     \eeq
  Hence the solution with a positive  $d F(\upsilon^*)/d \upsilon^* $ corresponds to $I(\upsilon^*) >0$ and is stable, as expected.

\subsection{Preemptive time-reversal symmetry breaking}

We now return to the full GL model for $\Delta_1$ and $\Delta_2$ and consider a possibility of a preemptive breaking of $Z_2$ symmetry associated with the
  relative phase, $\psi_1 - \psi_2 = \pm \pi/2$, between complex $\Delta^Q_1 = = |\Delta^Q_1|e^{i\psi_1}$
      and $\Delta^Q_2 = |\Delta^Q_2|e^{i\psi_2}$.
      We recall  that a nematic order is selected already within the hot spot model, while $Z_2$ part of the order parameter manifold associate with
        phase locking becomes relevant only once we go beyond hot spot approximation and include the interaction between CDW order parameters $\Delta^Q_k$ and $\Delta^Q_{-k}$.  Accordingly, the coupling constant associated with the nematic $Z_2$ symmetry is larger than the one associated with the phase $Z_2$ symmetry, and, hence $T_{n}$, at which a nematic order sets in, is larger than a temperature, $T_t$ at which the other $Z_2$ symmetry get broken. Still, it is essential to understand whether $T_t$ is larger than $T_{\rm cdw}$, i.e., whether $Z_2$ symmetry associated with $\psi_1 - \psi_2 = \pi/2$ or $-\pi/2$ gets broken
          at a temperature higher than the one when $U(1)$ symmetry of the common phase $\psi_1 + \psi_2$ gets broken.

  We assume that nematic order selects, say, ${\bf Q} = Q_x$ and consider GL model for $\Delta^x_1$ and $\Delta^x_2$.
 A preemptive instability with respect to the relative phase of $\Delta^x_1$ and $\Delta^x_2$
  would imply that  at some $T = T_t > T_{\rm cdw}$
     $\Delta^x_1$ and $\Delta^x_2$ form a bound state with zero total momentum.  In between $T_t$ and $T_{\rm cdw}$,  $\delta \rho(r) =  j_y
     (r)=0$, but
     $\Upsilon
     \propto \langle\delta \rho (r) j_y (r)\rangle$
      becomes non-zero. Under time reversal, $\Upsilon$ transforms into $-\Upsilon$, hence this order breaks $Z_2$ time reversal symmetry.

\subsubsection{Direct computation}

\begin{figure}
\includegraphics[width=0.8\columnwidth]{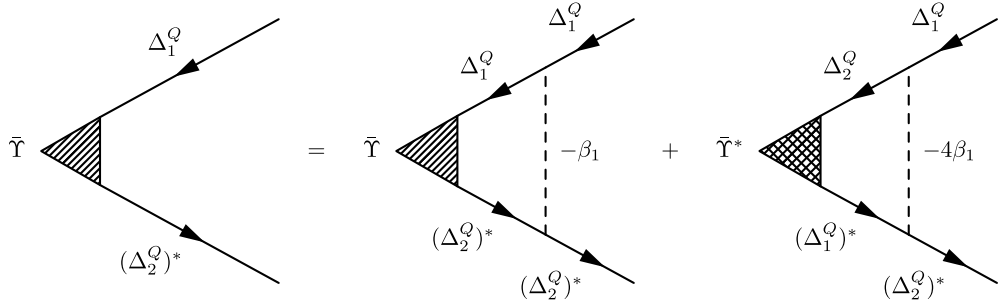}
\caption{Ladder equation for ${\bar \Upsilon} = \Delta_1^Q(\Delta_2^{Q})^*$. Of the two terms in the r.h.s., one contains ${\bar \Upsilon}$, another ${\bar
\Upsilon}^*$. For imaginary ${\bar \Upsilon}$, there is a sign change between these two terms.}
\label{Fig14}
\end{figure}

 One way to see that a preemptive transition is possible is to follow the same strategy as in the
  analysis of a spin-current order in anisotropic triangular antiferromagnets~\cite{starykh} and in Heisenberg-Kitaev model on a hyperhoneycomb
  lattice~\cite{arun_1,arun_2},  introduce a ``two-particle" collective variable
  ${\bar \Upsilon} =  \Delta_1^x(\Delta_2^{x})^*$, and solve for the emergence of a two-particle bound state instability in the
  same way as it is done for  superconductivity.
  For illustrative purposes we consider the effective action (\ref{ywth1_n}), although the actual calculation has to be performed for the more generic action (\ref{seff3}) as we will do below using HS approach.  We re-write (\ref{ywth1_n}) as
\begin{align}
{S}_{\rm eff}=&\alpha(|\Delta_1^x|^2+|\Delta_2^x|^2)+\beta(|\Delta_1^x|^4+|\Delta_2^x|^4)\nonumber\\
&+\beta(\Delta_{1}^{x}(\Delta_2^x)^*)(\Delta_{1}^{x}(\Delta_2^x)^*)+\beta((\Delta_1^x)^*\Delta_2^x)((\Delta_1^x)^*\Delta_2^x)+4
\beta(\Delta_1^x(\Delta_2^x)^*)(\Delta_2^Q(\Delta_1^x)^*).
\label{ch_5}
\end{align}
  The  ladder equation for ${\bar \Upsilon}$ is presented in Fig.\  \ref{Fig14} There are two terms in the r.h.s.\ of this graphic equation -- the first
  contains a ``direct" ${\bar \Upsilon} {\bar \Upsilon}$ interaction from the first term
   in the  second line of (\ref{ch_5}), and the second one contains the interaction between ${\bar \Upsilon}$ and ${\bar \Upsilon}^*$.  Both interactions
   are repulsive, hence no solution
    is possible if ${\bar \Upsilon}$ is real. However, if we search for a solution with a complex ${\bar \Upsilon}$, we obtain
     for infinitesimally small ${\bar \Upsilon}$
    \begin{align}
    {\bar \Upsilon} = -\beta P \left({\bar \Upsilon}+ 4 {\bar \Upsilon}^*\right)
   \label{ch_6}
   \end{align}
   where $P>0$ stand for convolution of the propagators of $\Delta_1^Q$ and $\Delta_2^Q$ fields.
    The only information about $P$ relevant to us at this stage is that it
  diverges at $T_{\rm cdw}$ when both propagators become massless. Hence, if Eq.\ (\ref{ch_6}) has a non-trivial solution, the corresponding $T$ is larger
  than $T_{\rm cdw}$.
     A simple analysis of  Eq.\ (\ref{ch_6}) shows  that the solution does exist if we set ${\bar \Upsilon}$ to be purely imaginary,
      ${\bar \Upsilon} = i \Upsilon$, because then the combination ${\bar \Upsilon}+ 4 {\bar \Upsilon}^*$ becomes equal to $-3 {\bar \Upsilon}$, and the
      minus sign compensates the overall minus sign in the r.h.s.\ of (\ref{ch_6}).  We emphasize that this is possible because the prefactor for ${\bar
      \Upsilon} {\bar \Upsilon}^*$ interaction term
   (the last term in (\ref{ch_5})) is 4 times larger than the direct ${\bar \Upsilon} {\bar \Upsilon}$ interaction term.  That ${\bar \Upsilon}$ is purely
   imaginary is entirely
 consistent with the fact that in mean-field approximation $\Delta_1^Q$ is real and $\Delta_2^Q$ is imaginary, hence in below $T_{\rm cdw}$,
  ${\bar \Upsilon} = \Delta_1^x (\Delta_2^{x})^*$ is also purely imaginary.

  \subsubsection{Hubbard-Stratonovich approach}

Another way to see  the emergence of a preemptive transition is to follow the same strategy as in the analysis of a pre-emptic nematic order
  and  apply HS transformation to the effective action (\ref{seff3}) by
introducing collective variables conjugated to quartic terms in  ${S}_{\rm eff}$. For this purpose, it is convenient to rescale
$\Delta^x_2$ as $\Delta^x_2 \to \Delta^x_2 (\beta_1/\beta_2)^{1/4}$, add the gradient terms and
re-write Eq.\ (\ref{seff3}) as
\begin{align}
{S}_{\rm eff}
\propto
&(\alpha_1 + q^2) |\Delta_1^x|^2+ ({\bar \alpha}_2 +q^2) |\Delta_2^x|^2+ \frac{\gamma_1}{2} \left(|\Delta_1^x|^2 +|\Delta_2^x|^2\right)^2 \nonumber\\
& -  \frac{\gamma_2}{2} \left(|\Delta_1^x|^2 - |\Delta_2^x|^2\right)^2 - \frac{\gamma_3}{2} \left(i \left((\Delta_1^x)^*\Delta_2^x -
\Delta_1^x (\Delta_2^x)^*\right)\right)^2
\label{seff3_1}
\end{align}
where $\alpha_1 =a (T-T_e)$, ${\bar \alpha}_2 = \alpha_2 (\beta_1/\beta_2)^{1/2} = a (T-T_o)$, where, we remind, $T_e$ and $T_o$ are (near identical)
 mean-field transition temperatures for even and odd in $k$ components of $\Delta^x$. Also
\beq
\gamma_1 = \frac{1}{2}\beta_1 + \frac{3}{2} \beta_3 \left(\frac{\beta_1}{\beta_2}\right)^{1/2},~~ \gamma_2 = \frac{3}{2} \beta_3
\left(\frac{\beta_1}{\beta_2}\right)^{1/2} -  \frac{1}{2} \beta_1,~~\gamma_3 = \beta_3 \left(\frac{\beta_1}{\beta_2}\right)^{1/2}
\label{apac_10}
\eeq
 The prefactors for the two $q^2$ terms in (\ref{seff3_1}) as well as the prefactors $a$ for $\alpha_1$ and ${\bar \alpha}_2$ do not have to be equal, but  this complication does not lead to new
 physics and we neglect it.

There are three quartic terms in (\ref{seff3_1}). Accordingly we introduce three HS bosonic fields  $\Upsilon$, congugated to
 $ i (\Delta^x_1 (\Delta^x_2)^{*} - (\Delta^x_1)^{*} \Delta^x_2)$,  $\Psi$, congugated to $(|\Delta^x_1|^2 +  |\Delta^x_2|^2)$,
  and $\Psi_1$, congugated to $(|\Delta^x_1|^2 - |\Delta^x_2|^2)$. The expectation value of each HS field is proportional to  the corresponding
  bilinear combination of $\Delta^x_1$ and $\Delta^x_2$. The field $\Psi$ describes Gaussian fluctuations of the modulus of a two-component order
   parameter and its expectation value is obviously non-zero at any $T$.  The field $\Psi_1$ describes fluctuations of a relative magnitude of
   $|\Delta^x_1|^2$ and $|\Delta^x_2|^2$.    For $\alpha_1 \neq {\bar \alpha_2}$, order parameters $\Delta^x_1$ and $\Delta^x_2$ are non-equal and
   $\langle|\Delta_1^x|^2 - |\Delta_2^x|^2\rangle$ is non-zero at any $T$, hence the expectation value of $\Psi_1$ is also non-zero for all $T$.  The field
   $\Upsilon$
    is different from the other two because the expectation value of $\langle(\Delta_1^x)^*\Delta_2^x - \Delta_1^x(\Delta_2^x)^*\rangle$ and hence of
    $\Upsilon$ becomes non-zero only due to spontaneous symmetry breaking.

We assume, without going into details, that the model is extended to large $M$, as in the case of a nematic transition, and analyze the effective action for composite HS fields within saddle-point approximation.
A similar HS approach has been recently used to study TRS breaking in Fe-pnictides~\cite{maiti_a,lara,alberto}
  We use, as before,
\begin{align}
\exp\[-\frac{\gamma_1(|\Delta_1^x|^2+|\Delta_2^x|^2)^2}2\]=&\int
\frac{d\Psi}{\sqrt{2\pi\gamma_1}}\exp\left({\frac{-{\Psi}^2}{2\gamma_1}}\right)\exp\left[{i{\Psi}(|\Delta_1^x|^2+|\Delta_2^x|^2)}\right]\nonumber\\
\exp\[-\frac{\gamma_2(|\Delta_1^x|^2-|\Delta_2^x|^2)^2}2\]=&\int
\frac{d\Psi_1}{\sqrt{2\pi\gamma_2}}\exp\left({-\frac{\Psi^2_1}{2\gamma_2}}\right)\exp\left[{\Psi_1(|\Delta_1^x|^2
-|\Delta_2^x|^2)}\right]\nonumber\\
\exp\left\{\frac{\gamma_3[i(\Delta_1^x(\Delta_2^x)^*-(\Delta_1^x)^*\Delta_2^x)]^2}2\right\}=&\int
\frac{d\Upsilon}{\sqrt{2\pi\gamma_3}}\exp\left({-\frac{\Upsilon^2}{2\gamma_3}}\right)\nonumber\\
&\times\exp\left\{i\Upsilon\[\Delta_1^x(\Delta_2^x)^*-(\Delta_1^x)^*\Delta_2^x\]\right\},
\end{align}
Substituting this transformation into (\ref{seff3_1}) and performing Gaussian  integration over the fields
$\Delta^x_1$ and $\Delta^x_2$ we obtain   the effective action in terms of collective variables $\Upsilon$, $\Psi$, and $\Psi_1$  in the form
\beq
{S}_{\rm eff} (\Upsilon,\Psi,\Psi_1) = T \int_q
\left\{\frac{\Upsilon^2} {2\gamma_3} + \frac{\Psi^2} {2\gamma_1} + \frac{\Psi^2_1} {2\gamma_2}
 + \log\[{(\alpha_1 + q^2 - i \Psi)^2 -\Psi_1^2 - \Upsilon^2}\]\right\}
\label{eq:4_1}
\eeq
where $\int_q =\int \frac{d^2 q}{4\pi^2}$.

 We analyze ${S}_{\rm eff} (\Upsilon,\Psi,\Psi_1)$ in the saddle-point approximation, by solving the coupled set of
  saddle-point
  equations
 \begin{align}
\Upsilon =&2 \gamma_3 \int_q \frac{\Upsilon}{(\alpha_+ - i \Psi +q^2)^2 - (\alpha_- + \Psi_1)^2 - \Upsilon^2} \nonumber \\
 \Psi =& 2 \gamma_1 \int_q \frac{(\alpha_+ - i \Psi + q^2)}{(\alpha_+ - i \Psi+q^2)^2 - (\alpha_- + \Psi_1)^2 - \Upsilon^2} \nonumber \\
 \Psi_1 =& 2 \gamma_2 \int_q \frac{(\alpha_- + \Psi_1)}{(\alpha_+ - i \Psi+q^2)^2 - (\alpha_- + \Psi_1)^2 - \Upsilon^2}
 \label{eq:5}
 \end{align}
 where $\alpha_{+} = (\alpha_1 + {\bar \alpha}_2)/2  = a (T-(T_e+T_o)/2)  \approx a(T-T_{\rm cdw})$ and $\alpha_{-} = ({\bar \alpha}_2-\alpha_1)/2 = (a/2) (T_e-T_o) >0$.

  Our goal is to verify whether a solution with $\Upsilon \neq 0$ emerges  before the primary CDW order sets in.
  In our 2D case, the primary order sets in when $\alpha_+ \to -\infty$, hence the emergence of $\Upsilon \neq 0$ at any finite $\alpha_+$ will be a preemptive instability.

  Introducing $\Psi = i \Psi_0$, $r_0 =
  \alpha_+ + \Psi_0$, and $r_1 = \alpha_- + \Psi_1$, we re-write the last two equations in (\ref{eq:5}) as
\bea
&&r_0 = \alpha_+ + 2 \gamma_1 \int_q \frac{r_0}{r^2_0 - r^2_1} \nonumber \\
&&r_1 = \alpha_- + 2 \gamma_2 \int _q \frac{r_1}{r^2_0 - r^2_1}
\label{apac_11}
\eea
Evaluating the integrals we obtain
\bea
&&r_0 = \alpha_+ + \frac{\gamma_1}{2\pi}\log\frac{\Lambda}{\sqrt{r_0^2-r_1^2}} \nonumber \\
&&r_1 = \alpha_- + \frac{ \gamma_2}{2\pi}\coth^{-1}\(\frac{r_0}{r_1}\)
\label{apyw_11}
\eea
($\coth^{-1} {x}$ is arc-hyperbolic-cotangent of $x$).

 The primary fields $\Delta^x_1$ and $\Delta^x_2$ get ordered when
$r^2_0 - r^2_1$ becomes equal to zero.  We see from (\ref{apac_11}) that this only happens at $\alpha_+ = -\infty$.
This is specific to $d=2$ and to systems with $M \geq 3$ components, as we already discussed.

\begin{figure}
\includegraphics[width=0.5\columnwidth]{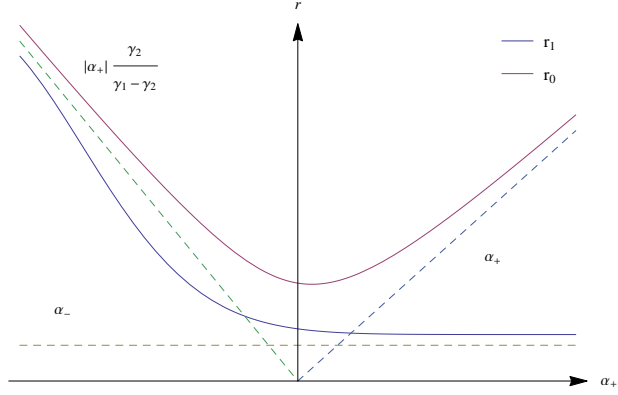}
\caption{the behavior of $r_1$ and $r_0$ as a function of $\alpha_+$.}
\label{r1alpha+}
\end{figure}

At high temperatures, $T \gg T_{\rm cdw}$, $\alpha_1 \approx \alpha_2 >0$, hence $\alpha_+ \gg \alpha_->0$. In this range,
the physically meaningful solution of \ref{apyw_11} is $r_0\approx\alpha_+$, $r_1 \approx \alpha_-$.
As temperature decreases, $\alpha_+$ and $r_0$ decrease, while $r_1$ increases.
Still, according to first equation in (\ref{apyw_11}), $r_0$ remains larger than $r_1$.
Eventually, $\alpha_+$ changes sign and becomes negative.  The quantities $r_0$ and $r_1$ evolve as shown in Fig.\ \ref{r1alpha+}.
   At finite but large negative $\alpha_+$, $r_0$ and $r_1$ are both large and Eqs.\ (\ref{apyw_11}) simplify to
\bea
&&r_0 = \alpha_+ + \frac{\gamma_1}{2\pi}\log\frac{\Lambda}{\sqrt{r_0^2-r_1^2}} \nonumber \\
&&r_1 =  \frac{ \gamma_2}{2\pi}\coth^{-1}\(\frac{r_0}{r_1}\).
\label{apyw_12}
\eea
From the second equation we obtain $r_0=r_1\coth (2\pi r_1/\gamma_2)$. Plugging this back to both sides of the first equation
 and introducing $\bar r_1=2\pi r_1/\gamma_2$ we find, at $|\alpha_+| \gg \log \Lambda$,
\begin{align}
\bar r_1\coth \bar r_1+\frac{\gamma_1}{\gamma_2}\log\frac{\bar r_1}{\sinh \bar
r_1}=\frac{2\pi}{\gamma_2}\alpha_+
\label{apyw_13}
\end{align}
Solving this equation we obtain
 \beq
 r_1 \approx
 |\alpha_+| \frac{\gamma_2}{\gamma_1 - \gamma_2}.
\label{apac_18}
\eeq
Note that, because  $r_0$ and $r_1$ are close to each other at large $|\alpha_+|$, the susceptibility of the primary fields
$\chi \propto 1/(r^2_0 - r^2_1)$ is strongly enhanced. Still, $r^2_0 > r^2_1$, i.e., the primary order does not develop.

We now look at the first equation in (\ref{eq:5}). Evaluating the integral, we find that the solution with infinitesimally small $\Upsilon$ emerges when
\beq
\frac{\gamma_3}{2\pi}\coth^{-1}\(\frac{r_0}{r_1}\)=r_1
\eeq
Using the second equation from (\ref{apyw_11}) we re-write this as the condition on a critical $r_{1,c}$,
\beq
r_{1,c} = \alpha_- \left(\frac{\gamma_3}{\gamma_3-\gamma_2}\right)
\label{apac_14}
\eeq
This critical $r_{1,c}$ is some positive number because $\alpha_- >0$ and $\gamma_3 > \gamma_2$ [see Eq.\ (\ref{apac_5}), $\gamma_3 > \gamma_2$ is equivalent to
 $J_{\beta_1} J_{\beta_2} > J^2_{\beta_3}$]. We next use the fact that
  $r_1$ monotonically increases as the temperature decreases. Hence,
 $r_1$ must reach $r_{1,c}$  at some finite $T=T_t$, and below this temperature the expectation value of $\Upsilon$ becomes
 non-zero.  A non-zero $\Upsilon = \pm \Upsilon_0$ in turn gives rise to a non-zero value of the  composite order parameter
 $\langle\Delta^x_1 (\Delta^x_2)^*\rangle \propto \pm i \Upsilon_0$.

In the consideration above we used the fact that $\alpha_->0$, in which case the expectation value of $\Phi_1$ is never zero (see Eq.\ (\ref{eq:5})), and only
 $\Upsilon$ field acquires a non-zero value due to spontaneous symmetry breaking.  In general $\alpha_- \propto (T_e-T_o) $ is non-zero and positive.
 However, we found earlier that it is quite small because both $T_e$ and $T_o$ are very close to the original $T_{\rm cdw}$, which, we remind, is a mean-field CDW transition temperature in the ``hot spot" approximation, when $\Delta^x_1$ and $\Delta^x_2$ are equivalent fields. If we set $T_e=T_o = T_{\rm cdw}$, i.e.,
  set $\alpha_- =0$, we immediately find from (\ref{eq:5}) that the field $\Phi_1$ can also order only due to symmetry breaking. The self-consistent equations
   for $\Phi_1$ and $\Upsilon$ now have equivalent kernels, and which of the two acquires a non-zero value depends on the ratio $\gamma_3\gamma_2$.
   Like we just said, in our case, $\gamma_3 > \gamma_2$, hence $\Upsilon$ field orders under proper conditions, but $\Phi_1 =0$. In this situation, only first two equations (\ref{eq:5}) matter, and solving them we obtain that $\Upsilon$ acquires a non-zero value when the two conditions are met:
   \bea
  r_0 &=& \alpha_+ + \frac{\gamma_3}{2\pi} \log{\frac{\Lambda}{r_0}} \nonumber\\
  1 &=& \frac{\gamma_3}{2\pi} \frac{1}{r_0}
  \eea
  Solving this set we obtain that, like in a more general case,  a non-zero $\Upsilon$ emerges at a negative but still finite
  \beq
  \alpha_+ = -\frac{\gamma_1}{2\pi} \left(\log{\frac{2\pi \Lambda}{\gamma_3}}  - \frac{\gamma_3}{\gamma_1}\right)
  \eeq
At larger negative $\alpha_+$, i.e., at smaller $T$, a non-zero $\Upsilon = \pm \Upsilon_0$  gives rise to a non-zero value of the  composite order parameter
 $\langle\Delta^x_1 (\Delta^x_2)^*\rangle \propto \pm i \Upsilon_0$.

When $\gamma_3 = \gamma_2$, or, equivalently, $\beta_3 =\beta_1 \beta_2$, the equations for $\Upsilon$ and $\Phi_1$ are identical, and one can immediately make sure that $\Upsilon$ and $\Phi_1$ in (\ref{eq:5}) can be cast as ``real" and ``imaginary"  components of the ``super-vector"
$\Theta = \sqrt{\Upsilon^2 + \Phi^2_1} e^{i\theta}$.  In the HS analysis, the magnitude of $\Theta$ becomes non-zero at some finite $T$, however neither $\Upsilon$ nor $\Phi_1$ order at any finite $T$ because of fluctuations between the directions of $\Upsilon$ and of $\Phi_1$.  In other words, in this situation, there will be no preemptive order which would break $Z_2$ TR symmetry. This is entirely consistent with the fact that without distinction between different $\alpha$ and $\beta$, the effective action decouples between $\Delta^Q_k$ and $\Delta^Q_{-k}$, such that both orders appear simultaneously at $T = T_{\rm cdw}$.  This last result shows that  non-equivalence of $\beta_i$ terms, namely the inequality $\beta_1 \beta_2 > \beta^2_3$, is the necessary condition
  for the existence of  a preemptive state with composite order which breaks TRS.

 \subsubsection{Preemptive order for state II}

 Before we proceed with the phase diagram, we briefly discuss potential preemptive orders for state II.
 As we found in the previous Section,  the CDW transitions into both versions of state II are first-order. In this situation,
  the analysis within the GL model is meaningless.
One can still argue, though, that because order parameter manifold in the CDW-ordered state has additional $Z_2$ component (ether $Q_x/Q_y$ or
 $\pm \pi/2$ for the relative phase between $\Delta_1$ and $\Delta_2$, depending on the realization of state II), there may be a preemptive transition into a state with a composite order parameter. However, to investigate this possibility, one has to go beyond GL expansion in powers of $\Delta$. We will not
  pursue this issue further.

\section{The phase diagram}
\label{sec:5}

\begin{figure}
\includegraphics[width=0.5\columnwidth]{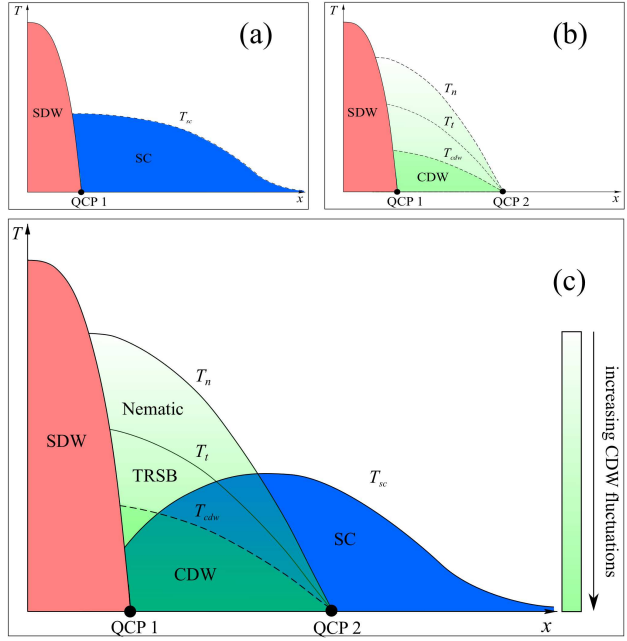}
\caption{Phase diagram for State I. { Panels (a) and (b)} -- the behavior of superconducting $T_{\rm sc}$ (panel a) and the onset temperatures for charge order $T_{n}$, $T_t$ and $T_{\rm cdw}$ (panel b), when superconducting and charge ordered are treated independent on each other. $T_{n}$ is the preemptive nematic transition temperature, and $T_t$ is the temperature below which a $q=0$ order emerges, breaking time-reversal symmetry. Panel (c) -- the full phase diagram, which includes the competition between superconductivity and charge order. QCP 1 and QCP 2 are quantum-critical points towards SDW  and CDW order,
 respectively.
}
\label{phases}
\end{figure}

For the rest of the paper we focus on state I, for which phase transitions are continuous ones.
 To construct the phase diagram for state I, we first consider how $T_{\rm cdw} (\xi)$ evolves when hole doping increases and magnetic correlation length decreases.
  We found  that at a finite $\xi$, the scale $ v^2_F\xi^{-2}/{\bar g} \sim {\bar g}/\lambda^2$ serves as the lower energy cut-off for the logarithm,
  i.e.,
   $T$ in (\ref{k0pilog}) gets replaced by, roughly, $(T^2 +{\bar g}^2/\lambda^4)^{1/2}$. As the consequence,  $T_{\rm cdw} (\xi)$ decreases with increasing $\xi$
   and vanishes when $\xi^{-1}_{\rm cr} \sim \bar g/v_F$, i.e. when the dimensionless coupling constant $\lambda \sim
   1$. We show this behavior in Fig.\ \ref{phases}(b).  The vanishing of $T_{\rm cdw} (\xi_{\rm cr})$ sets up a charge QCP at some distance away from a magnetic
   QCP.  The temperature $T_{n}$ at which
   composite nematic order sets in, and the temperature $T_t$ at which the preemptive TRSB order sets in,
    also gets smaller as $\xi$ increases.  We analyzed the emergence of the composite and CDW orders at $T=0$ using the same
   approach as in Ref.\ [\onlinecite{Fernandes_13}] (this requires one to include the dynamical term into $\alpha_q$ in equation (\ref{eq:4_1})) and found
   that the three lines, $T_{n}$, $T_t$, and $T_{\rm cdw}$ all terminate near the CDW quantum-critical point QCP 2, which actually becomes the point of weak first-order transition~\cite{Fernandes_13}.  It is possible, although  not proven yet, that a preemptive order survives down to $T=0$, in which case QCP 2 splits into
 two or even three quantum-critical points.
   We show the behavior of $T_{n} (\xi)$, $T_{t}(\xi)$ and $T_{\rm cdw} (\xi)$ in Fig.
   \ref{phases}(b).

The behavior of $T_{\rm cdw} (\xi)$ is different from that of superconducting $T_{\rm sc} (\xi)$. The latter does not vanish at a finite $\xi$ and just
interpolates between quantum-critical form $T_{\rm sc} \sim \omega_0$ at large $\xi$ and BCS form $T_{\rm sc} \propto \omega_0 e^{-1/\lambda}$ at smaller
$\xi$, when $\lambda$ becomes a small parameter.
This behavior of superconducting $T_{\rm sc}$  has been studied in Ref.\ [\onlinecite{acf}] and we also discuss it in detail in Appendix \ref{app:b} below.

   Panels (a) and (b) in Fig.\ \ref{phases} show the onset temperatures for superconducting order and for CDW and composite orders, when the CDW and SC  are
   considered independent on each other.
    In reality, charge and superconducting orders compete for hot fermions on the FS, and the
   competition implies that the order, which sets up first, tend to suppress the other one.
   In the spin-fluctuation approach, the value of $T_{\rm sc}$  is larger than $T_{\rm cdw}$, but the two are of the same order and comparable in magnitude
    The values of $T_{n}$ and $T_t$ are larger than $T_{\rm cdw}$, and we assume that  at large $\xi$, we have $T_{n},T_t > T_{\rm sc}$, i.e., the composite charge orders set up first
    upon the lowering of $T$.
   The composite order suppresses $T_{\rm sc}$ and gives rise to a non-monotonic behavior of $T_{\rm sc} (\xi)$ already in the paramagnetic phase.
   At the same time, it increases the correlation length for the primary CDW order parameter~\cite{rafael}, i.e, the composite order
    tends to increase $T_{\rm cdw}$. At larger $\xi$, $T_{\rm cdw}$ then well may become larger than the reduced $T_{\rm sc}$, in which case charge order
    develops prior to superconductivity.  At the lowest $T$, our calculations in Sec.\ IV show that  CDW and superconducting  orders
    co-exist. The  phase diagram for state I is shown in Fig.\ \ref{phases}(c).
     It has a number of features consistent with the experimental data on hole-doped
   cuprates.
   Namely, the theoretical phase diagram contains regions of SDW and $d$-wave superconductivity, and also a region with a nematic order, a region where time-reversal symmetry is broken, and a region of a true CDW order. The CDW order at $T=0$ co-exists with superconductivity and terminates at a CDW quantum-critical point QCP 2, distinct from the magnetic quantum-critical point QCP 1.  It is tempting to associate the $T_{n}$ line with the onset of nematic order
   seen in neutron
scattering~\cite{labacuo,stripes,hinkov,davis_1} and in Nernst experiments~\cite{nernst}, associate $T_t$ line with the onset temperature for the
 Kerr effect~\cite{kerr}, intra-unit cell magnetic order~\cite{bourges, greven},
  the magneto-electric birefringence~\cite{armitage}, and associate $T_{\rm cdw}$ with the onset temperature of  CDW order~
 \cite{ybco,ybco_1,X-ray,X-ray_1,wu,mark,ultra}, perhaps pinned by impurities\cite{mark_last,steve_last}. In our model calculations, the nematic transition temperature
 $T_{n}$ is larger than the onset temperature $T_t$ for time-reversal symmetry breaking. In general, the two temperatures are comparable, and the position of $T_{n}$ and $T_t$ lines on the phase diagram may depend on the type of material.
  The association of $T_t$ with these three experiments requires care because, as we said in the Introduction, Kerr effect does not change sign in a magnetic field
    over $10T$ (Ref.\onlinecite{kerr}) and linear birefringence is often associated with the breaking of a mirror symmetry rather than with breaking of time-reversal~\cite{armitage}. To address this issue in more detail one needs to study 3D systems, particularly the arrangements of the charge currents between neighboring layers.

\section{Comparison with ARPES data}
\label{sec:6}

In this section we discuss in some detail the comparison between our theory and ARPES data.
 The data
 on the fermionic spectral function in the pseudogap region all show~
 \cite{zxshen,arp,arp_1,arp_2,arp_3,mike_arc} that below a certain $T > T_{\rm sc}$,
  the spectral weight in the antinodal
 regions transforms from the FS to high frequencies, and the FS looks like a set of four disconnected
  Fermi arcs. We show below that this is an expected behavior for a
  system with strong CDW fluctuations, but without a true CDW order.

A generic charge order with an ordering momentum $Q$ introduces a new term $H'=\Delta_{k}^Q c^{\dagger}_{k+Q} c_{k-Q} +h.c.$ into the Hamiltonian. Then
fermions with momenta
$k \pm Q, k \pm 3Q, k \pm 5Q, ...$
 all become coupled. For commensurate
 $Q = \pi M/(N)$, where $M$ and $N$ are integers,
 the ``chain" of coupled momenta gets closed when after $N$ steps, for incommensurate $Q$ it is not closed, but for practical purposes
 one can approximate $Q$ by a close commensurate value.
To diagonalize such a Hamiltonian one has to solve a $N$-dimensional matrix equation
 ~\cite{subir_2}.
 The energy eigenstates with eigenvalues $E_1, E_2, \cdots, E_N$ are  linear combinations of the original fermions,
\begin{align}
\(\begin{array}{c}
d_1\\
d_2\\
\vdots\\
d_N\end{array}\)=
\(\begin{array}{c c c c}
u_{11}&u_{12}&\cdots&u_{1N}\\
u_{21}&u_{22}&\cdots&u_{2N}\\
\vdots&\vdots&\ddots&\vdots\\
u_{N1}&u_{N2}&\cdots&u_{NN}
\end{array}\)\(\begin{array}{c}
c_k\\
c_{k-2Q}\\
\vdots\\
c_{k-2
(N-1)
Q}\end{array}\),
\label{bb_2}
\end{align}
The ARPES spectral function measures the correlator of $c-$fermions and contains contributions from all eigenstates,  with different weights
\begin{align}
I(\omega,k)&\propto{\rm Im}(\langle c_k (\omega) c^{\dagger}_{k} (\omega)\rangle) \nonumber \\
&={\rm Im}\left(\sum_iu_{i1}^2\langle d_i (\omega)d_i^\dagger(\omega)\rangle\right) ={\rm Im}\left(\sum_i\frac{u_{i1}^2}{\omega-E_{i}-i\Gamma}\right),
\end{align}
We keep the damping term $\Gamma$  finite to model the state in which CDW fluctuations are well-developed but a true CDW order does not yet
occur~\cite{elihu}.

 We use this procedure to obtain the spectral function $I(\omega,k)$ at $\omega=0$, as a function of $k$ for ``damped" stripe CDW order with either $Q=Q_x$ or
 $Q=Q_y$.
   The position of the peak in this spectral function yields the location of the reconstructed FS in the CDW-ordered state, a $\Gamma$ gives a finite width to
    the peak.
    In a macroscopic system, there exist domains with stripes in both directions, and
     we assume that the measured
      ARPES intensity is the sum of $I(\omega,k)$ for $Q=Q_x$ and $Q=Q_y$.

\begin{figure}
\includegraphics[width=0.5\columnwidth]{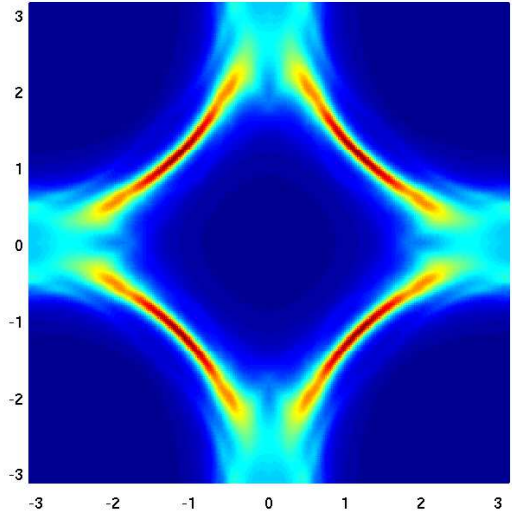}
\caption{The theoretical spectral function at  $\omega =0$ for a state with strong CDW fluctuations, which we model by introducing CDW orders
  $\Delta_x$ and $\Delta_y$, but keeping a finite lifetime of fermions on the FS.
    The Fermi arcs, terminating at hot spots,  are clearly visible.}
\label{FArc}
\end{figure}

 We show our result for the photoemission intensity $I(0,k)$ in Fig.\ \ref{FArc}.
 The Fermi arcs, terminating at hot spots,  are clearly visible. The actual FS's in the CDW-ordered state indeed cannot terminate inside the BZ,
but other pieces of the FS  have small spectral weights and are washed out by a finite $\Gamma$.  In the calculations we used
the dispersion from Ref.\ [\onlinecite{zxshen}]: $\epsilon(k_x, k_y)=-2t(\cos{k_x}+\cos{k_y})-4t^{'} (\cos{k_x} \cos{k_y})-2 t^{''}
(\cos{2k_x}+\cos{2k_y})-4 t^{'''}(\cos{2k_x} \cos{k_y}+\cos{k_x} \cos{2k_y})-\epsilon_0$, with
$t =0.22 {\rm eV}$, $t^{'} = -0.034315 {\rm eV}$, $t^{''} = 0.035977 {\rm eV}$, $t^{'''} = -0.0071637 {\rm eV}$ , and we took $\epsilon_0 = - 0.24327 {\rm
eV}$, slightly different
 from $-0.240577 {\rm eV}$ in~[\onlinecite{zxshen}], to get a commensurate  $2Q=0.2\pi$ instead of $2Q \approx 0.19 \pi$ in~[\onlinecite{zxshen}]. We then
 used $N=10, M=1$,  and set $\Gamma = 50 {\rm meV}$.

\begin{figure}
\includegraphics[width=0.5\columnwidth]{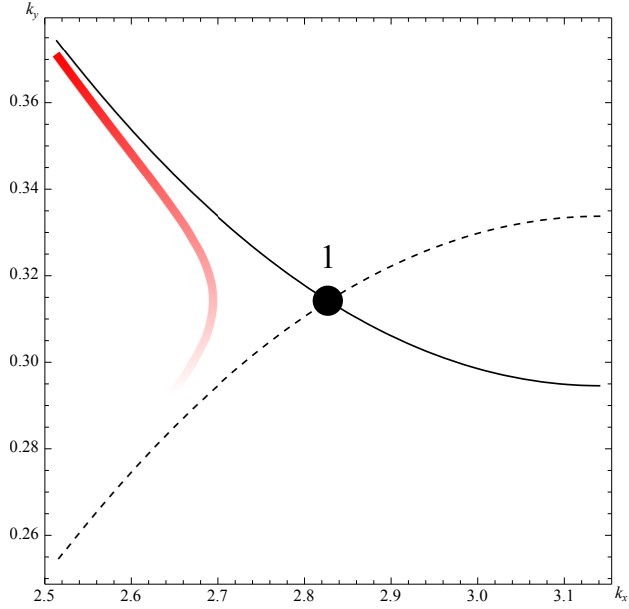}
\caption{The position of the peak in spectral function at $\omega=0$ around the hot spot 1 (red line).
The saturation of color indicates the spectral weight of the peak. The spectrum to the right of the hot spot 1 is pushed out of BZ boundary by large enough
CDW order parameter $\Delta$ used in the plot.}
\label{fig13}
\end{figure}

The appearance of the arcs can be understood analytically. Consider $Q=Q_y$ and focus on the region around hot spot 1 in Fig.\ \ref{Fig9}, with momenta
 near $(\pi-Q,Q)$.  One can easily verify that the most relevant momenta involved in CDW-induced mixing
 are  $(\pi-Q,Q)$ and $(\pi-Q,-Q)$, since for the other momenta in (\ref{bb_2}) either the gap is smaller, or the states are away from the FS.
  The effective $2\times 2$ Hamiltonian $H=H_0+H'$ can then be diagonalized by the standard Bogoliubov transformation. Defining $c_1=c_k$,
   $c_2=c_{k-2Q}$, $\epsilon_1=\epsilon_k$, $\epsilon_2=\epsilon_{k-2Q}$, and $\Delta=|\Delta_{k_0}^Q|$, with $k_0 = (\pi-Q,0)$
   we obtain
\begin{align}
\(\begin{array}{c}d_+\\d_-\end{array}\)=\(\begin{array}{c c}u&v\\-v&u\end{array}\)\(\begin{array}{c}c_1\\ c_2\end{array}\),
\end{align}
where
\begin{align}
u^2=&\frac{1}{2}\left[1+\frac{\epsilon_1-\epsilon_2}{\sqrt{(\epsilon_1-\epsilon_2)^2+4\Delta^2}}\right]\nonumber\\
v^2=&\frac{1}{2}\left[1-\frac{\epsilon_1-\epsilon_2}{\sqrt{(\epsilon_1-\epsilon_2)^2+4\Delta^2}}\right].
\label{uv}
\end{align}
The nergy eigenvalues are
\begin{align}
E_{\pm}=\frac{\epsilon_1+\epsilon_2}2\pm\sqrt{\(\frac{\epsilon_1-\epsilon_2}2\)^2+\Delta^2}.
\label{epm}
\end{align}
the ARPES spectral function at $\omega=0$ is
\begin{align}
I(\omega=0,k)\propto{\rm Im}(\langle c_1 c_1^\dagger\rangle) ={\rm Im}\left(\frac{u^2}{E_+-i\Gamma}+\frac{v^2}{E_--i\Gamma}\right).
\label{I}
\end{align}
The  peaks in the momentum distribution curves  are at $E_{\pm}=0$, which correspond to $\epsilon_1\epsilon_2=\Delta^2$.  This condition defines a
hyperbola
 in the momentum space around the hot spot 1, as shown in Fig.\ \ref{fig13}. The solid and dashed lines in this figure are
 the original FS $\epsilon_k =0$ and the ``shadow" FS $\epsilon_{k-2Q} =0$. At small $\Delta$, there is another part of the FS, to the right of point 1 in
 this figure,  but for large enough $\Delta$, used in the plot, this part is pushed out the BZ boundary.
 The spectral weight along the red line in Fig.\ \ref{fig13} depends on coherence factors and is much larger for the part which is close to the original FS
 than for the part close to the shadow FS.
 As a result, the only visible
 spectral peak at $\omega=0$ in the momentum space is along the former FS $\epsilon_k =0$, and it effectively terminates at the hot spot, as in Fig.
 \ref{fig13}.
 The contribution from the domain with $Q=Q_x$ is obtained in a similar manner, and the full result is the spectral function with the largest intensity at
  four Fermi arcs, as in Fig.\ \ref{FArc}.

  The Fermi arcs in the disordered CDW state and the Fermi pocket in the ordered CDW state, whose position and size are consistent with quantum oscillation measurements~\cite{suchitra,suchitra_last} have been recently obtained in the analysis~\cite{sad},
   similar to the one we presented here, but extended to the full Brillouin zone.

We next consider the dispersion along the BZ boundary in the antinodal (AN) regions. Experiments \cite{zxshen} performed on
$\rm{Pb_{0.55}Bi_{1.5}Sr_{1.6}La_{0.4}CuO_{6+\delta}}$ (Pb-Bi2201) have detected two prominent features:  (1) upon cooling below about the same temperature where arcs appear, the measured dispersion evolves into a band which comes towards a FS and then moves away from the FS, (2)
 the momentum, at with
 the reconstructed dispersion has a minimum, shifts from $k_F$ to a larger value $k_G$, (3)  once the system is further cooled down below  $T_{\rm sc}$,  a weak, ``shoulder"-like peak in appears in the energy
 distribution curve  at the binding energy $\omega\sim 25~{\rm meV}$.
  We find that all these features can be accounted for within our theory.

\begin{figure}
\includegraphics[width=0.5\columnwidth]{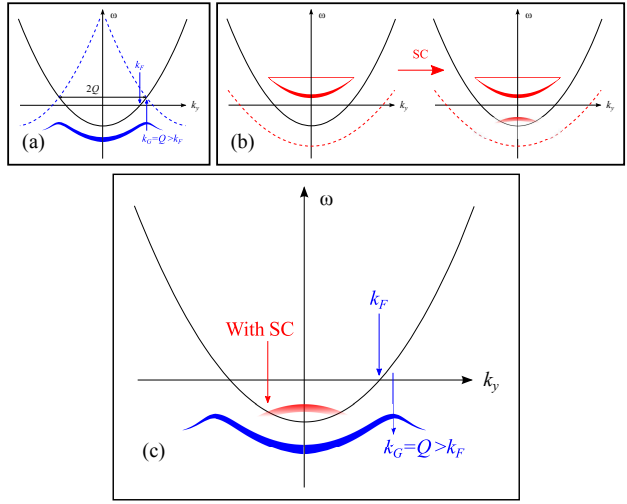}
\caption{Interpretation of the ARPES data around antinodal areas. Panel (a) -- contribution from the domain I with $Q=Q_y$.
 Panel (b) -- contribution from the domain II with $Q=Q_x$. Panel (c) -- The combined spectral peaks from both domains seen by ARPES. }
\label{ARPES}
\end{figure}

Because the features are at finite energy, we can safely neglect $\Gamma$ and compute ARPES dispersion assuming a true CDW order.
However, we still need to consider two domains: domain I with CDW order with ${\bf Q} =Q_y$ and domain II with CDW order with ${\bf Q} = Q_x$. For simplicity, we
will assume both CDW gaps  can be approximated by  constants, in which case  $\Delta_x=\mu\Delta_y$, with $\mu>1$. Because typical energy scale for
the fermionic dispersion in the AN region is much smaller than the bandwidth, we  again can neglect high-energy electronic
states. A simple analysis shows that for low-energy consideration it is sufficient to include three states with momenta $k, k+2Q_{y,x}, k-2Q_{y,x}$. We
show this in Fig.\ \ref{ARPES} (a) and \ref{ARPES} (b).

In domain I, the two states with momenta $k$ and $k+2Q_y$ cross at a small positive energy $\delta\epsilon$
  at $k_x = k_G = Q$, which is larger than the original $k_F$ simply because the distance between the two neighboring hot spots (one on top of the other) is larger  than
  the distance between the two points $(k_F,\pi)$ and $(-k_F,\pi)$, at which the FS crosses BZ boundary.
    The energy of the state with momentum $k-2Q_y$ is much larger in this
 region, so  we can further reduce the three-state system  to a two-state system. The energy eigenvalues at the crossing point are
  $E_{1,2} = \delta\epsilon\pm\Delta_y$. Once $\Delta_y$ exceeds $\delta \epsilon$, one of the energies, $E_1=\delta\epsilon-\Delta_y$, becomes negative, and the
  corresponding state becomes visible by ARPES.
  Evaluating $E_1 (k_x)$  at different $k_x$, we find that $E_1 (k_x)$ initially follows the original dispersion and moves towards zero,
  but deviates from the bare dispersion as $k_x$ approaches $k_F$,
   passes through a minimum at some finite negative energy, and then moves away from the Fermi level. We show this in Fig.
  \ref{ARPES} (a).
   The minimum of the reconstructed dispersion is right at $k_G> k_F$, where the two unreconstructed
   states with momenta $k$ and $k+2Q_y$ cross.  That the minimum of the reconstructed dispersion is at momentum larger than the original $k_F$
   is consistent with the experiment~\cite{zxshen}.
    One can easily make sure that the shift of the minimum to $k_G > k_F$
     is the consequence of the fact that the momentum of CDW order is along $Q_y$ or $Q_x$.  If CDW order parameter
   was with ${\bf Q}$ along the zone diagonal, the result would be the opposite -- the position of the local minimum would
   shift to a smaller momentum. This is yet another indication that CDW order does emerge with ${\bf Q} = (2Q,0)$ or $(0,2Q)$ rather than with the diagonal
  $(2Q, \pm 2Q)$.  Note in this regard that $k_G$ would remain equal to $k_F$ if the reconstruction of the fermionic dispersion was due to precursors of superconductivity.

We used the experimental value of $\delta\epsilon=5$ meV, and set $\Delta_y = 35$ meV to
 match the energy of the local minima at $k_G$ at $\omega=30$ meV, as in~[\onlinecite{zxshen}].

In domain II, with CDW order with $Q_x$, two out of three states are degenerate, and we define $\epsilon_{\pi+2Q,k_y}=\epsilon_{\pi-2Q,k_y} \equiv \epsilon_b(k_y)$. We also define
$\epsilon_a(k_y)=\epsilon_{\pi,k_y}$. Solving  3-by-3 matrix equation on eigenvalues and eigenfunctions,  we find that the dominant contribution
to the spectral function comes from the peak at $E= E_{11}=(\epsilon_a+\epsilon_b)/2+\sqrt{[(\epsilon_a-\epsilon_b)/2]^2+2\Delta_x^2}>\epsilon_a$. The value of $E_{11}$ is positive for the set of parameters which we used.
  Because ARPES can only detect filled states,  the peak at $E=E_{11} >0$ is invisible to ARPES.
In other words, in the normal state, the full dispersion, measured by ARPES, comes from domain I.

  Once superconductivity sets in at $T_{\rm sc}$,  electron and hole states get mixed up,  and the system develops a shadow image of $E_1$ at a negative energy,
  at $-\sqrt{E^2_{11} + \Delta^2_{\rm sc}}$. The superconducting gap $\Delta_{\rm sc}$ is rather small in Pb-Bi2201, hence the image is  approximately at
  $-E_{11}$.
  We show this in Fig.\ \ref{ARPES} (b).  The emergence of the new band below $T_{\rm sc}$
  is again consistent with  the
  experiment~\cite{zxshen}.
  To match the measured position of the new band at around $25$ meV, we use experimental values of $\epsilon_a(k_y=0)=-38$ meV, $\epsilon_b (k_y=0)=-59$ meV, and set $\Delta_x=51$ meV, larger than $\Delta_y =35$ meV. This is consistent with theoretical
 $\Delta_x = \mu \Delta_y$ and $\mu >1$.

In Fig.\ \ref{ARPES} (c) we show the combined peaks from both domains. This is our theoretical result for the spectral function for comparison with  ARPES.
In our view, the theoretical spectral function is quite consistent with the data.

The above analysis is valid in the vicinity of hot spots. More efforts are needed to see whether our CDW order is compatible with ARPES along the cuts
 away from hot regions, particularly near zone diagonals. The data in these regions have been successfully fitted by the theory based on pair-density-wave
 scenario~\cite{palee}.  To compare with the data within our CDW scenario one needs to go beyond what we did so far and to solve for CDW order parameter
 outside  hot regions. This would require full model calculations on a lattice.

 \section{Conclusions}
\label{sec:7}

The goal of this paper was
 two-fold. First, to understand whether spin-fluctuation approach, which describes $d$-wave superconductivity and non-Fl physics in the normal state, also
 allows one to understand the development of charge order in hole-doped cuprates. Second, to study the structure of charge order parameter and
 potential preemptive instabilities which break discrete symmetries but leave a continuous $U(1)$ phase symmetry intact.  We argued that
 magnetically-mediated interaction gives rise to  charge order with momenta
  $Q_x =(2Q,0)$ and $Q_y =(0,2Q)$, as seen in the experiments. The critical temperature for the onset of the charge order $T_{\rm cdw}$ is comparable to
  superconducting $T_{\rm sc}$ at large values of magnetic correlation length $\xi$ and the ratio $T_{\rm cdw}/T_{\rm sc}$ tends to one at $\xi = \infty$.
   At the same time, as $\xi$ decreases with increasing doping, the ratio $T_{\rm cdw}/T_{\rm sc}$ decreases, and at some finite doping
     $T_{\rm cdw}$ vanishes,
    setting up the second quantum critical point at some distance away from  the magnetic one.

   Our most significant observation is that CDW order parameter $\Delta^Q_k$ with a given ${\bf Q}$, say, $Q_y$  has two components, one ($\Delta^Q_{1}$) is an even function
    of the
    center of mass momentum $k$, another ($\Delta^Q_{2}$),  is an odd function of $k$. In real space, these two components describe, respectively, an incommensurate site and bond
   density variation $\delta \rho
   (r) \propto \cos 2Q y$, and an incommensurate current $j_x (r) \propto \sin 2Q y$.
    We derived and analyzed the full GL functional for four CDW components $\Delta^{Q_x}_1, \Delta^{Q_x}_2$, $\Delta^{Q_y}_1, \Delta^{Q_y}_2$
     first in the mean-field approximation and then beyond mean-field. Within mean-field, we found two CDW states -- state I and state II. Which of the two states is realized depends on the interplay between two system parameters, which are comparable to each other and which we only know approximately.
     The state I emerges via a continuous transition and is of stripe type:  non-zero CDW components have either ${\bf Q} =Q_x$ or $Q_y$. Both $\Delta_1$ and $\Delta_2$ are non-zero, and the relative phase between these two $U(1)$ fields is locked at $\pm \pi/2$. The full order parameter manifold for state I is $U(1) \times Z_2 \times Z_2$ where one $Z_2$
 is associated with the choice between $Q_x$ and $Q_y$, another $Z_2$ with the choice between $\pi/2$ and $-\pi/2$ for phase locking, and $U(1)$ is the symmetry with respect to the common phase of $\Delta_1$ and $\Delta_2$. To obtain a phase with this order parameter manifold it was essential to include the center-of-mass momentum dependence of $\Delta^Q_k$ for $k$ near the mid-point between neighboring hot spots.

  For state II, the CDW order emerges via a strong first-order transition, and in the ordered state
  $|\Delta^{Q_x}_1| = |\Delta^{Q_y}_2|$ and $|\Delta^{Q_y}_1| = |\Delta^{Q_x}_2|$. There are two realizations of state II and the choice is
   dictated by the interplay between parameters which become non-equal only due to $k-$dependence of $\Delta^Q_k$. One realization is the checkerboard
   order (all four $\Delta$'s are non-zero and equal by magnitude), another is a stripe state with only two non-zero components, say,
    $\Delta^{Q_x}_1$ and $\Delta^{Q_y}_2$. For both realizations, the order parameter manifold is $U(1) \times Z_2$, where in the first realization $Z_2$
     is associated with the phase locking at $\pm \pi/2$ between $\Delta_1$ and $\Delta_2$ (same locking for $Q_x$ and $Q_y$ components), and in the second
       $Z_2$ is associated with the symmetry between choosing $\Delta_1$ with $Q_x$ or $Q_y$.

   We focused on the state I because it emerges via a continuous transition and analyzed the GL action beyond mean-field. Our goal was to understand whether the two $Z_2$ Ising symmetries can be broken at higher temperatures than $T_{\rm cdw}$ at which $U(1)$ symmetry gets broken. We used HS approach, introduced composite fields conjugated to composite  order parameters, which order if the corresponding $Z_2$ symmetry gets broken,  integrated over the primary $\Delta$ fields, and analyzed the resulting effective action for the composite fields. We found that each of $Z_2$ degrees of freedom gets broken before a true CDW order sets in.
 We found that at a highest $T_{n}$ a nematic order sets in, i.e., the system selects $Q_x$ or $Q_y$, while $U(1)$ phase symmetry remains intact. Then, at $T_t \leq T_{n}$, another $Z_2$ symmetry gets broken, and the relative phase between $\Delta_1$ and $\Delta_2$ gets locked at $\pi/2$ or $-\pi/2$, while $U(1)$ symmetry of the common phase of $\Delta_1$ and $\Delta_2$ remains unbroken.
  In real-space picture, below $T_t$ both density and current components fluctuate, such that $\langle\delta \rho (r) \rangle  =  \langle
    j_x (r) \rangle =0$, however their fluctuations are correlated, and
      $\langle\delta \rho (r) j_x (r) \rangle$ is non-zero. Such an order breaks time-reversal symmetry. Finally, at $T_{\rm cdw} < T_t$, $U(1)$ symmetry gets broken and a true CDW order sets in.

The existence of the preemptive order is the crucial element in our scenario.  Without it, CDW instability would be subleading to $d$-wave superconductivity
 and also to bond order with diagonal ${\bf Q}_d = (2Q, \pm 2Q)$ as in the mean-field approximation both have slightly larger onset temperatures than $T_{\rm cdw}$ ($T_{\rm sc} \geq T_{\rm bo} \geq T_{\rm cdw}$).
 However, superconducting order parameter and  order parameter for bond charge order do not break $C_4$ lattice rotational symmetry and
  have only one, even in $k$, component. Accordingly, there are no preemptive
   instabilities for these orders. Because $T_{\rm sc}, T_{\rm bo}$ and $T_{\rm cdw}$ are close to each other at large $\xi$ and  $T_{n}, T_t > T_{\rm cdw}$, it is likely that
     they also exceed $T_{\rm sc}$ and $T_{\rm bo}$, in which case
        the first instability upon  lowering of $T$ is into a state with a composite CDW order.
          Once composite order forms, it reconstructs fermionic excitations and tends reduce the onset temperatures for superconductivity/bond-order because
   composite charge order and superconductivity/bond-order compete for the FS. At the same time, a composite CDW order
 increases the susceptibility for the primary CDW fields and hence increases $T_{\rm cdw}$,  much like a spin-nematic order in Fe-pnictides increases the
 Neel temperature of SDW order~\cite{rafael}.  An increase  of $T_{n}$ and $T_t$ compared to the onset of superconductivity/bond-order becomes even stronger once we
 include into consideration 2D fluctuation effects, because  near-degenerate $d$-wave
 superconductivity and bond order form weakly anisotropic $O(4)$ model, in which $T_{\rm sc}$ is strongly
 reduced by fluctuations from $O(4)$  manifold.

   The phase diagram resulting from our analysis is shown in Fig.\ \ref{phases}c.  It has numerous features consistent with the experiments on hole-doped
   cuprates.
   We performed a more detailed comparison with ARPES studies and found good quantitative agreement with the data.

   Overall, we believe that the most
   significant result of our theory is that it shows that pseudogap physics can be well understood within the same spin-fermion model which was earlier
   shown to yield $d$-wave superconductivity and non-Fermi liquid physics.  We believe that, with our result, the spin-fermion model re-emerges as the
   strong candidate for the theoretical model for the cuprates.

Several issues are not covered by this analysis and are left for further study.
One issue is
 the interplay between our spin-fluctuation scenario and the one based on microscopic analysis of charge fluctuations~\cite{castellani}.
 Another issue, specific to our model, is
to what extend $T_{\rm cdw}$ for a true CDW order and $T_{n}, T_t$ for preemptive transitions vary between different families of hole-doped
 cuprates. The third issue is the detailed analysis of the relation between our composite charge order which breaks time-reversal symmetry,  and Kerr effect~\cite{kerr} and
neutron scattering results from Refs.\ [\onlinecite{bourges,greven}]. And the fourth issue is the interplay between our incommensurate charge order and
incommensurate pair-density-wave (PDW) order discussed in Refs.\ [\onlinecite{fra_kiv,steve_k,palee}]. The two orders are ``cousins" in the same way as SC and diagonal bond order are.
Whether fluctuations between our CDW order and PDW order further complicate the phase diagram remains to be seen.

Note added (06/27/2014) The phenomenological GL model for PDW order parameter has been considered in a very recent preprint by D. Agterberg and M. Kashuap~\cite{dan}.
They introduced two order parameters ${\bar \Delta}^{\bar Q}_p = c^\dagger_{p+{\bar Q},\alpha}(i\sigma^y_{\alpha\beta}) c^\dagger_{-p+{\bar Q},\beta}$
with ${\bar Q}$ along $x$ and $y$ directions in momentum space, and argued that ${\bar \Delta}^{\bar Q}_p$ and ${\bar \Delta}^{-\bar Q}_p$ are not necessary the same.
The ``cousins" CDW and PDW order parameters transform into each other by changing one $c$ into $c^\dagger$ and replacing spin dependence $\delta_{\alpha\beta}$ for CDW into $i\sigma^y_{\alpha\beta}$ for PDW, but {\it without} changing the momentum. A cousin of our CDW order parameter
$\Delta^Q_k = c^\dagger_{k +Q,\alpha} \delta_{\alpha\beta} c_{k -Q,\beta}$ with the enter of mass momentum $k$ is PDW order $c^\dagger_{k+Q,\alpha}(i\sigma^y_{\alpha\beta}) c^\dagger_{k-Q,\beta} \equiv {\bar \Delta}^{2k}_{Q}$ with the total momentum ${\bar Q} =2k$. The orders ${\bar \Delta}^{\bar Q}_p$ and  ${\bar \Delta}^{-\bar Q}_p$  are then cousins of our $\Delta^Q_k$ and $\Delta^Q_{-k}$, and the non-equivalence of ${\bar \Delta}^{\bar Q}_p$ and  ${\bar \Delta}^{-\bar Q}_p$ explored in [\onlinecite{dan}]  is the  PDW
 analog of  the non-equivalence between $\Delta^Q_k$ and $\Delta^Q_{-k}$, which we explored in this paper.  Agterberg and Kashuap also identified an additional $Z_2$ component of the order parameter manifold, associated with time-reversal symmetry, and argued that $Z_2$ can be broken at a higher $T$ than the one at which a true PDW order develops.

\subsection{Acknowledgement}

We thank Ar.\ Abanov,  E.\ Abrahams, D.\ Agterberg, H.\ Alloul, P. \ Armitage, W.A.\ Atkinson, D.\ Basov, E.\ Berg, Ph.\ Bourges,  D.\ Chowdhury, C.\ Di\ Castro, P.\ Coleman,
A.\ Finkelstein, K.\ Efetov, R.\ Fernandes, P.\ Goswami, M.\ Greven, M.\ Grilli, M-H.\ Julien, A.\ Kapitulnik,
B.\ Keimer, S.\ Kivelson, D.\ Le\ Boeuf, P.\ A.\ Lee, S.\ Lederer, H.\ Meier, E.\ Mishchenko, M.\ Norman, A.\ Paramekanti, C.\ P\'epin,  N.\ Perkins, D.\ Pesin, D.\ Reznik, S.\ Sachdev, D.\ Scalapino, J.\ Schmalian, S.\ Sebastian, O.\ Starykh, L.\ Taillefer, A.\ Tsvelik, P.\ W\"olfle, and V.\ Yakovenko
 for fruitful discussions. The work was supported by the DOE grant DE-FG02-ER46900.

\appendix

\section{Gap equations for momentum and frequency-dependent CDW order parameters}
\label{app:a}

In this Appendix we present  the full linearized  equations for the CDW order parameters
 as integral equations in momentum and frequency.
We measure frequency and temperature in units of
$\omega_0$:  $\Omega^*_m = \Omega_m/\omega_0$, $\Sigma^* = \Sigma/\omega_0$, $T^* = T/\omega_0$, and measure momentum in units of $(\gamma \omega_0)^{1/2}
= 3{\bar g}/(2\pi v_F)$: $x^* = x/(\gamma \omega_0)^{1/2}, y^* =y/(\gamma \omega_0)^{1/2}$.
We recall that $\omega_0 = 9 {{\bar g}}/(16 \pi ) \times[(v_y^2-v_x^2)/v^2_F]$, where ${\bar g}$ is spin-fermion coupling.
 We keep the momentum dependence of $\Delta^Q_{k} (\Omega_m)$
along the FS and
 neglect the momentum dependence transverse to the FS. We consider the FS geometry as in Fig.\ \ref{fig3} and, to avoid too lengthy formulas,
   consider the limit $v_x =0$ when Fermi velocities at hot spots 1
 and 2 are anti-parallel and the ones at hot spots 3 and 4 are parallel.
  In this limit,  the momentum dependence along the FS is along $x$ axis for $\Delta^Q_{k_0}$ and
 along $y$ axis for $\Delta^Q_{k_\pi}$.      Integrating over momentum transverse to the FS we obtain
 \begin{align}
\Delta_{k_0}^{Q}(\Omega^*_{m}, x^*)=&-\frac{3 T^*}{8}\sum_{|\Omega^*_{m'}|<1} \int_{0}^{\infty}dy^*
\frac{\Delta_{k_\pi}^{Q}(\Omega^*_{m'},y^*)}{\sqrt{(y^*)^2+
|\Omega^*_{m}-\Omega^*_{m'}|}}\nonumber\\
&\times\frac{\left[x^* {\rm sgn}~(\Omega^*_{m'})+i\(\sqrt{(y^*)^2+|\Omega^*_{m}-\Omega^*_{m'}|}+\frac{3}{8}|{\tilde \Sigma}^*
(\Omega^*_{m'},y^*)|\)\right]^2}{\left[(x^*)^2 + \left(\sqrt{(y^*)^2+|\Omega^*_{m}-\Omega^*_{m'}|}+\frac{3}{8}|{\tilde \Sigma}^*
(\Omega^*_{m'},y^*)|\right)^2\right]^2} \label{Eq2aaa}\\
\Delta_{k_\pi}^{Q}(\Omega^*_{m},y^*)=&- T^*\sum_{|\Omega^*_{m'}|<1} \int_{0}^{\infty} dx^*
\frac{\Delta_{k_0}^{Q}(\Omega^*_{m'},x^*)}{\sqrt{(x^*)^2+|\Omega^*_m -\Omega^*_{m'}|}} \nonumber\\
&\times\frac{\sqrt{(x^*)^2+|\Omega^*_{m}-\Omega^*_{m'}|}+\frac{3}{8}|{\tilde \Sigma}^* (\Omega^*_{m'},x^*)|}{|{\tilde \Sigma}^*
(\Omega^*_{m'},x^*)|\[(y^*)^2 + \left(\sqrt{(x^*)^2+|\Omega^*_{m}-\Omega^*_{m'}|}+\frac{3}{8}|{\tilde \Sigma}^* (\Omega^*_{m'},x^*)|\right)^2\]}
\label{Eq2bbb}
\end{align}
where ${\tilde \Sigma}^* = \Omega^*_m + \Sigma^*$ and $G^{-1} (k, \omega) = i {\tilde \Sigma} (k,\omega) - \epsilon_k$. For the self-energy
we obtain
\begin{align}
\Sigma^* (\Omega^*_{m'},x^*) =&
T^*\sum_{|\Omega^*_{m'}|<\omega_0} \int_{0}^{\infty} dy^*  \frac{1}{\sqrt{(y^*)^2+
|\Omega_{m}-\Omega_{m'}|}} \nonumber \\
\times& \frac {{~{\rm sgn}}~(\Omega^*_{m'}) \left(\sqrt{(y^*)^2+|\Omega_{m}-\Omega_{m'}} +\frac{3}{8}|{\tilde \Sigma}^* (\Omega_{m'},y^*)|\right)+ix
}{(x^*)^2+\left(\sqrt{(y^*)^2+|\Omega_{m}-\Omega_{m'}|}+\frac{3}{8}|{\tilde \Sigma}^* (\Omega_{m'},y^*)|\right)^2}  \label{Eq2ccc}
\end{align}

The $ix$ term numerator is even in $\Omega^*_{m'}$ and renormalizes the Fermi velocity. We follow the standard procedure and incorporate this term into the bare dispersion. The expression for $\Sigma^* (\Omega^*_{m'},y^*)$ is obtained from (\ref{Eq2ccc}) by interchanging $x^*$ and $y^*$.

For large $\Sigma^* (\Omega^*_{m'},x^*) \propto (T \log \xi)^{1/2}$, the dependence on momentum in the self-energy and in the CDW order parameters can be
neglected, i.e.,  $\Delta_{k_0}^{Q}(\Omega^*_{m}, x^*) \approx \Delta_{k_0}^{Q}(\Omega^*_{m})$,
$\Delta^Q_{k_\pi}(\Omega^*_{m}, y^*) \approx \Delta_{k_\pi}^{Q}(\Omega^*_{m})$,  $\Sigma^* (\Omega^*_{m'},x^*) \approx  \Sigma^* (\Omega^*_{m'})$.
In this approximation, Eqs.\ (\ref{Eq2aaa}) - (\ref{Eq2ccc}) reduce to Eqs.\ (\ref{Eq1a}) - (\ref{frac_0})
  from the main text.  In general, however,
 the gap equations are integral equations in both momentum and frequency.  Moreover, at deviations from hot spots 1 and 2 along the FS $\Delta^Q_{k_0,
 x^*}$ acquires an imaginary part, which is also odd in frequency: $\Delta_{k_0}^{Q}(\Omega^*_{m}, x^*) = \Delta_{0,a} + i \Delta_{0,b} x^* {\rm sgn} \omega$, where $\Delta_{0,a}$
 and $\Delta_{0,b}$ are even functions of momentum and frequency.   To match this behavior, $\Delta^Q_{k_\pi}$ also acquires an imaginary part, odd in frequency,
 but at deviations from hot spots 3 and 4 transverse to the FS:
 $\Delta^Q_{k_\pi} (\Omega^*_{m}, x^*,y^*) = \Delta_{\pi,a} + i \Delta_{\pi,b}
    x^* {\rm sgn} \omega$, where $\Delta_{\pi,a}$ and $\Delta_{\pi,b}$ are again even functions of momentum and frequency.

In the same approximation, the linearized equation for $d-$wave superconducting order parameter $\Delta_{\rm sc} (\Omega^*_m, y^*)$ is
\begin{align}
\Delta_{\rm sc}(\Omega^*_{m},y^*)=&T^*\sum_{|\Omega^*_{m'}|<1} \int_{0}^{\infty} dx^* \frac{\Delta_{\rm sc} (\Omega^*_{m'},x^*)}{\sqrt{(x^*)^2+|\Omega^*_m
-\Omega^*_{m'}|}}\nonumber\\
&\times\frac{\sqrt{(x^*)^2+|\Omega^*_{m}-\Omega^*_{m'}|}+\frac{3}{8}|{\tilde \Sigma}^* (\Omega^*_{m'},x^*)|}{|{\tilde \Sigma}^*
(\Omega^*_{m'},x^*)|\[(y^*)^2 + \left(\sqrt{(x^*)^2+|\Omega^*_{m}-\Omega^*_{m'}|}+\frac{3}{8}|{\tilde \Sigma}^*
(\Omega^*_{m'},x^*)|\right)^2\]}\label{Eq2ddd}
\end{align}
The equation for rescaled gap function
${\bar \Delta}_{\rm sc} (\Omega_m^*) = \Delta_{\rm sc}(\Omega^*_{m}) /\Sigma^* (\Omega_m^*)$ for the most generic case, when
 (i) the gap depends on momentum along the FS and (ii) fermionic self-energy is not assumed to be larger than other terms in the pairing kernel, is
     \begin{align}
     {\bar \Delta}(\Omega^*_{m},x^*) =& T^* \sum_{|\Omega^*_{m'}|<1}
     \int_{0}^{\infty}\frac{dy^*}{\sqrt{(y^*)^2+
|\Omega^*_{m}-\Omega^*_{m'}|}}\left[\frac{{\bar \Delta}(\Omega^*_{m'},y^*)}{|\Omega_{m'}|} - \frac{{\bar \Delta}(\Omega^*_{m},x^*)}{\Omega_{m}}
\frac{\Omega^*_{m'}}{|\Omega_{m'}|}\right]\nonumber\\
&\times\frac{\sqrt{(y^*)^2+|\Omega^*_{m}-\Omega^*_{m'}|}+\frac{3}{8}|{\tilde \Sigma}^* (\Omega^*_{m'},y^*)|}{\left((x^*)^2 +
\left(\sqrt{(y^*)^2+|\Omega^*_{m}-\Omega^*_{m'}|}+\frac{3}{8}|{\tilde \Sigma}^* (\Omega^*_{m'},y^*)|\right)^2\right)^2}
\label{frac_4_2}
\end{align}
We see that the term with zero bosonic Matsubara does not cancel out completely. However, when ${\tilde \Sigma}^*$ is larger than other terms, the
dependence of
$ {\bar \Delta}(\Omega^*_{m},x^*)$ on $x^*$ and of $ {\bar \Delta}(\Omega^*_{m},y^*)$ on $y^*$ become weak,
 and Eq.\ (\ref{frac_4_2}) reduces to Eq.\ (\ref{frac_4}) in the main text.

\section{Evaluation of the terms $I_1-I_4$}
\label{app:extra}

In this Appendix we evaluate the terms $I_1$ - $I_4$, which we need to decide whether the system develops stripe or checkerboard order.
Each $I_i$ is a convolutions of four fermionic propagators:
\begin{align}
I_1\equiv&-\frac{1}{2}\int G_1^2G_2^2\nonumber\\
I_2\equiv&-\frac{1}{2}\int G_1^2G_5^2\nonumber\\
I_3\equiv&-\int G_1G_5^2G_6\nonumber\\
I_4\equiv&-\int G_1G_2G_5G_6.
\label{Is_1}
\end{align}
 The abbreviations for the Green's function are $G_1\equiv G(\omega_{m},{\bf k_1}+(k_x,k_y))$, etc, where $1,2$ and $5,6$ label hot spots in Fig.\ \ref{fig3}).
   The integrations are performed over running frequency $\omega_{m}$ and  momenta $k_x$ and $k_y$.  We use Green's functions for Free fermions and expand to linear order near hot spots. The Fermi velocities at relevant hot spots are  ${\bf v}_{F,{\bf k}_1} = (v_x, v_y)$ ${\bf v}_{F,{\bf k}_2} = (v_x, -v_y)$,
  ${\bf v}_{F,{\bf k}_5} =
   (-v_x, v_y)$, and ${\bf v}_{F,{\bf k}_6} = (-v_x, -v_y)$.

 For $I_1$ we obtain
\begin{align}
I_1=-\frac{T}{2}\sum_{m}\int_{-\Lambda}^{\Lambda}\frac{dk_x~dk_y}{(2\pi)^2}\left[\frac{1}{i\omega_{m}-(v_xk_x+v_yk_y)}\right]^2\left[\frac{1}{i\omega_{m}-(v_xk_x-v_yk_y)}\right]^2
\end{align}
We keep the upper cutoff $\Lambda$ in the momentum integrals. i.e., integrate over a finite momentum range around hot spots.
  We will take the limit $v_x\ll v_y$ and $v_x \Lambda \gg T$.
   The ratio  $v_x\Lambda/T$ can, in principle, be arbitrary for $T \sim T_{\rm cdw}$, but is definitely large for $T \to 0$. We will keep $v_x\Lambda \geq T$
 in our calculations.

Introducing the new parameters,
\begin{align}
x=v_xk_x,~~~y=v_yk_y,~~~\Lambda_x=v_x\Lambda,~~~\Lambda_y=v_y\Lambda \gg \Lambda_x.
\label{xy}
\end{align}
we re-write $I_1$  as
\begin{align}
I_1=-\frac{T}{8\pi^2v_xv_y}\sum_{m}\int_{-\Lambda_x}^{\Lambda_x}dx\int_{-\Lambda_y}^{\Lambda_y}dy\left(\frac{1}{y+x-i\omega_{m}}\right)^2\left(\frac{1}{y-x+i\omega_{m}}\right)^2.
\end{align}
 We separate the $y$-integral $\int_{-\Lambda_y}^{\Lambda_y}dy$ into $I_1=\int_{-\infty}^{\infty}dy-\int_{|y|>\Lambda_y}dy\equiv I_{1a}-I_{1b}$. For $I_{1a}$ we obtain
\begin{align}
I_{1a}=&-\frac{T}{8\pi^2v_xv_y}\sum_{m}\int_{-\Lambda_x}^{\Lambda_x}dx\int_{-\infty}^{\infty}dy\left(\frac{1}{y+x-i\omega_{m}}\right)^2\left(\frac{1}{y-x+i\omega_{m}}\right)^2\nonumber\\
=&-\frac{iT}{16\pi v_xv_y}\sum_{m}~{\rm sgn}(\omega_{m})\int_{-\Lambda_x}^{\Lambda_x}{dx}\left(\frac{1}{x-i\omega_{m}}\right)^3\nonumber\\
=&-\frac{iT}{16\pi v_xv_y}\sum_{m}\int_{-\Lambda_x}^{\Lambda_x}dx\left(\frac{1}{x-i|\omega_{m}|}\right)^3\nonumber\\
=&-\frac{iT}{32\pi v_xv_y}\sum_{m}\left[\left(\frac{1}{|\omega_{m}|+i\Lambda_x}\right)^2-\left(\frac{1}{|\omega_{m}|-i\Lambda_x}\right)^2\right]\nonumber\\
\approx&-\frac{i}{64\pi^2 v_xv_y}\int d\omega \left[\left(\frac{1}{|\omega|+i\Lambda_x}\right)^2-\left(\frac{1}{|\omega|-i\Lambda_x}\right)^2\right]\nonumber\\
=&-\frac{1}{16\pi^2v_xv_y}\frac{1}{\Lambda_x}.
\end{align}
 Note that the original integrand is singular in the infra-red, so it is important to keep the temperature {\it finite} as a regulator of the singularity and set $T\rightarrow0$ only at the very end of calculations.
  We will use the same procedure for the other integrals.

The contribution to $I_1$ from $|y|>\Lambda_y$ is
\begin{align}
I_{1b}=&-\frac{T}{8\pi^2v_xv_y}\sum_{m}\int_{-\Lambda_x}^{\Lambda_x}dx\int_{|y|>\Lambda_y}dy\left(\frac{1}{y+x-i\omega_{m}}\right)^2\left(\frac{1}{y-x+i\omega_{m}}\right)^2\nonumber\\
=&-\frac{T\Lambda_x}{2\pi^2v_xv_y}\sum_{m}\int_{\Lambda_y}^{\infty}\frac{1}{(y^2+\omega_{m}^2)^2}\nonumber\\
=&-\frac{1}{16\pi^2v_xv_y}\frac{\Lambda_x}{\Lambda_y^2}.
\end{align}
In the first line we used the fact that $\Lambda_y\gg\Lambda_x$. As we see, both contributions are small in $1/\Lambda$. The full result for $I_1$ is
\begin{align}
I_1=-\frac{1}{16\pi^2v_xv_y}\left(\frac{1}{\Lambda_x}-\frac{\Lambda_x}{\Lambda_y^2}\right) \approx - \frac{1}{16\pi^2v_x^2v_y}\frac{1}{\Lambda}
\end{align}

For $I_2$ we have
\begin{align}
I_2=-\frac{T}{8\pi^2v_xv_y}\sum_{m}\int_{-\Lambda_x}^{\Lambda_x}dx\int_{-\Lambda_y}^{\Lambda_y}dy\left(\frac{1}{x+y-i\omega_{m}}\right)^2\left(\frac{1}{x-y+i\omega_{m}}\right)^2.
\end{align}
 We again separate the integral over $y$ as $I_2=\int_{-\infty}^{\infty}dy-\int_{|y|>\Lambda_y}dy\equiv I_{2a}-I_{2b}$. This time, the integral over $y$ from $-\infty$ to $\infty$ vanishes because the poles are all located in the same half plane.
The integral $I_{2b}$ also vanishes:
\begin{align}
I_{2b}=&-\frac{T}{8\pi^2v_xv_y}\sum_{m}\int_{-\Lambda_x}^{\Lambda_x}dx\int_{|y|>\Lambda_y}dy\left(\frac{1}{x+y-i\omega_{m}}\right)^2\left(\frac{1}{x-y+i\omega_{m}}\right)^2\nonumber\\
=&-\frac{T\Lambda_x}{4\pi^2v_xv_y}\sum_{m}\int_{\Lambda_y}^{\infty}\left[\left(\frac{1}{y-i\omega_{m}}\right)^4+\left(\frac{1}{y+i\omega_{m}}\right)^4\right]\nonumber\\
\approx&-\frac{\Lambda_x}{8\pi^3v_xv_y}\int d\omega\int_{\Lambda_y}^{\infty}\left[\left(\frac{1}{y-i\omega}\right)^4+\left(\frac{1}{y+i\omega}\right)^4\right]\nonumber\\
=&0.
\end{align}
As a result, $I_2=0$.

Now we turn to $I_3$. We explicitly write it down as
\begin{align}
I_3=\frac{T}{4\pi^2v_xv_y}\sum_{m}\int_{-\Lambda_x}^{\Lambda_x}dx\int_{-\Lambda_y}^{\Lambda_y}dy\left(\frac{1}{x+y-i\omega_{m}}\right)^2\frac{1}{(x-y)^2+\omega_{m}^2}.
\end{align}
 As before, we write $I_3=\int_{-\infty}^{\infty}dy-\int_{|y|>\Lambda_y}dy\equiv I_{3a}-I_{3b}$. We evaluate $I_{3a}$ by extending the integral over $y$ onto the half plane where the integrand contains a single pole
\begin{align}
I_{3a}=&\frac{T}{4\pi^2v_xv_y}\sum_{m}\int_{-\Lambda_x}^{\Lambda_x}dx\int_{-\infty}^{\infty}dy\left(\frac{1}{x+y-i\omega_{m}}\right)^2\frac{1}{(x-y)^2+\omega_{m}^2}\nonumber\\
=&\frac{T}{16\pi v_xv_y}\sum_{m}~{\rm sgn}(\omega_{m})\int_{-\Lambda_x}^{\Lambda_x} dx \left(\frac{1}{x-i\omega_{m}}\right)^2\frac{1}{\omega_{m}}\nonumber\\
=&-\frac{T}{16\pi v_xv_y}\sum_{m}~{\rm sgn}(\omega_{m})~\frac{2\Lambda_x}{\Lambda_x^2+\omega_{m}^2}~\frac{1}{\omega_{m}}\nonumber\\
\approx&-\frac{1}{16\pi^2v_x v_y}\frac{1}{\Lambda_x}\log\frac{\omega_0}{T}.
\end{align}
We see that $I_{3a}$ is logarithmically singular  at $T \to 0$. The other part, $I_{3b}$, is regular at $T\to 0$. Therefore, to logarithmic accuracy,
\begin{align}
I_3 \approx - \frac{1}{16\pi^2v_x^2v_y}\frac{1}{\Lambda}\log\frac{\omega_0}{T}.
\label{ch_44}
\end{align}

Finally, for $I_4$ we have
\begin{align}
I_4=-\frac{T}{4\pi^2v_xv_y}\sum_{m}\int_{-\Lambda_x}^{\Lambda_x}dx\int_{-\Lambda_y}^{\Lambda_y}dy~\frac{1}{(x+y)^2+\omega_{m}^2}~\frac{1}{(x-y)^2+\omega_{m}^2}.
\end{align}
 The most straightforward  way to evaluate this integral is to first extend both $x$- and $y$-integrations to infinite limits and then check how the results change when we restore finite
  limits of integration.  To evaluate the integral in infinite limits, we introduce new variables $a=x+y$ and $b=x-y$ and re-express $I_4$  as
\begin{align}
I_4
\approx
&-\frac{T}{8\pi^2v_xv_y}\sum_{m}\int_{-\infty}^{\infty}da\int_{-\infty}^{\infty}db~\frac{1}{a^2+\omega_{m}^2}~\frac{1}{b^2+\omega_{m}^2}\nonumber\\
=&-\frac{1}{8 v_xv_y}T\sum_{\omega_m} \frac{1}{\omega_m^2}\nonumber\\
=&-\frac{1}{32 v_x v_y}\frac{1}{T}.
\label{ch_4}
\end{align}
We see that $I_4$ diverges as $1/T$. The divergence comes from momenta much smaller  than $\Lambda$, hence
 the prefactor for $1/T$ term does not depend on whether the limits of momentum integration are infinite or finite.  Integration in a finite limits gives rise to
   corrections to (\ref{ch_4}) of order $(1 + O(T/(v_x \Lambda)))$.

\section{On the stability of HS saddle-point in a complex plane}
\label{app:c1}

In the main text we discussed  the stability of the mean-field CDW solution.  We used HS transformation
   and obtained the effective action in terms of HS  collective variables ${\Delta}_+$ and ${\Delta}_-$, proportional to the two CDW order parameters
   $\Delta^Q_{k_0}$ and $\Delta^Q_{k_\pi}$.
    We found that the saddle-point solution below $T_{\rm cdw}$  is such that one variable is real and another is imaginary. Expanding near the
    saddle-point solution, we obtained the effective action which contains bilinear combination of fluctuations of ${\Delta}_+$ and ${\Delta}_-$ with
    imaginary prefactor.
    This form of the action persists also if we expand around ${\Delta}_{+} = {\Delta}_{-} =0$ at $T > T_{\rm cdw}$.   The complex form of the action
    requires extra care when one analyzes the convergence of the Gaussian integrals over fluctuations of ${\Delta}_+$ and ${\Delta}_-$

The same situation with two HS fields emerges in the analysis of a $Z_2$ spin-nematic order in Fe-pnictides~\cite{rafael}.
There, the solution of the saddle-point equations for collective nematic variables is again such that one variable is along real axis and the other is
along the imaginary axis. When one expands next the saddle-point solution, one faces the same issue of convergence of Gaussian integrals, taken by shifting
one of variables into a complex plane.

In this Appendix we discuss several generic issues, associated with the expansion near the saddle points in the complex plane and with the accuracy of using
saddle-point approximation for HS fields.  Specifically, we follow the suggestion put forward in  Ref.\ [\onlinecite{efetov_private}]
and consider the model with no momentum dispersion (formally, in dimension $D=0$).

  In this case the partition function can, in certain limits, be evaluated explicitly  with or without HS transformation. We use this fact to demonstrate
  that the computational procedure,  which we used in the main text, yields the results identical
  to the one which one obtains in a direct integration over the primary fields.

We consider two extensions of the original bosonic model, which allow us to put the calculation of the partition function under control.
 One is the extension to large $N$. It is obtained by extending the underlying fermionic model to $N \gg1$ fermionic flavors (for details see, e.g., Ref.
 \cite{rafael}). The net result of such an extension is the appearance of the overall factor of $N$ in the effective action for bosonic variables.  Another
 is the extension to the number of components of the bosonic fields to large $M$. This gives rise
 to more complex form of the action as quadratic and quartic terms in bosons change differently.  We argue below
   that HS  transformation is useful in the large $M$ limit, while at
    large $N$ it is easier to evaluate the partition function by  integrating directly over the original variables, without introducing HS collective
    variables.
    To put it differently, large $N$ and large $M$ limits are the two examples when one has to take care in choosing the variables whose fluctuations are
    weak.
     At large $N$, fluctuations of the original bosonic fields around mean-field solution are weak and there is no need to perform HS transformation. If
     one does transform to HS fields, one finds that fluctuations of HS fields around their mean-field solution are strong.  On the contrary, at large
     $M$,
     fluctuations of the original bose fields  are strong, while fluctuations of the HS fields around their saddle-point values are weak.  In this limit,
     HS transformation and the subsequent saddle-point analysis of the effective action in terms of the HS fields are fully justified.

     Which extension better describes the original model is a'priori unclear and requires complimentary analysis~\cite{rafael,tsvelik},
      particularly in the cases when the analysis in terms of HS variables yields a preemptive transition into a state with a composite order.
      In principle, such a transition may exist only at large enough $M$ and disappear when $M$ is reduced to the original value of order one.
      At the same time, we are not aware of the examples when a composite order, detected in the model, extended to large $M$,
       does not exist in the original model with $M = O(1)$.

 \subsection{The model with one real field}

We assume that HS transformation from the original fermionic variables to collective bosonic  variables is already performed and consider the action in
terms of bosonic fields.  As a warm-up, consider the  effective action for one  real single-component bosonic field $\Delta$:
\be
S [\Delta] = \alpha \Delta^2 + \frac{1}{2} \Delta^4
\label{a1}
\ee
The partition function is
\be
I =  \frac{1}{\sqrt{\pi}} \int e^{-S[\Delta]} d \Delta
\label{a2}
\ee
At a mean-field level, $\langle\Delta\rangle =0$ at $\alpha >0$ and $\langle\Delta\rangle = \pm (-\alpha)^{1/2}$ at $\alpha <0$.

\subsubsection{Large $N$.}

Consider first the extension of the model to large $N$ (large number of flavors of original fermions).
This extension adds $N$ as the overall factor to the action~\cite{rafael}
\be
 S [\Delta] = N\left[\alpha \Delta^2 + \frac{1}{2} \Delta^4\right]
\label{a11}
\ee

We first compute the partition function directly and  by using HS transformation to composite bose fields.

  For $\alpha >0$ and $\alpha^2 N \gg 1$, the quartic term can be neglected and we immediately obtain
\be
I = \frac{1}{\sqrt{N \alpha}},~~~ \alpha >0
\label{a3}
\ee
For $\alpha <0$ and $N \alpha^2 \gg1$,  we expand near $\Delta =  \pm  (-\alpha)^{1/2} = (|\alpha|)^{1/2}$ and after simple integration obtain
\be
I = 2 \frac{e^{\frac{N\alpha^2}{2}}}{\sqrt{2 N |\alpha|}},~~~ \alpha <0
\label{a4}
\ee
(the overall factor $2$ comes from summing up contributions from positive and negative $\Delta$).

The crossover between the two results for $I$ occurs in the range $N \alpha^2 \leq 1$.  In this range, more accurate analysis is needed
 to compute the partition function.

We now use HS transformation
 \be
 e^{-N \Delta^4/2} = \frac{1}{\sqrt{2\pi N}} \int d \psi ~e^{-\frac{\psi^2}{2N}} e^{i\psi \Delta^2}
 \label{a5}
 \ee
 Substituting this into (\ref{a2}) and integrating over $\Delta$, we obtain
 \bea
 &&I = \frac{1}{2\pi} \int d \psi ~e^{-{S}_{\rm eff} [\psi]} \nonumber \\
 && {S}_{\rm eff} [\psi] = \frac{1}{2} \ln(\alpha + i \psi) + \frac{N \psi^2}{2}
 \label{a6}
 \eea
 So far, this is exact (the argument of the log is perfectly well defined for real $\psi$), and the integration over $\psi$
  indeed gives the correct I, as one can easily check.  However, the reason we use HS transformation is that we hope that
   the integration over the field $\psi$ can be done by expanding around a saddle point.  Let's see what we get if we do this.

 First, let's obtain the saddle-point solution. Differentiating ${S}_{\rm eff} [\psi]$ from (\ref{a6}) over $\psi$, we obtain
 \be
 N\psi = \frac{i}{2} \frac{1}{\alpha - i \psi}
 \label{a7}
 \ee
 The solution is along the imaginary axis: $-i\psi = \psi_0$. Introducing $\alpha + \psi_0 = r_0$, we re-express (\ref{a7}) as
 \be
 r_0 = \alpha + \frac{1}{2N r_0}
 \label{a8}
 \ee
 There are two solutions of this equation. For one, $r_0>0$, for the other $r_0 <0$.
 At large positive $\alpha$, $\psi_0$ is obviously small and hence $r_0 \approx \alpha >0$. Because $r_0$ never crosses zero (see (\ref{a8})),
  only the solution with a positive $r_0$ is physically relevant. We have
  \be
  r_0 = \frac{\alpha}{2} + \sqrt{\frac{\alpha^2}{4} + \frac{1}{2N}},~~~\psi_0 = -\frac{\alpha}{2} + \sqrt{\frac{\alpha^2}{4} + \frac{1}{2N}},
   \label{a9}
 \ee
Now let's expand around this saddle-point.  Introducting ${\tilde \psi}$ via $\psi = i \psi_0 + {\tilde \psi}$ and substituting into (\ref{a6}), we
obtain, without making any approximation,
 \be
 {S}_{\rm eff} [\psi] = S_0 + N \frac{{\tilde \psi}^2}{2} + \frac{1}{2} \left[\ln{\left(1 - \frac{i {\tilde \psi}}{r_0}\right)}+  \frac{i {\tilde
 \psi}}{r_0}\right]
 \label{a10}
 \ee
 where
 \be
 S_0 = N \frac{\psi^2_0}{2} - \frac{1}{2} \ln{r_0}.
 \label{a111}
 \ee
 Then
 \be
  I = \frac{e^{-\frac{N \psi^2_0}{2}}}{\sqrt{2\pi r_0}} ~ {\tilde I}
\label{a12}
 \ee
 where
 \be
 {\tilde I} = \int_{-\infty}^{\infty} d{\tilde \psi} e^{-\left[N \frac{{\tilde \psi}^2}{2} + \frac{1}{2} \left(\ln{\left(1 - \frac{i {\tilde
 \psi}}{r_0}\right)}+  \frac{i {\tilde \psi}}{r_0}\right)\right]}
 \label{a14}
 \ee

 So far, all transformations were exact.  Now let's see whether we can treat ${\tilde \psi}$ as small and expand near the saddle point.
 Let's start with positive $\alpha$.  At $\alpha >0$ and $\alpha^2 N \gg1$ we have
 $r_0 \approx \alpha$ and $\psi_0 \approx 1/(2 \alpha N)$. Expanding in (\ref{a14}) to second order in ${\tilde \psi}$, we obtain
 \be
 {S}_{\rm eff} [\psi] = S_0 +  N\frac{{\tilde \psi}^2}{2} \left(1 + \frac{1}{2N \alpha^2}\right)
 \label{a15}
 \ee
The integral over ${\tilde \psi}$ in (\ref{a14}) is perfectly convergent, and evaluating it we obtain
 \be
I = \frac{1}{\sqrt{N \alpha}},~~~ \alpha >0
\label{a16}
\ee
which is the same result as in (\ref{a3}).

For $\alpha <0$ (and, still, $\alpha^2 N \gg 1$)  the situation is more complex as now $\psi_0 \approx |\alpha|$ and $r_0 \approx 1/(2 \alpha| N)$.
Substituting these forms into (\ref{a12}), (\ref{a14}), and rescaling, we find
\be
I = \sqrt{\frac{|\alpha| N e}{\pi}} e^{\frac{N \alpha^2}{2}} {\bar I}
\label{a17}
\ee
where in rescaled variables
\be
{\bar I} = r_0 \int_{-\infty}^{\infty} d u ~e^{-\frac{u^2}{8 \alpha^2 N}} e^{-\frac{1}{2} \left(\ln{(1 - iu)} + iu\right)}
\label{a18}
\ee
and $u = {\tilde \psi}/r_0$.
The exponent $e^{\frac{N \alpha^2}{2}} $ comes from the saddle point and is the same as in (\ref{a4}).  However, the
prefactor cannot be obtained by expanding around the saddle point -- we clearly see from (\ref{a18}) that typical $u$ are of order one, hence one cannot
approximate the logarithmical term in the exponent in (\ref{a18}) by expanding to order $u^2$. Rather, one has to evaluate the full integral.
This shows that saddle-point approximation is only partially valid at large $N$ -- the exponent in $I$ comes out right, but the prefactor cannot be
obtained by expanding near the saddle point to order ${\tilde \psi}^2$.

The integrand in (\ref{a18}) converges at $u \to \pm \infty$, and the integral can be easily evaluated by closing the integration contour in the lower
half-plane
 ($u = a-ib$, $b >0$).
 There is a branch cut along negative imaginary axis, at $b >1$. Integrating over the boundaries of the branch cut we obtain after simple algebra
 \bea
{\bar I} &=& -ir_0 \int_{1}^{\infty} d b ~e^{-b/2} \left[e^{-\frac{1}{2} \left(\ln{b-1} -i\pi\right)}-e^{-\frac{1}{2} \left(\ln{b-1}
+i\pi\right)}\right]\nonumber \\
&& = 2 r_0 \int_{1}^{\infty} \frac{d b ~e^{-b/2}}{\sqrt{b-1}} = 2 r_0 \sqrt{\frac{2\pi}{e}}
\label{a19}
\eea
Substituting this into (\ref{a17}) we obtain
\be
I = 2 \frac{e^{\frac{N\alpha^2}{2}}}{\sqrt{2 N |\alpha|}},~~~ \alpha <0
\label{a20}
\ee
This expression coincides with (\ref{a4}), as it indeed should.

The message from this analysis is that, at large $N$, there is no advantage of using HS transformation and expanding around a saddle point -- it is more
straightforward to
 compute $I$ by directly integrating over $\Delta$ and expanding around its mean-field solution along the real axis.  For $\alpha <0$, one has to expand
 around the minimum of $S[\Delta]$ at a non-zero $\Delta = \pm |\alpha|^{1/2}$, and this expansion is controlled by $1/N$. Still, one can get the correct
 result for $I$ even from HS analysis. The exponent at $\alpha <0$ comes from the saddle point, but
   to get the prefactor right one has to do full integration, without expanding to quadratic order in the deviations from the saddle point.

\subsubsection{Large $M$.}

Let's now consider different extension of  Eq.\ (\ref{a1}). Suppose that the field $\Delta$ has $M$ components, and $M \gg1$.

At large $M$, it is convenient to rescale the prefactor for $\Delta^4$ term to 1/M and analyze the action
\be
S[\Delta] = \alpha \sum_{i=1}^M \Delta^2_i + \frac{1}{2M} \left( \sum_{i=1}^M \Delta^2_i\right)^2.
\label{a21}
\ee
The partition function is
\be
I = \frac{1}{\pi^{M/2}} \prod_{i=1}^M \int d \Delta_i e^{-S[\Delta]}
\label{a22}
\ee

As in the previous Section, we compute $I$ in two ways: (i) by directly integrating over $\Delta_i$ and (ii) by using HS transformation

We begin with  direct computation. Introducing M-dimensional spherical variables, one can re-write (\ref{a22}) as
\bea
I &=& (A_M/\pi^{M/2}) \int_0^\infty r^{M-1} dr e^{-(\alpha r^2 + r^4/(2M))} \nonumber \\
&& = A_M M^{M/2} \int_0^\infty \frac{dx}{x}  e^{M[\ln x -\alpha x^2 -(1/2) x^4]}
\label{a23}
\eea
where $A_M = 2 \pi^{M/2}/\Gamma(M/2)$ is the area  of a $M-$dimensional sphere with unit radius.
At large $M$, $\Gamma(M/2) \approx \sqrt{4\pi/M} (M/2)^{M/2} e^{-M/2} $.

Because of prefactor $M$ in the exponent in (\ref{a23}), the integral over $x$ can be evaluated by expanding around the minimum of $\ln x -\alpha x^2
-(1/2) x^4$.
The position of the minimum is at $x =x_0$, where
\be
x_0 =\[-\frac{\alpha}{2} + \sqrt{\frac{\alpha^2}{2} +1/2} \]^{1/2}.
\label{a24}
\ee
Introducing $x = x_0 + {\tilde x}$ and  expanding around saddle point
 we obtain perfectly convergent integral over ${\tilde x}$  with $M$ in the exponent, which justifies the expansion.
    Evaluating the integral over $\delta x$ and adding the contribution from the saddle point, we obtain
\be
I = e^{M\left(S_0 + \frac{1+ \ln 2}{2}\right)}  \frac{1}{(\alpha^2 +2)^{1/4} (\sqrt{\alpha^2+2} -\alpha)^{1/2}}
\label{a25}
\ee
where $S_0 = \ln x_0 -\alpha x^2_0 - (1/2) x^4_0$.
Note that only a portion of the exponential prefactor in (\ref{a25}) comes from the saddle point, another portion comes from  $A_M$ (the area of
$M-$dimensional sphere).

We next evaluate the partition function using HS transformation.
 Using a generic formula for $e^{-X^2/2M}$ for real $x$:
  \be
 e^{-X^2/2M} = \sqrt{\frac{M}{2\pi}} \int d \psi e^{-M \frac{\psi^2}{2}} e^{i\psi X},
 \label{a5_1}
 \ee
 applying it to $X = \sum_{i=1}^M |\Delta_i|^2$  and integrating over the components of the $\Delta$ field, we obtain
 \be
 I =  \sqrt{\frac{M}{2\pi}} \int d \psi e^{-M\left[\frac{\psi^2}{2} + \frac{1}{2} \ln(\alpha - i \psi)\right]}
 \label{a26}
\ee
 The saddle point is at $\psi = i \psi_0$, where
 \be
 \psi_0 = -\frac{\alpha}{2} + \sqrt{\frac{\alpha^2}{4} +\frac{1}{2}}
 \label{a27}
\ee
Like before, we can re-write the equation for the saddle point as
 \be
 r_0 = \alpha + \frac{1}{2r_0}
 \label{a28}
\ee
where $r_0 =\alpha + \psi_0$.
This equation formally has two solutions, for one $r_0 >0$, for the other $r_0 <0$. However, only the solution with $r_0 >0$ is meaningful
   because (i) at large positive $\alpha$, $\psi_0$ is small and $r_0 \approx \alpha >0$ and (ii) $r_0$ doesn't change sign as a function of $\alpha$
   because
     $r_0 =0$ is not a solution of (\ref{a28}).

Introducing $\psi = i \psi_0 + {\tilde \psi}$ and expanding the exponent around the saddle point
  we obtain
 \be
 I =  \sqrt{\frac{M}{2\pi}} e^{M (\psi^2_0/2 - \ln{r_0})}  \int d {\tilde \psi} e^{-M\left[\frac{{\tilde \psi}^2}{2} \left(1 + 1/(2r^2_0)\right)\right]}
 \label{a29}
\ee
The integration is elementary and yields
 \be
 I = \frac{1}{\sqrt{1 + \frac{1}{2r^2_0}}}  e^{M (\psi^2_0/2 - \ln{r_0})}
 \label{a30}
\ee
Using Eq.\ (\ref{a27}) and the fact that $r_0 = \alpha + \psi_0 = \frac{\alpha}{2} + \sqrt{\frac{\alpha^2}{4} +\frac{1}{2}}$, one can easily verify that the
expressions for the partition functions obtained directly and using HS transformation,  Eqs.\ (\ref{a25}) and (\ref{a30}), are identical.

The message here is  that  the partition function in the large $M$ limit can computed directly by integrating over $\Delta$ field, but it can also be
  obtained by using the HS transformation and expanding around the saddle point.  This expansion is perfectly well justified at large $M$ and, moreover,
    the exponent in $I$ in (\ref{a30}) contains the action taken right at the saddle-point. In the direct integration, the exponent in $I$  comes partly from the
    action at the minimum and partly from $A_M$.

Note that in both calculations  we computed the partition function by expanding around the
 extremal value of the action (first derivative of the action vanishes).
 In the direct integration over $\Delta$, the point for which $d S/d \Delta =0$ is along the real axis,  and the integration over fluctuations of $\Delta$
 is also along the real axis.   Within the HS approach, the extremum of the action is along the imaginary axis, and by writing
 $\psi = i \psi_0 + {\tilde \psi}$ we shift the integration contour into the complex plane. For one-component model this is not a dangerous procedure as
   the only requirement on the integration over ${\tilde \psi}$ in the HS approach is that the integration contour should merge with the real axis at
   infinite $\psi$.  Still,  the agreement between $I$ obtained via HS transformation and by direct integration over $\Delta$ along the real axis tells us
   that the shift into the complex plane, used in the HS calculation, is perfectly legitimate procedure. For one-field case, there is little doubt that
   this is true,
    but we will see below that the analogy between direct and HS calculations helps us to justify the integration over HS fields over the contour in the
    complex plane in a more involved  case of two bose fields.

 \subsection{The model with two order parameters}

For definiteness, consider the two-field model discussed in Ref.\ [\onlinecite{rafael}] in connection with a preemptive spin-nematic order
\begin{align}
S [\Delta_1,\Delta_2] =\alpha(\Delta_1^2+\Delta_2^2)+  \frac{1}{2} \left(\Delta_1^2+\Delta^2_2\right)^2 -\frac{\beta}{2}
\left(\Delta_1^2-\Delta_2^2\right)^2
\label{a31}
\end{align}
where $0<\beta<1$.

As before, we extend the model separately to large $N$ and to large $M$.

\subsubsection{Large $N$.}

The extension to large $N$ is straightforward -- one just has to multiply the effective action in Eq.\ (\ref{a31}) by $N$. We have

\be
I = \frac{1}{\pi} \int d \Delta_1\ d \Delta_2\ e^{-N S[\Delta_1,\Delta_2]}
\label{a32}
\ee

We begin with a direct computation of $I$. Introducing $\Delta_1 = \Delta \cos{\phi}, \Delta_2 = \Delta \sin{\phi}$ and substituting into (\ref{a32}), we
obtain
\be
I = \frac{2}{\pi} \int_0^\infty dx  \int_{-\pi/4}^{\pi/4} d \phi\ e^{\left[-N \left(\alpha x + \frac{x^2}{2} \left(1 - \beta
\cos^2{2\phi}\right)\right)\right]}
\label{a33}
\ee
For $\alpha >0$ and $\alpha^2 N \gg 1$, the $x^2$ term in the exponent is irrelevant and we get
\be
I = \frac{1}{N\alpha}
\label{a340}
\ee
For negative $\alpha$ and, again, $\alpha^2 N \gg 1$, one can complete the square in the exponential term, introduce $y = x -|\alpha|/(1-\beta \cos^2
{2\phi})$ as a new variable, and integrate over $y$ in infinite limits. This yields
\be
I = \frac{4}{\sqrt{2\pi N}} \int_{-\pi/4}^{\pi/4} \frac{d\phi}{\sqrt{1-\beta \cos^2 {2\phi}}} e^{\frac{N\alpha^2}{2 (1-\beta \cos^2 {2\phi})}}
\label{a34}
\ee
The exponent has a maximum at $\phi =0$. Expanding near the maximum, we obtain
\be
I = \frac{2 \sqrt{2 \pi}}{\sqrt{N (1-\beta)}} e^{\frac{N\alpha^2}{2 (1-\beta)}} \int_{-\infty}^\infty dz e^{-\frac{2 N \alpha^2 \beta z^2}{(1-\beta)^2}}
\label{a35}
\ee
Evaluating then the integral over $z$, we obtain
\be
I = \frac{2}{N |\alpha|} \sqrt{\frac{1-\beta}{\beta}} e^{\frac{N\alpha^2}{2 (1-\beta)}}
\label{a36}
\ee

Now let's see whether we can reproduce this result using the HS analysis.
We use
\begin{align}
e^{-\frac{N \beta}{2}(\Delta_1^2+\Delta_2^2)^2}&= \sqrt{\frac{N}{2\pi \beta}}\int d\psi~e^{\left(\frac{-N{\psi^2}}{2\beta}\right)}e^{iN \psi
(\Delta_1^2+\Delta_2^2)}\nonumber\\
e^{\frac{N}{2}(\Delta_1^2-\Delta_2^2)^2}&=\sqrt{\frac{N}{2\pi}} \int d\upsilon~e^{\left(-\frac{N \upsilon^2}{2\beta}\right)}e^{N
\upsilon(\Delta_1^2-\Delta_2^2)}
\label{a37}
\end{align}
Substituting these integrals into (\ref{a32}) and integrating over $\Delta_1$ and $\Delta_2$, we obtain
\be
I = \frac{1}{2\pi \sqrt{\beta}} \int d \psi d \upsilon e^{-{S}_{\rm eff} [\psi, \upsilon]}
\label{a38}
\ee
where
\be
{S}_{\rm eff} [\psi, \upsilon] = N \left(\frac{\psi^2}{2} + \frac{\upsilon^2}{2\beta} \right) + \frac{1}{2} \ln{\left[(\alpha-i\psi)^2 -
\upsilon^2\right]}
\label{a39}
\ee
The equations on the extremum of the action  are
\begin{align}
&\upsilon \left[1 -\frac{\beta}{N} \frac{1}{(\alpha-i\psi)^2 - \upsilon^2}\right] =0 \nonumber \\
& -i \psi = \frac{1}{N} \frac{\alpha-i\psi}{(\alpha-i\psi)^2 - \upsilon^2}
\label{a40}
\end{align}
One obvious solution is $\upsilon =0$, $\psi = i \psi_0$, where, like in the previous case,
\be
\psi_0 = -\frac{\alpha}{2} + \sqrt{\frac{\alpha^2}{4} + \frac{1}{N}}
\label{a41}
\ee
and
\be
r_0 = \alpha + \psi_0 = \frac{\alpha}{2} + \sqrt{\frac{\alpha^2}{4} + \frac{1}{N}}
\label{a42}
\ee

Introducing $\psi = i \psi_0 + {\tilde \psi}$ and expanding around this saddle point, we obtain
\be
{S}_{\rm eff} [\psi,\upsilon] = S_0 + N \frac{\upsilon^2}{2\beta} \left(1 - \frac{\beta}{N r^2_0}\right) +   N \frac{{\tilde \psi}^2}{2}\left(1 +
\frac{1}{N r^2_0}\right)
\label{a43}
\ee
where
\be
S_0 = - \frac{N}{2} \psi^2_0  + \ln{r_0}
\label{a44}
\ee
For $\alpha >0$ and $\alpha^2 N \gg 1$, $r_0 \approx \alpha$ and $\psi_0 \approx 1/(N \alpha)$.
In this case, saddle point is a minimum along real  $\upsilon$ and real ${\tilde \psi}$.  The effective action can be approximated by
 \be
{S}_{\rm eff} [\psi,\upsilon] \approx  \ln{\alpha} + N \left(\frac{\upsilon^2}{2\beta} + \frac{{\tilde \psi}^2}{2}\right)
\label{a45}
\ee
Substituting this into the integral for $I$ and integrating over $\upsilon$ and over ${\tilde \psi}$, we obtain
\be
I = \frac{1}{N\alpha}
\label{a46}
\ee
which coincides with (\ref{a34}).

For $\alpha <0$ and $\alpha^2 N \gg 1$, the situation is different.  Now $r_0 \approx 1/(N |\alpha|)$ and the prefactor for the $\upsilon^2$ term in
 (\ref{a43}) becomes
 negative: $1-\beta/(N r^2_0) \approx - \beta N \alpha^2 <0$.  This obviously implies that the extremum at $\upsilon =0$ is a maximum rather than a
 minimum, and one has to search for a solution of the saddle-point equations with $\upsilon \neq 0$. Such solution is a ``nematic" solution in the current
 nomenclature, although a true nematic order is indeed impossible in zero-dimensional case.

 The solution of (\ref{a40}) for $\upsilon = \pm \upsilon_0 \neq 0$ is:
 \bea
 &&\psi_0 = \frac{|\alpha|}{1-\beta},~~ \upsilon^2_0 = \left(\frac{\alpha \beta}{1-\beta}\right)^2 - \frac{\beta}{N} \nonumber \\
 && r_0 = \alpha + \psi_0 =  \frac{|\alpha|\beta }{1-\beta},~~ r^2_0 - \upsilon^2_0 = \frac{\beta}{N}
 \label{a470}
\eea
Such a solution is possible when $\alpha^2 N > (1-\beta)^2/\beta$.

Expanding near  $\psi_0$ and $\pm \upsilon_0$, we obtain,
\be
I = \frac{1}{\pi \sqrt{\beta}} e^{\frac{N \alpha^2}{2(1-\beta)}} \left(\frac{ N e}{\beta}\right)^{1/2} {\tilde I}
 \label{a47}
\ee
where
\be
{\tilde I} = \int_{-\infty}^\infty d x \int_{-\infty}^\infty d y\ e^{-\frac{N}{2} \left(\frac{x^2}{\beta} + y^2\right)} e^{-\frac{1}{2} \left[\ln{\left(1 -
\frac{2N}{\beta}
 \left(i y r_0 + x \upsilon_0\right) - \frac{N(x^2 +y^2)}{\beta}\right)} + \frac{2N}{\beta}
 \left(i y r_0 + x \upsilon_0\right)\right]}
 \label{a48}
\ee
where $x = \upsilon - \upsilon_0$ and $|x|$ is assumed to be small.
As in the case of one field, we cannot expand under the logarithm as typical $x$ and $y$ are such that the argument of the logarithm is of order one.
The integrals over $x$ and over $y$ are, however, fully convergent, and the integration can be performed in any order.  We notice that the integrand
vanishes for all large  $y$ in the lower half-plane and integrate over $y$ by closing the contour in the lower half-plane of $y$. There is again a branch
but along the negative imaginary axis of $y$.  Closing the contour such that it doesn't cross the imaginary axis of $y$ in the range where the branch cut
exists, and integrating over $y$, we obtain, after straightforward algebra,
\be
{\tilde I} = 2 \frac{(1-\beta)}{|\alpha| N} \left(\frac{2\pi}{e}\right)^{1/2} \int_{-\infty}^\infty d x\ e^{-\frac{x^2 N}{2\beta} (1-\beta)}
 \label{a49}
\ee
In writing (\ref{a49}) we used the fact along the branch cut $i y r_0 + x \upsilon_0 = O(1/N)$ and $r_0 = \upsilon_0 + O(1/N)$. To leading order in $1/N$
we then have
$y^2 \approx -x^2$, such that $x^2/\beta + y^2$ in the exponent in (\ref{a48}) can be approximated by  $x^2 (1-\beta)/\beta$.

Integrating finally over $x$ in (\ref{a49}) and substituting the result into (\ref{a47}), we obtain
\be
I = \frac{2}{N |\alpha|} \sqrt{\frac{1-\beta}{\beta}} e^{\frac{N\alpha^2}{2 (1-\beta)}}
\label{a500}
\ee
This is exactly the same result as Eq.\ (\ref{a36}).

Furthermore, can easily check that the direct calculation and the one using the HS transformation yield the same values of average quantities. In
particular,
the direct evaluation of
\be
Q = \frac{|\Delta^2_1-\Delta^2_2|}{\Delta^2_1+\Delta^2_2}
\label{a50}
\ee
 yields $Q = O(1/N)$ for $\alpha >0$ and $Q \approx 1 - O(1/N)$ for $\alpha <0$.  In both cases, calculations are under control at $\alpha^2 N \gg 1$.
 The analysis based on HS transformation yields the same result.  The crossover from small $Q$ to $Q \approx 1$ occurs in a narrow range $\alpha^2 N \leq
 1$. In principle, one can compute $I$ in this range and obtain the full crossover behavior of $Q$ and related quantities. This, however, requires more
 computational efforts.

The conclusion of large $N$ analysis is that we can reproduce the result for the partition function at large $N$ by using the HS transformation.
 At $\alpha <0$, to do so we need to expand near the ``nematic" solution $\upsilon = \pm \upsilon_0$.  Typical deviations from the saddle-point solution at
 non-zero $\upsilon_0$ are small in $1/N$.  We cannot expand under the logarithm in (\ref{a48}) because typical values of the argument are of order one,
 but the
  integrand, viewed as a function of $y = {\tilde \psi} = \psi -i \psi_0$,  is nicely convergent and can be evaluated using standard means. Once we
  integrate over $y$, the remaining integral over $x = \upsilon - \upsilon_0$ is a conventional gaussian integral with large prefactor $N$ in the exponent.
  Obviously, typical $x^2$ are of order $1/N$ and are small.

\subsubsection{Large $M$.}

 Let's now extend the original model of two scalar field to the model of two $M-$component fields, and take the limit $M \gg1$.
 We have
 \be
 S[\Delta_1,\Delta_2] =\alpha \sum_{i=1}^M (\Delta^2_{1,i} +\Delta^2_{2,i}) +  \frac{1}{2M} \left( \sum_{i=1}^M (\Delta^2_{1,i} +\Delta^2_{2,i}) \right)^2
- \frac{\beta}{2M} \left( \sum_{i=1}^M (\Delta^2_{1,i} -\Delta^2_{2,i}) \right)^2
\label{a510}
\ee
 and
 \be
 I = \frac{1}{\pi^M} \prod_{i=1}^M \int d \Delta_{1,i} d \Delta_{2,i} e^{-S[\Delta_1,\Delta_2]}
\label{a51}
\ee

We again compute $I$ in two ways -- directly and via HS transformation.  We will see that HS approach is
 advantageous because the part of the action associated with the deviations from the saddle point contains large $M$ in the prefactor.
 At the same time, the action written in terms of $\upsilon-\upsilon_0$ and $\phi -i \phi_0$  has cross-term with imaginary coefficient.
   The validity of the evaluation of the gaussian integral over fluctuations in this situation has been
    questioned in Ref.\ [\onlinecite{efetov_private}].  We will see that
  the computation of I by direct integration over $\Delta_1$ and $\Delta_2$ is free from such complications as the integrals do not have to be shifted
   from the real axis.  We argue that the way how the gaussian integration has to be done in HS approach is set by the necessity to obtain the same I as in
   the direct calculation.  In this respect, zero-D case is a blessing, as for any $D >0$ there is no way to check HS calculation by directly integrating
   over
    $\Delta$  (the $\Delta^4$ term contains components with four different momenta, subject to momentum conservation).

We begin with the direct calculation of $I$. Using $m-$dimensional spherical coordinates for each of the two M -component fields, we re-write (\ref{a51})
as
 \be
 I = \frac{A^2_m}{\pi^M} \int_0^\infty d \Delta_{1} \int_0^\infty d \Delta_{2} \left(\Delta_1 \Delta_2\right)^{M-1} e^{-S[\Delta_1,\Delta_2]}
\label{a52}
\ee
where, as before $A_M = 2 \pi^{M/2}/\Gamma(M/2)$  is the area of a unit sphere in M-dimensions.  At large $M$, $\Gamma(M/2) \approx 2 (M/2)^{M/2} e^{-M/2}
\sqrt{\pi/M}$.

Introducing
\be
\Delta_1 = \sqrt{z} \cos{\phi/2}, \Delta_2 = \sqrt{z} \sin{\phi/2}, ~~0<\phi < \pi
 \label{a53}
\ee
we re-express $I$ as
\be
I = \frac{A^2_M}{\pi^M 2^{M+1}} \int \frac{dz}{z} \int_0^\pi \frac{d \phi}{\sin{\phi}} e^{M \ln[z\sin{\phi}] -\alpha z - \frac{z^2}{2M} \left( 1-\beta
\cos^2{\phi}\right)}
 \label{a54}
\ee
Introducing $z = xM$ and $u = \cos{\phi}$, we re-write Eq.\ (\ref{a54}) as
\be
I = \frac{A^2_M M^M}{\pi^M 2^{M+1}} \int \frac{dx}{x} \int_{-1}^1 \frac{d u}{1-u^2} e^{-MS[x,u]}
\label{a55}
\ee
where
\be
S[x,u] = -\ln{x} - \frac{1}{2} \ln{(1-u^2)} +\alpha x + \frac{x^2}{2} \left( 1-\beta u^2\right)
 \label{a56}
 \ee
Because the exponent in (\ref{a55}) contains an overall factor of $M$, we search for the extreme of $S[x,u]$ at $x=x_0$ and $u = u_0$. Differentiating over
$x$ and over $u$, we obtain
\bea
&& \frac{1}{x_0} -\alpha -x_0  \left( 1-\beta u^2_0\right) =0 \nonumber \\
&& u_0\left[\beta x^2_0- \frac{1}{1-u^2_0}\right] =0
  \label{a57}
 \eea

The second equation in (\ref{a57}) has two solutions: $u_{0,1}=0$ and $u^2_{0,2} = 1-1/(\beta x^2_0)$.
For the first solution we have from the first equation in (\ref{a57}) $x=x_{0,1}$, where
\be
x_{0,1} = -\frac{\alpha}{2} + \sqrt{\frac{\alpha^2}{4} +1}
\label{a570}
 \ee
 (we recall that, by construction, $x>0$).
 For the  second solution, we have
 \be
 x_{0,2} = - \alpha/(1-\beta)~~ {\text{and}}~~ u^2_{0,2} = 1 - \alpha^2_c/\alpha^2
\label{a58}
 \ee
where $\alpha_c = (1-\beta)/\sqrt{\beta}$.  Obviously, the solution with a non-zero $u_{0,2}$ exists only
 for $\alpha <0$, when $|\alpha| > \alpha_c$.  At the critical value $\alpha = - \alpha_c$, $x_{0,1} = x_{0,2} = 1/\sqrt{\beta}$.

We now expand the action near each of the solutions. Expanding near $u_0=0$, $x= x_{0,1}$ we obtain
 \be
S[x,u] = S[x_{0,1}, 0] + \frac{u^2}{2} \left(1 -\beta x^2_{0,1} \right) + \frac{1}{2} (x-x_{0,1})^2 \left(1 + \frac{1}{x^2_{0,1}}\right)
 \label{a59}
 \ee
 We see that the prefactor for the $(x-x_{0,1})^2$ term is definitely positive, but the one for $u^2$ term may have either sign.
 The solution with $u_0 =0$ is the minimum of the effective action in the region where $\beta x^2_{0,1} <1$. An elementary calculation shows that
  this holds when $\alpha > - \alpha_c$.  We checked the second solution for these $\alpha$ and found that it corresponds to the maximum of the action and
  is  therefore irrelevant. Evaluating the Gaussian integrals over $u$ and over $x-x_{0,1}$, we obtain
 \be
I = \frac{A^2_M M^{M-1}}{\pi^{M-1} 2^{M}} \frac{1}{\sqrt{1-\beta x^2_{0,1}}} \frac{1}{\sqrt{1+x^2_{0,1}}}
\label{a60}
\ee

The case $\alpha > \alpha_c$ is more interesting for our purposes. Now the solution with $u_{0,1} =0$ becomes a maximum with respect to variations of $u$,
and we need to look at another extremal solution $ u = \pm u_{0,2} = \pm (1 - \alpha^2_c/\alpha^2)^{1/2}$ and $x = x_{0,2}$.
 Expanding in ${\tilde u} = u - u_{0,2}$ and ${\tilde x} = x - x_{0,2}$, we obtain
  \be
S[x,u] = S[x_{0,2}, u_{0,2}] + A {\tilde x}^2 + B {\tilde u}^2 - 2C {\tilde x} {\tilde u}
 \label{a61}
 \ee
where
 \bea
 A &=& \frac{1}{2} (1-\beta) + \beta (1-u^2_{0,2}) \nonumber \\
 B &=& \frac{u^2_{0,2}}{(1-u^2_{0,2})^2} \nonumber \\
 C^2 &=& \frac{\beta u^2_{0,2}}{1-u^2_{0,2}}
 \label{a62}
 \eea
 One can immediately make sure that
 \be
 AB-C^2 = \frac{1-\beta}{2} \frac{u^2_{0,2}}{(1-u^2_{0,2})^2} >0
  \label{a63}
 \ee
The integral
\be
J = \int d{\tilde x}\ d {\tilde u}\ e^{-M \left[A {\tilde x}^2 + B {\tilde u}^2 - 2C {\tilde x} {\tilde u}\right]}
  \label{a64}
 \ee
 then perfectly converges, no matter in what order we integrate. There is indeed no need to shift the integration contour from the real axis.
 Integrating in (\ref{a64}), we obtain
 \be
 J = \frac{\pi}{M} \frac{1}{\sqrt{AB-C^2}}
   \label{a65_1}
 \ee
 Substituting this result into the expression for $I$ and multiplying the result by $2$ because there are two extremal points
  $+ u_{0,2}$ and $-u_{0,2}$ and one has to expand near both, we obtain
  \be
  I = \frac{\sqrt{2}}{\beta^{M/2}} e^{\frac{M}{2} \left[\frac{\alpha^2}{1-\beta} +1\right]} \frac{\sqrt{(1-\beta)}}{\sqrt{\alpha^2 -\alpha^2_c}}
     \label{a65}
 \ee
 This result is valid as long as $u_{2,0}$ exceed typical $|\tilde u|$. The corresponding condition is $\alpha^2 -\alpha^2_c \geq 1/\sqrt{M}$.

   We next compute $I$ by applying HS transformation. The computational steps are the same as at large $N$, and the expression for $I$
    is

    \be
    I = \frac{M}{2\sqrt{\beta}} \int d \psi\ d\upsilon\ e^{-M {S}_{\rm eff} [\psi, \upsilon]}
      \label{a66}
 \ee
where
\be
{S}_{\rm eff} [\psi, \upsilon] = \frac{\psi^2}{2\beta} + \frac{\upsilon^2}{2} + \frac{1}{2} \log\left[(\alpha - i \psi)^2 - \upsilon^2\right]
    \label{a67}
 \ee
 The saddle-point equations have the same form as at large N: $\psi = i \psi_0$ and $\upsilon = \upsilon_0$, where
 \bea
&&\psi_0 = \frac{\alpha + \psi_0}{(\alpha + \psi_0)^2 - \upsilon^2_0} \nonumber \\
&&\upsilon_0 \left(1 - \frac{\beta}{(\alpha + \psi_0)^2 - \upsilon^2_0}\right) =0
     \label{a68}
 \eea
 One solution is obviously
 \be
 \upsilon_{0,1} =0, ~~\psi_{0,1} = -\frac{\alpha}{2} + \sqrt{\frac{\alpha^2}{4} +1}
      \label{a69}
 \ee
 The other solution is
\be
\upsilon^2_{0,2} = \frac{\beta^2}{(1-\beta)^2} \left(\alpha^2 - \alpha^2_c\right),~~\psi_{0,2} = -\frac{\alpha}{1-\beta},
      \label{a70}
 \ee
 where $\alpha_c = (1-\beta)/\sqrt{\beta}$ is the same as the one introduced after Eq.\ (\ref{a59}).

 For $\alpha > - \alpha_c$, one can easily show that the solution with $\upsilon_{0,1}=0$ corresponds to the minimum of ${S}_{\rm eff} [\psi,
 \upsilon]$.
  Expanding near this point and evaluating the (fully convergent) gaussian integrals over $\upsilon$ and over ${\tilde \psi} =\psi -i\psi_0$, we
  immediately
   reproduce Eq.\ (\ref{a60}).

 For $\alpha < - \alpha_c$, we need to consider the second solution and expand around a non-zero $\upsilon_{0,2}$ and $\psi_{0,2}$. There are two
  solutions: $+ \upsilon_{0,2} = \pm (\beta/(1-\beta)) (\alpha^2 - \alpha^2_c)^{1/2}$ and $-\upsilon_{0,2}$. We expand near one of them, say,
  $+\upsilon_{0,2}$, assume that we are in the region where $(\upsilon -\upsilon_{0,2})^2 \ll \upsilon^2_{0,2}$, and multiply the result by $2$.
   Expanding near $\upsilon_{0,2}$ and $\psi_{0,2}$, we obtain
   \be
S[\psi,\upsilon] = S[\psi_{0,2}, \upsilon_{0,2}] + {\bar A} {\tilde \psi}^2 - {\bar B} {\tilde \upsilon}^2  + 2i C {\tilde \psi} {\tilde \upsilon}
 \label{a71}
 \ee
where
 \bea
 {\bar A}  &=& \frac{1}{2} \left(1 + \frac{\beta + 2 \upsilon^2_{0,2}}{\beta^2}\right) \nonumber \\
 {\bar B} &=& \frac{\upsilon^2_{0,2}}{\beta^2} \nonumber \\
 {\bar C}^2 &=& \frac{\upsilon_{0,2} \sqrt{\upsilon^2_{0,2}+\beta}}{\beta^2}
 \label{a72}
 \eea
 One can immediately make sure that all three pre-factors are positive,  but now
 \be
 {\bar A} {\bar B}-{\bar C}^2 = -2 \frac{(1-\beta) \upsilon^2_{0,2}}{\beta^3} <0
  \label{a73}
 \ee
The quadratic form (ref{a71}) has exactly the same form as the one in our analysis of the stability of the CDW solution, see Eq.\ (\ref{y_3_2}) in the main
text.
 Here, however, we have a benchmark -- the result for $I$ must agree with Eq.\ (\ref{a65}).

  We first follow the analysis in the main text and combine the last three terms in the r.h.s.\ of (\ref{a71}) in the same way as we did there,  into
  \be
  {\bar A} \left({\tilde \psi} + i \frac{{\bar C}}{{\bar A}}{\tilde \upsilon}\right)^2 + \frac{{\bar C}^2 -{\bar A}{\bar B}}{{\bar A}} {\tilde \upsilon}^2
  \label{a74}
 \ee
We then integrate first over ${\tilde \psi}$ by shifting the integration variable by adding an imaginary constant, and then over
${\tilde \upsilon}$.  Both integrals are fully convergent, and integrating over ${\tilde \psi}$ and ${\tilde \upsilon}$ and assembling the prefactors, we
obtain
  \be
  I = \frac{\sqrt{2}}{\beta^{M/2}} e^{\frac{M}{2} \left[\frac{\alpha^2}{1-\beta} +1\right]} \frac{\sqrt{(1-\beta)}}{\sqrt{\alpha^2 -\alpha^2_c}}
 \label{a76}
 \ee
This is exactly the same result as Eq.\ (\ref{a65}).  The agreement justifies the integration procedure we used in the main text.
Like we said there, we could alternatively  integrate over ${\tilde \upsilon}$ first by
combining the last three terms in the r.h.s.\ of (\ref{a71}) into
  \be
  -{\bar B} \left({\tilde \upsilon} - i \frac{{\bar C}}{{\bar B}}{\tilde \psi}\right)^2 - \frac{{\bar C}^2 -{\bar A} {\bar B}}{{\bar B}} {\tilde \psi}^2,
  \label{a75}
 \ee
and then integrate over ${\tilde \psi}$.  In this integration procedure both integrals are formally  divergent if we integrate in infinite limits.
However, as we said in the main text, if we integrate in finite limits and set the limit of integration to infinity only after the integration, we
 do reproduce the same result as in Eqs.\ (\ref{a65}) and (\ref{a76}).

The conclusion of large $M$ analysis
is that it is perfectly legitimate to expand near the saddle point in which one variable is along the real axis and the other is
 along the imaginary axis, and the way to obtain the correct result is to combine variables such that we get two convergent integrals. Here we explicitly
 verified this by comparing the answer, Eqn (\ref{a76}),  with the one obtained by integrating over real axis, without shifting the contour, Eqn.
 (\ref{a65}).

 Another way to see that the integration around the saddle point in HS scheme is non-controversial is to consider the last three terms in the r.h.s.\ of
 (\ref{a71}) as a matrix and evaluate its two eigenvalues. This is a simple exercise, and the result is that both eigenvalues have positive real parts when
 ${\bar C}^2 > {\bar  A} {\bar  B}$, which implies that Gaussian integrals over fluctuations are convergent.

Indeed, in zero-D case there is no true  nematic order.  Still,  the analysis in this Appendix shows that one can successfully apply
   the HS procedure and reproduce the exact results for the partition function by integrating in the near vicinity of the  saddle point. The key message is
   that the need to integrate over fluctuations  along the contour in the complex plane is not an obstacle -- Gaussian integrals over fluctuations
   near the saddle point nicely converge.

\section{An alternative HS analysis, with saddle points along real axis}
\label{app:c2}

In Sec.\ \ref{ivb} of the main text we represented the 4-fermion interaction in the CDW channel as in Eq.\ (\ref{aa_1}), i.e. as
 \begin{align}
 H'=\bar\chi(\bar \rho_{k_0}\rho_{k_\pi}+\bar\rho_{k_\pi}\rho_{k_0})=\frac{\bar\chi}{2}\(\bar\rho_{k_0}+\bar\rho_{k_\pi}\)\(\rho_{k_0}+\rho_{k_\pi}\)-\frac{\bar\chi}{2}\(\bar\rho_{k_0}-\bar\rho_{k_\pi}\)\(\rho_{k_0}-\rho_{k_\pi}\).
\label{eq1}
 \end{align}
 In this Appendix we consider a more general representation
 \begin{align}
 H'=\bar\chi(\bar \rho_{k_0}\rho_{k_\pi}+\bar\rho_{k_\pi}\rho_{k_0})=&\frac{\bar\chi}{2\sqrt{a_1a_2}}\(\sqrt{a_2}\bar\rho_{k_0}+\sqrt{a_1}\bar\rho_{k_\pi}\)\(\sqrt{a_2}\rho_{k_0}+\sqrt{a_1}\rho_{k_\pi}\)\nonumber\\
 &-\frac{\bar\chi}{2\sqrt{a_1a_2}}\(\sqrt{a_2}\bar\rho_{k_0}-\sqrt{a_1}\bar\rho_{k_\pi}\)\(\sqrt{a_2}\rho_{k_0}-\sqrt{a_1}\rho_{k_\pi}\),
 \end{align}
 in which we initially treat $a_1$ and $a_2$ are arbitrary positive parameters.
 We see that we have two Hermitian interactions, the first one is repulsive and the second is attractive. We now introduce two HS fields $\Delta'$ and $\Delta$ for these interactions and perform HS transformation. We introduce $\tilde \chi\equiv \bar \chi/\sqrt{a_1a_2}$, and we use the same identities as Eq.\ (\ref{ywfri1}),
 \begin{align}
\exp{\(\frac{\tilde\chi}{2}{\bar z_+} z_-\)} &= \int \frac{d {\Delta'} d {\bar\Delta'}}{2\pi\tilde\chi} \exp{\[-\frac{|{\Delta}'|^2}{2\tilde\chi} + \frac{i}{2}\( {\Delta'} z_+ + {\bar\Delta'} {\bar z_+}\)\]}
\nonumber\\
\exp{\(\frac{\tilde\chi}{2}{\bar z_-} z_-\)} &=  \int \frac{d {\Delta} d {\bar\Delta}}{2\pi\tilde\chi} \exp{\[-\frac{|{\Delta}|^2}{2\tilde\chi} +\frac{1}{2}\( {\Delta} z_- + {\bar\Delta} {\bar z_-}\)\]}.
\label{ywfri2}
\end{align}
All integrals converge along the real axis. We apply these identities to  $z_+ = \sqrt a_2{\rho}_{k_0} + \sqrt a_1{\rho}_{k_\pi}$
 and $z_- = \sqrt a_2{\rho}_{k_0} - \sqrt a_1 {\rho}_{k_\pi}$
  and obtain the effective action
 \begin{align}
 S_{\rm {\rm eff}}=S_0&+\frac{1}{2\tilde\chi}\bar\Delta'\Delta'-\frac{i}{2}\bar\Delta'\(\sqrt{a_1}\rho_{k_\pi}+\sqrt{a_2}\rho_{k_0}\)-\frac{i}{2}\(\sqrt{a_1}\bar\rho_{k_\pi}+\sqrt{a_2}\bar\rho_{k_0}\)\Delta' \nonumber\\
 &+\frac{1}{2\tilde\chi}\bar\Delta\Delta-\frac{1}{2}\bar\Delta\(\sqrt{a_2}\rho_{k_0}-\sqrt{a_1}\rho_{k_\pi}\)-\frac{1}{2}\(\sqrt{a_2}\bar\rho_{k_0}-\sqrt{a_1}\bar\rho_{k_\pi}\)\Delta,
 \label{seff}
\end{align}
where $S_0$ contains the fermionic part of the action.

\subsection{Fluctuations at $T > T_{\rm cdw}$}

Now, let's integrate out the fermions and expand the effective action in powers of $\Delta$ and $\Delta'$.  To quadratic
  order in the HS fields we obtain
\begin{align}
S_{\rm {\rm eff}}[\bar\Delta',\Delta',\bar\Delta,\Delta]=&\frac{1}{2\tilde\chi}|\Delta'|^2+\frac{1}{2\tilde\chi}|\Delta|^2+\frac{1}{4}(a_1{A_2}+a_2{A_1})(|\Delta'|^2-
|\Delta|^2)\nonumber\\
&-\frac{i}{4}(a_2{A_1}-a_1{A_2})(\bar\Delta'\Delta+\bar\Delta\Delta').
\end{align}
where $A_1$ and $A_2$ are defined in Eq.\ (\ref{ch_10}).
We now choose
\begin{align}
a_1={A_1} ~~{\rm and}~~ a_2={A_2}.
\label{aa12}
\end{align}
With this choice, the effective action becomes Hermitian and we the HS fields $\Delta$ and $\Delta'$ decouple:
\begin{align}
S_{\rm {\rm eff}}[\bar\Delta',\Delta',\bar\Delta,\Delta]=\left (\frac{1}{2\tilde\chi}+\frac{A_1A_2}{2}\right) |\Delta'|^2+\left(\frac{1}{2\tilde\chi}-\frac{A_1A_2}{2} \right)|\Delta|^2.
\end{align}
 We see that the prefactor for $|\Delta'|^2$ is always positive, i.e. $\langle\Delta'\rangle=\langle\tilde\chi\(\sqrt{{A_1}}\rho_{k_\pi}+\sqrt{{A_2}}\rho_{k_0}\)\rangle=0$. Using this condition,  we find the relation between the average values of $\rho_{k_\pi}$ and $\rho_{k_0}$:
\begin{align}
\mu\equiv-\frac{\langle\Delta_{k_\pi^Q}\rangle}{\langle\Delta_{k_0}^Q\rangle}=-\frac{\langle\chi\rho_{k_\pi}\rangle}{\langle\chi\rho_{k_0}\rangle}=\sqrt{\frac{A_2}{A_1}},
\label{67}
\end{align}
 The prefactor for $|\Delta|^2$ term becomes negative at $T_{\rm cdw}$, for which, in original parameters,
  \begin{align}
 \bar \chi \sqrt{A_1 A_2}=1.
 \label{68}
 \end{align}
  Restoring the frequency and momentum dependence of $\bar \chi$,  we find the same CDW instability condition as in Eq.\ (\ref{tuac_6}). One can easily make sure
   that Eq.\ (\ref{67}) is also equivalent to the condition on $\langle\Delta_{k_\pi^Q}\rangle/\langle\Delta_{k_0}^Q\rangle$, which we obtained by
   solving the linearized gap equations (\ref{k0pilog}).

\subsection{Fluctuations at $T < T_{\rm cdw}$}

\subsubsection{Near $T_{\rm cdw}$}

  Next suppose that we at a temperature $T=T_{\rm cdw}-\delta$ and $\delta$ is small. In this range, the order parameter $|\Delta|^2\sim\delta$ is also small, and one can restrict with only fourth-order terms in the expansion in powers of $|\Delta|$. Expanding in $S_{\rm eff}$ we ontain
   \begin{align}
S_{\rm {\rm eff}}[\bar\Delta',\Delta',\bar\Delta,\Delta]=&\frac{1}{2\tilde\chi}(|\Delta'|^2+|\Delta|^2)+\frac{1}{4}(a_1{A_2}+a_2{A_1})(|\Delta'|^2-|\Delta|^2)\nonumber\\
&-\frac{1}{16}(a_1^2I_2+a_2^2I_1)\[(\bar\Delta^2-\bar\Delta'^{2})(\Delta^2-\Delta'^{2})-4|\Delta|^2|\Delta'|^2\]\nonumber\\
&-\frac{i}{4}(a_2{A_1}-a_1{A_2})(\bar\Delta'\Delta+\bar\Delta\Delta')\nonumber\\
&+\frac{i}{8}(a_2^2I_1-a_1^2I_2)\[\bar\Delta\bar\Delta^{'}(\Delta^2-\Delta'^{2})+(\bar\Delta^2-\bar\Delta'^{2})\Delta\Delta'\],
\end{align}

 We first find the modified relation between parameters $a_1$ and $a_2$, which will keep $\Delta'=0$ as an extremum of the action and then show that  this
 extremum  is a local minimum. The need to modify $a_1/a_2$  ratio comes from the fact that in the CDW-ordered state
   fourth-order terms bring $O(|\Delta|^2)=O(\delta)$ Gaussian corrections to  quadratic terms  in $\Delta'$ and $\bar\Delta'$.
A simple experimentation shows that we need to keep $\Delta'=0$  as an extremum we need to choose
\begin{align}
\frac{a_2}{a_1}=\frac{A_2}{A_1}+\frac{1}{2}\(|I_2|-\frac{A_2^2}{A_1^2}|I_1|\)|\Delta|^2,
\end{align} From the fact that $\langle\Delta'\rangle=\langle\tilde\chi\(\sqrt{a_1}\rho_{k_\pi}+\sqrt{a_2}\rho_{k_0}\)\rangle=0$ we find that
\begin{align}
\mu^2\equiv\frac{\langle\tilde\chi\rho_{k_\pi}\rangle^2}{\langle\tilde\chi\rho_{k_0}\rangle^2}=\frac{A_2}{A_1}+\frac{1}{2}\(|I_2|-\frac{A_2^2}{A_1^2}|I_1|\)|\Delta|^2.
\end{align}
One can straightforwardly check that this equation is consistent with Eq.\ \ (\ref{sat4}).

Next we verify that $\Delta'=0$ and $\Delta=O(\sqrt\delta)$  correspond to a local minimum of the effective action. We  write down the effective action
 as
 \begin{align}
S_{\rm eff}[\bar\Delta',\Delta',\bar\Delta,\Delta]=&\(\frac 1{\tilde\chi}+a\delta\)|\Delta'|^2-a\delta|\Delta|^2\nonumber\\
&+\frac{1}{16}(A_1^2|I_2|+A_2^2|I_1|)\[|\Delta'|^4+|\Delta|^4-(\bar\Delta^2\Delta'^2+\bar\Delta'^2\Delta^2)-4|\Delta|^2|\Delta'|^2\],
\label{y_1}
\end{align}
where $a>0$ is a number of order one. Expanding around
\begin{align}
|\Delta'|=&0\nonumber\\
|\Delta|=&\sqrt\frac{8\delta}{A_1^2|I_2|+A_2^2|I_1|}
\end{align}
 we immediately find that this solution is the local minimum of $S_{\rm eff}$.

It we now neglect the non-critical fields $\bar\Delta'$ and $\Delta'$, we
 obtain the effective action in terms of the order parameter
$\Delta $ {\it along the real axis}.  The action has the form
\begin{align}
S_{\rm eff}= \alpha |\Delta|^2 + \beta |\Delta|^4 + ...,
\label{y_2}
\end{align}
where $\alpha=-a\delta=a(T-T_{\rm cdw})$ and $\beta=({1}/{16})(A_1^2|I_2|+A_2^2|I_1|) >0$.  This agrees with Eq.\ (\ref{sat5}) (we recall that
 $(1+\lambda)/(1-\lambda)=\sqrt{A_1/A_2}$ in (\ref{sat5})).
 If we wouldn't neglect $\Delta'$ but rather integrated over it, (assuming that fourth order term $|\Delta'|^4$ is irrelevant) we obtained the same
 effective action as in (\ref{y_2}) but with slightly modified prefactors.

\subsubsection{Smaller temperatures, full non-linear analysis}

When $\delta$ is not small we can no longer expand in $\Delta$.  We go back to the original effective action Eq.\ (\ref{seff})
\begin{align}
S_{\rm eff}=&\frac{1}{2\tilde\chi}(|\Delta'|^2+|\Delta|^2)+\Psi_\pi^\dagger\mathcal{G}_{\pi}^{-1}\Psi_\pi+\Psi_0^\dagger\mathcal{G}_{0}^{-1}\Psi_0,
\end{align}
where we defined $\Psi_{\pi}^{\dagger}=(c_{k_\pi+Q}^\dagger,c_{k_\pi-Q}^\dagger)$, $\Psi_{0}^{\dagger}=(c_{k_0+Q}^\dagger,c_{k_0-Q}^\dagger)$, and
\begin{align}
\mathcal{G}_{\pi}^{-1}=\(
\begin{array}{c c}
G^{-1}_{k_\pi+Q} & \frac{\sqrt{a_1}}{2}(\bar\Delta+i\bar\Delta')\\
-\frac{\sqrt{a_1}}{2}(\Delta+i\Delta') & G^{-1}_{k_\pi-Q}
\end{array}
\),~\mathcal{G}_{0}^{-1}=\(
\begin{array}{c c}
G^{-1}_{k_0+Q} & \frac{\sqrt{a_2}}{2}(\bar\Delta-i\bar\Delta')\\
-\frac{\sqrt{a_2}}{2}(\Delta-i\Delta') & G^{-1}_{k_0-Q}
\end{array}
\)
\end{align}
Explicitly  integrating out fermionic degrees of freedom we obtain
\begin{align}
S_{\rm eff}[\bar\Delta',\Delta',\bar\Delta,\Delta]=&\frac{1}{2\tilde\chi}(|\Delta'|^2+|\Delta|^2)\nonumber\\
&-\log\left\{G^{-1}_{k_\pi+Q}G^{-1}_{k_\pi-Q}-\frac{a_1}{4}\[|\Delta|^2-|\Delta'|^2+i(\bar\Delta\Delta'+\bar\Delta'\Delta)\]\right\}\nonumber\\
&-\log\left\{G^{-1}_{k_0+Q}G^{-1}_{k_0-Q}-\frac{a_2}{4}\[|\Delta|^2-|\Delta'|^2-i(\bar\Delta\Delta'+\bar\Delta'\Delta)\]\right\}.
\end{align}
The summations over frequency and momentum are assumed.

Like before, we must tune  $a_1$ and $a_2$ such that $\Delta'=0$ remains an extremum of the action, and then to show that this extremum
 is actually a local minimum.

First we differentiate the effective action with respect to $\bar\Delta'$, and we find,
\begin{align}
\left.\frac{\partial S_{\rm eff}}{\partial \bar\Delta'}\right|_{\Delta'=0}=&\frac{ia_1}{4}\sum_{k,\omega}\frac{\Delta}{G_{k_\pi+Q}^{-1}G_{k_\pi-Q}^{-1}-\frac{a_1}{4}|\Delta|^2}-\frac{ia_2}{4}\sum_{k,\omega}\frac{\Delta}{G_{k_0+Q}^{-1}G_{k_0-Q}^{-1}-\frac{a_2}{4}|\Delta|^2}\nonumber\\
=&-\frac{i}{4}(a_1\bar{A_2}-a_2 \bar A_1)\Delta,
\end{align}
where in the last line we have defined
\begin{align}
\bar A_2=\sum_{k,\omega}\frac{1}{-G_{k_\pi+Q}^{-1}G_{k_\pi-Q}^{-1}+\frac{a_1}{4}|\Delta|^2}\nonumber\\
\bar A_1=\sum_{k,\omega}\frac{1}{-G_{k_0+Q}^{-1}G_{k_0-Q}^{-1}+\frac{a_2}{4}|\Delta|^2}.
\end{align}
Diagrammatically, $\bar A_1$ and $\bar A_2$ are nothing but the polarization bubbles with fully dressed normal and anomalous propagators.
 Imposing the condition on the extremum of the action, we obtain
\begin{align}
a_1=\bar A_1 ~~{\rm and}~~ a_2=\bar A_2,
\end{align}
which is to be compared with Eq.\ (\ref{a12}). Taking derivative with respect to $\Delta$, we find that the CDW instability sets in when
\begin{align}
\tilde\chi \bar A_1 \bar A_2=1.
\label{Tcdw}
\end{align}

Next we show that $\Delta'=0$ is a local minimum. We define the real part and imaginary parts of $\Delta$ and $\Delta'$ as
\begin{align}
\Delta&=x+iy\nonumber\\
\Delta'&=x'+iy'.
\end{align}
Substituting this into the action we obtain
\begin{align}
S_{\rm eff}[x,y,x',y']=&\frac{1}{2\tilde\chi}(x^2+y^2+x'^2+y'^2)\nonumber\\
&-\log\left\{G^{-1}_{k_\pi+Q}G^{-1}_{k_\pi-Q}-\frac{\bar A_1}{4}\[x^2+y^2-x'^2-y'^2+i(xx'+yy')\]\right\}\nonumber\\
&-\log\left\{G^{-1}_{k_0+Q}G^{-1}_{k_0-Q}-\frac{\bar A_2}{4}\[x^2+y^2-x'^2-y'^2+i(xx'+yy')\]\right\}.
\end{align}
Differentiating $\mathcal {S}_{\rm eff}$ twice with respect to $x'$ and using Eq.\ (\ref{Tcdw}) we obtain
\begin{align}
\left.\frac{\partial^2S_{\rm eff}}{\partial x'^2}\right|_{x'=0,y'=0}=2\bar A_1\bar A_2-\frac{x^2}{16}(\bar A_2^2|\bar I_1|+\bar A_1^2|\bar I_2|),
\label{24}
\end{align}
where
\begin{align}
\bar I_2=&-\frac{1}{2}\sum_{k,\omega}\frac{1}{(-G_{k_\pi+Q}^{-1}G_{k_\pi-Q}^{-1}+\frac{1}{4}\bar A_1 (x^2+y^2))^2},\nonumber\\
\bar I_1=&-\frac{1}{2}\sum_{k,\omega}\frac{1}{(-G_{k_0+Q}^{-1}G_{k_0-Q}^{-1}+\frac{1}{4}\bar A_2 (x^2+y^2))^2}
\end{align}
These two are given by the same square diagrams from Fig.\ \ref{2}, which we used before, but for the case when CDW order is already developed
Evaluating the integrals we find that $\partial^2S_{\rm eff}/{\partial x'^2}$ is positive no matter what $x$ is.
 The same holds for differentiation over $y'$. Hence, $\Delta'=0$ is a local minimum.

The rest of the analysis proceeds the same way as near $T=T_{\rm cdw}$. Namely, if we neglect $\Delta'$, the effective action in terms of $\Delta$ has a conventional form,
 and $|\Delta|^2$ increases as $T$ decreases. This still holds even if we perform Gaussian integration over fluctuations of $\Delta'$.

\section{An alternative method to go beyond hot spot treatment}
\label{app:d}
\begin{figure}
\includegraphics[width=0.5\columnwidth]{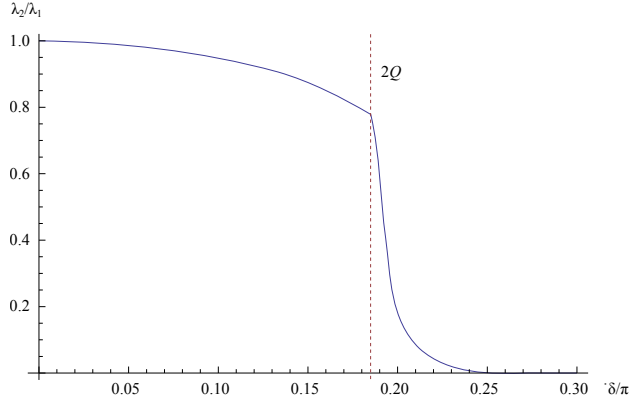}
\caption{The ratio of the eigenvalues $\lambda_1$ and $\lambda_2$ for even and odd in $k$ CDW order parameters, $\Delta_1^Q$  and $\Delta_2^Q$,
respectively,
 as a function as the momentum integration range $\delta$ of integration around the hot spots.
 Note the change of the behavior at $\delta=2Q$. We set  $T=1 {\rm ~meV}$.}
\label{TRSB}
\end{figure}

In this Appendix we present a complimentary approach to go beyond hot spot treatment of the CDW order parameters $\Delta_1^Q$ and $\Delta_2^Q$. The
conclusion we reach here is the same -- $\Delta_1^Q$ has a stronger instability. For GL coefficients in Eq.\ (\ref{seff3}), this corresponds to
$\alpha_2>\alpha_1$. In the main text we assumed that even and off components of $\Delta^Q_k$ behave as $\cos k$ and $\sin k$, respectively, along the direction of the center-of-mass momentum ${\bf k}$. Here we assume a more simple momentum dependence of $\Delta^Q_{1} (k)$ and $\Delta^Q_{2} (k)$, namely,
assume that the CDW gap is concentrated around anti-nodal regions, and even component is a constant and the odd component by a linear dependence on momentum in a hot region:
\begin{align}
\Delta_1^Q({ k})&\approx{\rm const.} = \Delta^Q_1 (k_0) \nonumber\\
\Delta_2^{Q}({ k})& \approx |\Delta^Q_2 (k_0)| \frac{\pi-k_x -k_y}{\pi-k_0},
\end{align}
where $k_0$ is the center-of-mass momenta when $\Delta^Q_k$ connects two fermions right at hot spots.
Obviously, $\Delta_1^Q$ and $\Delta_2^Q$ are symmetric and antisymmetric about the $(\pi,0)$ point, respectively.
We approximate the interaction $\bar\chi({\bf k},{\bf k'})$ by a constant within some momentum window
 $\pi-\delta<k_x'-k_x<\pi+\delta$ and $\pi-\delta<k_y'-k_y<\pi+\delta$ and set it to zero outside this window. We then explicitly compute  the eigenvalues
 for even and odd in $k$ solutions
   and compare them.  The two eigenvalues are given by ($\lambda_1$ is for even solution)
   \begin{align}
\lambda_1=&\bar\chi^2\int_{-\delta}^{\delta}\frac{d{ p}~f(\epsilon_{{ k}_\pi+{ p}+{ Q}})-f(\epsilon_{{ k}_\pi+{ p}-{ Q}})}{\epsilon_{{ k}_\pi+{ p}-{
Q}}-\epsilon_{{ k}_\pi+{ p}+{ Q}}}\int_{-\delta}^{\delta}\frac{d{ q}~f(\epsilon_{{ k}_0+{ p}+{ q}+{ Q}})-f(\epsilon_{{ k}_0+{ p}+{ q}-{ Q}})}{\epsilon_{{
k}_0+{ p}+{ q}-{ Q}}-\epsilon_{{ k}_0+{ p}+{ q}+{ Q}}}\frac{\Delta_1^{Q}({ k}_0+{ p+q})}{\Delta_1^{Q}({ k}_0)}\nonumber\\
\lambda_2=&\bar\chi^2\int_{-\delta}^{\delta}\frac{d{ p}~f(\epsilon_{{ k}_\pi+{ p}+{ Q}})-f(\epsilon_{{ k}_\pi+{ p}-{ Q}})}{\epsilon_{{ k}_\pi+{ p}-{
Q}}-\epsilon_{{ k}_\pi+{ p}+{ Q}}}\int_{-\delta}^{\delta}\frac{d{ q}~f(\epsilon_{{ k}_0+{ p}+{ q}+{ Q}})-f(\epsilon_{{ k}_0+{ p}+{ q}-{ Q}})}{\epsilon_{{
k}_0+{ p}+{ q}-{ Q}}-\epsilon_{{ k}_0+{ p}+{ q}+{ Q}}}\frac{\Delta_2^{Q}({ k}_0+{ p+q})}{\Delta_2^{Q}({ k}_0)},
\label{alpha12}
\end{align}
where $f(\epsilon)$ is the Fermi function.
We set ${\bf Q} = Q_y$ and ${\bf k}_0 = (\pi-Q,0)$ and  evaluated the two integrals numerically  using the dispersion relation $\epsilon_k$  for $\rm{Pb_{0.55}Bi_{1.5}Sr_{1.6}La_{0.4}CuO_{6+\delta}}$ (see
 Ref.\ [\onlinecite{zxshen}]).
 In  Fig.\ \ref{TRSB} we show  our results for $\lambda_2/\lambda_1$ as a function of the size of the momentum window $\delta$.

 We see from the plot  that, as expected, $\lambda_2<\lambda_1$, i.e., the even solution emerges at a higher $T$. At the same time, the values of
 $\lambda_2$ and $\lambda_1$ are quite close as long as  $\delta <2Q$.  That $\lambda_1$ is larger, but $\lambda_2$ is close second is consistent with
  the analysis in the main text and with
 Refs.~[\onlinecite{subir_2,subir_4}].

\section {Bond order  with diagonal momenta $(\pm 2Q, \pm 2Q)$}
\label{app:b}

   \begin{figure}
\includegraphics[width=0.5\columnwidth]{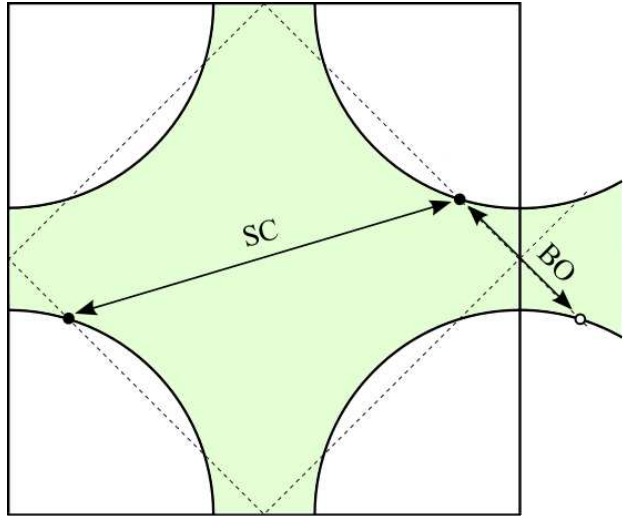}
\caption{Fermions which contribute to diagonal bond order and $d-$wave superconducting order. Filled and empty circles denote particle and hole states,
respectively.}
\label{bond order}
\end{figure}

For completeness and for comparison with our results on CDW order with $Q_x$ or $Q_y$, we also consider charge order with momenta ${\bf \bar Q}=(2Q, \pm
2Q)$, as depicted in Fig.\ \ref{bond order}. A charge order with diagonal momentum
 has been studied in Refs.\ [\onlinecite{ms, efetov}]. The critical temperature for the instability towards such order
 is exactly the same as superconducting $T_{\rm sc}$ if one neglects the curvature of the FS near the hot spots.
   The gap function for a diagonal charge order has a $d$-wave structure $\langle{c}^{\dagger}(k+\bar Q){c}(k-\bar Q)\rangle=\Delta(\cos k_x-\cos
   k_y)$, the same as a $d-$wave  superconducting order parameter. A $d-$wave charge order
   does not create a charge density modulation $\langle c^{\dagger}(r)c(r)\rangle$, but it introduces modulations of the correlation  function between
   neighboring sites:
\begin{align}
\langle c^{\dagger}({\bf r})c({\bf r+a})\rangle=2\Delta\cos {\bf \bar Q}\cdot\left({\bf r}+\frac{{\bf a}}{2}\right)(\delta_{{\bf a},
{\bf x}}-\delta_{{\bf a},{\bf y}})
\end{align}
where
${\bf x}$ and ${\bf y}$ are vectors along $x$ and $y$ directions, in units of interatomic spacing ${\bf a}$.
 A charge order of this kind is called bond order (Bo)

To obtain the onset temperature for BO, $T_{\rm bo}$,  and compare it with SC $T_{\rm sc}$ in the presence of FS curvature,
  we add to the spin-fermion action two
infinitesimal
 vertices $\Phi_0(k) c_{k,\alpha}(i\sigma^{y}_{\alpha\beta})c_{-k,\beta}$ and $\Psi_0(k)c_{k,\alpha}\delta_{\alpha\beta}c^{\dagger}_{k+\bar  Q,\beta}$,
  where $k$ stands for 2+1 momentum $(\omega_m, {\bf k})$. These vertices gets renormalized by spin-fermion interaction, and the critical temperature
  ($T_{\rm bo}$
   or $T_{\rm sc}$)  is obtained when the corresponding susceptibility diverges, i.e., the solution for fully renormalized $\Phi(k)$ or $\Psi(k)$ exists
   even when the
    bare vertices are set to zero.

\begin{figure}
\includegraphics[width=0.5\columnwidth]{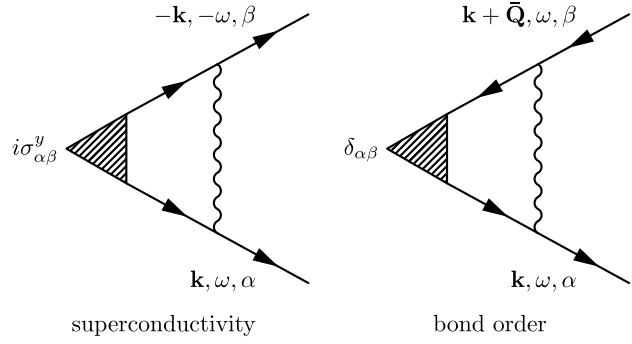}
\caption{The diagrammatic expressions for the fully renormalized vertices in superconducting and bond order channels.}
\label{fig2}
\end{figure}

   The authors of~[\onlinecite{ms,efetov}] have demonstrated that a superconducting instability and an instability towards bond order come from the
   fermions located in the same
   hot regions, only for bond order one of the regions is shifted by $(2\pi,0)$. The ladder renormalizations of $\Phi_0(k)$ and $\Psi_0(k)$ are shown in
    in Fig.\ \ref{fig2}, where the wavy line is the spin-fermion interaction. In analytical form we have, at the corresponding critical temperatures,
\begin{align}
\Phi(k)=&-3{\bar g} \int G(k')G(-k')\chi(k-k')\Phi(k'+\pi)\nonumber\\
\Psi(k)=&3{\bar g} \int G(k')G(k+\bar Q)\chi(k-k')\Psi(k'+\pi),
\label{ch_11}
\end{align}
where the spin-fermion coupling ${\bar g}$ and the dynamical spin susceptibility $\chi (k-k')$ are defined in Eqs.\ (\ref{as_1}) and (\ref{as_1_1}) in
the main text. The difference in the overall sign in the r.h.s is due to different Pauli algebra -- for superconducting vertex
 ${\sigma}^{i}_{\alpha'\alpha}(i\sigma^{y}_{\alpha\beta}){\sigma}^{i}_{\beta\beta'}=-3i\sigma^{y}_{\alpha'\beta'}$, while for bond vertex
 ${\sigma}^{i}_{\alpha'\alpha}(\delta_{\alpha\beta}){\sigma}^{i}_{\beta\beta'}=3\delta_{\alpha'\beta'}$. One can easily verify~\cite{ms} that,
  if one neglects the curvature of the FS, one finds $\epsilon_{k+\bar Q} = -\epsilon_k$, and hence  $G(-k)=-G(k+\bar Q)$.
    In this approximation, the kernels in the two equations (\ref{ch_11}) become identical, hence, $T_{\rm sc} =
    T_{\rm bo}$. Once the curvature of the FS is included, the
    degeneracy is lifted and $T_{\rm sc} > T_{\rm bo}$.  The reasoning is that superconductivity involves fermions with
    strictly opposite $k$, and the momentum integration can still be replaced by the integration over $\epsilon_k$, with a constant prefactor, even in the
    presence of the FS curvature. For BO, the relation $\epsilon_{k +\bar Q} = - \epsilon_k$ no longer holds in the presence of the FS curvature,
    and this reduces the kernel in the Eq.\ (\ref{ch_11}) for $\Psi$.

Explicitly, expanding near a hot spot, we obtain
 \begin{align}
 \epsilon(k)=&- v_F(k_{\perp}+{\tilde \kappa} k_{\|}^2/k_F) \nonumber \\
 \epsilon(k+\pi)=& v_F(k_{\perp}-{\tilde \kappa} k_{\|}^2/k_F)
 \end{align}
  where $k_\|$ and $k_\perp$ are momenta parallel and perpendicular to the FS, respectively, and ${\tilde \kappa}$ is a dimensionless parameter
  characterizing the curvature of the FS.  We will  use dimensionless parameters
\begin{align}
\tilde g\equiv &\frac{\bar g}{v_F k_F}\nonumber\\
\tilde \xi\equiv &\xi k_F.
\end{align}
where $\xi$ is the magnetic correlation length, present in $\chi (k-k')$.
The dimensionless coupling  $\tilde g$, dimensionless correlation length $\tilde \xi$, and dimensionless FS curvature
 $\tilde\kappa$  are three input parameters for the consideration in this Section.
  An additional parameter, set by the FS geometry,  is the angle between Fermi velocities at hot spots separated by $(\pi,\pi)$.
  To simplify the presentation, we assume that these two velocities are orthogonal to each other.

In this Appendix we follow earlier works~\cite{acf,acs,ms,efetov,senthil,norman} and assume that the spin-fermion interaction can be well approximated by its value between
fermions
 on the FS. Integrating  over momenta transverse to the FS in the fermionic propagators, we obtain integral equations for $\Phi(\omega_m,k_\|)$ and
 $\Psi(\omega_m,k_\|)$, which depend on frequency and on momenta along the FS. The equations are
\begin{align}
\Phi(\omega_m,k_\|)=\frac{3 {\tilde g} k_F}{2} \int_{m'k_\|'}K(\omega_m,k_\|,\omega_m',k_\|';0)\Phi(\omega_m',k_\|'), \label{Phi}
\\
\Psi(\omega_m,k_\|)=\frac{3 {\tilde g} k_F}{2}  \int_{m'k_\|'}K(\omega_m,k_\|,\omega_m',k_\|';{\tilde \kappa})\Psi(\omega_m',k_\|'), \label{Psi}
\end{align}
where $\int_{m'k_\|'}$ stands for $T\sum_{m'}\int dk_\|'/2\pi$ and
\begin{align}
K(\omega_m,k_\|,\omega_m',k_\|';{\tilde \kappa})=\frac{|\omega_m'+\Sigma (\omega_m', k_\parallel')|}
{[\omega_m'+\Sigma (\omega_m', k_\parallel')]^2+v_F^2{\tilde \kappa}^2k_\|^{\prime 4}/k_F^2}~\frac{1}{k_\|^2+k_\|^{\prime
2}+\gamma|\omega_m-\omega_m'|+{\tilde \xi}^{-2}k_F^2}
\label{kernel}
\end{align}

\subsection {$T_{\rm sc}$ and $T_{\rm bo}$  at the onset of SDW order, $\xi^{-1} =0$}
\label{tc}

It is convenient to introduce the set of rescaled variables
 \begin{align}
\bar T &=\frac{\pi T}{\omega_0},~~~\bar \omega_m=\frac{\omega_m}{\omega_0},~~~\bar k_\|=\frac{k_\|}{\sqrt{\gamma\omega_0}}.
\end{align}
where $\omega_0 = 9 {{\bar g}}/(16 \pi ) \times[(v_y^2-v_x^2)/v^2_F]$ was introduced in the main text.
In these notations, the linearized gap equation for BO becomes
\begin{align}
\Psi(\bar \omega_m,\bar k_\|)=\frac{1}{4\pi}\int_{\bar T}\frac{d\bar\omega_m~d\bar k_\|'}{\bar k_\|^2+\bar
k_\|^{'2}+|\bar\omega_m-\bar\omega_m'|}\frac{|\bar\omega_m'+\bar\Sigma(\bar\omega_m',\bar k_\|')|}{|\bar\omega_m'+\bar\Sigma(\bar\omega_m',\bar
k_\|')|^2+16{\tilde g}^2{\tilde \kappa}^2\bar k_\|^{'4}/\pi^2}\Psi(\bar \omega_m',\bar k_\|'),
\label{ch_8}
\end{align}
where the rescaled self-energy is\cite{acs,ms}
\begin{align}
\bar \Sigma(\bar\omega_m,\bar k_\|)=\sqrt{|\bar\omega_m|+\bar k_\|^2}-\left|\bar k_\|\right|.
\end{align}
 We verified, using the same strategy as in our earlier work~\cite{wang} on superconducting $T_{\rm sc}$ at ${\tilde \xi}^{-1}=0$,
  that the leading contribution to the r.h.s of Eq.\ (\ref{ch_8}) comes from the region where $\bar \Sigma>\bar \omega_m'$ and $\bar k_\|^2>\bar \omega$. In
  this region, the momentum dependence of $\Psi$ is more relevant than its frequency dependence.
   Keeping only the momentum dependence in $\Psi$ and introducing $x=\bar k_\|^2$ and $y=\bar k$, we re-write (\ref{ch_8}) as
\begin{align}
\Psi(y)=\frac{1}{2\pi}\int_{\bar T_{\rm bo}}^{1}\frac{dx}{x+y}\log\frac{x^2+64{\tilde g}^2{\tilde \kappa}^2x^3}{\bar T^2_{\rm bo}+64{\tilde g}^2{\tilde
\kappa}^2x^3}\Psi(x).
\label{Psixy}
\end{align}
For superconductivity, the same procedure yields
\begin{align}
\Phi(y)=\frac{1}{\pi}\int_{\bar T_{\rm sc}}^{1}\frac{dx}{x+y}\log\frac{x}{\bar T_{\rm sc}}\Phi(x).
\label{Phixy}
\end{align}
Comparing Eqs.\ (\ref{Psixy}) and (\ref{Phixy}),  we find  that extra terms in the logarithm in (\ref{Psixy}) make it smaller than the logarithm in (\ref{Phixy}), hence in the presence of a FS curvature $T_{\rm bo}$ gets smaller than $T_{\rm sc}$. Specifically, the curvature term couples to $x^3$ and provides a soft {\it upper} cutoff to the integral over $x$, at $x \sim 1/({\tilde g}{\tilde \kappa})^2$.
At the same time, $T_{\rm bo}$ remains finite, no matter how large $\kappa$ is. Indeed, at large ${\tilde g}{\tilde \kappa}$ we have $T_{\rm bo} \propto  T_{\rm sc}/({\tilde g}{\tilde \kappa})^2 \ll 1$.  In other words, at ${\tilde \xi}^{-1} =0$,  there is no threshold value of ${\tilde \kappa}$ above which BO would not develop.

\begin{figure}
\includegraphics[width=0.5\columnwidth]{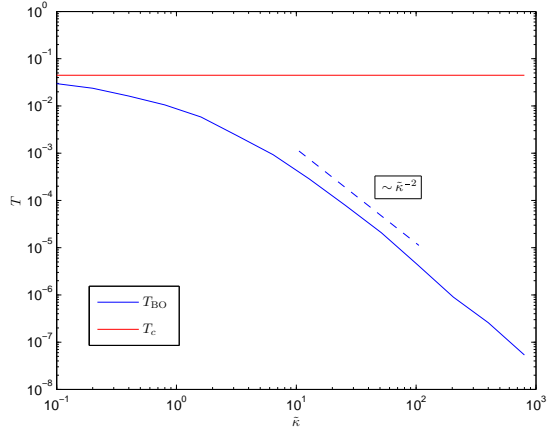}
\caption{The onset temperatures for SC  and BO $T_{\rm sc}$ and $T_{\rm bo}$, respectively as functions of dimensionless FS
curvature
${\tilde \kappa}$ at the onset of SDW order (magnetic $\xi^{-1} = 0$).  We  set ${\tilde g}=0.1$.
 Superconducting $T_{\rm sc}$ is not affected by the curvature, while $T_{\rm bo}$ decreases but remains finite. In analytical consideration $T_{\rm
 bo}/T_{\rm sc}$ was found to scale as ${\tilde \kappa}^{-2}$ at large enough curvature. We show this functional behavior by a dashed line.}
\label{fig3_1}
\end{figure}

To check our analytical reasoning, we solved Eqs (\ref{Phi}) and (\ref{Psi})  numerically and obtained the same result, namely $T_{\rm bo}$ decreases with
increasing ${\tilde \kappa}$ but remains finite. We show the results in Fig.\ \ref{fig3_1}. We set  ${\tilde g}=0.1$, ${\tilde \xi}=\infty$, and varied
${\tilde \kappa}$.

\subsection{$T_{\rm sc}$ and $T_{\rm bo}$  at a finite ${\tilde \xi}$}
\label{app:b2}

When the system moves away from the magnetic QCP, it eventually recovers a conventional FL behavior in the
 normal state. Indeed, as the correlation length ${\tilde \xi}$ decreases, $\omega_{\rm sf} = {\tilde \xi}^{-2}k_F^2/\gamma$
 becomes the upper energy cutoff for the pairing~\cite{acs,maslov_1}.
  Below this scale, the spin susceptibility can be treated as frequency-independent and the fermionic self-energy is linear in frequency:
\begin{align}
\Sigma(\omega_m,k_\|)=&\frac{3{\tilde g}
k_F}{2\pi\sqrt{\gamma}}\left(\sqrt{|\omega_m|+k_\|^2/\gamma+\omega_{\rm sf}}-\sqrt{k_\|^2/\gamma+\omega_{\rm sf}}\right){\rm sgn}{(\omega_m)}\nonumber\\
\approx&\frac{3{\tilde g {\tilde \xi}\sqrt{\omega_{\rm sf}}}}{2\pi}\frac{\omega_m}{2\sqrt{\omega_{\rm sf}}}\nonumber\\
=&\lambda\omega_m.
\end{align}
In the last line we have defined $\lambda=\frac{3{\tilde g} {\tilde \xi}}{4\pi}$.
Plugging this into Eq.\ (\ref{Phi}) for superconducting $T_{\rm sc}$ and using the condition that typical $\omega,\omega_m',k_\|^{2},k_\|^{'2}/\gamma$ are
all small in this limit compared to $\omega_{\rm sf}$, we  integrate over momentum $k_\|'$ and  obtain
\begin{align}
\Phi=\frac{\lambda}{1+{\lambda}}\log\frac{\omega_{\rm sf}}{T_{\rm sc}}\Phi.
\end{align}
For superconducting $T_{\rm sc}$ we then have a conventional
BCS-McMillan result~\cite{mcmillan}
\begin{align}
T_{\rm sc}\sim\omega_{\rm sf}\exp\left(-\frac{1+\lambda}{\lambda}\right).
\end{align}
 Hence, as the system moves away from the QCP, it crosses over to a BCS behavior, and $T_{\rm sc}$ gradually decreases as ${\tilde \xi}$ decreases and
 $\lambda$ gets smaller.

For BO,  the gap equation in the rescaled variables becomes, in this limit,
\begin{align}
\Psi=\frac{1}{4\pi}\int_{{\bar T}_{\rm bo}}\frac{d\omega_m~d\bar k_\|'}{\bar
k_\|^{'2}+\tilde\omega_{\rm sf}}\frac{(1+\lambda)|\bar\omega_m'|}{(1+\lambda)^2|\bar\omega_m'|^2+16{\tilde g}^2{\tilde \kappa}^2\bar k_\|^{'4}/\pi^2}\Psi,
\label{ch_9}
\end{align}
where we  defined
\begin{align}
\tilde\omega_{\rm sf} \equiv\frac{\omega_{\rm sf}}{\omega_0} =\left(2\lambda\right)^{-2}.
\end{align}

Typical $\bar k_\|^{'2}$ are of order ${\tilde \omega}_{\rm sf}$,
 and in the second term in the denominator we can safely replace $\bar k_\|^{'4}$ by $\tilde\omega_{\rm sf}^2$.
   We see that the curvature ${\tilde \kappa}$ now appears in a combination with a constant term and
     provides a {\it lower} cutoff for the BCS-like logarithmic behavior.
      This is qualitatively different from the behavior at the magnetic QCP, where the curvature was coupled to the running variable $x^3$.
       Because of the cutoff, the frequency integral in (\ref{ch_9})  no longer diverges at $T=0$. Hence, at some critical ${\tilde \xi}$, the linearized gap equation for
       BO vertex $\Psi$ has a solution at $T=0$.  Setting $T=0$ in
  (\ref{ch_9})   and integrating over $\bar k_{\|}$, we obtain the condition when $T_{\rm bo} =0$:
\begin{align}
\Psi=\frac 12~\frac{\lambda \Psi}{1+\lambda}\int_{-\omega_{\rm sf}}^{\omega_{\rm sf}}\frac{d\omega~ |\omega|}{|\omega|^2+16\tilde g^2\tilde
\kappa^2\omega_{\rm sf}^2/\pi^2}.
\end{align}
Canceling out $\Psi$ and integrating over frequency, we find
\begin{align}
1=\frac{\lambda}{1+\lambda}\left(\log{\frac{\pi}{4{\tilde g}{\tilde \kappa}}}\right).
\label{txi}
\end{align}
This defines a critical ${\tilde \xi}$ at which BO vanishes:
\begin{align}
{\tilde \xi}_{\rm cr}^{-1}\sim\tilde{g}\log\left(\frac{\pi}{4 {\tilde g}{\tilde \kappa}}\right).
\label{xcr}
\end{align}
At smaller  ${\tilde \xi}<{\tilde \xi}_{\rm cr}$, the equation on $\Psi$ only allows a trivial solution $\Psi=0$, hence BO does not develop at any
$T$.

\begin{figure}
\includegraphics[width=0.5\columnwidth]{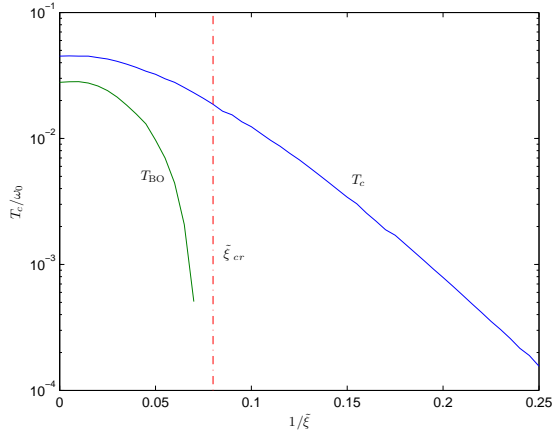}
\caption{$T_{\rm sc}$ and the onset temperature $T_{\rm bo}$ for BO with diagonal ${\bf Q}$ as functions of the  magnetic
correlation length.
We set ${\tilde \kappa}=0.14$ and ${\tilde g}=0.1$. The red dashed line is ${\tilde \xi}_{\rm cr}$, given by Eq.\ (\ref{xcr}).}
\label{figg}
\end{figure}
\begin{figure}
\includegraphics[width=0.5\columnwidth]{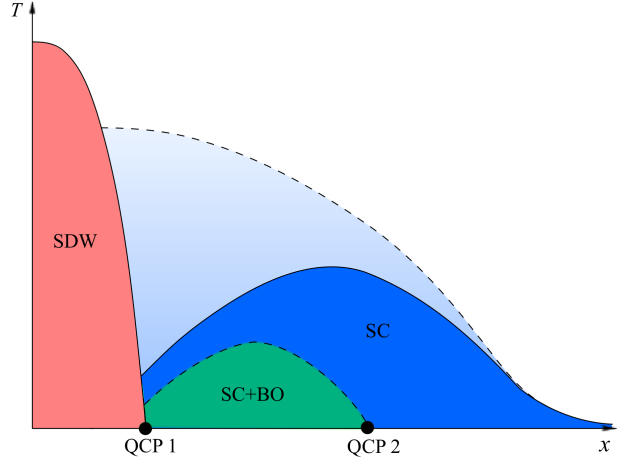}
\caption{A schematic phase diagram for the (artificial) case when the only two competing states are SC and BO with a diagonal ${\bf
Q}$.
 The light blue region is where the pseudogap phase, which combines SC fluctuations at large $x$ (smaller $\xi$) and fluctuations between SC  and
 BO at small $x$ (large $\xi$). }
\label{fig4}
\end{figure}

To verify this, we solved for $T_{\rm sc}$ and $T_{\rm bo}$ numerically. We set  ${\tilde g}=0.1$ and ${\tilde \kappa}=0.14$ and varied ${\tilde \xi}$.
 We found that  superconducting $T_{\rm sc}$  crosses over  to BCS behavior at small enough ${\tilde \xi}$ and that for BO there exists a
 critical
  ${\tilde \xi}$ at which $T_{\rm bo}=0$.  We show the results in Fig.\ \ref{figg}.

 Although the behavior of $T_{\rm bo}$  resembles that of $T_{\rm cdw}$ for CDW order with $Q_x/Q_y$,
   the physics behind the reduction of these temperatures with decreasing magnetic $\xi$ is different.  For BO with diagonal ${\bar Q} = (2Q, \pm 2Q)$, the
   reduction of $T_{\rm bo}$  compared to $T_{\rm sc}$ and its eventual vanishing is solely due to the FS curvature. If we set ${\tilde \kappa}$ to zero,
$T_{\rm sc}$ and $T_{\rm bo}$ remain identical at any $\xi$.  For CDW order with $Q_x/Q_y$, the reduction and eventual vanishing
 of $T_{\rm cdw}$ upon decreasing of ${\tilde \xi}$ is not related to curvature and holds even if we set curvature to zero.  At a small curvature then,
 $T_{\rm bo} > T_{\rm cdw}$, but which temperature is larger at ${\tilde \kappa} = O(1)$ depends on numbers.  We emphasize again in this regard  that
  for CDW with $Q_x/Q_y$ the first instability upon lowering $T$ is at $T> T_{\rm cdw}$, towards the state which breaks
  an Ising $Z_2$  symmetry. No such transition holds for BO with a diagonal ${\bf Q}$.

\subsection{Interplay between superconductivity and bond order}

In Fig.\  \ref{fig4} we present the phase diagram for the (artificial) case when the only two competing states are superconductivity and BO with a diagonal
${\bf Q}$, i.e. when CDW order with $Q_x/Q_y$ is just neglected.
  Because $T_{\rm sc}$ is larger than $T_{\rm bo}$, the leading instability upon lowering $T$ is always into a
  superconducting state, BO may appear only at a lower $T$.  At the same time, at large $\xi$, $T_{\rm sc}$ by itself is reduced because over some
  range of $T$ the system fluctuates between superconductivity and BO~\cite{efetov,subir_2} (a light blue region in Fig.\ \ref{fig4}).
 The phase diagram in Fig.\ \ref{fig4} is similar to that in Ref.\ [\onlinecite{efetov}], but differs in that in our analysis BO only emerges at
 ${\tilde \xi} > {\tilde \xi}_{\rm cr}$, i.e., there exists a  quantum-critical point  towards BO (QCP 2) at some distance away from a magnetic quantum-critical
 point QCP 1.


\begin{thebibliography} {99}
\bibitem{labacuo}  M. H\"ucker, M. V. Zimmermann, G. D. Gu, Z. J. Xu, J. S. Wen, Guangyong Xu, H. J. Kang, A. Zheludev, and J. M. Tranquada, {Phys. Rev. B} {\bf 83,  }104506 (2011).
\bibitem{stripes} J. M. Tranquada, G. D. Gu, M. H\"ucker, Q. Jie, H.-J. Kang, R. Klingeler, Q. Li, N. Tristan, J. S. Wen, G. Y. Xu, Z. J. Xu, J. Zhou, and M. v. Zimmermann, {Phys. Rev. B} {\bf  78,  }174529 (2008).
\bibitem{hinkov} V. Hinkov, P. Bourges, S. Pailh\`es, Y. Sidis, A. Ivanov, C. D. Frost, T. G. Perring, C. T. Lin, D. P. Chen, and B. Keimer, {Nat. Phys.} {\bf  3}, 780 (2007).
\bibitem{davis} K. Fujita, A. R. Schmidt, E-A Kim, M.\ J. Lawler, D.-H. Lee, J. C. Davis, Hiroshi Eisaki, S. Uchida, {J. Phys. Soc. Jpn.}  {\bf 81,} (2012) 011005.
\bibitem{davis_1}K. Fujita, M. H. Hamidian, S. D. Edkins, C. K. Kim, Y. Kohsaka, M. Azuma, M. Takano, H. Takagi, H. Eisaki, S. Uchida, A. Allais, M. J. Lawler, E.-A. Kim, S. Sachdev, and J. C. Séamus Davis,  arXiv:1404.0362.
\bibitem{kerr} H. Karapetyan, M. Hücker, G. D. Gu, J. M. Tranquada, M. M. Fejer, Jing Xia, and A. Kapitulnik, {Phys. Rev. Lett.} {\bf 109,  }147001 (2012) and references therein.
For the discussion on how Kerr effect can be understood without time-reversal symmetry breaking see  P. Hosur, A. Kapitulnik, S. A. Kivelson, J. Orenstein, and S. Raghu, {\it Phys. Rev. B} {\bf  87,  }115116 (2013).
\bibitem{bourges} for the latest review see Y. Sidis and P. Bourges, arXiv:1306.5124.
\bibitem{greven} Yuan Li, V. Bal\'edent, G. Yu, N. Bari\v{s}i\'c, K. Hradil, R. A. Mole, Y. Sidis, P. Steffens, X. Zhao, P. Bourges, and M. Greven, {Nature} {\bf 468,  }283 (2010).
\bibitem{armitage} Y. Lubashevsky, LiDong Pan, T. Kirzhner, G. Koren, and N. P. Armitage, Phys. Rev. Lett. 112, 147001, (2014).
\bibitem{ybco} G. Ghiringhelli, M. Le Tacon, M. Minola, S. Blanco-Canosa, C. Mazzoli, N.B. Brookes, G.M. De Luca, A. Frano, D. G. Hawthorn, F. He, T. Loew, M. Moretti Sala, D.C. Peets, M. Salluzzo, E. Schierle, R. Sutarto, G. A. Sawatzky, E. Weschke, B. Keimer, and L. Braicovich, {Science}, {\bf 337}, 821 (2012).
\bibitem{ybco_1} A. J. Achkar, R. Sutarto, X. Mao, F. He, A. Frano, S. Blanco-Canosa, M. Le Tacon, G. Ghiringhelli, L. Braicovich, M. Minola, M. Moretti Sala, C. Mazzoli, Ruixing Liang, D. A. Bonn, W. N. Hardy, B. Keimer, G. A. Sawatzky, and D. G. Hawthorn, Phys. Rev. Lett., {\bf 109}, 167001 (2012).
\bibitem{X-ray} R. Comin, A. Frano, M. M. Yee, Y. Yoshida, H. Eisaki, E. Schierle, E. Weschke, R. Sutarto, F. He, A. Soumyanarayanan, Y. He, M. Le Tacon, I. S. Elfimov, J. E. Hoffman, G. A. Sawatzky, B. Keimer, and A. Damascelli, Science {\bf 343}, 390-392 (2014)
\bibitem{X-ray_1} E. H. da Silva Neto, P. Aynajian, A. Frano, R. Comin, E. Schierle, E. Weschke, A. Gyenis, J. Wen, J. Schneeloch, Z. Xu, S. Ono, G. Gu, M. Le Tacon, A. Yazdani,  Science 343, 393-396 (2014).
\bibitem{mark_last}T. Wu, H. Mayaffre, S. Kr\"amer, M. Horvati\'c, C. Berthier, W.N. Hardy, R. Liang, D.A. Bonn, and M.-H Julien, arXiv:1404:1617.
\bibitem{steve_last} L. Nie,  G. Tarjus, and S. A. Kivelson, Proc. Nat. Acad. Sci. {\bf 111}, 7980 (2014).
\bibitem{wu}Tao Wu,	 Hadrien Mayaffre, Steffen Kr\"amer, Mladen Horvati\'c, Claude Berthier, W. N. Hardy, Ruixing Liang, D. A. Bonn, and Marc-Henri Julien, {Nature} {\bf 477}, 191-194 (2011).
\bibitem{mark}Tao Wu, Hadrien Mayaffre, Steffen Kr\"amer, Mladen Horvati\'c, Claude Berthier, Philip L. Kuhns, Arneil P. Reyes, Ruixing Liang, W. N. Hardy, D. A. Bonn, and Marc-Henri Julien, {Nat. Commun.}, {\bf 4} 2113 (2013).
\bibitem{ultra} David LeBoeuf, S. Kr\"amer2, W. N. Hardy, Ruixing Liang, D. A. Bonn, and Cyril Proust, {Nat. Phys.} {\bf 9}, 79 (2013).
\bibitem{suchitra}  N. Harrison, and S. E. Sebastian, {  Phys. Rev. Lett.} {\bf 106}, 226402 (2011);
S. E. Sebastian, N. Harrison, and G. G. Lonzarich, Rep. Prog. Phys. {\bf 75} 102501 (2012).
\bibitem{taillefer}  N. Doiron-Leyraud and L. Taillefer, {  Physica C} {\bf 481}, 161 (2012).
\bibitem {zxshen} R.-H. He, M. Hashimoto, H. Karapetyan, J. D. Koralek, J. P. Hinton, J. P. Testaud, V. Nathan, Y. Yoshida, Hong Yao, K. Tanaka, W. Meevasana, R. G. Moore, D. H. Lu, S.-K. Mo, M. Ishikado, H. Eisaki, Z. Hussain, T. P. Devereaux, S. A. Kivelson, J. Orenstein, A. Kapitulnik, and Z.-X. Shen, { Science} {\bf 331,  }1579 (2011).
\bibitem{zxshen_0} I. M. Vishik, M. Hashimotoc, R.-H. He, W.-S. Lee, F. Schmitt, D. Lu, R. G. Moore, C. Zhang, W. Meevasan, T. Sasagawa, S. Uchida, Kazuhiro Fujita, S. Ishida, M. Ishikado, Y. Yoshida, H. Eisaki, Z. Hussain, T. P. Devereaux, and Z.-X. Shen, {  Proc. Natl. Acad. Sci.} {\bf 110}, 17774 (2013).
 \bibitem{Kaminski} A. Kaminski, T. Kondo, T. Takeuchi, and G. Gu,  arXiv:1403.0492 and references therein.
 \bibitem{basov} D. N. Basov, R. D. Averitt, D. van der Marel, M. Dressel, and K. Haule, {Rev. Mod. Phys.} {\bf  83}, 471 (2011).
\bibitem{alloul} H. Alloul, arXiv:1302.3473; C. R. Physique 15 (2014).
\bibitem{grilli} C. Castellani, C. Di Castro, and M. Grilli, Phys. Rev. Lett. {\bf 75}, 4650 (1995);  A. Perali, C. Castellani, C. Di Castro, and M. Grilli, Phys. Rev. B {\bf 54}, 16216 (1996).
\bibitem{kontani} see S. Onari and H. Kontani, Phys. Rev. Lett. {\bf 109}, 137001 (2012) and references therein.
\bibitem{castellani} see C. Castellani et al., J. Phys. Chem. Sol. {\bf 59}, 1694 (1998).
\bibitem{extra_gr} S. Andergassen, S. Caprara, C. Di Castro, and M. Grilli, Phys. Rev. Lett. 87, 056401 (2001).
\bibitem{gr_raman} S. Caprara et al, Phys. Rev. Lett. {\bf 95}, 117004 (2005); S. Caprara et al,  Phys. Rev. B {\bf 84}, 054508 (2011).
\bibitem{gr_stm} G. Seibold, M. Grilli, and J. Lorenzana,  Phys. Rev. Lett. {\bf 103}, 217005 (2009).
\bibitem{gr_arpes} G. Mazza, M. Grilli, C. Di Castro, and S. Caprara, Phys. Rev. B {\bf 87}, 014511 (2013).
\bibitem{review_features} for a review, see G. Seibold et al, Physica C {\bf 481}, 132 (2012) and references therein.
\bibitem{varlamov} A. Perali {\it el al}, Phys. Rev. B {\bf 62}, R9295(R) (2000).
\bibitem{scalapino} D. J. Scalapino, Rev. Mod. Phys. {\bf 84}, 1383 (2012).
\bibitem{pines} P. Monthoux, D. Pines, and G. G. Lonzarich, {Nature} {\bf 450}, 1177-1183 (2007)
\bibitem{acf} Ar. Abanov, A. V. Chubukov, and M. A. Finkelstein, {  Europhys. Lett.} {\bf 54,} 488 (2001).
\bibitem{acs} Ar. Abanov, A. V. Chubukov, and J. Schmalian, {  Adv. Phys.} \textbf{52,} 119 (2003).
\bibitem{hartnol} S. A. Hartnoll, D. M. Hofman, M. A. Metlitski, and S. Sachdev, {  Phys. Rev. B} {\bf  84,  }125115 (2011).
\bibitem{maslov} A. V. Chubukov, D. L. Maslov, and V. I. Yudson, Phys. Rev. B {\bf 89}, 155126 (2014).
\bibitem{tremblay} D. B. Kyung, S. S. Kancharla, D. Sénéchal, A.-M. S. Tremblay, M. Civelli, and G. Kotliar, {  Phys. Rev. B} {\bf  73}, 165114 (2006).
\bibitem{joerg} J. Schmalian, D. Pines, and B. Stojkovich, {  Phys. Rev. Lett.} {\bf 80},  3839 (1998).
\bibitem{sedrakyan} T. A. Sedrakyan and A. V. Chubukov, {  Phys. Rev. B} {\bf  81}, 174536 (2010).
\bibitem{wang_el} Y. Wang and A. V. Chubukov, Phys. Rev. B {\bf 88}, 024516 (2013).
\bibitem{el_prec} see, e.g., N.P. Armitage, P. Fournier, and R.L. Greene, Rev. Mod. Phys. 82, 2421-2487 (2010)
\bibitem{varma} C. M. Varma, {  Phys. Rev. B} 55 14554 (1997).
\bibitem{kotliar} Z. Wang, G. Kotliar, and X.-F. Wang, {  Phys. Rev. B} {\bf  42}, 8690 (1990).
\bibitem{sudip} S. Chakravarty, R. B. Laughlin, D. K.  Morr, and C. Nayak, {  Phys. Rev. B} {\bf  63}, 094503 (2001).
\bibitem{lee} P. A. Lee, N. Nagaosa, and X.-G. Wen, {  Rev. Mod. Phys.} {\bf 78}, 17 (2006).
\bibitem{millis} E. Gull, O. Parcollet, and A. J. Millis, {  Phys. Rev. Lett.} {\bf 110},  216405 (2013).
\bibitem{tremblay_1} A.-M. S. Tremblay, arXiv:1310.1481.
\bibitem{ph_ph} T.-P. Choy and Ph. Phillips, {  Phys. Rev. Lett.} {\bf 95}, 196405 (2005).
\bibitem{rice} T. M. Rice,  K.-Y. Yang, and F. C. Zhang, {   Rep. Prog. Phys.} {\bf 75},  016502 (2012).
\bibitem{emery} V. J. Emery and S. A. Kivelson, {  Nature} {\bf 374}, 434 (1994).
\bibitem{randeria} L. Benfatto, S. Caprara, C. Castellani, A. Paramekanti, and M. Randeria, {  Phys. Rev. B} {\bf 63}, 174513 (2001).
\bibitem{mike_last} V. Mishra, U. Chatterjee, J. C. Campuzano, and M. R. Norman,  Nature Physics {\bf 10}, 357-360 (2014).
\bibitem{ms} M. A. Metlitski and S. Sachdev, {  Phys. Rev. B} {\bf 82}, 075128 (2010).
\bibitem{efetov}  K. B. Efetov, H. Meier and C. P\'epin, {  Nat. Phys.} {\bf 9} 442, (2013).
\bibitem{subir_3} L. E. Hayward, D. G. Hawthorn, R. G. Melko, and S. Sachdev, arXiv:1309.6639.
\bibitem{efetov_2} H. Meier, M. Einenkel, C. P\'epin, and K. B. Efetov {  Phys. Rev. B} {\bf  88,  }020506(R)(2013).
\bibitem{finn} Ar. Abanov, A. V. Chubukov, and A. M. Finkel'stein, Europhys. Lett. {\bf 54}, 488 (2001).
\bibitem{wang} Y. Wang and A. V. Chubukov, {  Phys. Rev. Lett.} {\bf 110}, 127001 (2013).
\bibitem{X-ray_last}R. Comin, R. Sutarto, F. He, E. da Silva Neto, L. Chauviere, A. Frano, R. Liang, W. N. Hardy, D. Bonn, Y. Yoshida, H. Eisaki, J. E. Hoffman, B. Keimer, G. A. Sawatzky, and A. Damascelli, arXiv:1402.5415.
\bibitem{efetov_3}  H. Meier, C. P\'epin, M. Einenkel, and K. B. Efetov, Phys. Rev. B 89, 195115 (2014).
\bibitem{subir_2} S. Sachdev and R. La Placa, {  Phys. Rev. Lett.} {\bf 111}, 027202 (2013).
\bibitem{subir_4} J. D. Sau and S. Sachdev, Phys. Rev. B 89, 075129 (2014).
\bibitem{bill} W. A. Atkinson, A.P. Kampf, and S. Bulut, arXiv:1404.1335.
\bibitem{rafael} R. M. Fernandes, A. V. Chubukov, J. Knolle, I. Eremin, and J. Schmalian, Phys. Rev. B {\bf  85},  024534 (2012).
\bibitem{lederer}S. Lederer, Y. Schattner, E. Berg, and S. A. Kivelson, arXiv:1406.1193
\bibitem{tsvelik}A. M. Tsvelik, Phys. Rev. B {\bf 89}, 184515 (2014).
\bibitem{ccm}  C. Castellani, C. Di Castro, and W. Metzner, Phys. Rev. Lett. {\bf 72}, 316 (1994).
\bibitem{varma_2}  C. M. Varma, Phys. Rev. B {\bf 73}, 155113 (2006).
\bibitem{Fernandes_13}R. M. Fernandes, S. Maiti, P. W\"olfle, and A. V. Chubukov, {Phys. Rev. Lett.}, {\bf 111,} 057001 (2013).
\bibitem{umbrella}  E. G. Moon and S. Sachdev, Phys. Rev. B {\bf 82}, 104516 (2010).
\bibitem{vvc} B. A. Vorontsov, M. G. Vavilov and A. V. Chubukov,  {Phys. Rev. B} \textbf{81}, 174538 (2010).
\bibitem{fs} R. M. Fernandes and J. Schmalian, {Phys. Rev. B} \textbf{82},014521 (2010).
\bibitem{fra_kiv} E. Fradkin and S. A. Kivelson,  Nature Physics {\bf 8}, 865-866 (2012).
\bibitem{steve_k}E. Berg, E. Fradkin, S. A. Kivelson,and J. Tranquada, {  New J. Phys.} {\bf 11}, 115004 (2009).
\bibitem{mike_arc} M. R. Norman, H. Ding, M. Randeria, J. C. Campuzano, T. Yokoya, T. Takeuchi, T. Takahashi, T. Mochiku, K. Kadowaki, P. Guptasarma, and D. G. Hinks, {Nature} {\bf 392}, 157 (1998).
\bibitem{raghu_a} A. V. Maharaj, P. Hosur, and S. Raghu, arXiv:1406.4154.
\bibitem{arun_new} Mark H. Fischer, Si Wu, Michael Lawler, Arun Paramekanti, and Eun-Ah Kim, arXiv:1406.2711.
\bibitem{ar_mike} Ar. Abanov, A. V. Chubukov,and M. R. Norman,{  Phys. Rev. B} {\bf  78}, 220507(R) (2008).
\bibitem{moon} E-G. Moon and A.V. Chubukov, J. of Low Temp. Phys. {\bf 161}, p 263 (2010).
\bibitem{comm_A} This result holds when fermionic self-energy $\Sigma (k, \Omega_m)$ is approximated by its value at hot spots. When the  $k-$dependence along the FS is included, the summation of the leading logarithms still does not lead to the instability~\cite{wang}, but the form of $\Delta_{\rm SC} (\Omega_m)$ is more complex~\cite{ms,wang}.
\bibitem{msv} A. J. Millis, S. Sachdev, and C. M. Varma, Phys. Rev. B {\bf 37}, 4975 (1988).
\bibitem{senthil} D. F. Mross, J. McGreevy, H. Liu, and T. Senthil, Phys. Rev. B {\bf 82}, 045121 (2010).
\bibitem{max_very_last} M. A. Metlitski, D. F. Mross, S. Sachdev, and T. Senthil,  arXiv:1403.3694.
\bibitem{son} D. T. Son, Phys. Rev. D \textbf{59}, 094019 (1999).
\bibitem{efetov_private} K. B. Efetov, private communication.
\bibitem{private}  K.B. Efetov and C. P\'epin, private communication.
\bibitem{Fern_13_a} G.-W. Chern, R. M. Fernandes, R. Nandkishore, and  A. V. Chubukov, Phys. Rev. B {\bf 86}, 115443 (2012).
\bibitem{deban} D. Chowdhury and  S. Sachdev,  arXiv:1404.6532.
\bibitem{maxim} M. G. Vavilov, A. V. Chubukov, and  A. B. Vorontsov, Supercond. Sci. Technol. {\bf 23}, 054011 (2010).
\bibitem{inc_1}S. Tewari, C. Zhang, V. M. Yakovenko, and S. Das Sarma, {  Phys. Rev. Lett.} {\bf 100,  }217004 (2008).
\bibitem{inc_2} C.-H. Hsu, S. Raghu, and S. Chakravarty, {  Phys. Rev. B} {\bf 84}, 155111 (2011).
 \bibitem{polyakov}  A.M. Polyakov,  Phys. Lett. B {\bf 59} (1), 79-81 (1975).
\bibitem{starykh} A.V. Chubukov and O. A. Starykh, Phys. Rev. Lett. {\bf 110}, 217210 (2013).
 \bibitem{arun_1} A. Dhar {\it et al.}, {Phys. Rev. A} {\bf 85}, 041602 (2012).
 \bibitem{arun_2} A. Dhar {\it et al.}, {  Phys. Rev. B} {\bf 87}, 174501 (2013).
 \bibitem{maiti_a}S. Maiti and A. V. Chubukov, Phys. Rev. B {\bf 87}, 144511 (2013).
\bibitem{lara}  M. Marciani, L. Fanfarillo, C. Castellani, and L. Benfatto, Phys. Rev. B 88, 214508 (2013).
\bibitem{alberto} A. Hinojosa, R. Fernades, and A. V. Chubukov, arXiv:1405.7077.
\bibitem{nernst} R. Daou {\it et al}, Nature {\bf 463}, 519-522 (2010).
\bibitem{arp} A. A. Kordyuk {\it et al}, Eur.Phys.J. {\bf188}, 153 (2010).
\bibitem{arp_1}T. Kondo {\it et al} Phys. Rev. Lett. {\bf 111}, 157003 (2013).
\bibitem{arp_2} U. Chatterjee {\it et al}, Proc. Nat. Acad. Sci., {\bf 108} 9346 (2011).
\bibitem{arp_3} H.-B. Yang {\it et al}, Phys. Rev. Lett. {\bf 107}, 047003 (2011).
\bibitem{elihu} A. V. Chubukov, M. R. Norman, A. J. Millis, and E. Abrahams, Phys. Rev. B {\bf 76}, 180501(R) (2007).
\bibitem{suchitra_last} N. Harrison and  S. E. Sebastian,  arXiv:1401.6590.
\bibitem{sad} A. Allais, D. Chowdhury, and S. Sachdev,  arXiv:1406.0503.
\bibitem{palee} P. A. Lee,  arXiv:1401.0519.
\bibitem{norman} Ar. Abanov, A. V. Chubukov, and M. R. Norman, Phys. Rev. B {\bf 78}, 220507(R) (2008).
\bibitem{maslov_1} A. V. Chubukov and D. L. Maslov, Phys. Rev. B 86, 155136 (2012); Phys. Rev. B {\bf 86}, 155137 (2012).
\bibitem{mcmillan}  W. L. McMillan, Phys. Rev. \textbf{167}, 331 (1968).
\bibitem{dan} D. Agterberg and M. Kashuap, arXiv:1406.4959.

\end{thebibliography}
\end{document}